\documentclass[twocolumn, amsmath]{aastex631}
\usepackage{amsmath}
\usepackage{hyperref}
\usepackage{comment}

\newcommand{\Lagr}{\mathcal{L}}

\begin{document}

\title{Low frequency radio continuum imaging and SED modeling of 11 LIRGs: radio-only and FUV to radio bands}

%\title{Radio-only and FUV to radio SED modeling of 11 local luminous infrared galaxies }
\author[0000-0002-4679-0525]{Subhrata~Dey}
\affiliation{Astronomical Observatory of the Jagiellonian University, Orla 171, 30-244 Krak\'{o}w, Poland }

\author[0000-0002-2224-6664]{Arti~Goyal}
\affiliation{Astronomical Observatory of the Jagiellonian University, Orla 171, 30-244 Krak\'{o}w, Poland }

\author[0000-0003-3080-9778]{Katarzyna~Ma{\l}ek}
\affiliation{National Centre for Nuclear Research, ul. Pasteura 7, 02-093 Warsaw, Poland}

\author[0000-0002-2801-766X]{Timothy~J.~Galvin}
\affiliation{International Centre for Radio Astronomy Research, Curtin University, Bentley, WA 6102, Australia}
\affiliation{CSIRO Space \& Astronomy, PO Box 1130, Bentley WA 6102, Australia}

\author{Nicholas~Seymour}
\affiliation{International Centre for Radio Astronomy Research, Curtin University, Bentley, WA 6102, Australia}

\author{Tanio~D\'iaz~Santos}
\affiliation{Institute of Astrophysics, Foundation for Research and Technology-Hellas, GR-71110, Heraklion, Greece}

\author{Julia~Piotrowska}
\affiliation{Astronomical Observatory of the Jagiellonian University, Orla 171, 30-244 Krak\'{o}w, Poland }

\author{Vassilis Charmandaris}
\affiliation{Institute of Astrophysics, Foundation for Research and Technology-Hellas, GR-71110, Heraklion, Greece}
\affiliation{Department of Physics, University of Crete, Heraklion, 71003, Greece}
\affiliation{European University Cyprus, Diogenes street, Engomi, 1516 Nicosia, Cyprus}
\begin{abstract}
We present the detailed analysis of 11 local luminous infrared galaxies (LIRGs) from ultraviolet through far-infrared to radio ($\sim$70\, MHz to $\sim$15\,GHz) bands. We derive the astrophysical properties through spectral energy distribution (SED) modeling using the Code Investigating GALaxy Emission (CIGALE) and UltraNest codes. The radio SEDs include our new observations at 325 and 610\,MHz from the GMRT and the measurements from public archives. Our main results are (1) radio SEDs show turnovers and bends, (2) the synchrotron spectral index of the fitted radio spectra ranges between $-$0.5 and $-$1.7, and (3) the infrared luminosity, dust mass, dust temperature, stellar mass, star-formation rates (SFRs) and AGN fraction obtained from CIGALE falls in
the range exhibited by galaxies of the same class. The ratio of 60$\mu$m infrared and 1.4\,GHz radio luminosity, the 1.4\,GHz thermal fraction, and emission measure range between 2.1 and 2.9,  0.1\% and 10\%,  0.02 and 269.5$\times$10$^{6}$\,cm$^{-6}$\,pc, respectively. We conclude that the turnovers seen in the radio SEDs are due to free-free absorption; this is supported by the low AGN fraction derived from the CIGALE analysis. The decomposed 1.4\,GHz thermal and nonthermal radio luminosities allowed us to compute the star formation rate (SFR) using scaling relations.
A positive correlation is observed between the SFR$_{IR}$ obtained 10\,Myr ago (compared to 100\,Myr ago) and 1.4\,GHz radio (total and nonthermal) because similar synchrotron lifetimes are expected for typical magnetic field strengths observed in these galaxies ($\approx$50$\mu$G).
\end{abstract}


\section{Introduction}\label{sec:intro}

\begin{table}
\centering
\scriptsize
%\small
\caption{Basic information on our sample of LIRGs\label{tab:sample}}
\begin{tabular}{ccccc}\\
\hline
Name  &  R.A.(J2000)  &  Dec.(J2000)  &  $z$ & log$_{10}$($L_{IR}$)   \\
      & (h\,m\,s) & ($^\circ$\,$^\prime$\,$\arcsec$)&  & ($L_\odot$)  \\ 
(1)  & (2)  & (3)          & (4)            &  (5)         \\ \hline
 ESO 500-G034 & 10 24 31.4   & $-$23 33 10    &  0.0122    &  10.77  \\
 NGC 3508     & 11 02 59.7   & $-$16 17 22    &  0.0128   &   10.65  \\
ESO 440-IG058 & 12 06 51.9   & $-$31 56 54    &  0.0232   &   11.18   \\
ESO 507-G070  & 13 02 52.3   & $-$23 55 18    &  0.0217   &   11.34   \\
NGC 5135      & 13 25 44.0   & $-$29 50 01    &  0.0136   &   11.12   \\
 IC 4280      & 13 32 53.4   & $-$24 12 26    &  0.0162   &   10.85   \\
 NGC 6000     & 15 49 49.6   & $-$29 23 13    &  0.0070   &   10.92  \\
IR 16164-0746 & 16 19 11.8   & $-$07 54 03    &  0.0271   &   11.29  \\
ESO 453-G005  & 16 47 31.1   & $-$29 21 22    &  0.0209   &   11.69  \\
IR 18293-3413 & 18 32 41.1   & $-$34 11 27    &  0.0181   &   11.62  \\
ESO 593-IG008 & 19 14 31.1   & $-$21 19 09    &  0.0485   &   11.77   \\
\hline
\end{tabular}

Columns : (1) source name, (2) right ascension, (3) declination, (4) spectroscopic redshift from the NASA/IPAC Extragalactic Database (NED)\footnote{\url{https://ned.ipac.caltech.edu/}}; (5) absolute far-$IR$ luminosity from Table\,1 of \citet[][]{Condon96}, except for NGC\,3508 which is taken from Table\,1 of \citet[][]{Condon90}.
\end{table}

 Characterized by a prodigious amount of emission at infrared ($IR$) wavebands, luminous and {\it ultra}-luminous infrared galaxies ((U)LIRGs) dominate the infrared sky. LIRGs and ULIRGs have infrared luminosity in the wavelength range 8$\mu$m $<$ $\lambda$ $<$ 1000$\mu$m, $L_{IR}$, $\sim> 10^{11}$\,$L_\odot$ and $\sim> 10^{12}$\,$L_\odot$, respectively, where $L_\odot$ is the solar luminosity \citep[][]{Helou88}. As the name suggests, these galaxies bridge the gap between underlying astrophysical processes contributing to emission in normal star-forming galaxies and the active galactic nuclei (AGN) activity \citep[see, for a review][]{Sanders96, Perez-Torres21}. Their spectral energy distribution (SED), although dominated by emission at infrared wavebands, ranges from radio to UV/X-rays frequencies \citep[][]{Yun01, Pereira-Santaella2011}, containing imprints of different astrophysical processes such as star formation, stellar evolution, chemical enrichment, processes in the interstellar medium, and the AGN \citep[e.g.,][]{Conroy13}. Therefore, a detailed and broadband SED modeling provides not only important constraints on astrophysical properties shaping SEDs, but also the evolutionary history of a galaxy, providing insights into the cosmic evolution of the galaxy population \citep[][]{Lonsdale06}.

As the radio waves remain unaffected by dust, the study of the radio continuum offers a promising approach for studying the astrophysical properties of galaxies. The q$_{IR}$ parameter, defined as the ratio of infrared (60-100\,$\mu$m) to radio (1.4\,GHz) luminosities, shows a surprisingly tight correlation for normal galaxies because emission at these wavebands is ascribed to a common origin and interpreted as calorimetric models \citep[][]{Helou88, Condon92, Yun01, Murphy09}. In this framework, the galaxies are optically thick to UV radiation from young massive stars that are absorbed by the dust in the interstellar medium and reradiated in the far-infrared (FIR) regime. Later on, these stars explode to form type II supernovae and accelerate cosmic-ray electron (CRe) that produces radio emission via a synchrotron process before escaping the galaxy \citep[][]{Voelk89}. \citet[][]{Helou85} suggested that galaxies can be optically thin to both UV photons and cosmic rays, but coupling between gas and magnetic field should exist to maintain the radio-IR correlation \citep[see also,][]{Lacki10, Tabatabaei13}. Furthermore, a secondary component of the radio continuum emission due to free-free interactions between charged particles, i.e., free-free radiation, is also produced by ionization of gas in H$_\textsc{II}$ regions. In general, synchrotron emission is characterized by a power-law emission spectrum, $f_{\nu}\,\propto \nu^{\alpha}$  $(\alpha \simeq -0.8)$ dominating the 1--10 GHz frequency range, whereas free-free emission has almost a flat spectrum with flux proportional to $\nu^{-0.1}$, dominating in frequency range $\geq$ 10 GHz \citep[][]{Condon92}. The radio spectra of galaxies bend (or flatten) at lower frequencies $<$1GHz due to absorption processes such as free-free absorption (FFA), synchrotron self-absorption (SSA) or the Tsytovitch-Razin effect \citep[][]{Israel90, Condon92, Clemens10, Marvil15}.

 Therefore, the exact shape of the radio spectra between the MHz to GHz range depends on either the quantity and distribution of ionized (thermal) gas in galaxies \citep[][]{Vardoulaki15, Clemens08, Chyzy18, Galvin2018} or due to the presence of the AGN \citep[][]{Clemens10}. Typically, the thermal fraction, $TF$, defined as the ratio of thermal to total emission, ranges between 0.1 and 10\% at 1.4\,GHz for normal star-forming galaxies and LIRGs  \citep[see,][]{Galvin2018}. Therefore, constraining the radio spectrum to low frequencies is essential to understand the absorption models for these galaxies.

Multiwavelength SED modeling, from ultraviolet (UV) to IR, provides information about the light emitted by stars, either directly or through reprocessing by the gas (emission and absorption features in the SED) and dust in the interstellar medium, while radio SED probes the nonthermal and thermal processes in galaxies \citep[see, for a review,][]{Walcher10, Tabatabaei17, Perez-Torres21}. Therefore, different regimes of the broad SED provide critical insight into the nature, origin of emission, and factors that establish the energy balance. 
The astrophysical properties of galaxies using the Code Investigating GALaxy Emission (CIGALE) model set for main-sequence (normal) star-forming galaxies \citep[][]{Ciesla2017, Vika2017, Pearson2018, ricco2021, shirley2021}, LIRGs, and ULIRGs have been characterized \citep[][]{Malek2017, Malek2018, Pasparialis2021}. As these galaxies can host AGNs, the CIGALE code includes the AGN component in the modeling. In particular, the AGN fraction (defined as the ratio of $IR$ luminosity due to AGN and a sum of $IR$ luminosity due to AGN and starburst), stellar mass (M$_\star$), star formation rate (SFR$_{IR}$), dust luminosity ($L_{IR}$), dust temperature ($T_{\rm dust}$) have been obtained. The most significant finding of these studies is that LIRGs are characterized by a relatively higher SFR$_{IR}$, $L_{IR}$, $T_{\rm dust}$, AGN fraction compared to normal star-forming galaxies and lower than those obtained by ULIRGs \citep[][]{Malek2017}. On the other hand, detailed radio-SED analyses of a large sample of galaxies are rare, primarily due to the lack of wide-area multifrequency radio surveys and targeted follow-up of suitable samples.

\begin{figure*}
    \hspace{0cm}{\includegraphics[width =  0.30\textwidth]{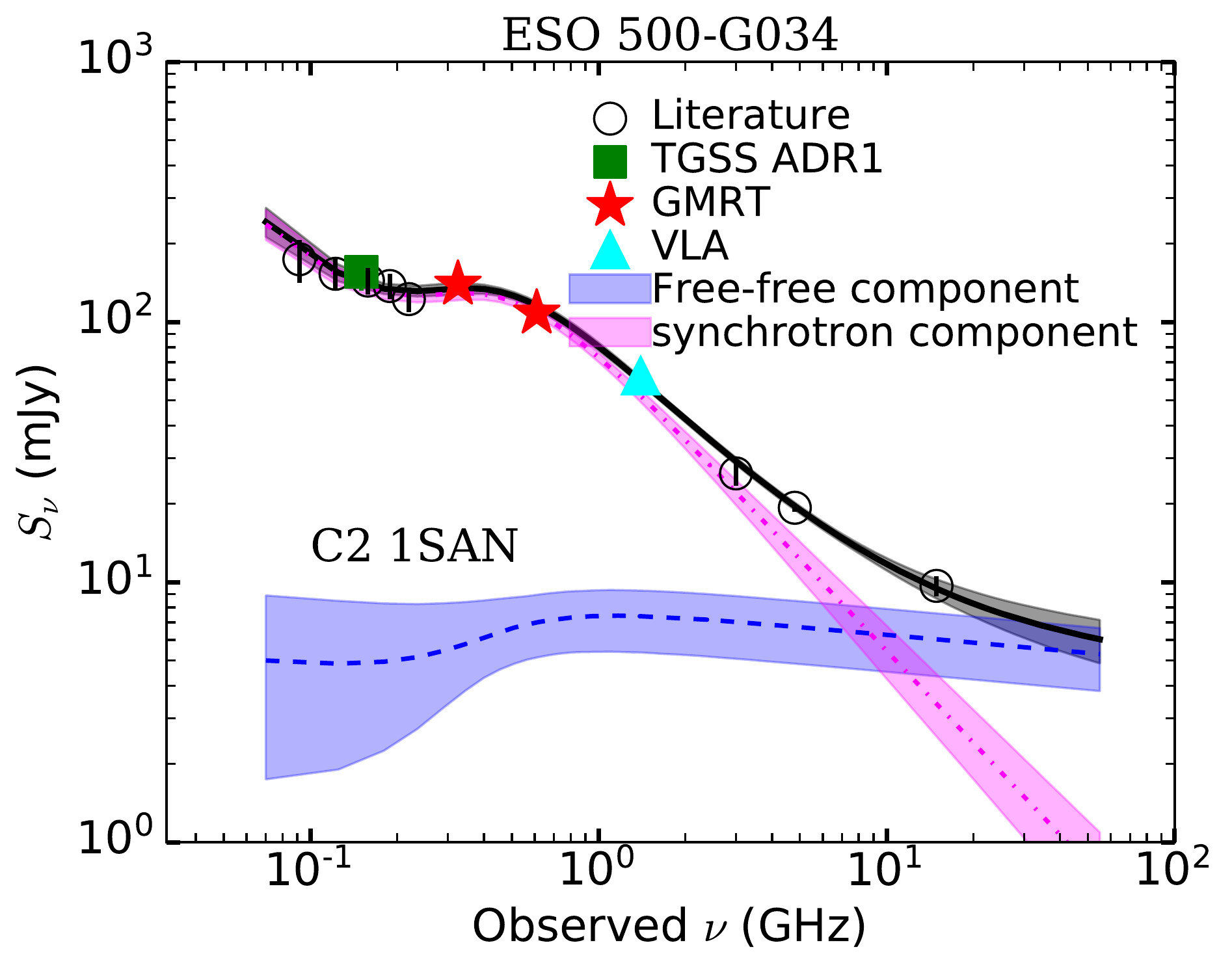}}
    \hspace{0cm}{\includegraphics[width =  0.30\textwidth]{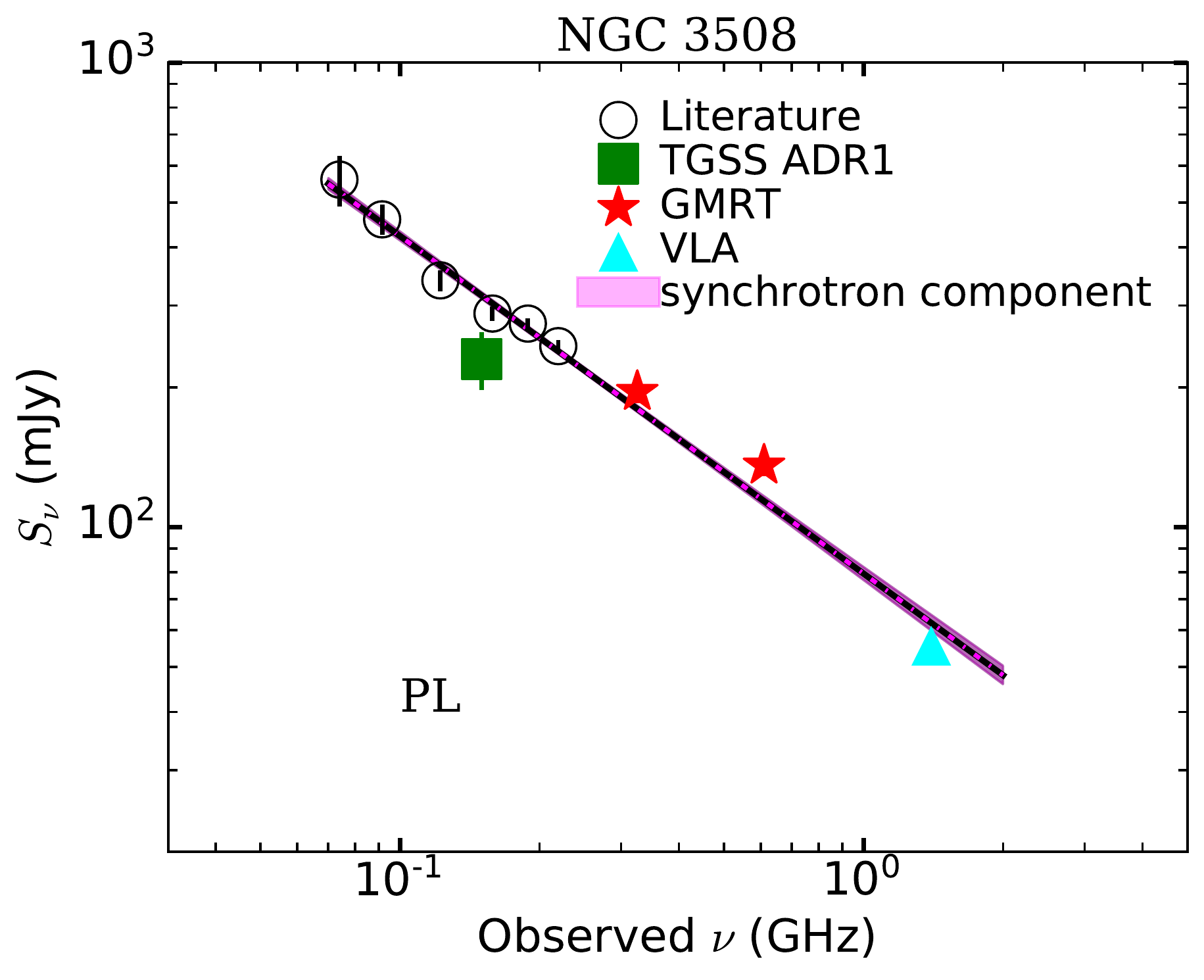}}
    \hspace{0cm}{\includegraphics[width =  0.30\textwidth]{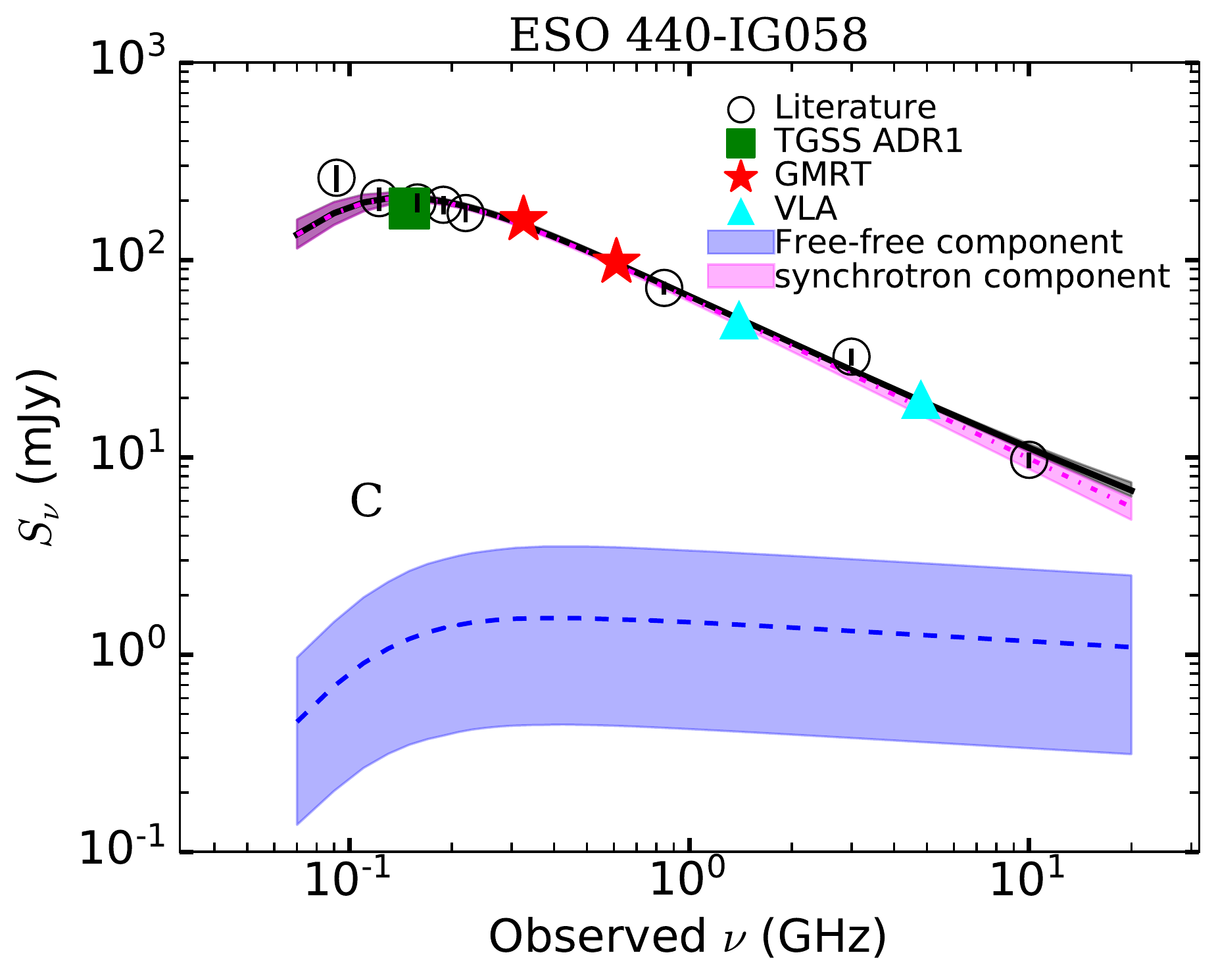}}
    \hspace{0cm}{\includegraphics[width =  0.30\textwidth]{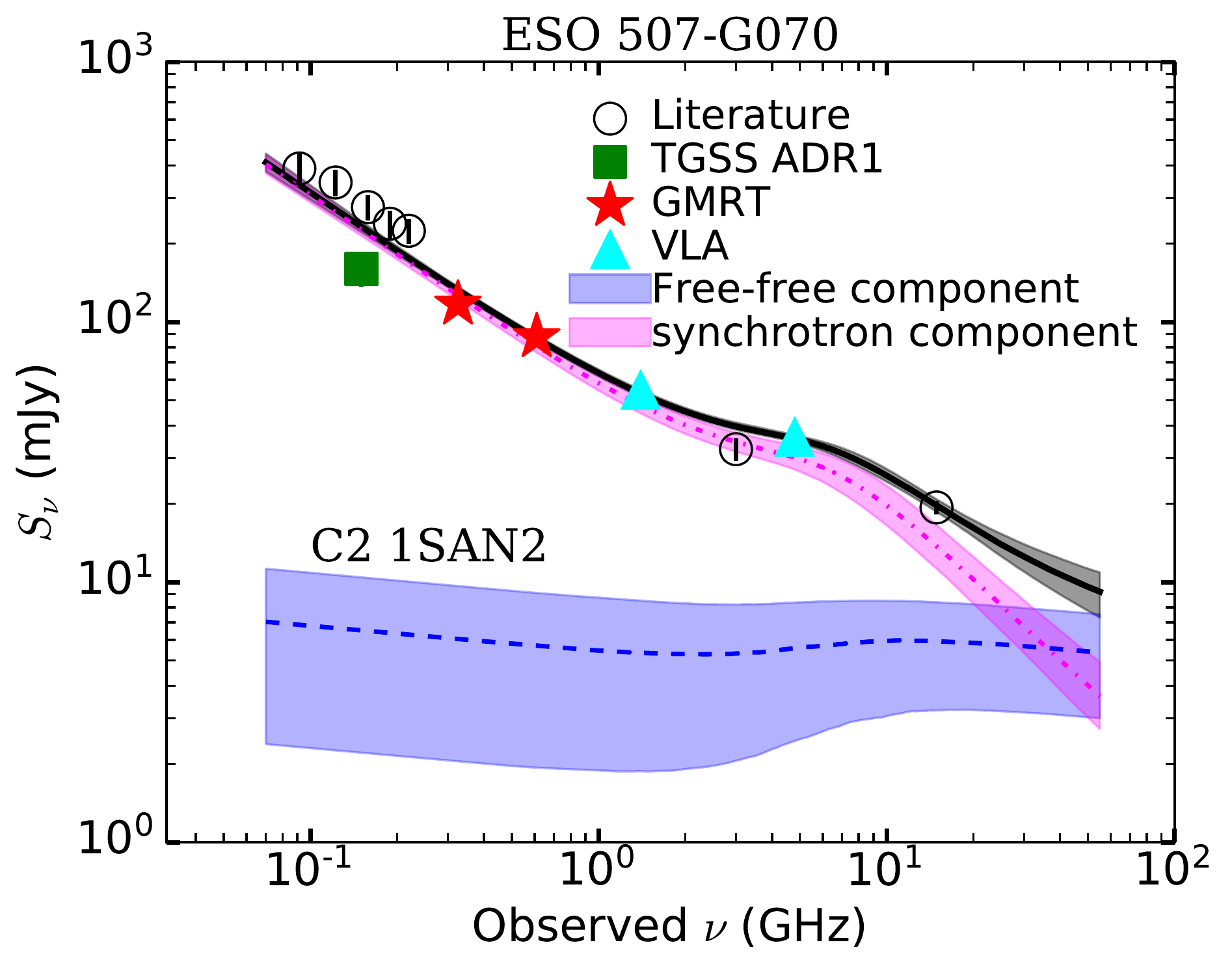}}
    \hspace{0cm}{\includegraphics[width =  0.30\textwidth]{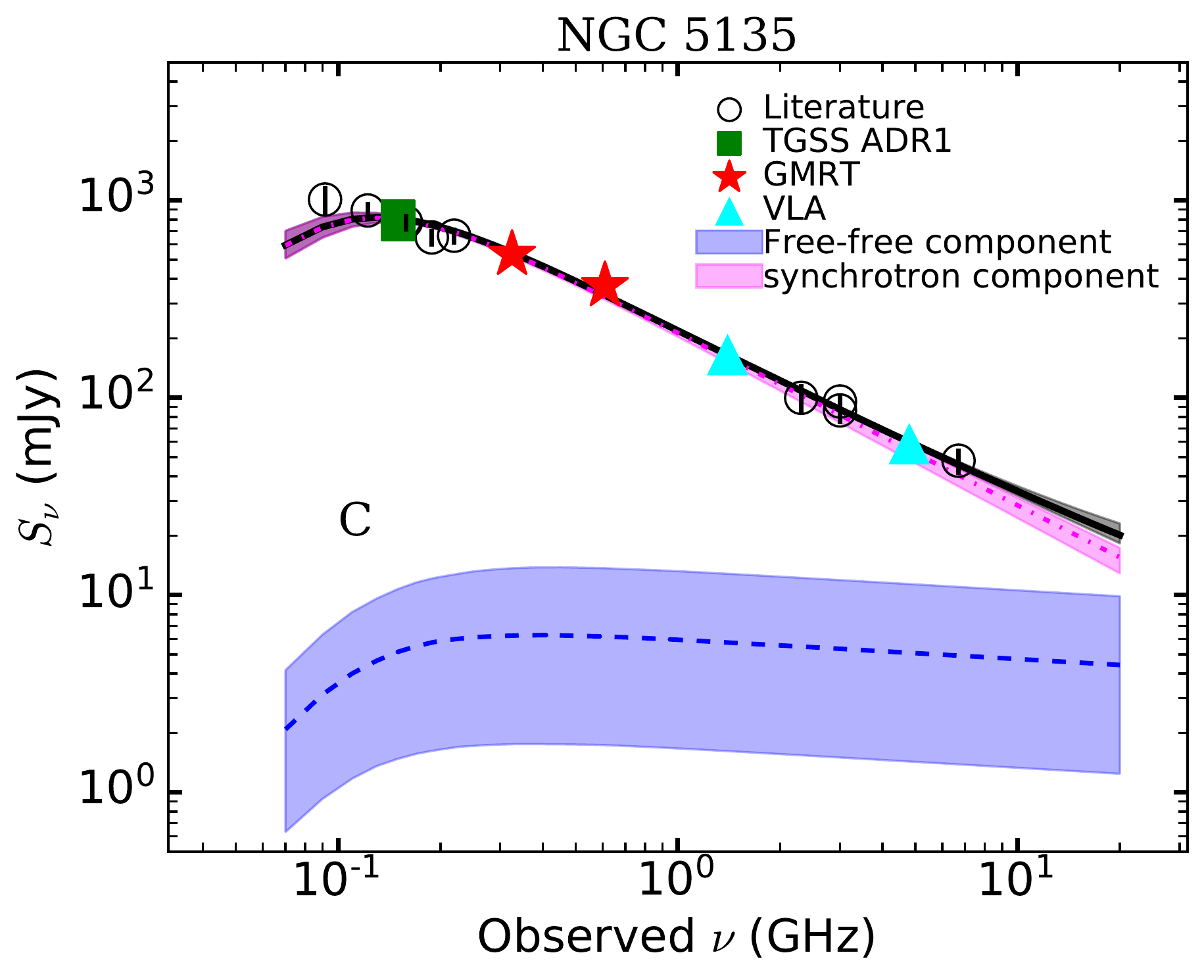}}
    \hspace{0cm}{\includegraphics[width =  0.30\textwidth]{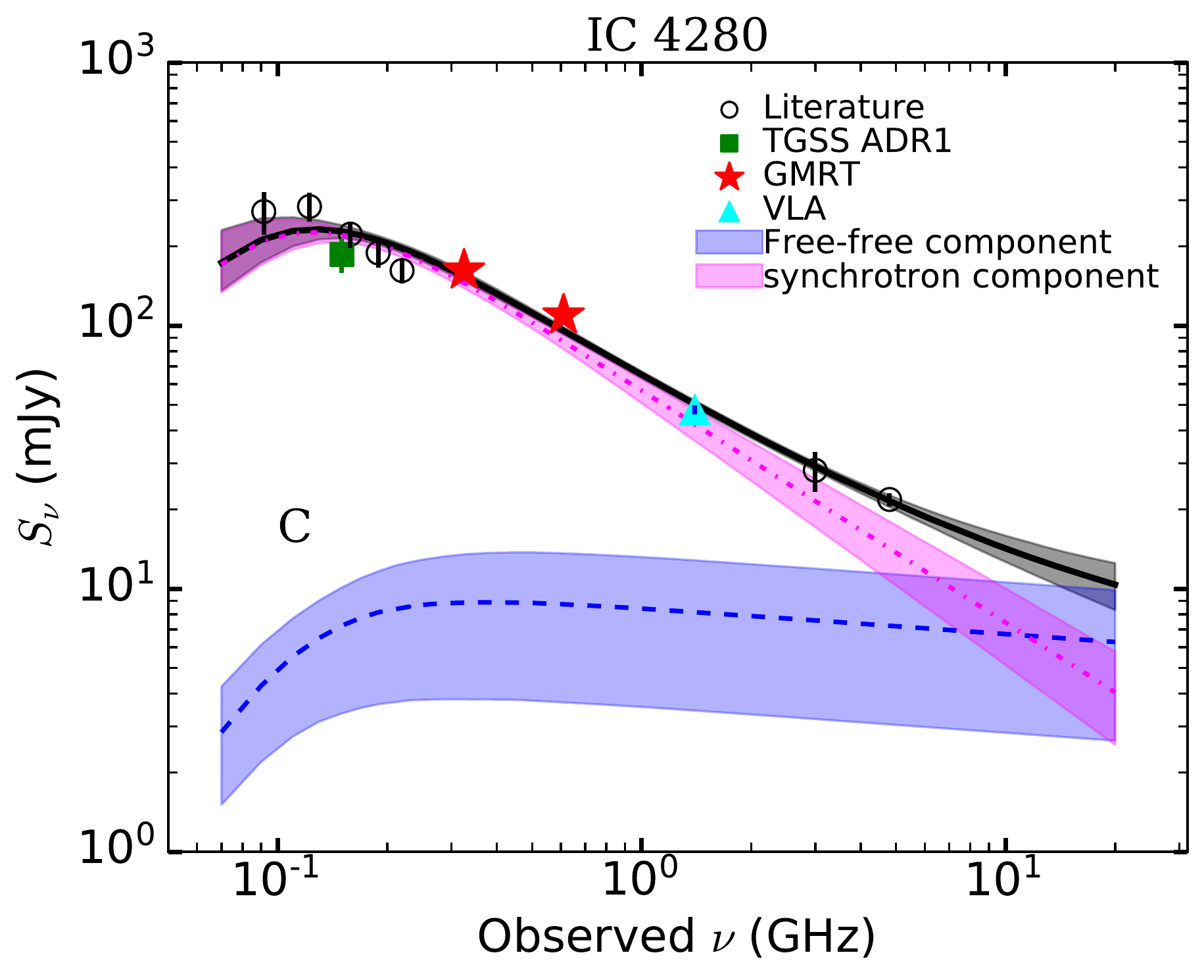}}
   \hspace{0cm}{\includegraphics[width =  0.30\textwidth]{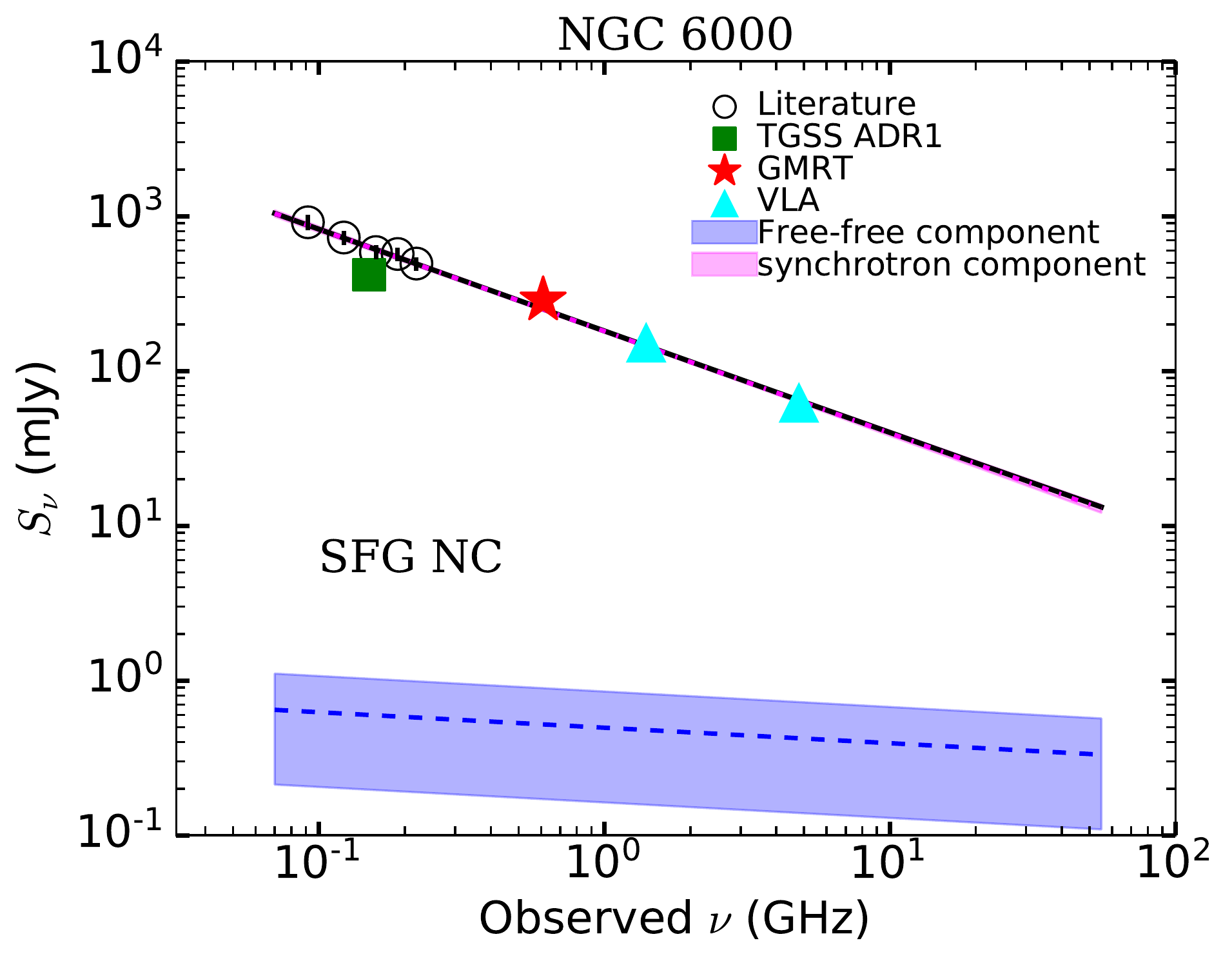}}
    \hspace{0cm}{\includegraphics[width =  0.30\textwidth]{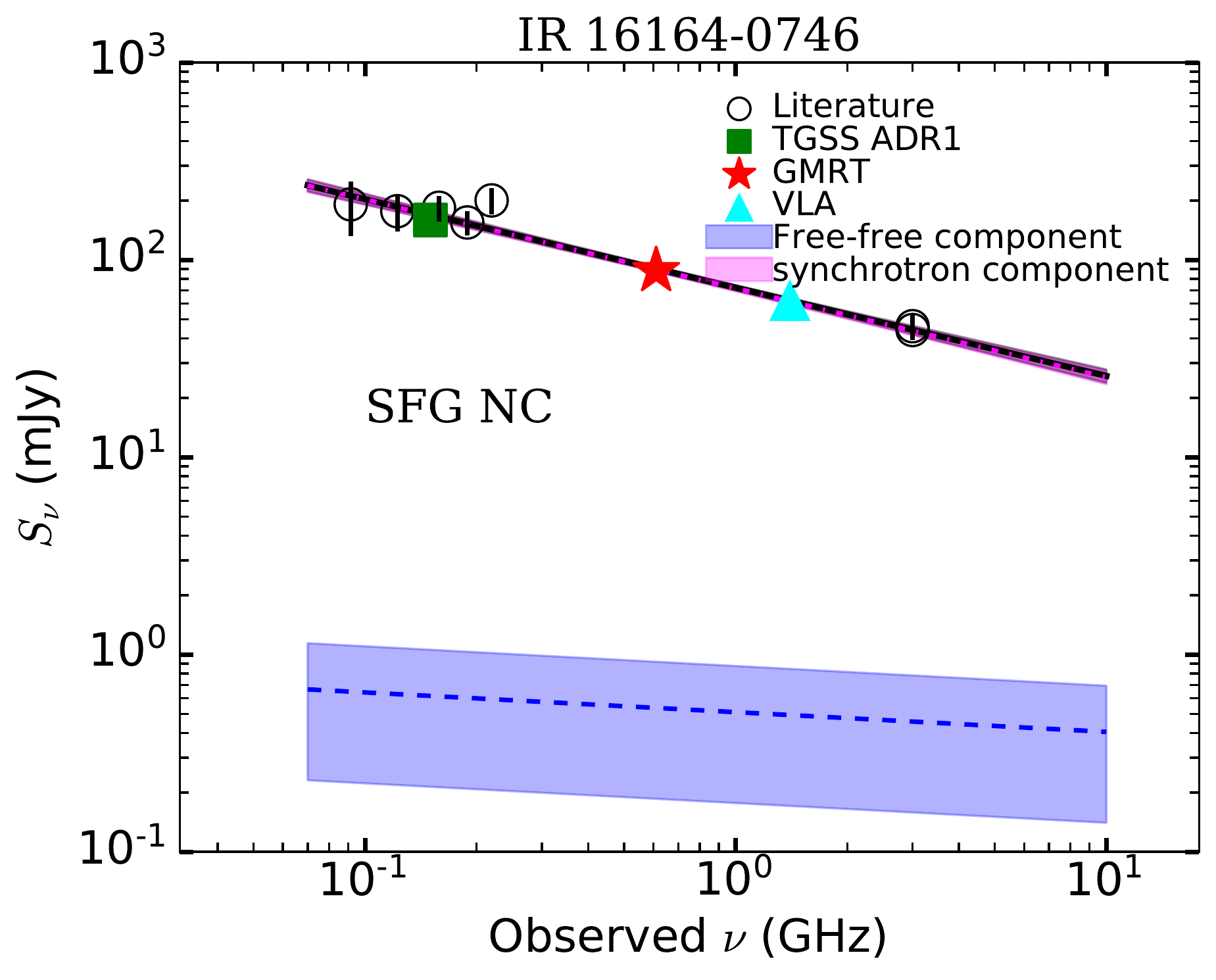}}
    \hspace{0cm}{\includegraphics[width =  0.30\textwidth]{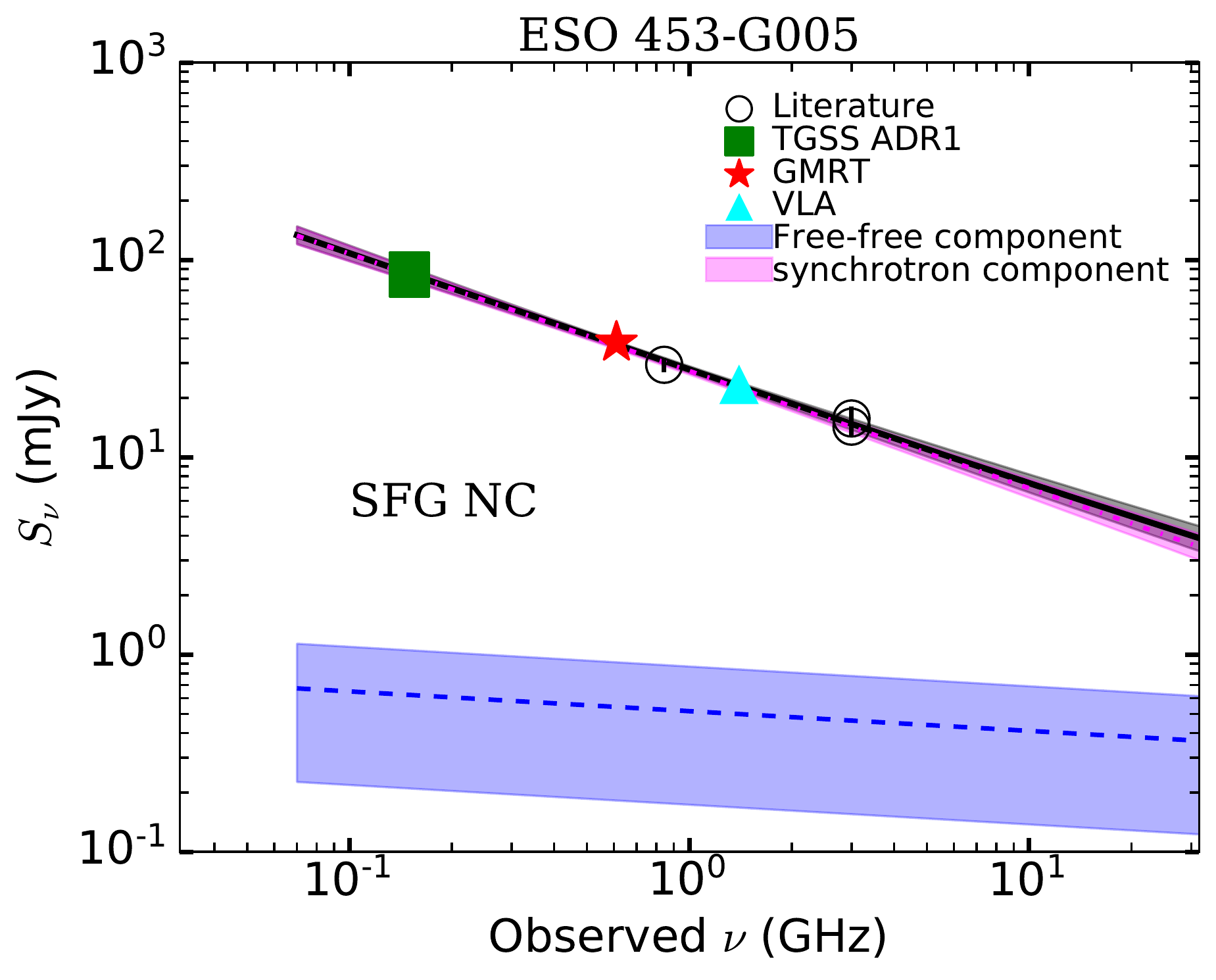}}
    \hspace{0.7cm}{\includegraphics[width =  0.30\textwidth]{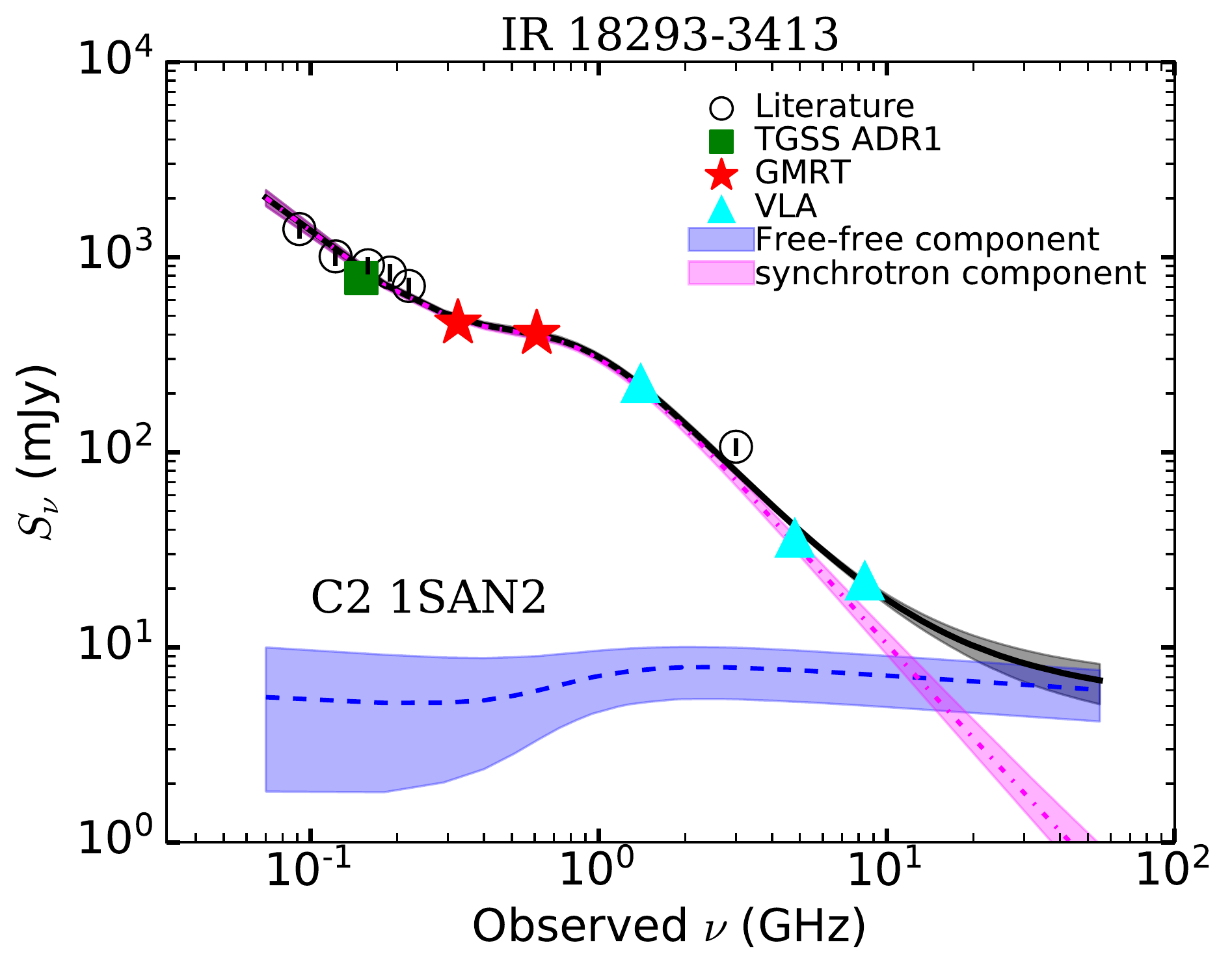}}
    \hspace{0.5cm}{\includegraphics[width =  0.30\textwidth]{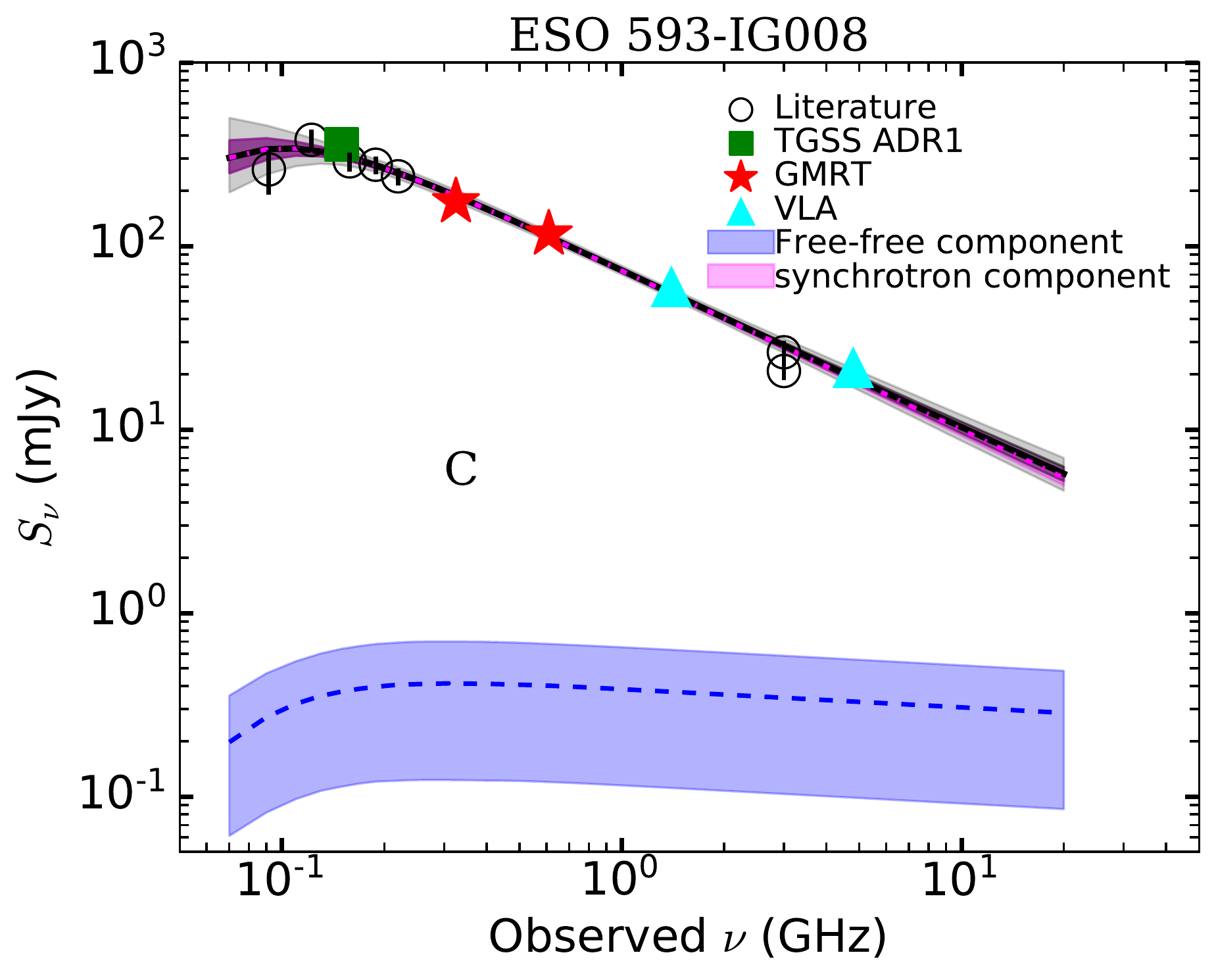}}
\caption{Radio SED for our sample of LIRGs. Galaxy name and the best-fit radio SED model name are given at the top and inside each panel. The solid black line represents the best-fit model, while the gray shaded region represents the 1$\sigma$ uncertainties sampled by the UltraNest package. The dot-dashed magenta and dotted blue lines show the decomposed synchrotron and free-free components. The pink and blue shaded regions represent the 1$\sigma$ uncertainties in the synchrotron and free-free components, respectively. For galaxies fitted with two-component models, the free-free and synchrotron components correspond to the sum of individual free-free and synchrotron components. The flux density measurements obtained with GMRT (365 and 610\,MHz) are shown by the filled red star symbol, while the archival VLA data that were reduced and analyzed by us (1.4, 4.8, 8.4, and 14.9\,GHz) are shown by the filled cyan triangle. The filled green square indicates the TGSS\,ADR1 150\,MHz fluxes, and the open circle represents the flux density measurements taken from the literature (GLEAM, SUMSS, NVSS, VLASS, ATCA, and Effelsberg telescope) (see Section~\ref{sec:datarad}, Table~\ref{tab:radiosummary}). }\label{fig:radiosed}
\end{figure*}

 With this motivation, we performed a SED modeling of radio-only and far-ultraviolet (FUV) to radio bands of a sample of 11 LIRGs, including our new measurements at 325 and 610\,MHz frequencies using the Giant Metrewave Radio Telescope \citep[GMRT; ][]{Swarup90}\footnote{\url{http://www.gmrt.ncra.tifr.res.in/}}, operated by the National Centre of Radio Astronomy-Tata Institute of Fundamental Research, India. In this paper, we report the results of detailed radio SED modeling covering $\sim$80\,MHz to $\sim$15\,GHz and CIGALE SED modeling covering GHz band radio frequencies up to UV frequencies. As the CIGALE modeling cannot capture the complex shapes of low-frequency radio spectra exhibited by LIRGs and ULIRGs \citep[e.g.,][]{Clemens08, Clemens10, Galvin2018}, therefore, we include $>$1.4~GHz band radio fluxes to cover the five decades of the spectral range. Furthermore, an essential aspect of CIGALE modeling is that it works on the energy balance between $UV$ and $IR$, which is ultimately related to the radio luminosity, hence including radio fluxes in CIGALE modeling is essential for better constraining the model parameter space (in particular, the $L_{\rm dust}$).

    One of the primary goals of this paper is to compare the astrophysical properties resulting from SED modeling at radio-only and FUV to radio bands. The shapes of SEDs reflect the radiation laws and their parameters (such as power-law energy index or emissivity) and the physical processes affecting those parameters, such as cooling or heating mechanisms in the medium. Moreover, integrated SEDs provide total energetics from different frequency regimes, and comparing those offers key information on the nature of emission and general factors that determine their energy balance principle. Furthermore, our radio SED modeling allowed us to decompose the nonthermal and thermal emission components and estimate the star formation rate (SFR) using the basic nonthermal and thermal radio SFR calibration presented in \citet[][]{Murphy11} and compare them with the SFR$_{IR}$ estimated from CIGALE SED modeling.

 We perform detailed SED modeling of a sample of 11 nearby LIRGs (median redshift equal to 0.0181), focusing separately on radio-only and FUV-radio spectral bands. The paper is organized as follows. Sample selection is given in Section~\ref{sec:sample} while Section~\ref{sec:data} describes data collection and analysis. Section~\ref{sec:sedmodeling} gives details of the radio-only and panchromatic (FUV to radio) SED modeling procedures. Section~\ref{sec:results} gives the results of model fitting, i.e., characterization of astrophysical properties of our sample. The discussion of the results obtained is given in Section~\ref{sec:discussion} while the conclusion is given in Section~\ref{sec:conclusion}.

\section{sample selection}\label{sec:sample}

In this study, we have assembled a sample of 10 LIRGs with the selection criteria log$_{10}(L_{IR})$ $>$ 10.75 $L_\odot$ from the 1.425\,GHz Atlas of the IRAS Bright Galaxy Sample catalog \citep[][]{Condon96} and one galaxy, NGC\,3508 from the 1.49 GHz Atlas of the IRAS Bright Galaxy Sample \citep[][]{Condon90} of with flux densities greater than 5.24\,Jy at 60$\mu$m. These galaxies were selected based on the availability of 150\,MHz data in the original release of the TIFR GMRT Sky Survey (TGSS) Data Release (DR)~4, covering 1\,sr of the sky, using the GMRT \citep[see, for a basic description of the survey;][]{Sirothia14}. We note that \citet[][]{Intema17} presented the entire TGSS data in their alternate data release (TGSS\,ADR1). Although, we selected our sample from the original TGSS DR\,4 release, which had slightly better sensitivity for extended emission (rms error $\sim$7-9 mJy\,beam$^{-1}$) as compared to the TGSS\,ADR1 \citep[][]{Intema17}, we use TGSS\,ADR1 integrated fluxes for radio SED fitting as its reduction and calibration methodology is fully described. The basic properties of the LIRG sample are listed in Table~\ref{tab:sample}.

\begin{table*}
%\centering
%\scriptsize
%\small
\caption{An overview of the natural log of the Bayes odds ratio from the UltraNest fitting of each model to each source. In boldface, we present the most preferred model with highest evidence value (see Eq.~\ref{eq:evidence}). \label{tab:bayesodds}}
\begin{tabular}{ccccccccc}\\
\hline
Name  &  PL  & SFG   &  C & SFG    & C2 & C2 & C2 & C2 \\
      &  & NC &   & NC2 & 1SA & 1SAN & 1SAN2 &  \\ 
\hline
 %ESO 500-G034 & $-$60.7 & $-$58.2 & $-$34.8 & \textbf{0}  & $-$1.9 & $-$5.4  \\
ESO 500-G034 & $-$68.1 &	$-$67.5 &	$-$6.3	& $-$76.3 &	$-$1.2 &	\textbf{0} &	$-$1.5 &	$-$4.2 \\

 NGC 3508     &  \textbf{0}	&	$-$3.5 	&	$-$7.2	&	$-$8.7	&	$-$19.3 &	$-$4.6	&	$-$2.8	&	$-$7.1\\
 
ESO 440-IG058 &  $-$17.2 &  $-$14.9 & 	\textbf{0} &  $-$25.7 &  $-$3.9	&  $-$9.1	&  $-$6.8	&  $-$2.3
\\
ESO 507-G070 &  $-$74.2	&  $-$69.9	&  $-$159.1
&  $-$21.8	&  $-$235.1	&  $-$1.9	& \textbf{0}	&  $-$453.1
  \\
NGC 5135      &  $-$6.5	&  $-$3.7	&  \textbf{0}	&  $-$11.8	&  $-$13.0	& $-$2.9	&  $-$35.4	&  $-$6.8
 \\
 IC 4280  &  $-$3.2	&  $-$6.1
 & \textbf{0}	&  $-$12.4	&  $-$8.8	&  $-$1.6	&  $-$4.4	&  $-$0.7
  \\
 NGC 6000    &  $-$3.4
& \textbf{0}	&  $-$21.4	&  $-$6.1	&  $-$11.2	&  $-$1.6	&  $-$3.1
 &  $-$408.8
\\
IR 16164-0746 &   $-$3.4 &	\textbf{0}	&  $-$1.2	&  $-$1.3	&  $-$4.1	&  $-$0.9	&  $-$0.8	&  $-$3.0
  \\
ESO 453-G005  &  $-$2.9	 & \textbf{0}	&  $-$0.6	&  $-$3.5	&  $-$4.9	&  $-$2.1	&  $-$2.3	&  $-$1.5
 \\
IR 18293-3413 &  $-$88.7	&  $-$86.8	&  $-$90.4	&  $-$99.5	&  $-$4.3	&  $-$5.1 &	\textbf{0}	&  $-$5.1
  \\
ESO 593-IG008   &  $-$3.0	&  $-$5.9 &	\textbf{0}	&  $-$7.5	&  $-$4.3	&  $-$6.8	&  $-$3.8	&  $-$2.1
\\
\hline
\end{tabular}
\end{table*}

\section{multi-wavelength datasets: Radio to FUV}\label{sec:data}

\subsection{Radio data and AIPS analysis}
\label{sec:datarad} 

To construct the integrated radio SED for modeling, we use data in the $\sim$70 MHz to 15 GHz frequency range. These include our continuum observations at 325 and 610\,MHz conducted using the GMRT (ID: 23\_051, PI: Arti Goyal) and the publicly available archival datasets at 4.8, 8.4, and 14.5 GHz from the Karl G. Jansky Very Large Array (VLA\footnote{\url{https://public.nrao.edu/telescopes/vla/}}), operated by the National Radio Astronomical Observatory, USA.

\subsubsection{New GMRT data at 325 and 610\,MHz} 

We carried out radio continuum imaging of our sample using the GMRT at 325\,MHz and 610\,MHz. The primary and phase calibrators used in our observations are provided in Table~\ref{tab:gmrtobs}. We observed each target with 32\ MHz bandwidth divided into 256 spectral channels. We observed standard flux calibrators at the beginning and end of the observation to calibrate the antenna gains. Phase calibrators were selected from the NRAO VLA calibrator manual list and were within 20 deg. Phase calibrators were observed every 30 minutes for a typical duration of 4-5 minutes to correct for antenna gain drifts, atmospheric and ionospheric gain, and phase variations. Each source was observed for a total duration of 32 minutes, in two scans consisting of 16 minutes each to enable better U-V coverage.  

\subsubsection{Archival VLA data at 1.4, 4.8, 8.4, and 14.9\,GHz}
We analyzed the archival VLA data for galaxies wherever possible. Most galaxies in our sample were observed for a few minutes of integration time with different array configurations and different central frequencies. The dataset with the largest on-source integration time was reduced when several observations were available with the same configuration and central frequency. 

\subsubsection{AIPS analysis}

 The interferometric observations from both the GMRT and the VLA were analyzed using NRAO AIPS\footnote{The National Radio Astronomy Observatory is a facility of the National Science Foundation operated under cooperative agreement by Associated Universities, Inc.}. Data reduction was carried out in a standard fashion. The flux density scale of \citet[][]{Baars77} was used to obtain the flux densities of the primary (flux) calibrator, the secondary (phase) calibrator, and the target source. Antennas and baselines affected with strong radio frequency interference (RFI), nonworking antennas, were edited out after visual inspection. For the GMRT datasets, bandpass calibration was determined using the phase calibrator and the spectral channels were collapsed to generate the continuum database. Usually, spectral channels below ten and above 200 were discarded before collapsing the data. Images were produced using task \textsc{imagr} on the channel collapsed data. To correct for distortions in the imaging, the large field of view with noncoplanar baselines (GMRT at frequencies $<$1 GHz), polyhedron imaging was employed where the field of view was divided into smaller fields (facets). These were $5\times5$ facets that covered the entire field of view up to the half power beam width (HPBW). Usually, 3-5 rounds of phase-based self-calibration were performed iteratively by choosing point sources in the field such that the flux density is $>$3$\sigma$ with one synthesized beam. The final images were made with full {\it UV}-coverage and robust weighting of 0 to weight the UV data \citep{Briggs95}. Facets were combined using the task \textsc{flatn}. The same steps were followed for the VLA observations except that the data were obtained in two intermediate frequency (IF) channels calibrated for antenna gains before averaging them together for imaging. We did not apply any band-pass calibration since the data were obtained in a single spectral channel of 50\,MHz bandwidth (BW). The final images were corrected for the reduction in sensitivity due to the shape of the antenna beam using task \textsc{pbcor} with the specified parameters for the GMRT\footnote{\url{http://www.ncra.tifr.res.in:8081/~ngk/primarybeam/beam.html}} and the VLA\footnote{\url{http://www.aips.nrao.edu/cgi-bin/ZXHLP2.PL?PBCOR}}.

 Integrated flux densities (and uncertainty) were obtained using the task \textsc{TVSTAT} in \textsc{AIPS} for the GMRT and VLA images. We note that the synthesized beam sizes range from $\sim$0.6-38${^{\prime\prime}}$. Assuming a typical resolution of 5${^{\prime\prime}}$, the linear scale is be 0.5--4\,kpc at the galaxy distance ($z$ =  0.007--0.048). Therefore, it is reasonable to state that we obtain emissions from extended regions in most galaxies. Furthermore, we note in GMRT observations, all the proposed galaxies are detected at\,610 MHz observations while LIRGs NGC\,6000, IR\,16164$-$0746 and ESO\,453-G005 could not be detected at 325\,MHz because the data could not be calibrated by weak phase calibrator.

\subsubsection{Radio fluxes from the literature}
We searched for flux measurements at other frequencies along with the observations described above. In particular, we obtained measurements at the central frequencies of 74\,MHz from the VLA Low-Frequency Sky Survey \citep[VLSSr;][]{Cohen07}, 74-231\,MHz GaLactic, and Extragalactic MWA Survey \citep[GLEAM;][]{Wayth15}, 150\,MHz TGSS ADR, 843\,MHz Sydney University Molonglo Sky Survey \citep[SUMSS;][]{Mauch03, Mauch13}, 3.0\,GHz VLA Sky Survey \citep[VLASS;][]{Lacy20, Gordon21} within the positional uncertainties provided by the survey parameters. For LIRGs ESO440-IG080 and ESO500-G034, we also included 10.0\,GHz flux densities from the Australia Telescope Compact Array, published in \citet[][]{Hill01}. For NGC\,5135, we included 2.3\,GHz measurements from the $S-$band Polarization All-Sky Survey \citep[][]{Meyers17}, and 6.7\,GHz measurements from the Effelsburg telescope  \citep[][]{Impellizzeri08}. In addition, we also included 1.4\,GHz\,NRAO\,VLA\,Sky\,Survey \citep[NVSS;][]{Condon98} flux for the galaxy IR\,18293$-$3413. The GLEAM survey frequency bands are 72-103\,MHz, 103-134\,MHz, 139-170\,MHz, 170-200\,MHz, and 200-231\,MHz where each band is divided into 7.68\,MHz sub-channels for imaging purposes \citep[][]{Walker17}. The survey provides flux measurements into subbands of 7.68\,MHz BW (each band has a frequency resolution of 30.72\,MHz), which are not independent of each other; therefore, we averaged the fluxes within each band for spectral modeling.

Table~\ref{tab:radiosummary} summarizes the GMRT and VLA datasets analyzed by us, while Figure~\ref{fig:overlays} provides total intensity radio maps overlaid on Digital Sky Survey (DSS)2-$R$-band images in Appendix~\ref{app:A}. For radio SED modeling, we further upscaled the errors to account for variations in uncalibrated system temperature. In particular, we added in quadrature to the flux density errors an additional 10\% for 150\,MHz TGSS ADR1 \citep[][]{Intema17}, 5\% for 325 and 610\,MHz GMRT \citep[][]{Zywucka14, Mhlahlo21}, 3\% for 1.4, 4.8, 8.4, and 14.9\,GHz VLA \citep[][]{Perley17}, 3\% for NVSS \citep[][]{Condon98}, 10\% for VLASS \citep[][]{Lacy20}, 10\% for GLEAM \citep[][]{Mhlahlo21}, 5\% for SUMSS \citep[][]{Mauch03, Galvin2018}, 15\% for Effelsberg measurements \citep[][]{Impellizzeri08}, 10\% for $S$-PASS \citep[][]{Meyers17}, and 5\% for ATCA \citep[][]{Galvin2018} measurements, respectively, as calibration error \citep[see Eq.\,1 of ][]{Zywucka14}. As can be seen from Table~\ref{tab:radiosummary} that the synthesized beam sizes for our continuum images range between a few arcseconds and 10$\times$ a few arcseconds. Because we analyzed datasets taken mostly at the B, C, and D array configurations of the VLA, it is unlikely that our galaxies are missing flux due to the lack of short \textsc{U-V} spacing data. To establish this, we compared the 1.4\,GHz fluxes from Table~\ref{tab:radiosummary} with the NVSS fluxes, obtained at a resolution of 45$^{\arcsec}$. The two measurements are comparable within the measurement uncertainties. Table~\ref{tab:radioflux} provides the list of flux measurements used for the radio SED fitting.

\subsection{UV, optical, and infrared fluxes}\label{sec:datauv}
We collected photometric measurements from several instruments from both ground and space-based facilities for SED modeling using CIGALE. Specifically, we obtained the fluxes from the literature by cross-matching the optical positions of our galaxies to public databases such as NED \citep[][]{ned}, SIMBAD \citep[][]{Wenger00}, VizieR \citep[][]{Vizier}, and NASA/IPAC Infrared Science Archive (IRSA)\footnote{\url{https://irsa.ipac.caltech.edu/}} using a matching radius of 5$^{\arcsec}$--15$^{\arcsec}$. This matching radius was chosen to ensure that there exists a counterpart, depending on the instrument's resolution. Specifically, the $UV$ and optical data were collected from the NED and VizieR Photometry viewer, while the infrared data were obtained from the IRSA database.

About 20--30 bands of $UV$--$IR$ broadband photometry are available for these sources. They include measurements from Galaxy Evolution Explorer (GALEX), Optical/UV monitor of XMM-{\it Newton} telescope (XMM-OM), \textit{Swift} ultraviolet/optical telescope, SkyMapper Southern Sky Survey (SMSS), Sloan Digital Sky Survey (SDSS) DR\,16, the Two-Micron All-Sky Survey (2MASS), the Wide-field Infrared Survey Explorer (WISE), {\it Spitzer} space telescope, AKARI, and {\it Herschel} Space Observatory. Table~\ref{tab:uvdata} gives the instrument characteristics. In the case of the availability of multiple flux measurements at a given wavelength, we chose the one that contained the entire galaxy. Moreover, five galaxies in our sample, namely, ESO\,440-IG058, ESO\,507-G070, ESO\,593-IG008, IR\,16164-0746, and IR\,18293-3413 are either interacting, merging, or post-merging type; for these, the fluxes used in modeling include emission from the companion, too. Table~\ref{tab:UV-IRflux} gives the integrated fluxes along with the integration area per band for each source used for SED modeling. For galaxies ESO 440-IG058 and NGC 5135, the IRAC apertures were optimized on a source by source basis to cover individual components when measuring galaxies in merger systems and to contain all the integrated flux in the case of isolated galaxies (Mazzarella et al., in preparation)

\section{SED modeling}\label{sec:sedmodeling}

\subsection{Radio SED}\label{sec:radiosed}

Integrated radio SEDs are modeled with physically motivated scenarios in which the radio continuum originates from either single or two emission regions characterized by the same or different populations of CRe and optical depths. Most of our adopted models are presented in \citet[][]{Galvin2018}. All modeling was performed in the observers' frame with a reference frequency, $\nu_0$ = 1.4\,GHz. We consider the following models:\\ 
\subsubsection{Single power-law (PL)}\label{model:pl}

A single power-law model with emission characterized by synchrotron processes, following the form:

\begin{equation}
   S_\nu =  A \left(\frac{\nu}{\nu_0}\right)^\alpha
\end{equation}
where, A and $\alpha$ are the normalization and the spectral index, respectively, treated as free parameters.

\subsubsection{Synchrotron and Free-Free Emission (SFG NC) }\label{model:sfgnc}
A radio continuum as a sum of optically thin (no curvature) synchrotron and free--free emission processes, following the form:

\begin{equation}
 S_\nu  =   A\left(\frac{\nu}{\nu_0}\right)^{\alpha}+B\left(\frac{\nu}{\nu_0}\right)^{-0.1} 
\end{equation}
where A and B are the nonthermal and thermal normalization components, respectively, treated as free parameters. The free parameter, $\alpha$, is the synchrotron spectral index.

\subsubsection{Synchrotron and Free-Free Emission with free-free absorption (C) }\label{model:c}

A radio continuum as a sum of optically thick synchrotron and free--free emission processes where the synchrotron emission can be suppressed due to FFA. If the frequency of this turnover due to FFA is parametrized by $\nu_{t,1}$  then the optical depth, $\tau$, can be defined as $(\nu/\nu_{t,1})^{-2.1}$. This model has the following form:
\begin{equation}
S_\nu  =  \left(1-e^{-\tau}\right)\left[B+A\left(\frac{\nu}{\nu_{t,1}}\right)^{0.1+\alpha}\right]\left(\frac{\nu}{\nu_{t,1}}\right)^2
\end{equation}

where A and B are the nonthermal and thermal normalization components and treated as free parameters. The free parameters, $\nu_{t,1}$ and $\alpha$ are the turnover frequency and the synchrotron spectral index, respectively. To reduce the degeneracy of the model, we replace the term $\nu_0$ with the turnover frequency parameter for each component. A key point here is that the models assume that the synchrotron emission is completely commingled within extended plasma, which causes the free-free absorption (a plasma will exhibit FFA regardless of the physical origin of the radio photon - i.e. it does not matter what causes a radio photon to be there; it will get absorbed by the plasma). 
%It is likely that some electrons will probably make it to less dense regions of the galaxy before losing all their energy, so there will be some `leakage’ which we do not have the observations to fully constrain here.} 

\subsubsection{Multiple Free-Free absorption (FFA) components}\label{model:ffa}

Multiple (two) components with emission and absorption representing two star-forming regions with different orientations or compositions. The radio continuum could be complex in these cases, as these regions are integrated by large synthesized beams for unresolved galaxies. We distinguish five scenarios in the multiple components framework which are described below:\\
(a) radio continuum originating from two different relativistic electron populations, characterized by synchrotron spectral indices, $\alpha$ and  $\alpha_2$ in two distinct star-forming regions without FFA turnovers, labelled as SFG NC2, following the form: 

\begin{eqnarray}
S_\nu  =   A\left(\frac{\nu}{\nu_0}\right)^{\alpha} + B\left(\frac{\nu}{\nu_0}\right)^{-0.1} \nonumber \\ + C\left(\frac{\nu}{\nu_0}\right)^{\alpha2} + D\left(\frac{\nu}{\nu_0}\right)^{-0.1}
\end{eqnarray}

where, A and C are the nonthermal normalization components, respectively, while B and D are the thermal normalization components, respectively. A, B, C, D,  $\alpha$, and $\alpha_2$ are treated as free parameters.  

(b) radio continuum characterized by the same (single) relativistic electron population, $\alpha$, in two distinct star-forming regions having different optical depths, $\tau_1$ and $\tau_2$, respectively, labelled as ``C2 1SA'':

\begin{eqnarray}
S_\nu  =   \left(1-e^{-\tau_1}\right)\left[B+A \left(\frac{\nu}{\nu_{t,1}}\right)^{0.1+\alpha}\right]\left(\frac{\nu}{\nu_{t,1}}\right)^2 \nonumber\\ +
\left(1-e^{-\tau_2}\right)\left[D+C	\left(\frac{\nu}{\nu_{t,2}}\right)^{0.1+\alpha}\right]\left(\frac{\nu}{\nu_{t,2}}\right)^2
\end{eqnarray}

where, A and C are the nonthermal normalization components, respectively, while B and D are the thermal normalization components, respectively. A, B, C, D,  $\alpha$, $\tau_1$ and $\tau_2$ are treated as free parameters.

(c) radio continuum characterized by the same (single) relativistic electron population, $\alpha$, in two distinct star-forming regions, one without a turnover (optically thin) while the other characterized by optical depth, $\tau_2$ (optically thick), respectively, labelled as ``C2 1SAN'': 

\begin{eqnarray}
	S_\nu  =   \left(\frac{\nu}{\nu_0}\right)^{-2.1}\left[B+A	\left(\frac{\nu}{\nu_0}\right)^{0.1+\alpha}\right]\left(\frac{\nu}{\nu_0}\right)^2 + \nonumber \\
	\left(1-e^{-\tau_2}\right)\left[D+C	\left(\frac{\nu}{\nu_{t,2}}\right)^{0.1+\alpha}\right]\left(\frac{\nu}{\nu_{t,2}}\right)^2
\end{eqnarray}

where A and C are the nonthermal normalization components, respectively, while B and D are the thermal normalization components, respectively. A, B, C, D,  $\alpha$, and $\tau_2$ are treated as free parameters. This model is most suited to explain the high frequency kinks, elaborated in \citealt{Condon_yin90} and \citet[][]{Clemens10}.

(d) radio continuum characterized by the two different relativistic electron populations, $\alpha$ and $\alpha_2$, in two distinct star-forming regions, one without a turnover while the other characterized by optical depth, $\tau_2$, respectively, labelled as ``C2 1SAN2'':

\begin{eqnarray}
S_\nu  =   \left(\frac{\nu}{\nu_0}\right)^{-2.1}\left[B+A	\left(\frac{\nu}{\nu_0}\right)^{0.1+\alpha}\right]\left(\frac{\nu}{\nu_0}\right)^2 + \nonumber\\
	\left(1-e^{-\tau_2}\right)\left[D+C	\left(\frac{\nu}{\nu_{t,2}}\right)^{0.1+\alpha2}\right]\left(\frac{\nu}{\nu_{t,2}}\right)^2
\end{eqnarray}

where A and C are the nonthermal normalization components, respectively, while B and D are the thermal normalization components, respectively. A, B, C, D,  $\alpha$, $\alpha_2$, and $\tau_2$ are treated as free parameters.

(e) radio continuum characterized by the two different relativistic electron populations, $\alpha$ and $\alpha_2$, in two distinct star-forming regions, characterized by optical depths, $\tau_1$ and $\tau_2$, respectively, labelled as ``C2''.

\begin{eqnarray}
S_\nu  =   \left(1-e^{-\tau_1}\right)\left[B+A	\left(\frac{\nu}{\nu_{t,1}}\right)^{0.1+\alpha}\right]\left(\frac{\nu}{\nu_{t,1}}\right)^2 + \nonumber \\
    \left(1-e^{-\tau_2}\right)\left[D+C	\left(\frac{\nu}{\nu_{t,2}}\right)^{0.1+\alpha_2}\right]\left(\frac{\nu}{\nu_{t,2}}\right)^2
\end{eqnarray}

where A and C are the nonthermal normalization components, respectively, while B and D are the thermal normalization components, respectively. A, B, C, D,  $\alpha$, $\alpha_2$, $\tau_1$ and $\tau_2$ are treated as free parameters.

Models labeled as SFG NC2, C2 1SAN2 and C2 are motivated by galaxy merger scenario where two distinct systems with distinct electron populations drives a new burst of star formation. For fitting the models and model comparison, we applied the Bayesian inference package called UltraNest \citep[][]{Buchner21}. UltraNest works on the principle of the Monte Carlo technique called nested sampling \citep[][]{Skilling04}. The advantage of nested sampling is that it allows Bayesian inference on arbitrary user-defined likelihood and provides posterior probability distributions on model parameters and marginal likelihood (``evidence'') $Z$. The likelihood function used in UltraNest is given as:

\begin{equation}
\mathrm{ln}\,\Lagr\left(\theta\right)  =   -\frac{1}{2}\sum_{n} \left[\frac{\left(D_n-f\left(\theta\right)\right)^2}{\sigma_n^2} + \mathrm{ln}\left(2\pi\sigma_n^2\right)\right],
\label{eq:like}
\end{equation}

where $D_n$ and $\sigma_n$ are the vectors at $n$ different frequencies containing flux densities and uncertainties. $f(\theta)$ represents the model fitted with the data and the parameter vector $\theta$. For model fitting, we assume independent flux measurements with normally distributed errors, which is a prerequisite for the likelihood function used by the UltraNest. Within the Bayesian framework, the posterior distribution of any model parameter requires a prior distribution along with a likelihood function, which gives the confidence interval on the derived parameter. In our analysis, we consider a uniform prior distribution of model parameters. We constrain the priors for the normalization parameters, A, B, C, and D as positive, spectral index parameters $\alpha$ and  $\alpha_2$ in the range $-$0.2 and $-$1.8, and turnover frequencies are between 1\,MHz and 50\,GHz \citep[see, for details ][]{Galvin2018}.

The evidence value is used to predict the most preferred model by calculating the Bayes odds ratio as follows:

\begin{equation}\label{eq:evidence}
\Delta\mathcal{Z}  =   e^{{\mathcal{Z}}_1-{\mathcal{Z}}_2}
\end{equation}

where ${\mathcal{Z}}_1$ and ${\mathcal{Z}}_2$ are the evidence values for models $M_1$ and $M_2$, respectively. The strongest evidence supporting $M_1$ over $M_2$ is when $\Delta\mathcal{Z}>$150 while for 150$>$ $\Delta\mathcal{Z}$ $>$20 and 20$>$ $\Delta\mathcal{Z}$ $>$3, respectively, it is either strong or positive evidence. For $\Delta\mathcal{Z}$ $<$3, the models are considered indistinguishable. Table~\ref{tab:bayesodds} summarizes the natural logarithm of the Bayes odds ratio between each model and the most preferred model \citep[][]{Kass95}. This means that for a given model, log$_e$(1)  =   0.0 indicates the most preferred model. The least preferred model will have the most negative value in this representation. For each source, we provide the natural logarithm of the Bayes odds ratio for the most preferred model in boldface.    

The best-fit radio SEDs are presented in Figure~\ref{fig:radiosed} while Table~\ref{tab:radiosed_pars} gives the corresponding model parameters, along with the 1$\sigma$ uncertainties corresponding to the 16th and 84th percentile of the posterior distribution of the parameter. Our radio-only modeling separates the thermal and nonthermal components from the total emission. Using the best-fit model, we estimated total thermal(nonthermal) emission by setting the normalization of the nonthermal(thermal) component(s) to zero. Table~\ref{tab:rad_flux} gives the total, thermal, and nonthermal fluxes at 1.4\,GHz, derived from the radio-only SED modeling, along with the 1$\sigma$ uncertainties corresponding to the 16th and 84th percentile of the posterior distribution of the parameter. To assess the degeneracy caused by the number of free parameters in the best-fit model, we give corner plots for all our galaxies in Appendix (Figure~\ref{fig:corner_radio}). The complete figure set (11 images) is available in the online journal. Figure~\ref{fig:corner_radio}. The corner plot gives the one- and two- dimensional projections of the posterior probability distribution of parameters. Most of our corner plots show poorly constrained thermal components, most likely due to weaker thermal emission in the frequency range covered by our data. In Table~\ref{tab:rad_flux}, we also provide the 1.4\,GHz thermal fraction ($TF_{1.4}$). The thermal fraction ($TF$) at a given frequency is the ratio of thermal radio emission to total radio emission (synchrotron and free-free emission). NGC\,3508 shows an excellent fit to the single power law (PL) shape. Three other galaxies, NGC\,6000, IR\,16164$-$0746, and ESO\,453-G005, are fitted with a single-component model without a turnover (SFG NC). The galaxies ESO\,440-IG058, NGC\,5135, IC\,4280, and ESO\,593-IG008 are fitted with the single-component model with low-frequency turnover (C). For galaxies ESO\,500$-$G034 is best-fitted with a multiple component model characterized with a single relativistic electron population in two different star-forming regions (one region with no turnover at low frequencies and the other characterized by a turnover (C2 1SAN). For galaxies ESO\,507$-$G070 and IR\,18293$-$3413, the best fit model turned out to be multiple components, one characterized by different relativistic electron populations in two different star-forming regions (one region with no turnover at low frequencies and the other characterized by a turnover (C2 1SAN2). None of our galaxies fits with models described by a single relativistic electron population in two different star-forming regions with different optical depths (C2 1SA) and a multiple component model representing two different electron populations in the two distinct star-forming galaxies with different optical depths (C2).
    
\begin{table*}
\caption{List of input parameters for CIGALE modeling}\label{tab:sedpar}

  \begin{center}
 
    \begin{tabular}{l c c}
     \hline\hline
    \textbf{\textit{Parameters}} & & \textbf{\textit{Values}}\\
     \hline\hline
     \multicolumn{3}{c}{\textbf{delayed star formation history + additional burst} \citep{ciesla2015}}   \\
     \hline
     e-folding time of the main stellar population model [Myr] &$\tau_{main}$ & 300-15000 by a bin of 300\\
    e-folding time of the late starburst population model [Myr] &$\tau_{burst}$   & 50-1000 by a bin of 50\\
     mass fraction of the late burst population & $f_{burst}$ &  0.05, 0.1, 0.3, 0.6, 0.9\\
   Age of the main stellar population in the galaxy [Myr]   & age   & 1000, 2000, 3000, 4500, 5000, 6500, 10000, 12000\\
    Age of the late burst [Myr] & age$_{burst}$  & 10.0, 40.0, 80.0, 110, 150, 170\\
    \hline\hline
    \multicolumn{3}{c}{\textbf{stellar synthesis population} \citep{bruzandcharlot2003}}   \\
    \hline
    Initial mass function & IMF & \citep{Salpeter1995}\\
    Metallicity & $Z$ &  0.02 \\
    Separation age &    &  1\,Myr\\
    \hline\hline
    \multicolumn{3}{c}{\textbf{dust attenuation laws} \citep{calzettii2000}}   \\
    \hline
    \hline
    Colour excess of young stars & E(B-V) & 0.1-2 by a bin of 0.2 \\
    Reduction factor$^{(iii)}$ & f$_{att}$ & 0.3, 0.44, 0.6,0.7 \\
    \hline
    \hline
    \multicolumn{3}{c}{\textbf{dust grain model; THEMIS} \citep{Jones2017}}   \\
    \hline
    fraction of small hydrocarbon solids & q$_{hac}$ &  0.02, 0.06, 0.1,0.17, 0.24 \\
    Minimum radiation field & U$_{min}$ & 1, 5, 10, 15, 20, 30 \\
    power law index of the radiation & $\alpha$ & 2 \\
    fraction illuminated from $U_\mathrm{min}$ to $U_\mathrm{max}$ & $\gamma$ & 0.02,0.06,0.1,0.15,0.2 \\
    \hline\hline
    \multicolumn{3}{c}{\textbf{active nucleus model; Skirtor} \citep{Stalevski2012,Stalevski2016}} \\
    \hline
    optical depth at 9.7$~\mu$m & $\tau_{9.7}$ & 3.0, 7.0 \\
    torus density radial parameter  & pl & 1.0 \\
    torus density angular parameter & q & 1.0 \\
    angle between the equatorial plan and edge of the torus [deg] & oa & 40 \\
    ratio of outer to inner radius & R & 20 \\
    fraction of total dust mass inside clumps [\%] & Mcl & 97 \\
    inclination (viewing angle) [deg] & i & 30 (type 1), 70 (type 2) \\
    AGN fraction & & 0.0-0.4 by a bin of 0.05\\
    extinction law of polar dust && SMC \\
    E(B-V) of polar dust && 0.01-0.7 by a bin of 0.5\\
    Temperature of the polar dust & K & 100 \\
    Emissivity index of the polar dust && 1.6 \\
    
   \hline \hline
   \multicolumn{3}{c}{\textbf{synchrotron emission}}   \\
    \hline
    FIR/radio parameter$^*$ & $q_{IR}$ &  2.3 - 2.9 by a bin of 0.1 \\
    Power-law slope (Flux $\propto$ Frequency$^{\alpha_{synch}}$) & $\alpha_{synch}$ & $-$1.8 to $-$0.2 by a bin of 0.1 \\
    \hline
    \end{tabular}
    \end{center}
    $^*$ computed as $\log_{10}L_{IR(8-1000\mu\rm m)}$ -$\log_{10}L_{1.4\,GHz}$ where $L_{1.4\,GHz}$ is the radio luminosity at 1.4\,GHz.
    \end{table*}

\begin{figure}
    \centering
    \hspace{0cm}{\includegraphics[width =  0.48\textwidth]{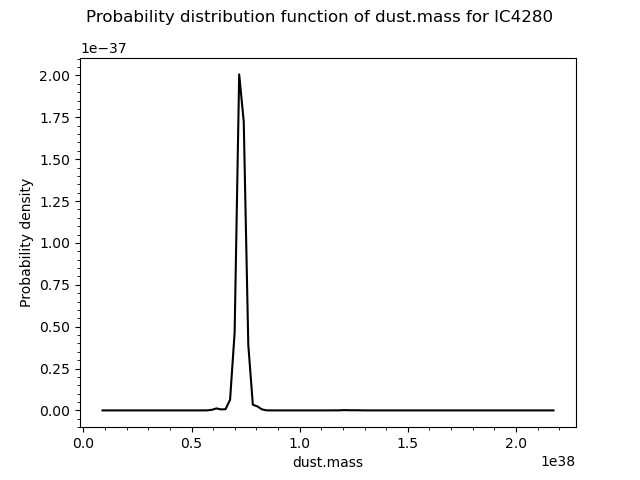}}
\caption{Probability distribution function (PDF) of the estimated dust mass for galaxy IC4280 using Bayesian inference.}\label{fig:pdf} 

\end{figure}

\begin{figure*}

    \hbox{
    \hspace{0cm}{\includegraphics[width =  0.33\textwidth]{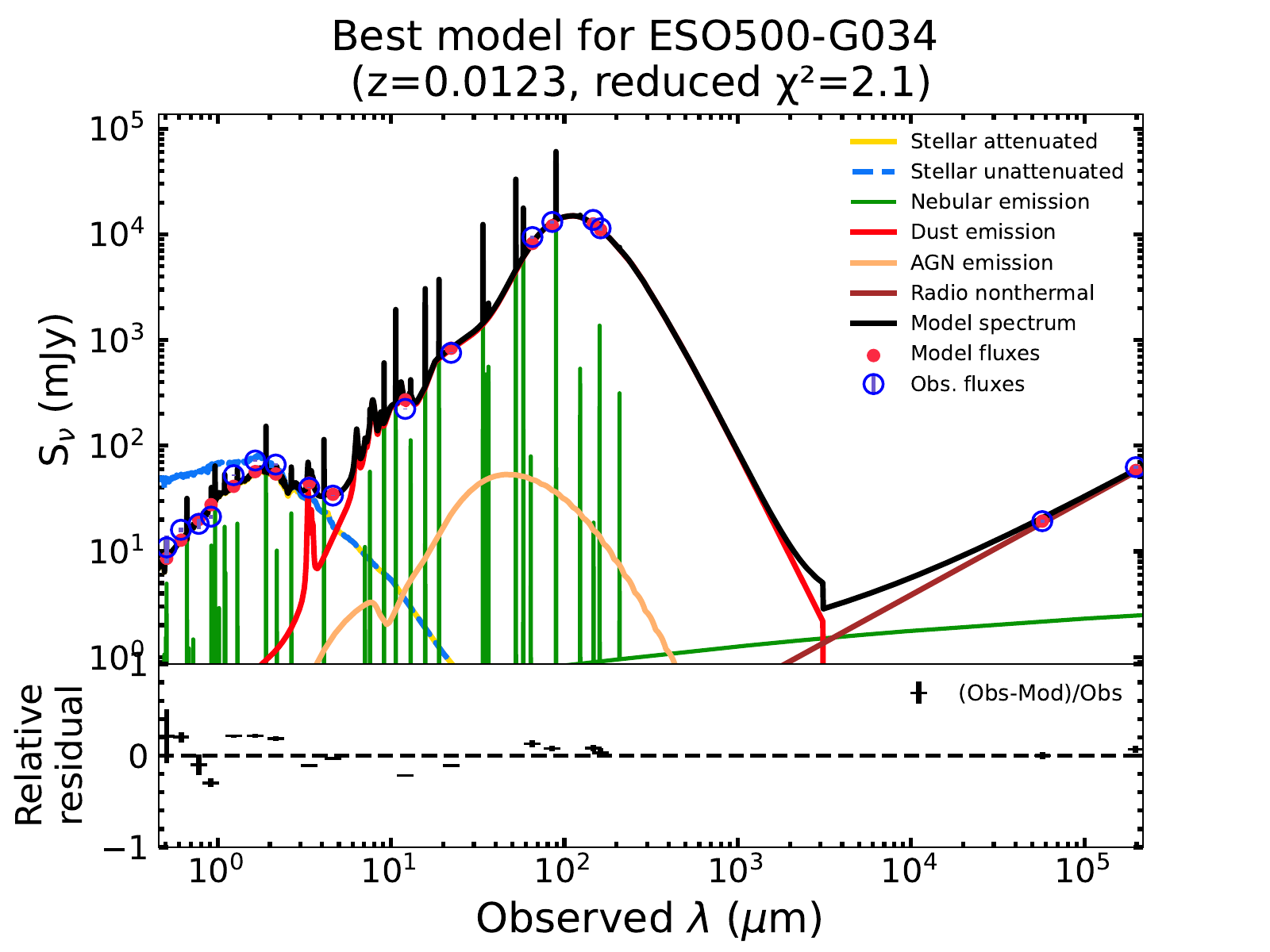}}
    \hspace{0cm}{\includegraphics[width =  0.33\textwidth]{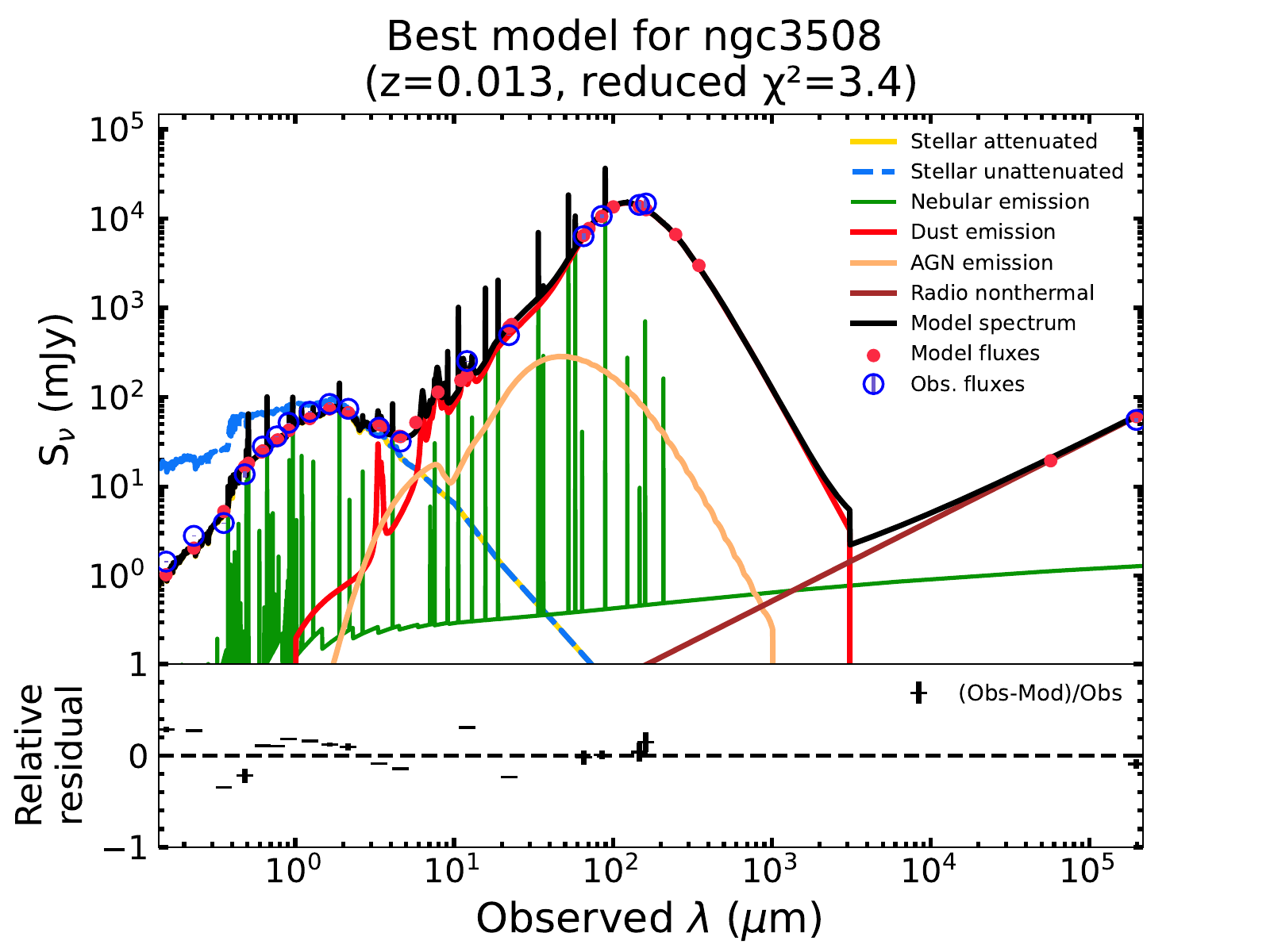}}
    \hspace{0cm}{\includegraphics[width =  0.33\textwidth]{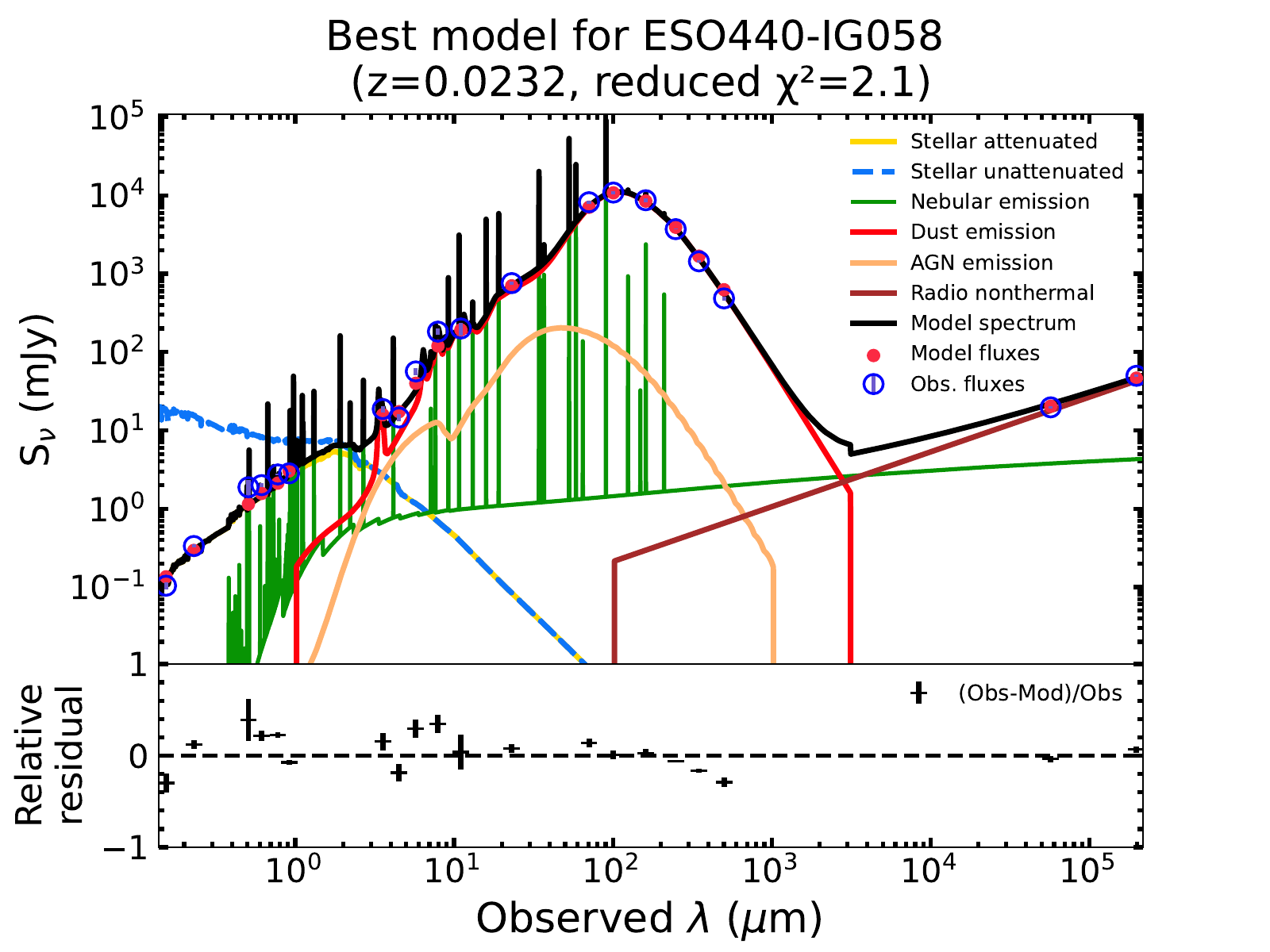}}
    }
    \hbox{
    \hspace{0cm}{\includegraphics[width =  0.33\textwidth]{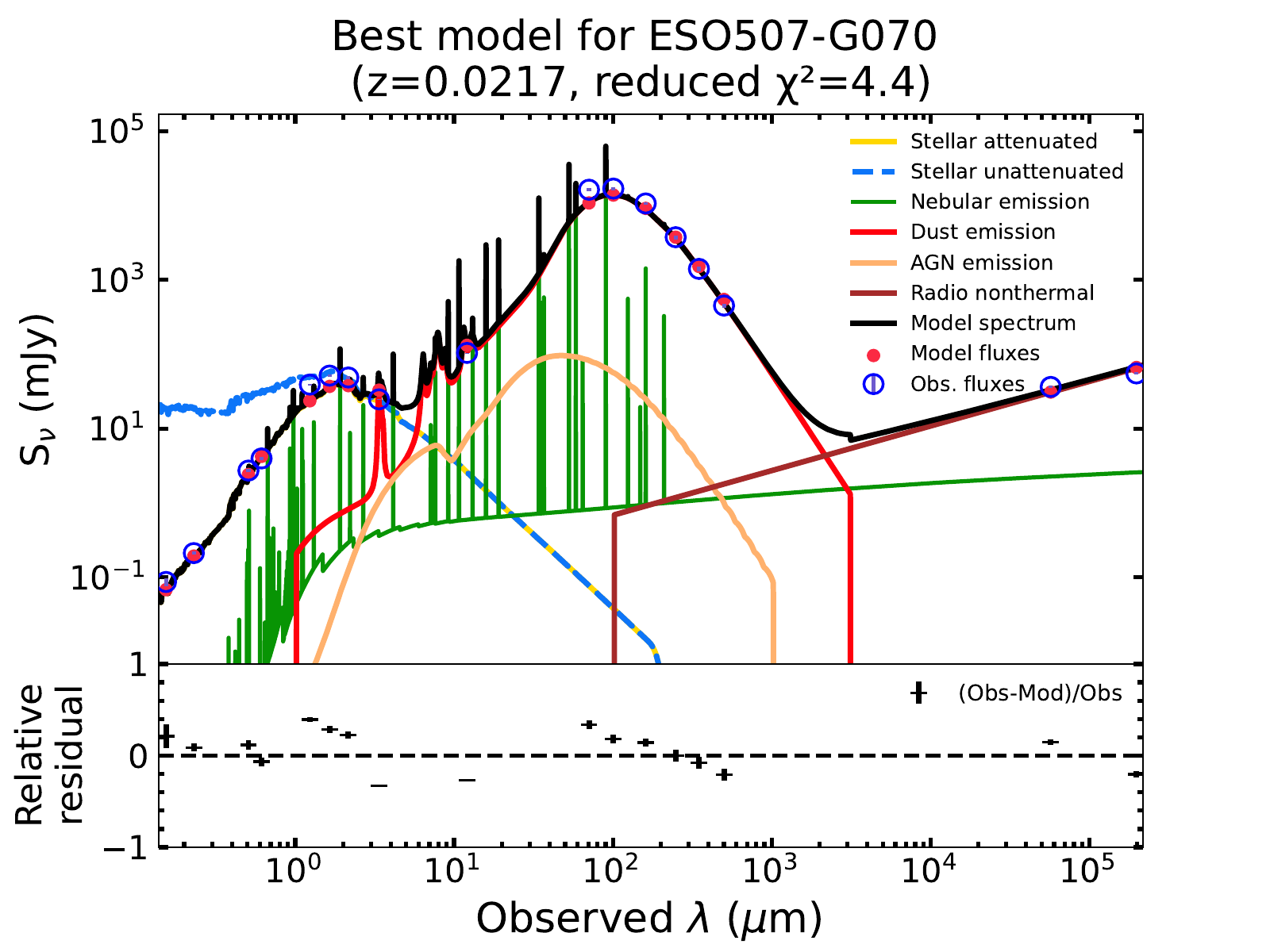}}
     \hspace{0cm}{\includegraphics[width =  0.330\textwidth]{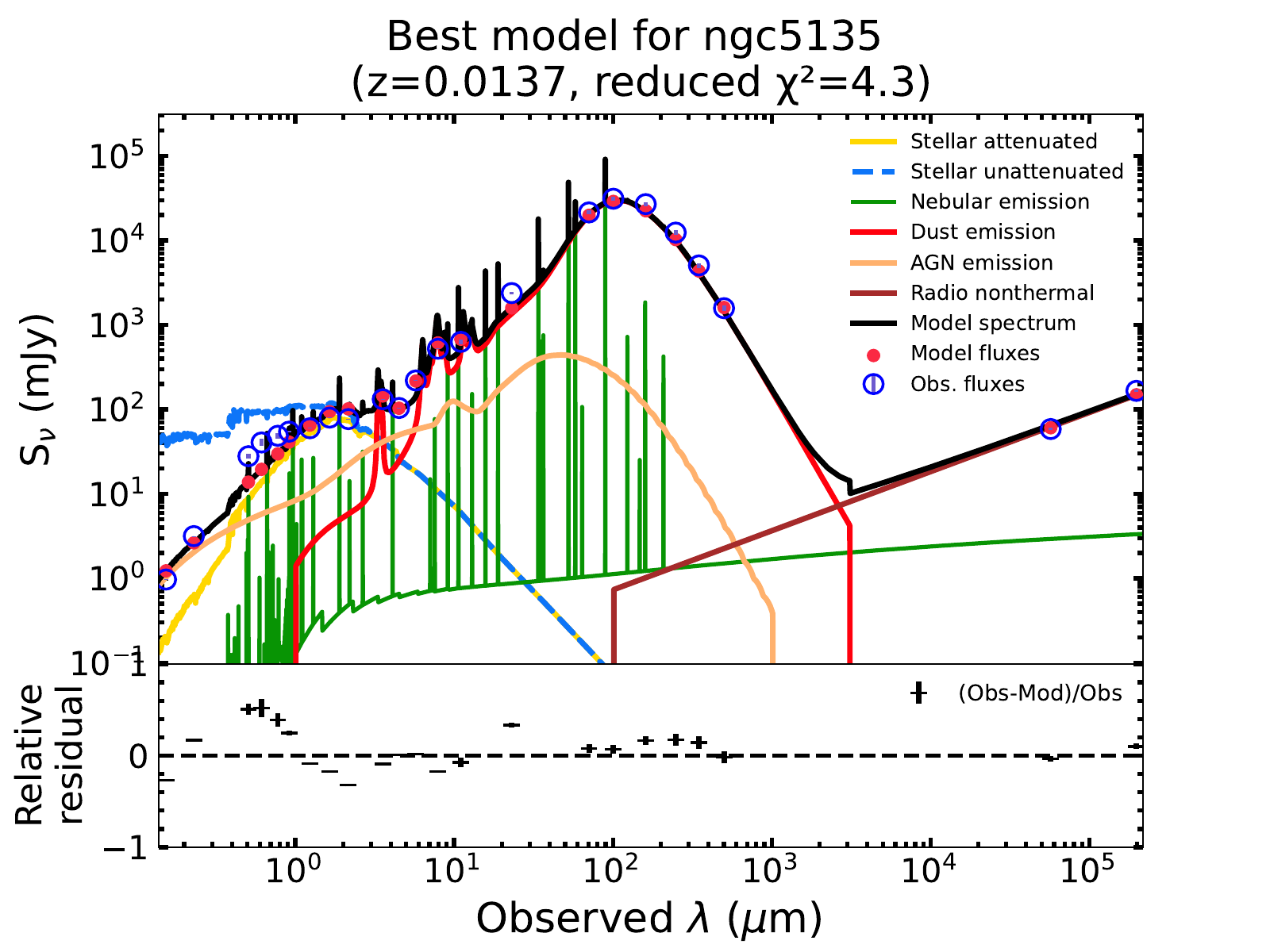}}
      \hspace{0cm}{\includegraphics[width =  0.330\textwidth]{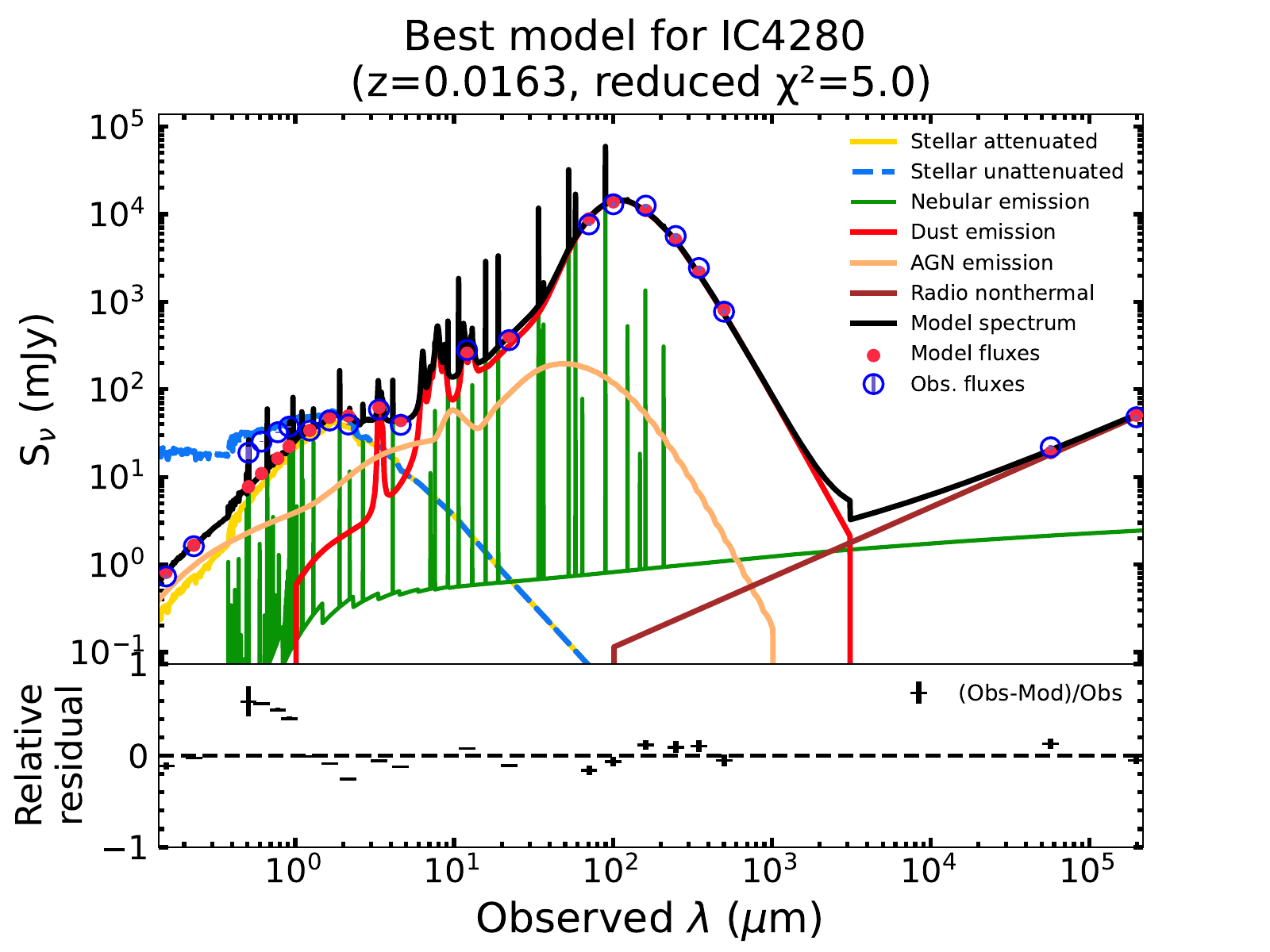}}
    }
    \hbox{
     \hspace{0cm}{\includegraphics[width =  0.330\textwidth]{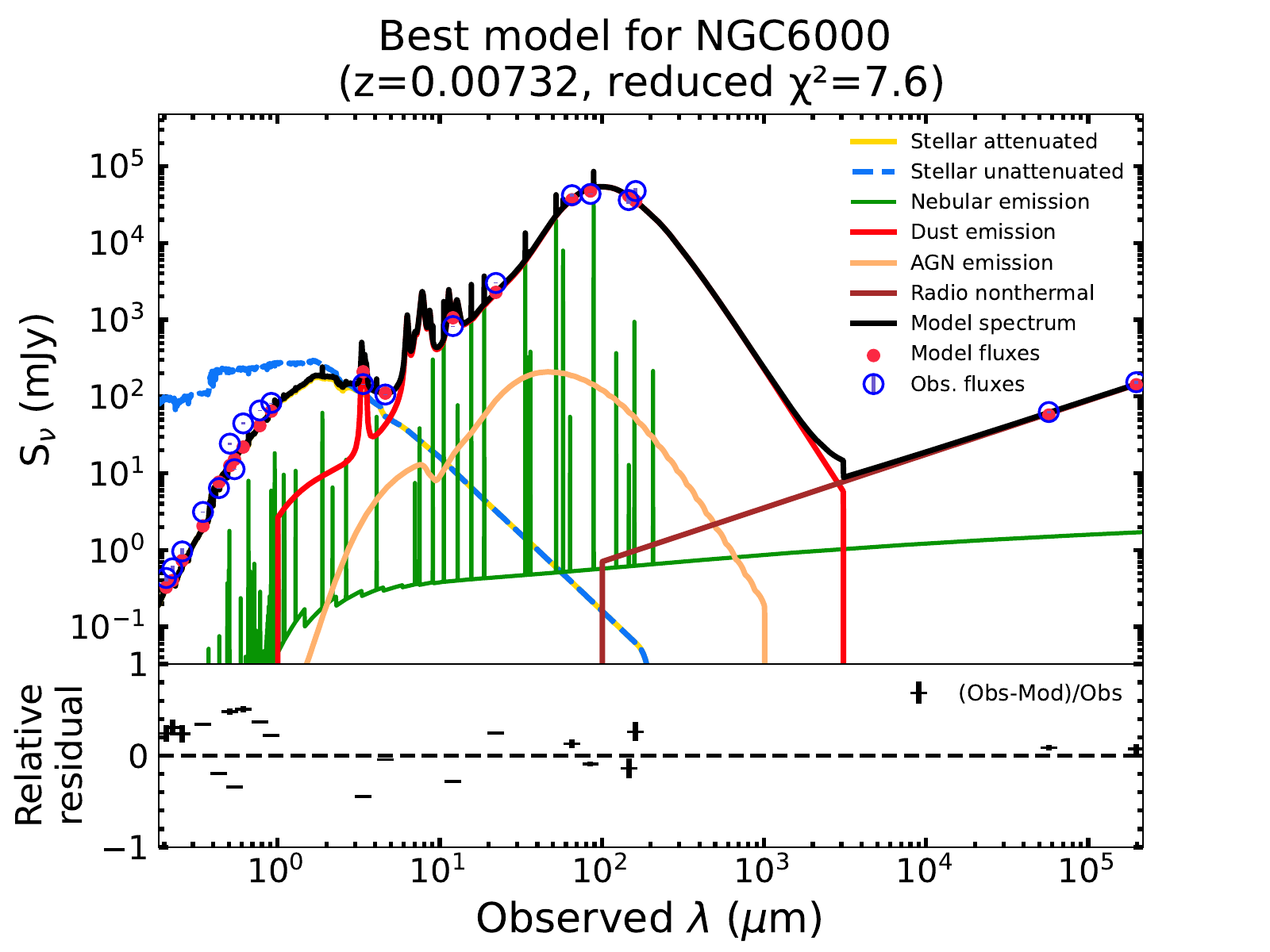}}
    \hspace{0cm}{\includegraphics[width =  0.330\textwidth]{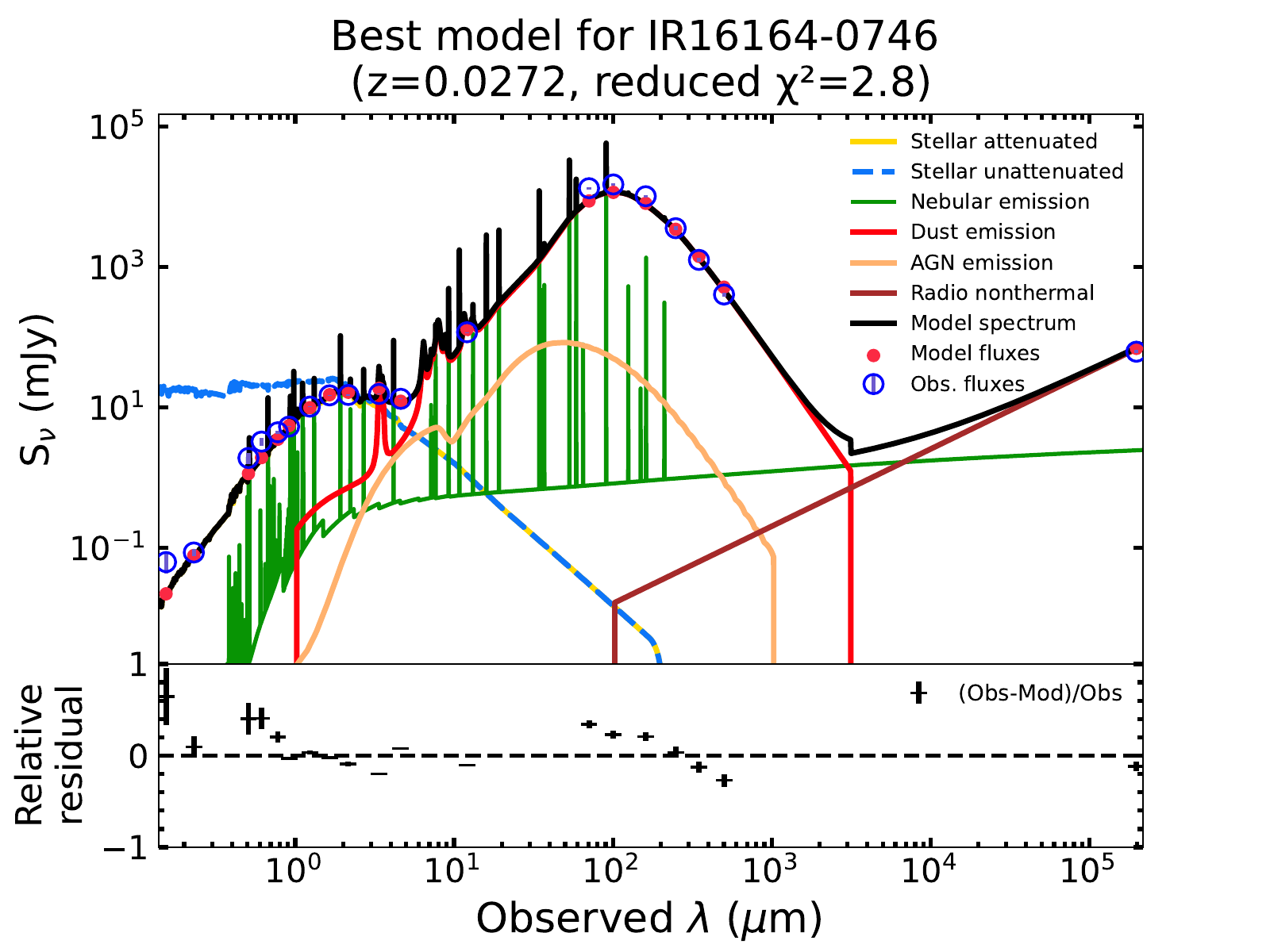}}
    \hspace{0cm}{\includegraphics[width =  0.330\textwidth]{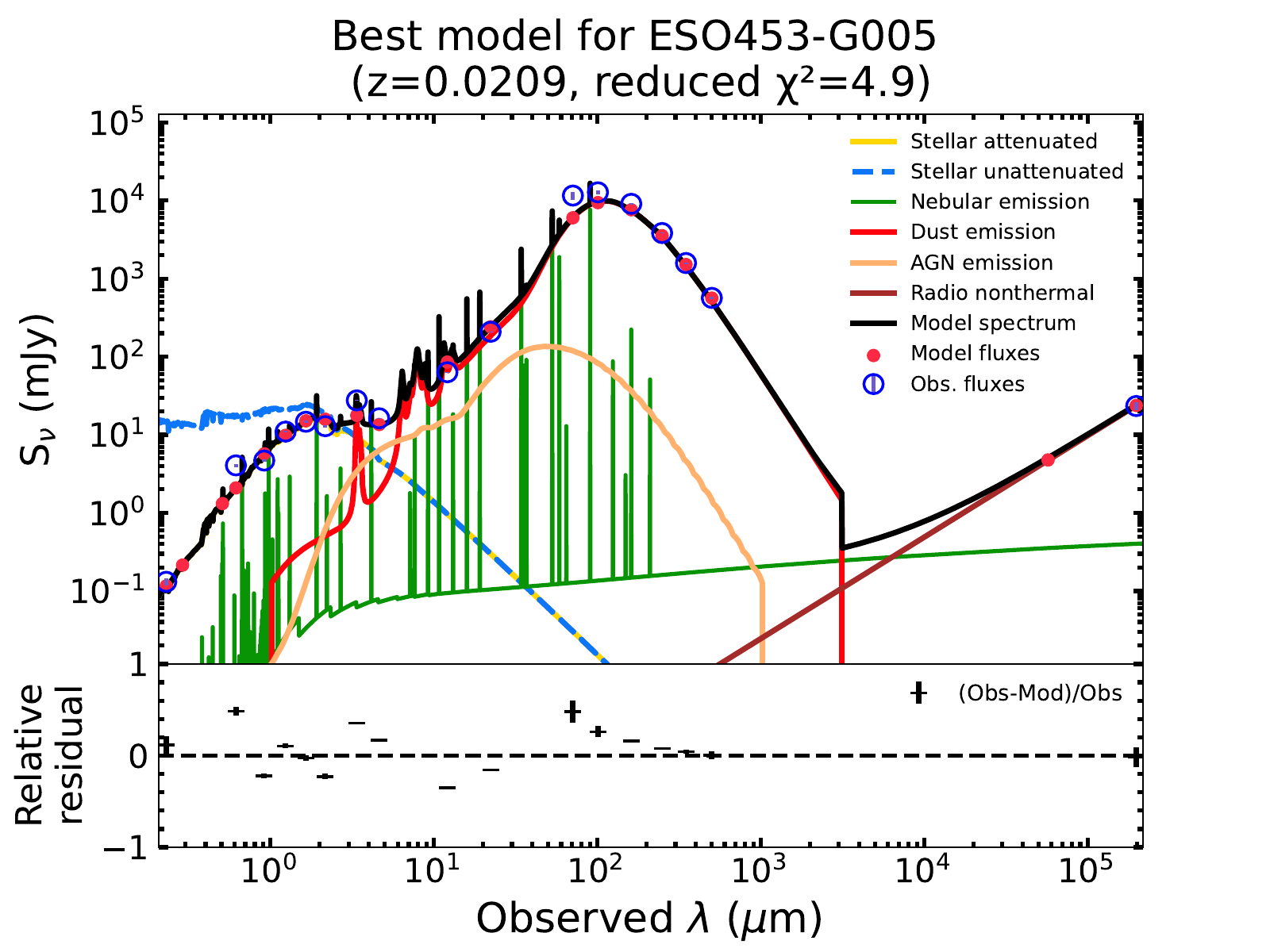}}
    }
 \hbox{
    \hspace{0cm}{\includegraphics[width =  0.330\textwidth]{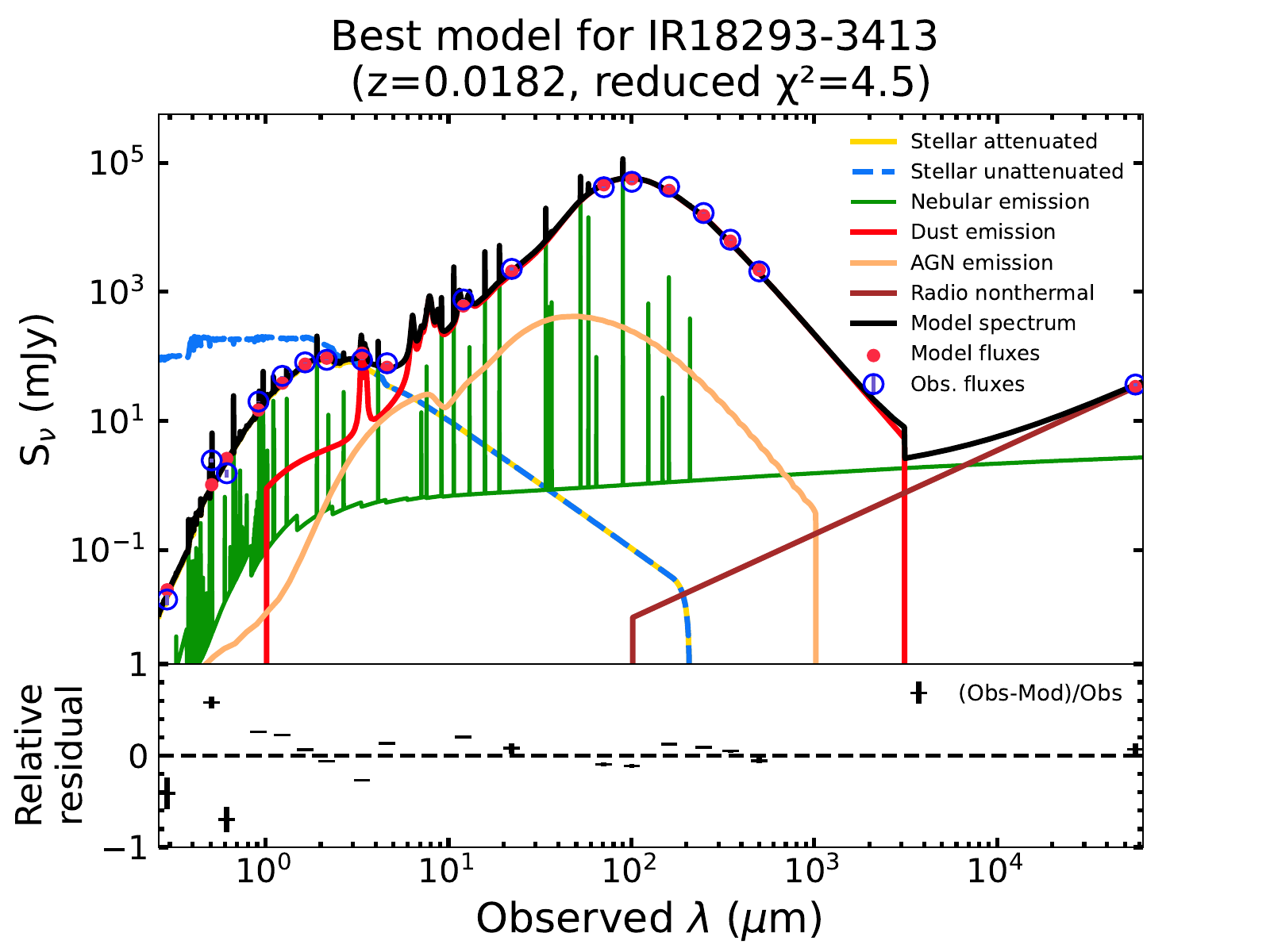}}
    \hspace{0cm}{\includegraphics[width =  0.330\textwidth]{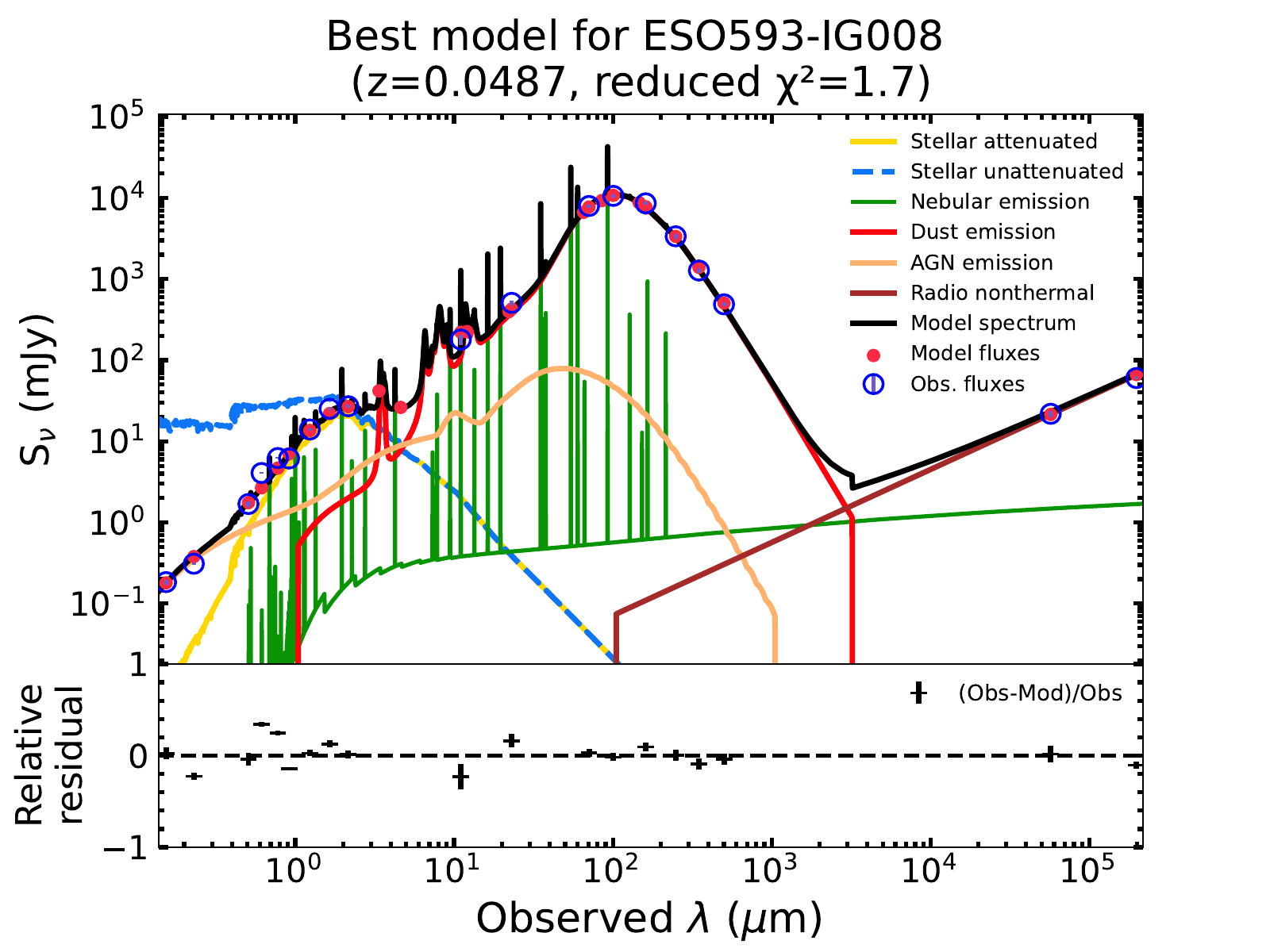}}    
   
    }  
\caption{Best-fit CIGALE SEDs of the 11 LIRGs studied here. The open and filled symbols give the observed and modeled flux densities. The goodness of fit is estimated by reduced $\chi^2$ shown at the top of each panel, along with the name and redshift of the galaxy. In almost all cases, SEDs are well modeled, giving reasonable estimates of astrophysical properties.}\label{fig:cigalefitting}
\end{figure*}

\subsection{UV to radio SED modeling with CIGALE}\label{sec:cigale}

We estimate astrophysical properties of our galaxies using the SED fitting technique with the CIGALE version 2020.0 \footnote{\url{https://cigale.lam.fr/}} \citep{Noll2009, Boquien2019}. The CIGALE modeling works on the energy balance principle, i.e., the energy emitted by dust in the mid and far-$IR$ corresponds to the energy absorbed by dust in the $UV$ to optical range \citep[][]{Efsthathiou2003}. Moreover, it uses a Bayesian-like approach to model the SED and obtains the model parameters efficiently. CIGALE is parallelized and modular, which makes it user-friendly and efficient, as it does not solve the computationally demanding radiative transfer equation \citep[][]{Boquien2019}. It provides the best fit model for the SED by selecting a suitable set of parameters. For this, a large grid of models is fitted to the data. The grid dimension is set by the number of input parameters used to define the different galaxy components, such as stellar emission spectra, star formation history, AGN emission, dust attenuation, nebular emission, and the slope of the radio synchrotron spectrum.

 The interpretation of the observed SED is based on a comparison of all the modeled SEDs obtained from the fixed grid of parameters used for the modeling. First, each model is scaled and normalized to the data by minimizing $\chi^2$. Then the probability that a given model matches the data is quantified by the likelihood taken as e$^{-\chi^2/2}$. These likelihoods can then be used as weights to estimate the physical parameters (the likelihood-weighted means of the physical parameters) and the related uncertainties (the likelihood-weighted standard deviation of the physical parameters). Finally, models with low probability are discarded, leaving only the best models to determine the physical parameters. Due to this process, the calculated uncertainties are symmetric \citep[see also, Section 4.3 of ][]{Boquien2019}. This method of choosing the best-fit model also takes care of the degeneracies in the model parameters, as only one parameter value (one with the highest probability in the pdf) can result in best-fit SED. 
 
 Next, we provide a brief account of the selected models used in our study.The first step towards obtaining the SED model requires building the stellar emission, which consists of selecting the single stellar population model \citep[in our case;][]{Bruzual2003}, assuming \citet{Salpeter1995} initial mass function. Next, to model the star formation history (SFH), we adopted a delayed SFR with an additional burst profile \citep[in accordance to][]{Malek2018}.
This form of SFH provides good estimates of the SFR-$M_{\star}$ relation compared to observations \citep{ciesla2015}. The SFH is defined as:

\begin{equation}
    \rm SFR \left( t \right) \propto 
    \begin{cases}
\rm SFR_{delayed} \left( t \right) : \text{if} \,\, t < t_0.\\
\rm SFR_{delayed} \left( t \right)\, +\, \rm SFR_{burst} \left( t \right), : \text{if} \,\, t \geq t_0.
\end{cases}
\end{equation}

where $t_{0}$ is the age since the onset of the second episode (burst) of star formation.

To model starlight attenuation by dust, we chose the extinction law of \citet{calzettii2000}, and to model dust emission, we selected The Heterogeneous Evolution Model for Interstellar Solids  \citep[THEMIS;][]{Jones2017} model.
THEMIS successfully explains the observed Far-UV to Near-IR extinction and the shape of the IR to mm dust thermal emission. Furthermore, it predicts the observed relationship between the E(B-V) color excess and the inferred sub-mm opacity derived from {\it Planck} observation \citep[see for information][]{Jones2017, Nersesian2019}.
To incorporate the AGN component in the SED, we selected the SKIRTOR module \citep[][]{Stalevski2012, Stalevski2016}. SKIRTOR is based on the 3D radiative transfer code SKIRT \citep{Baes2011}. It includes obscuration by dusty torus and obscuration by dust settled along with the polar directions.
Since our galaxies have a rich amount of radio data, we used 1.4\,GHz and 4.8\,GHz fluxes to model the nonthermal synchrotron emission taking into account the power law of the synchrotron spectrum and the ratio of the far-IR/radio correlation in the CIGALE fitting.\\

Table~\ref{tab:sedpar} gives the set parameters used to build the SEDs of our galaxies. We adopted parameters similar to those used in \citet{Malek2017, Pasparialis2021} and modified them accordingly to suit the galaxies in the current sample.
The SED fitting was performed with a very careful adjustment of the fitting parameters, module by module.
We performed the pdf analysis method to calculate the likelihood function ($\chi^2$) for all possible combinations of parameters \citep[see Section 4.3 of ][]{Boquien2019}. We generate the marginalized pdf for every parameter and assess the suitability of the SED model by visual inspection. An example of this method is given in Figure~\ref{fig:pdf} for the galaxy IC4280 for dust mass. Figure~\ref{fig:cigalefitting} provides the CIGALE fit for our LIRGs. We note that some of the photometric data and the best model vary and, for them, the residuals exceed 20-30\%. We want to stress that those residuals are often related to strong emission lines, visible for the nebular model. Moreover, this kind of catalog, which includes data from different surveys and instruments, sometimes performed more than 10--20 years ago (GALEX, XMM-OM), reduced with different reduction procedures, can result in heterogeneous photometric coverage of the spectra. In the SED fitting procedure, all measurements are treated with the same weight, and it can happen that the residuals of some of them are not as small as we would expect. The most significant residuals can be found for the $g^\prime$ and $r^\prime$ bands for our galaxies. As fits for the UV part of the spectra look very good, we can assume that the young stellar population was fitted with very good precision. The same observation can be made for $i^\prime$, $z^\prime$, and near-IR measurements, which assumes a good estimation of the stellar mass and properties of older stellar populations.

Another issue we want to address here is effect of different aperture sizes taken for flux measurements at different wavelengths for the SED modeling. Indeed, the apertures do not match for galaxies in our sample. The resulting SED fitting can be influenced by the decrement in the flux, especially nearby galaxies, where the beam size can be smaller than the galaxy itself. Also, a too large aperture in the case of a smaller galaxy can result in measuring a partial flux from another nearby galaxy.
In our case, we tried to use an aperture size that was adequate to the size of a galaxy. Of course, it is impossible to match it perfectly without re-measuring fluxes, as was done in \citet[][]{Ramos-Padilla20}. As expected, this uncertainty in flux measurements results in errors for different estimated physical properties. In addition, CIGALE adds a 10\% error to account for all these deviations. Problems with aperture, data reduction, and normalization can be seen as ``jumps'' in the SED instead of a smooth transition between different measurements. However, in the case of galaxies studied in this paper, the spectral coverage with broadband photometry is dense enough to obtain realistic physical properties. For the SED fitting procedure, fluxes were used together with the accompanying error. In our CIGALE analysis, we have also checked all pdfs for estimated physical properties such as stellar masses, star formation rate, dust mass, and dust temperature. We did not find any strange or suspicious behavior that incorrect photometric matching can trigger. We note that  \citet[][]{Ramos-Padilla20} showed, recalculating flux measurements can increase the quality of estimated astrophysical properties, however we did not remeasure the fluxes in our analysis.

\clearpage
\newpage
\begin{table*}
\setlength{\extrarowheight}{5pt}
\centering
%\scriptsize
%\small
\caption{The most preferred models were selected on the basis of the Bayes odds ratio and the constrained value of the free parameters. The nominal value is taken as the 50$^{th}$ percentile of the posterior distribution of the samples, and the 1$\sigma$ uncertainties are provided by the 16$^{th}$ and 84$^{th}$ percentile. Parameters not included in the models are \textbf{indicated as} ``-''.\label{tab:radiosed_pars}}
\begin{tabular}{cccccccccc}\\
\hline
Name  &  Model  & A   & B  & $\alpha$ & $\nu_{t,1}$ & C & D & $\alpha_2$ & $\nu_{t,2}$ \\
 &  & (mJy) & (mJy) &  & (GHz) & (mJy) & (mJy) &  & (GHz) \\
 (1) & (2) & (3) & (4) & (5) & (6) & (7) & (8) & (9) & (10)\\
\hline
 ESO 500-G034 & C2 1SAN & $6.21^{+2.23}_{-1.83}$ & $3.58^{+3.13}_{-2.48}$ & $-1.18^{+0.11}_{-0.12}$ & - & $154.37^{+11.53}_{-10.90}$ & $3.80^{+3.20}_{-2.64}$  & - & $0.52^{+0.07}_{-0.06}$ \\
 NGC 3508  & PL   &  $62.20^{+2.55}_{-2.31}$ &  - & $-0.73^{+0.02}_{-0.02}$  & - & -  &  -  & - & -  \\
ESO 440-IG058 & C & 325.81$^{+19.13}_{-16.75}$  & $1.79^{+2.33}_{-1.28}$   & $-0.82^{+0.03}_{-0.04}$ &  $0.14^{+0.02}_{-0.02}$ & - & - &- & -\\
ESO 507-G070  & C2 1SAN2 & $39.95^{+4.46}_{-4.30}$ & $5.10^{+3.25}_{-3.30}$   &  $-0.77^{+0.05}_{-0.06}$ & - & $21.07^{+3.98}_{-4.20}$ & $2.25^{+1.77}_{-1.53}$  & $-1.50^{+0.38}_{-0.22}$ & 7.49$^{+1.85}_{-2.01}$\\
NGC 5135 & C     & $1294.29^{+80.51}_{-70.09}$   & $7.34^{+8.88}_{-5.28}$   & $-0.88^{+0.03}_{-0.05}$ & $0.13^{+0.02}_{-0.02}$   & - & - & - & - \\
 IC 4280  & C   & $359.44^{+43.22}_{-30.17} $ & $10.43^{+5.69}_{-5.93}$ &  $-0.89^{+0.09}_{-0.12}$ & $0.13^{+0.03}_{-0.03}$  & - & - & - & -  \\
 NGC 6000     & SFG NC   & $144.60^{+3.21}_{-3.26}$   & $0.48^{+0.34}_{-0.32}$ & $-0.66^{+0.01}_{-0.02}$ & - & - & - & - & - \\
IR 16164-0746 & SFG NC & $61.52^{+2.27}_{-2.33}$ & $0.49^{+0.36}_{0.33}$ & $-0.45^{+0.03}_{-0.03}$ & - & - & - & - & -\\
ESO 453-G005  & SFG NC & $22.34^{+1.06}_{-0.99}$ & $0.52^{+0.32}_{-0.35}$ & $-0.59^{+0.04}_{-0.04}$  & - & - & - & - & -\\
IR 18293-3413 & C2 1SAN2 & $34.13^{+12.20}_{-9.91}$ & $4.09^{+3.28}_{-2.75}$ & $-1.33^{+0.12}_{-0.14}$ & - & $373.64^{+42.37}_{-41.82}$ & $4.19^{+3.08}_{-2.79}$ & $-1.74^{+0.08}_{-0.05}$ & $1.06^{+0.08}_{-0.09}$ \\
ESO 593-IG008 &  C  & $540.79^{+73.47}_{-47.65}$    & $0.49^{+0.34}_{-0.34}$  & $-0.87^{+0.03}_{-0.03}$ & $0.10^{+0.02}_{-0.02}$ & - & - & - & -\\
\hline
\end{tabular}

Columns: (1) galaxy name, (2) best-fit model, (3) and (7) nonthermal normalization components, (4) and (8) thermal normalisation components, (5) and (9) synchrotron spectral indices, (6) and (10) turnover frequencies.
\end{table*}

\begin{table}[htp!]
\centering
\scriptsize
\caption{Total, thermal, and nonthermal fluxes estimated at 1.4 GHz from the radio-only SED fitting and the corresponding thermal fraction estimated at 1.4 GHz. The nominal value is taken as the 50$^{th}$ percentile of the posterior distribution of the samples, and the 1$\sigma$ uncertainties are provided by the 16$^{th}$ and 84$^{th}$ percentile.
\label{tab:rad_flux}}
\begin{tabular}{ccccc }
\hline
Name & $S^{\rm total}_{1.4}$ & $S^{th}_{1.4}$ & $S^{\rm nth}_{1.4}$& $TF_{1.4}$ \\
  & (mJy) & (mJy) & (mJy) &   \\
 (1)  & (2) & (3) & (4)& (5) \\
\hline
ESO 500-G034 &59.74$^{+1.94}_{-1.74}$ & 7.32$^{+1.82}_{-2.22}$ & 52.57$^{+3.06}_{-2.74}$ & 0.11$^{+0.02}_{-0.02}$ \\
NGC 3508 & 61.84$^{+2.55}_{-2.42}$ & $-$ & 61.84$^{+2.55}_{-2.42}$ & 0 \\
ESO 440-IG058 & 49.98$^{+1.03}_{-1.04}$ &  1.59$^{+1.79}_{-1.13}$ & 48.39$^{+1.66}_{-2.24}$ & 0.03$^{+0.02}_{-0.02}$ \\
ESO 507-G070 & 53.02$^{+1.59}_{-1.59}$ & 4.75$^{+3.41}_{-3.11}$ & 48.24$^{+3.31}_{-3.52}$ & 0.09$^{+0.04}_{-0.04}$\\
NGC 5135 & 164.33$^{+3.58}_{-3.74}$ & 6.04$^{+7.28}_{-4.43}$ & 158.29$^{+6.20}_{-9.60}$ & 0.04$^{+0.02}_{-0.03}$  \\
IC 4280 & 50.32$^{+1.52}_{-1.73}$ & 8.53$^{+5.04}_{-5.17}$ & 41.84$^{+6.12}_{-6.24}$ & 0.12$^{+0.04}_{-0.04}$ \\
NGC 6000 & 146.98$^{+3.07}_{-3.21}$ & 0.49$^{+0.35}_{-0.33}$ & 146.00$^{+4.00}_{-3.00}$ & 0.003$^{0.002}_{0.002}$ \\
IR 16164-0746 & 62.04$^{+2.45}_{-2.59}$  & 0.51$^{+0.34}_{-0.33}$ & 61.50$^{+2.50}_{-2.60}$ & 0.008$^{+0.004}_{-0.004}$   \\
ESO 453-G005 & 23.04$^{+1.35}_{-1.24}$ & 0.52$^{+0.33}_{-0.35}$ & 22.50$_{+1.40}^{-1.40}$ & 0.02$^{+0.01}_{-0.01}$ \\
IR 18293-3413 & 220.59$^{+7.18}_{-7.23}$ & 7.76$^{+2.10}_{-2.15}$ & 212.84$^{+7.95}_{-7.75}$ & 0.04$^{+0.01}_{-0.01}$\\
ESO 593-IG008 & 58.59$^{+1.72}_{-1.67}$ & 0.36$^{+0.25}_{-0.26}$ & 58.22$^{+1.70}_{-1.71}$ & 0.006$^{+0.003}_{-0.003}$\\
\hline
\end{tabular}

Columns: (1) galaxy name, (2--4) total, thermal, and nonthermal fluxes at 1.4\,GHz, (5) thermal fraction at 1.4 GHz 
\end{table}

\begin{table*}
\scriptsize
\centering
\caption{CIGALE SED fit results.\label{tab:sedfitresults}}
\begin{tabular}{cccccccccc}
\hline
Name & $\log_{10}(M_{\star})$  & $\log_{10}(M_{\rm dust})$ & $T_{\rm dust}$  & $\log_{10}(\rm SFR_{\it IR})$ & AGN fraction & $\alpha_{\rm synch}$ & $\log_{10}(L_{\rm dust}$) & $q_{IR}$  & $\chi^2$\\

 & ($M_\odot$) & ($M_\odot$) & (K) & ($M_\odot\,\rm yr^{-1}$) & (\%) & & ($L_\odot$)   & \\

(1)  & (2)  & (3)  & (4) &  (5) & (6) & (7) & (8) & (9)  \\
\hline
ESO 500-G034  & 10.12 $\pm$ 0.07 & 7.37 $\pm$ 0.04 & 27.55 $\pm$ 0.42 & 0.72 $\pm$ 0.14 & 1.11 $\pm$ 0.4    & $-$0.90$\pm$0.04 & 11.86 $\pm$ 0.02 & 2.30 $\pm$ 0.01 & 2.1 \\
NGC 3508      & 10.32 $\pm$ 0.02 & 7.68 $\pm$ 0.02   & 24.16 $\pm$ 0.02  & 0.59 $\pm$ 0.02  & 6.00 $\pm$ 0.001 & $-$0.61 $\pm$0.03 & 11.25 $\pm$ 0.02  & 2.43 $\pm$ 0.04 & 3.4\\
ESO 440-IG058 & 9.51 $\pm$ 0.08 & 7.82 $\pm$ 0.02  & 27.24 $\pm$ 0.002 & 1.63 $\pm$ 0.02  &4.53 $\pm$ 0.02    & $-$1.20$\pm$0.09 & 10.99 $\pm$ 0.02 & 2.60 $\pm$ 0.05 & 2.1\\
ESO 507-G070  & 10.79 $\pm$ 0.02 & 7.61 $\pm$ 0.02   & 30.70 $\pm$ 0.003 & 1.35 $\pm$ 0.02 & 2.00 $\pm$ 0.0004  & $-$0.74$\pm$0.08 & 10.93 $\pm$ 0.02 & 2.67 $\pm$ 0.05 & 4.4\\
NGC 5135      & 10.34 $\pm$ 0.03 & 7.89 $\pm$ 0.08   & 25.77 $\pm$ 1.52 & 1.05 $\pm$ 0.02 & 6.00 $\pm$ 0.04  & $-$0.76$\pm$0.09 & 11.21 $\pm$ 0.02 & 2.50 $\pm$ 0.03 & 4.3\\
IC 4280       & 10.44 $\pm$ 0.06 & 7.66 $\pm$ 0.04 & 27.18 $\pm$ 0.50 & 1.07 $\pm$ 0.02 &  6.10 $\pm$ 0.60   & $-$0.96$\pm$0.15 & 11.36 $\pm$ 0.02 & 2.30 $\pm$ 0.02 & 5.0\\
NGC 6000      & 10.32 $\pm$ 0.05 & 7.39 $\pm$ 0.05   & 28.91 $\pm$ 0.50 & 0.31 $\pm$ 0.12 & 1.06 $\pm$ 0.41   & $-$0.90$\pm$0.03 & 10.79 $\pm$ 0.02 &2.50 $\pm$ 0.02 & 7.6\\
IR 16164-0746 & 10.12 $\pm$ 0.06 & 7.82 $\pm$ 0.02   & 29.21 $\pm$ 0.003 & 1.52 $\pm$ 0.05 & 2.24 $\pm$ 0.01 & $-$0.80$\pm$0.10 & 10.77 $\pm$ 0.02 &2.45 $\pm$ 0.06 & 2.8 \\
ESO 453-G005  & 9.62 $\pm$ 0.10 & 7.69 $\pm$ 0.02    & 27.24 $\pm$ 0.05 & 0.52 $\pm$ 0.20 & 4.72 $\pm$ 0.88   &  $-$0.82 $\pm$ 0.04 & 10.97 $\pm$ 0.02 & 2.50 $\pm$ 0.02 & 4.9\\
IR 18293-3413 & 10.67 $\pm$ 0.06 & 8.06 $\pm$  0.02  & 30.70 $\pm$ 0.06 & 1.19 $\pm$ 0.32 & 2.55 $\pm$ 1.16  & $-$1.19 $\pm$0.09 & 11.72 $\pm$ 0.02 & 2.60 $\pm$ 0.004 & 4.5 \\
ESO 593-IG008 & 10.91 $\pm$ 0.19 & 8.30 $\pm$ 0.03   & 29.55 $\pm$ 0.40 & 1.71 $\pm$ 0.16 & 3.93 $\pm$ 2.7   & $-$0.73 $\pm$0.68  & 11.18 $\pm$ 0.02 & 2.40 $\pm$ 0.001 & 1.7\\
\hline

\end{tabular}

Columns: (1) galaxy name, (2) stellar mass, (3) dust mass, (4) dust temperature, (5) instantaneous star formation rate, (6) AGN fraction, (7) slope of power-law synchrotron emission, $\alpha$, (8) dust luminosity, (9) $q_{IR}$, (10) reduced $\chi^2$ value for the best-fit model. 
\end{table*}

\begin{figure}
    \includegraphics[width =  0.45\textwidth]{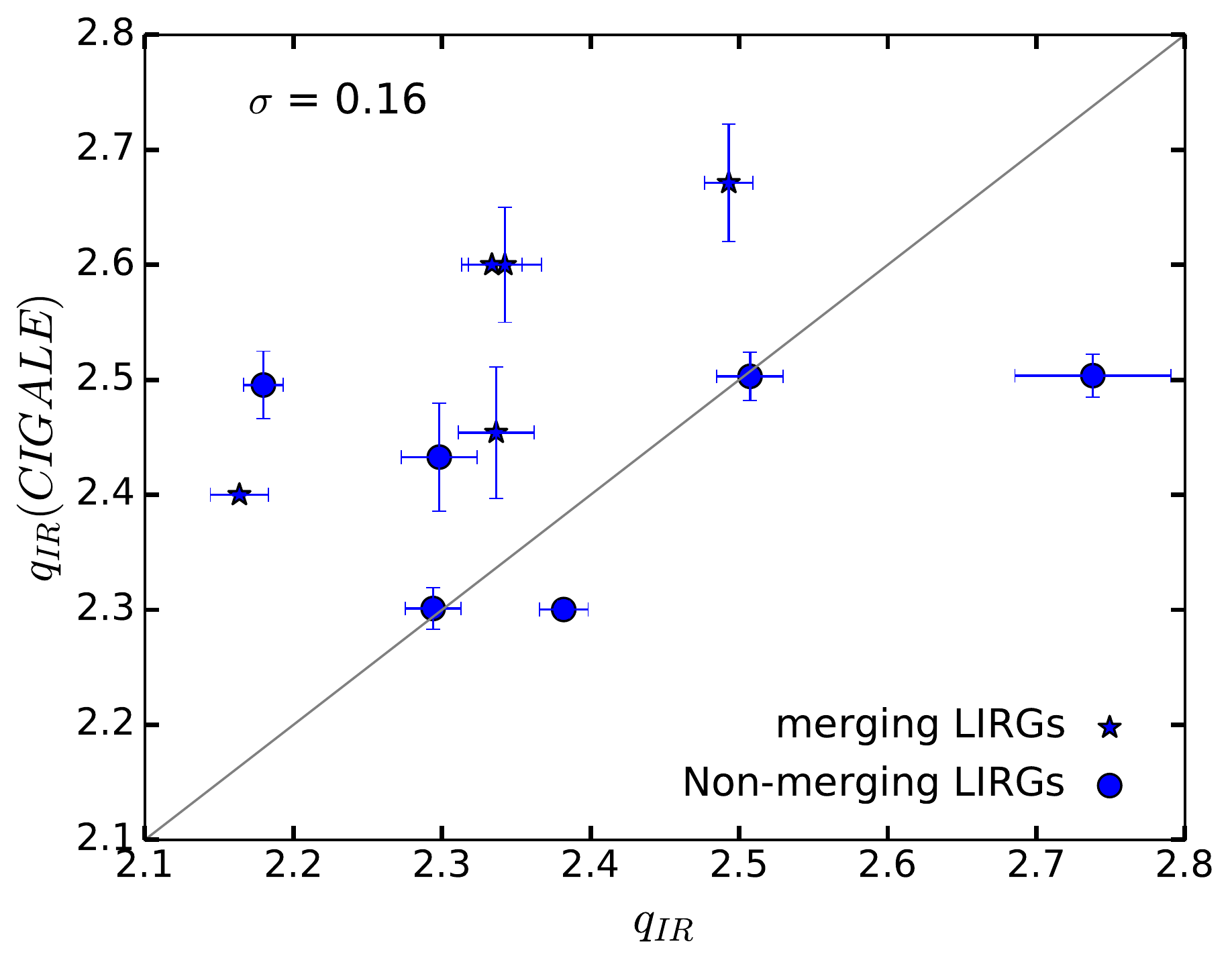}
    \caption{Comparison of $q_{IR}$ estimated by CIGALE modeling and  Eq.~\ref{eq:qir}.  The grey line represents the one-to-one relation and the scatter, $\sigma$, from unity is given. }\label{fig:qir}
\end{figure}

\begin{figure}
    \includegraphics[width =  0.45\textwidth]{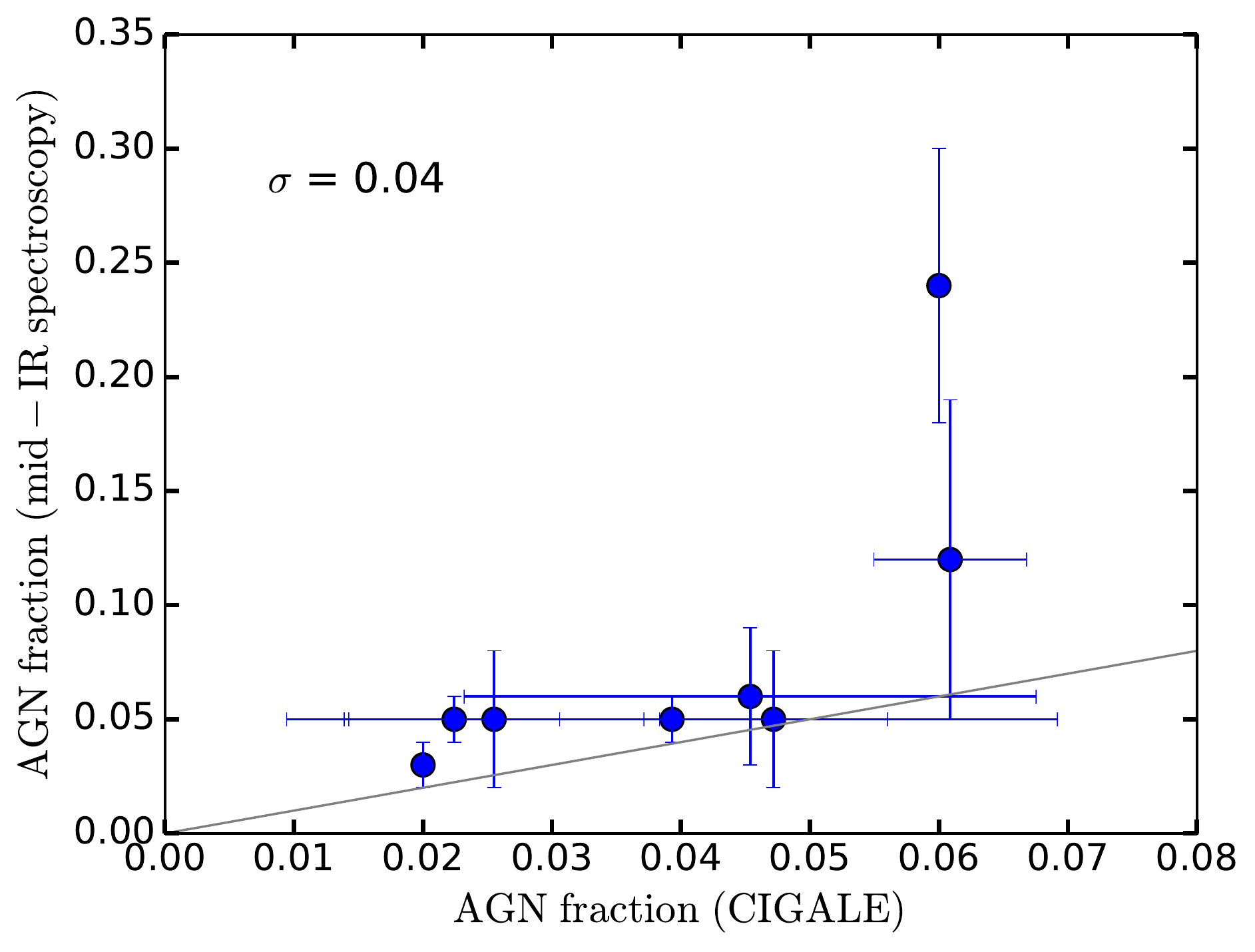}
    \caption{Comparison of bolometric AGN fractions obtained in this work and \citet[][]{Diaz-Santos17} for 8 LIRGs in our sample. The grey line represents the one-to-one relation and the scatter, $\sigma$, from unity is given. }\label{fig:agnfrac}
\end{figure}

\begin{table}
\centering
\caption{Emission Measure (EM) is derived from the model most supported by the evidence value for each source. Sources without constrained EM are marked by `--'\label{tab:em}}
\begin{tabular}{ccc}
\hline
Name & EM$_1$ & EM$_2$  \\
  & ($10^6$\,cm$^{-6}$\,pc) & ($10^6$\,cm$^{-6}$\,pc) \\
\hline
ESO 500-G034 & $-$ & 0.77 \\
NGC 3508 & $-$ & $-$ \\
ESO 440-IG058 & 0.05 & $-$\\
ESO 507-G070 & $-$ & 269.5 \\
NGC 5135 & 0.04 & $-$ \\
IC 4280 & 0.05 & $-$ \\
NGC 6000 & $-$ & $-$ \\
IR 16164-0746 & $-$ & $-$ \\
ESO 453-G005 & $-$ & $-$ \\
IR 18293-3413 & $-$ & 3.44 \\
ESO 593-IG008 & 0.02 & $-$ \\
\hline
\end{tabular}
\end{table}

\section{Results}\label{sec:results}

\subsection{Radio-only and FUV to radio SED modeling}\label{sec:resultmodel}

The densely sampled radio-only and panchromatic (FUV to radio) SEDs for our sample are fitted with UltraNest and CIGALE modeling tools, respectively. Figures~\ref{fig:radiosed} and ~\ref{fig:cigalefitting} present radio-only and FUV to radio spectral fits, respectively. ~\ref{tab:radiosed_pars} provides Bayesian estimates of the model parameters while and Table~\ref{tab:sedfitresults} provides frequentist estimates of the model parameters. Our main results are as follows:

\begin{enumerate}

 \item{The radio continuum SEDs of our galaxies covering $\sim$70\,MHz-$\sim$15\,GHz frequency range show a variety of features: (1) single PL model arising from synchrotron emission without thermal contribution, (2) single-component PL models due to synchrotron and free-free emission with and without turnover at low frequencies, and (3) two-component models with distinct emission regions, optical depths, and sometimes different spectral index of the CRe (Figure~\ref{fig:radiosed}). The derived radio synchrotron spectral index ranges between $-$0.45$_{-0.03}^{+0.03}$ and $-$1.74$_{-0.08}^{+0.04}$. The turnover frequencies for SEDs fitted with models with absorption features fall in the range between 100$_{-20}^{+20}$\,MHz and 7.55$_{-1.97}^{+1.94}$\,GHz (Table~\ref{tab:radiosed_pars}).}

\item {Figure~\ref{fig:cigalefitting} provides the CIGALE fit for our LIRGs.  CIGALE-modeling of the galaxies provides the rest-frame infrared luminosity in the wavelength range of 8-1000\,$\mu$m (log$_{10}L_{IR(8-1000\mu\,\rm m)}$) as 10.77$\pm$0.02 -- 11.86$\pm$0.02 $L_{\odot}$. The $q_{IR}$ parameter is $\sim$2.30$\pm$0.02 -- $\sim$2.67$\pm$0.05 while the radio spectral index estimated between 1.4\,GHz and 5\,GHz is in the range of $\sim -$0.61$\pm$0.03 -- $-$1.20$\pm$0.09. The range of estimated stellar mass, dust mass, and dust temperature are $\log_{10}M_\star$ =  9.51$\pm$0.08 -- 10.91$\pm$ 0.19\,$M_\odot$, $\log_{10}M_{\rm dust}$ = 7.37$\pm$0.04 -- 8.30$\pm$0.03 \,$M_\odot$, and $T_{\rm dust}$ = 24.16$\pm$0.02 -- 30.70$\pm$0.06\,K, respectively. The star formation rate $\log_{10}$SFR$_{IR}$ is 0.31$\pm$0.12 -- 1.71$\pm$0.16 ($M_\odot\,\rm yr^{-1}$). The AGN fraction that contributes to the optical and $IR$ emission ranges between 1.06$\pm$0.41\% and 6.10$\pm$0.60\% (Table~\ref{tab:sedfitresults}).}

\item{Including the radio measurements in CIGALE modeling allowed us to obtain better constrained $L_{\rm dust}$ values as compared to CIGALE modeling results in the literature which did not include radio measurements. Although the dust luminosity of our galaxies falls in the range exhibited by galaxies of the same class, we note that our $L_{\rm dust}$ estimates have $\sim$1 order of magnitude lower uncertainties (0.02\,dex, Table~\ref{tab:sedfitresults}) than those provided in the literature \citep[$\sim$0.2\,dex from][]{Malek2017}.}

\item{Figure~\ref{fig:qir} shows the comparison between the $q_{IR}$ estimated from the CIGALE modeling and Eq.~\ref{eq:qir}. The relation between these values is closer to unity  and agree within errors, with a scatter (or average deviation), $\sigma$ = 0.16.}

\item{To cross-check the robustness of the CIGALE analysis, we present a comparison of the AGN fractions obtained for eight galaxies in our sample with the AGN fractions derived by \citet[][]{Diaz-Santos17} in Figure~\ref{fig:agnfrac}. CIGALE computes the AGN fraction using an AGN template that contributes between $UV$ and $IR$ wavelengths. In contrast, the \citet[][]{Diaz-Santos17} computes the AGN fraction using the mid-$IR$ (MIR) spectral diagnostics, including emission-line ratios, the equivalent width of 6.2\,$\mu$m polycyclic aromatic hydrocarbon (PAH), the dust continuum slope at 30 and 15 $\mu$m as well as spectral template-based diagrams. The relation between the these values is closer to unity and they agree within error, with a scatter, $\sigma$=0.04. A close correspondence between our and \citet[][]{Diaz-Santos17} values validates the analysis procedure.}

\end{enumerate}

\renewcommand{\thefigure}{\arabic{figure}}

\begin{figure}
    \includegraphics[width =  0.5\textwidth]{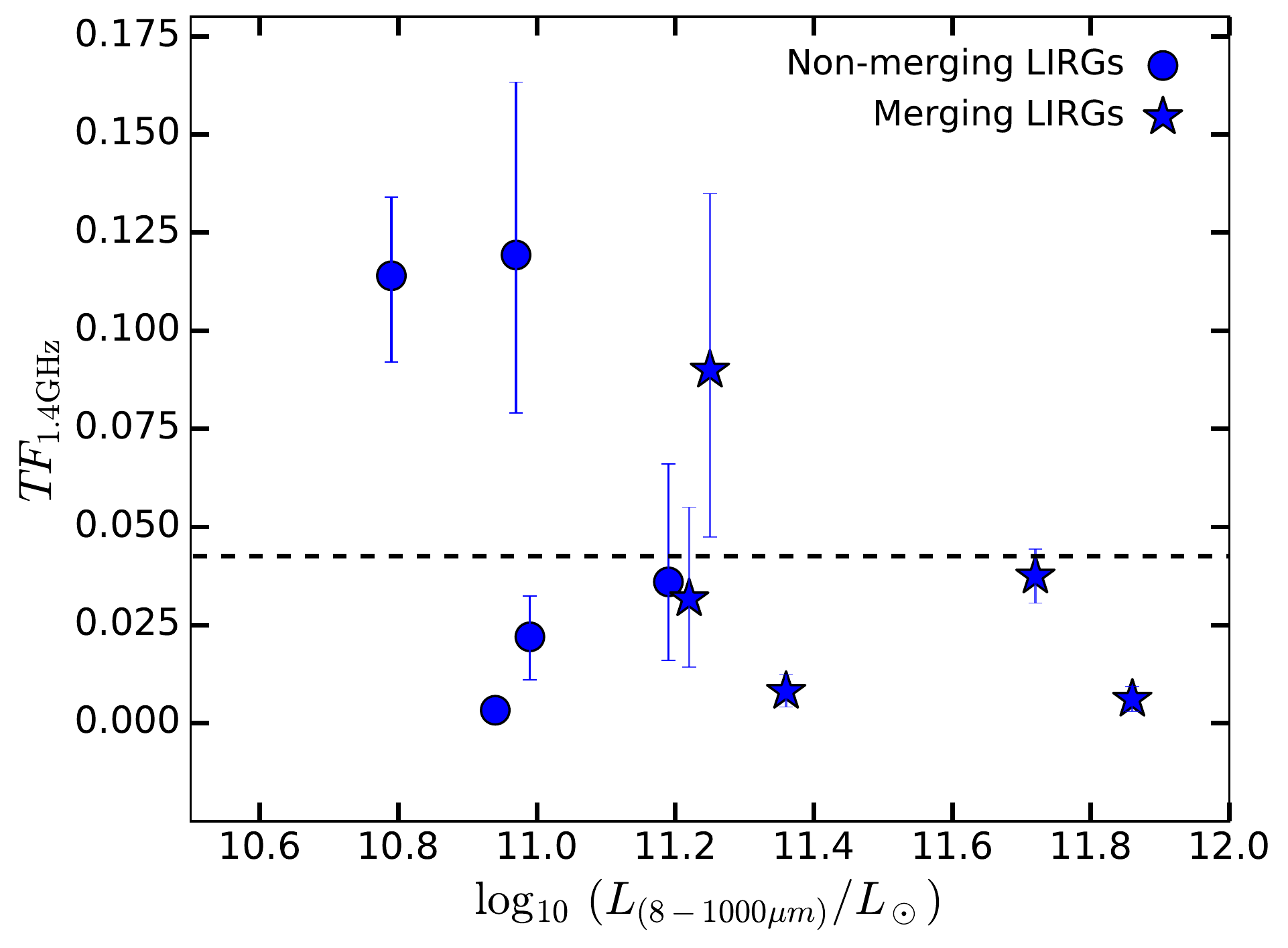}
    \caption{Thermal fraction ($TF$) at 1.4\,GHz plotted against the rest-frame infrared luminosity obtained from our CIGALE modeling. The dashed line marks the average thermal fraction = 0.042 for our sample.}\label{fig:tf}
\end{figure} 

\begin{figure}
    \includegraphics[width =  0.5\textwidth]{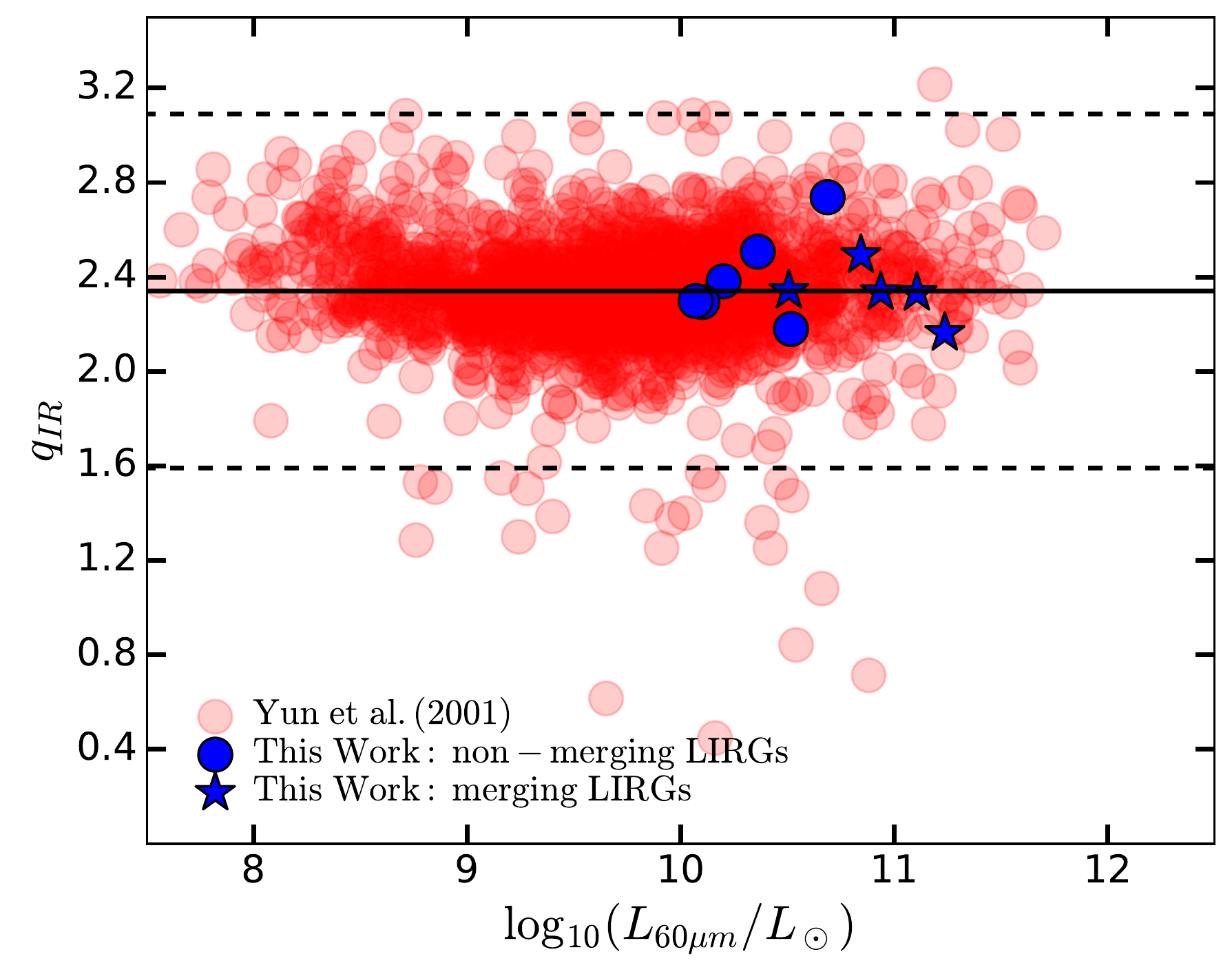}
    \caption{Distribution of $q_{IR}$ with 60\,$\mu$m infrared luminosity for our sample of LIRGs. The solid horizontal line marks the mean $q_{IR}$ =  2.34 and the dashed lines represent upper and lower 3$\sigma$ bounds \citep[][]{Yun01}.}\label{fig:lirqir}
\end{figure} 

\begin{figure}
    \includegraphics[width =  0.5\textwidth]{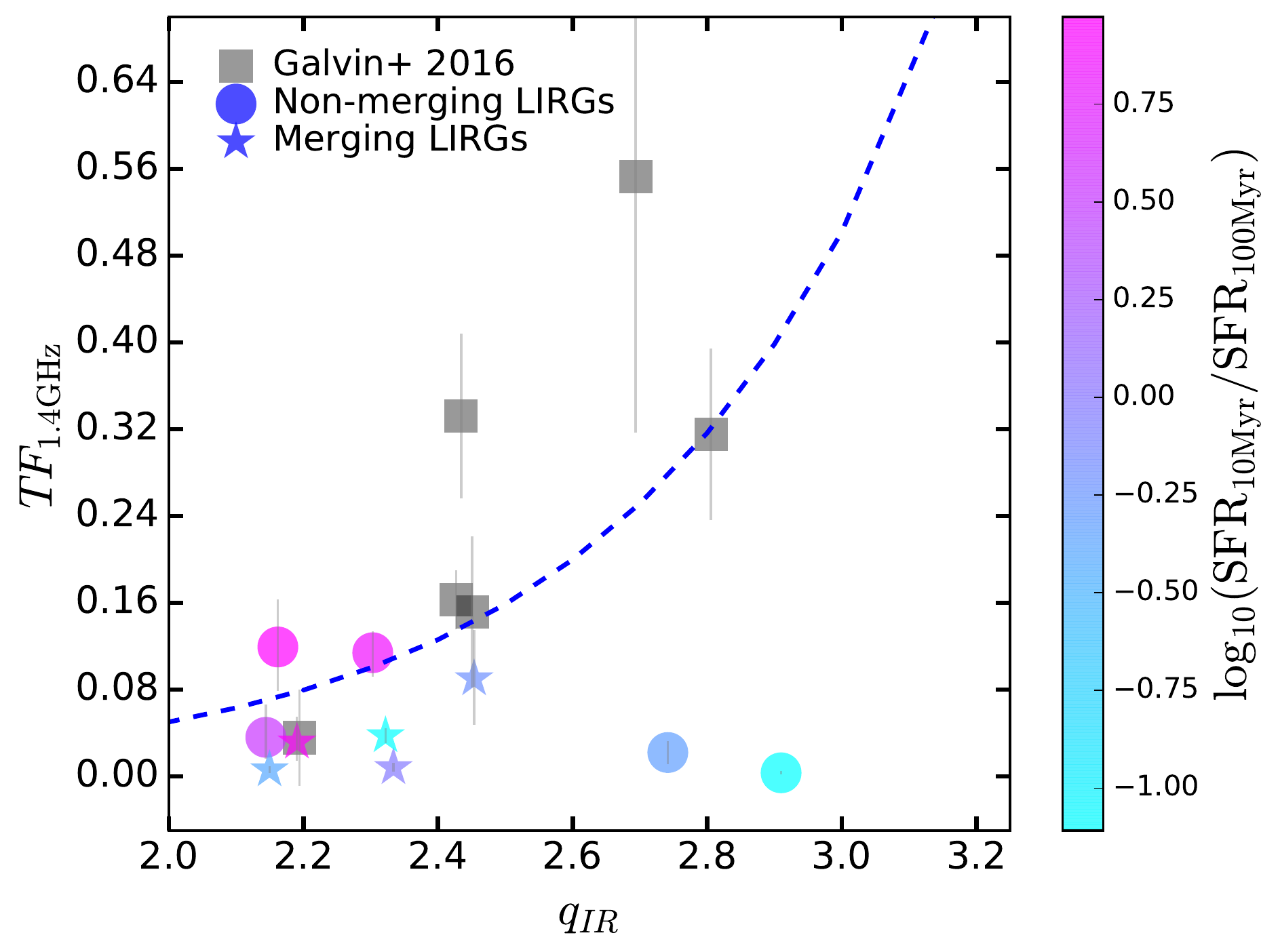}
    \caption{Distribution of thermal fraction with $q_{IR}$ (Eq.~\ref{eq:qir}) for our sample of LIRGs. The color bar indicates the SFR$_{IR}$ obtained for a time interval of 10\,Myr ago from the CIGALE analysis. The dashed line marks the relation between thermal fraction and $q_{IR}$ given by $TF =  $\,$1.7\times$10$^{q_{IR}-3.53}$ \citep[][]{Marvil15}.}\label{fig:tfsfh}
\end{figure}

\begin{figure}
    \centering
    \hspace{0cm}{\includegraphics[width =  0.50\textwidth]{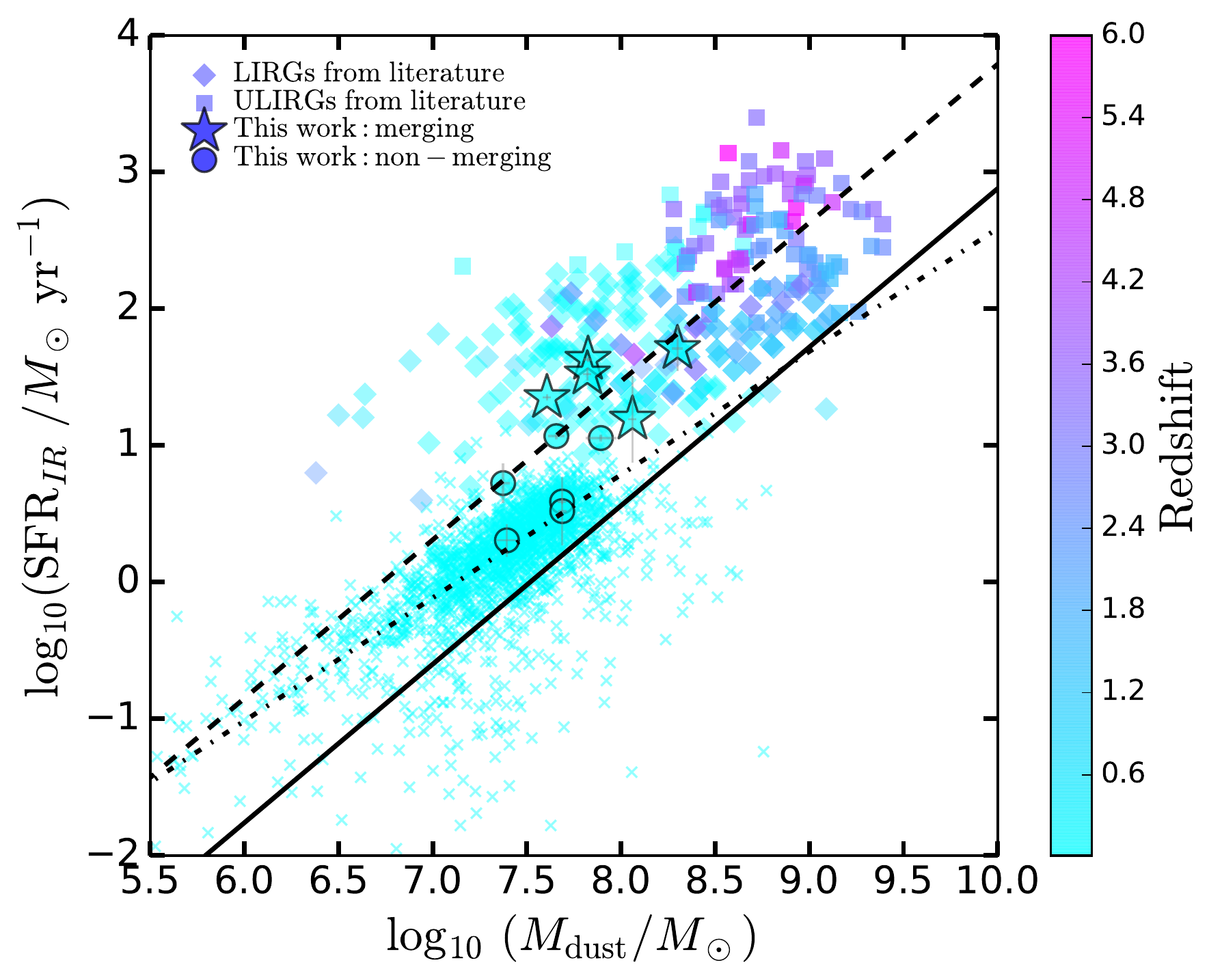}}
\caption{SFR$_{IR}$ vs M$_{\rm dust}$ for our sample of LIRGs obtained from CIGALE modeling. The solid and dashed lines, respectively, are the relations derived from low$-z$ dusty galaxies (H-ATLAS) and starburst galaxies (local ULIRGs and $z>$ 2 submillimeter galaxies) by \citet{Rowlands2014}. The dot-dashed line represents the best-fit line to the SDSS-IRAS selected local star-forming galaxies of infrared luminosities (10$\leq$\rm log$_{10}$($L_{\rm dust}$/ $L_{\odot}$ $<$ 12) given by \citet{Dacunha2010b}\label{fig:sfrmdust} }

\end{figure}    
    
\begin{figure}
    \centering
    \hspace{0cm}{\includegraphics[width =  0.50\textwidth]{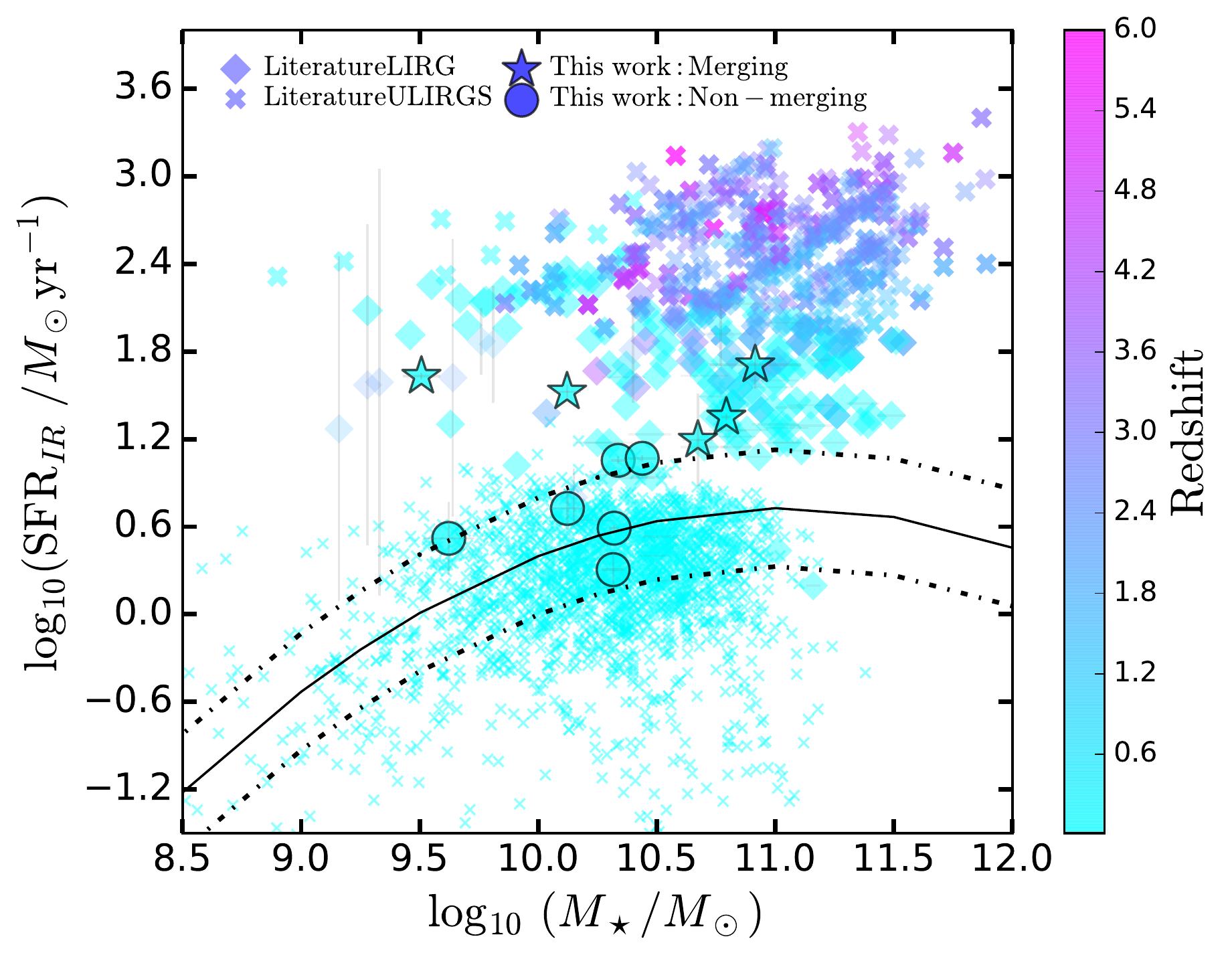}}
\caption{SFR$_{IR}$ vs M$_{\star}$   for our sample of LIRGs obtained from CIGALE modeling. The solid line represents the relation between M$_{\star}$- SFR$_{IR}$ for the main-sequence galaxies given by \citet{Schreiber15} for a redshift range of 0.02. The dotted and dash-dotted lines show the range of the main sequence spreading $\pm$0.4 dex above and below the solid line.} \label{fig:stellarmasssfr}

\end{figure}

\begin{figure}
    \centering
    \hspace{0cm}{\includegraphics[width =  0.50\textwidth]{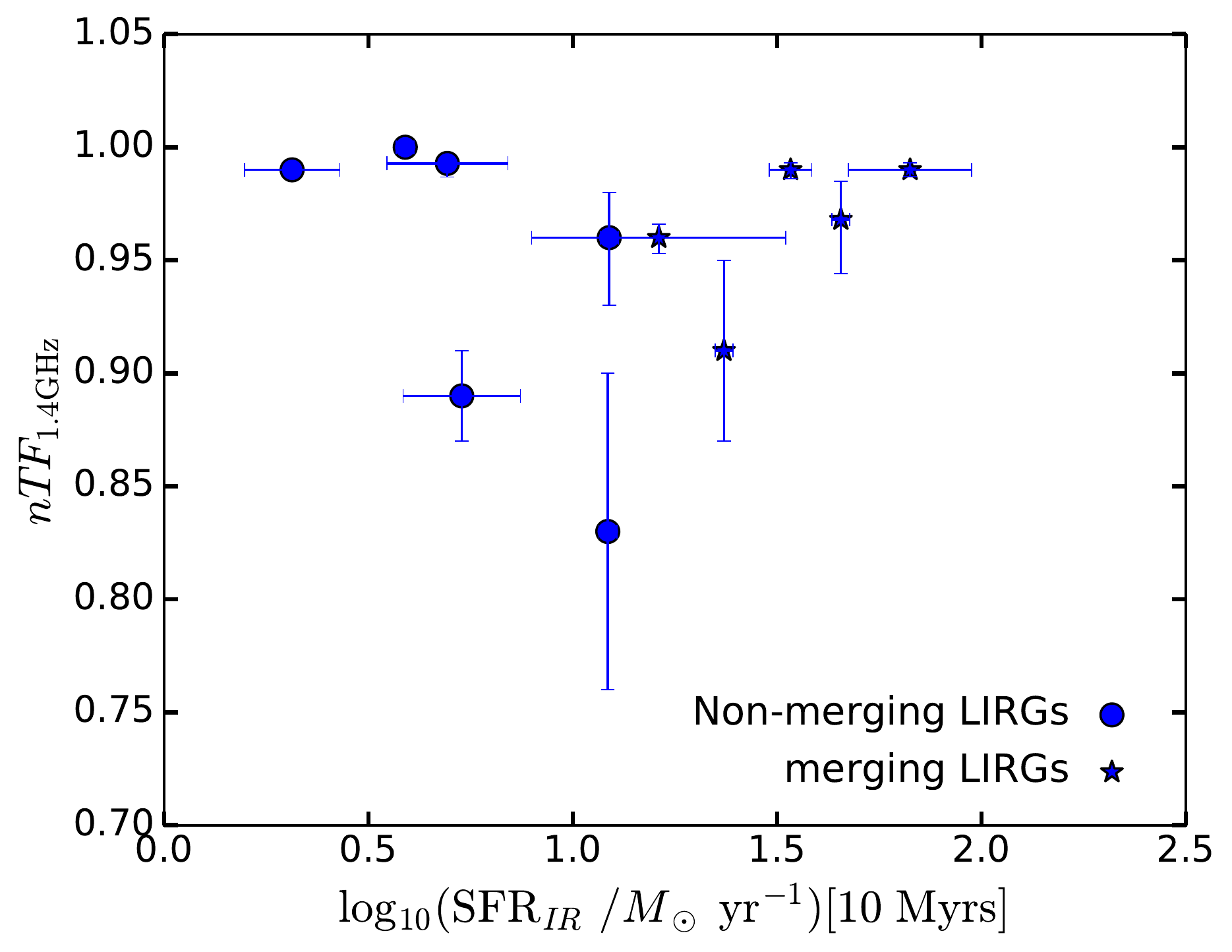}}
\caption{1.4\,GHz non-thermal fraction, $nTF$, as a function of SFR$_{IR}$ for a time interval of 10\,Myr. The nonthermal fraction is estimated in the same manner as that of the thermal fraction (see Section ~\ref{sec:tf}).} \label{fig:ntf}

\end{figure}

\subsection{Specific results on individual galaxies}\label{sec:resultgal}

\begin{enumerate}

\item{{\bf ESO\,500-G034}: classified as an intermediate between the Seyfert type AGN nucleus and a starburst galaxy \citep[][]{Hill99}, the densely sampled radio spectrum covering $\sim$2.3 decades in frequency (70\,MHz-14.9\,GHz) is fitted with a two-component model representing two distinct star-forming regions where the first component is without turnover. The turnover frequency is $\sim$0.52\,GHz. The synchrotron spectral index, $\alpha$ = $-$1.15. The $TF$ at 1.4\,GHz is $\sim$11.4\%. The CIGALE modeling estimates a %high 
SFR$_{IR}$ = 5.72\,$M_\odot$\,yr$^{-1}$, high $L_{\rm dust}$ = 10$^{11.86}$\,$L_\odot$, and low AGN fraction = 1.11\%, respectively. The low AGN fraction obtained by our analysis contradicts the previous results by \citet{Hill99} who classified this source as an intermediate between an AGN and a starburst galaxy \citep[its low is AGN fraction is also confirmed by][]{Diaz-Santos17}.}

\item{{\bf NGC\,3508}: classified as a spiral galaxy \citep[][]{Veilleux95}, its densely sampled radio spectrum covering 2.9 decades in frequency (70\,MHz-3\,GHz) is fitted with a single component model characterized by synchrotron emission only. The synchrotron spectral index, $\alpha$ = $-$0.73. The $TF$ is 0 as this emission is characterized by synchrotron emission (i.e. no free-free component). The CIGALE modeling gives SFR$_{IR}$ = 3.90\,$M_\odot$\,yr$^{-1}$, $L_{\rm dust}$ =  10$^{11.25}$\,$L_\odot$, and AGN fraction = 6.00\%, respectively, for this galaxy.}

\item{{\bf ESO\,440-IG058}: classified as a galaxy merger with a LINER-type AGN northern neighbor \citep[][]{Corbett03}, this galaxy is known to show shock-dominated emission for the southern system \citep[][]{Zaurin11}. Its radio SED covering 2.1 decades (70\,MHz-10.0\,GHz) is fitted with single-component star-forming region properties (including absorption) with a turnover at $\sim$140\,MHz. Synchrotron spectral index, $\alpha$ = $-$0.82. The computed $TF$ is 3.17\%. The CIGALE-estimated SFR$_{IR}$, $L_{\rm dust}$, AGN fraction are 42.66\,$M_\odot$\,yr$^{-1}$, 10$^{10.99}$\,$L_\odot$, $\sim$4.53\%, respectively. Our SFR$_{IR}$ value is comparable within uncertainty to that obtained in literature \citep[$\sim$36$\,M_\odot$\,yr$^{-1}$;][]{Miluzio13} and \citep[$\sim$48$\,M_\odot$\,yr$^{-1}$;][]{Illana17}. 
%Additionally, the companion galaxy at the north-west does not seem to contribute to radio frequencies.
}

\item{{\bf ESO\,507-G070}: classified as a Seyfert-type AGN nucleus \citep[][]{Condon96} in a post-merger system \citep[][]{Stierwalt13, Pasparialis2021}, its radio SED covering $\sim$2.3\,decades in frequency (70\,MHz-14.9\,GHz) is best-fitted with a two-component model representing two distinct star-forming regions with different ultrarelativistic electron populations and high frequency turnover. The synchrotron spectral indices for the two populations, $\alpha$ and $\alpha_2$, are $-$0.77 and $-$1.50, respectively. The $TF$ is $\sim$9\%. The SFR$_{IR}$, $L_{\rm dust}$, AGN fraction are 22.40\,$M_\odot$\,yr$^{-1}$, 10$^{10.93}$\,$L_\odot$, $\sim$2.0\%, respectively. Its UV-$IR$ SED has previously been modeled with CIAGLE by \citet[][]{Pasparialis2021} who reported comparable $L_{\rm dust}$ (=10$^{11.1}$\,$L_\odot$), and SFR$_{IR}$ (45.1\,$M_\odot$\,yr$^{-1}$), and lower AGN fraction (0\%) as compared to our results. 
}

\item{{\bf NGC\,5135}: classified as having a Seyfert-type AGN nucleus \citep[][]{Condon96} with strong starburst along the spiral arms \citep[][]{Marin07}, its radio SED covering $\sim$2.0\,decades in frequency (70\,MHz-6.7\,GHz) is fitted with single-component star-forming region properties (including absorption) with a turnover at $\sim$0.130\,GHz. The synchrotron spectral index, $\alpha$ = $-$0.88. The $TF$ is $\sim$3.6\%. The CIGALE-estimated SFR$_{IR}$, $L_{\rm dust}$, AGN fraction are 11.22\,$M_\odot$\,yr$^{-1}$, 10\,$^{11.21}$\,$L_\odot$, $\sim$6\%, respectively. 

}

\item{{\bf IC\,4280}: classified as a spiral galaxy \citep[][]{Fairall89, Jin19}, its radio SED covering $\sim$1.6\,decades (70\,MHz-3.0\,GHz) is best-fitted with a single-component star-forming region properties (including absorption) with a turnover at $\sim$140\,MHz. The synchrotron spectral index, $\alpha$ =  $-$0.9. The $TF$ is 11.9\%. The CIGALE-estimated SFR, $L_{\rm dust}$, AGN fraction are 11.75\,$M_\odot$\,yr$^{-1}$, 10$^{11.36}$\,$L_\odot$, $\sim$6\%, respectively. 

}

\item{{\bf NGC\,6000}: classified as a starburst galaxy \citep[][]{Carollo02}, its radio SED covering $\sim$1.6\,decades in frequency (70\,MHz-3.0\,GHz) is best-fitted with emission from a single star-forming region without a low frequency turnover. The synchrotron spectral index is $-$0.66. The $TF$ is 0.3\%. The CIGALE-estimated SFR, $L_{\rm dust}$, and AGN fraction are 2.04\,$M_\odot$\,yr$^{-1}$, 10\,$^{10.79}$\,$L_\odot$, $\sim$1.06\%, respectively.}

\item{{\bf IR\,16164-0746}: classified as a late stage merger \citep[][]{Stierwalt13} with a single LINER-type nucleus and a tidal-tail \citep[][]{Dixon11, Hung15}, its radio SED covering $\sim$1.6\,decades in frequency (70\,MHz-3.0\,GHz) is best-fitted with emission from a single star-forming region without low frequency turnover. The synchrotron spectral index is $\sim$-0.5. The $TF$ is 0.8\%. The CIGALE-estimated SFR, $L_{\rm dust}$, AGN fraction are 33.11\,$M_\odot$\,yr$^{-1}$, 10\,$^{10.77}$\,$L_\odot$, $\sim$2.2\%, respectively. }

\item{{\bf ESO\,453-G005}: a galaxy pair that appears to be dominated by star-forming activity, although it does not show signs of interaction \citep[][]{Rich15}, its poorly sampled radio SED covering $\sim$1.3\,decades in frequency (150\,MHz-3.0\,GHz) is best-fitted with emission from a single star-forming region without a low-frequency turnover. The synchrotron spectral index is $\sim$-0.6. The $TF$ is 2\%. The CIGALE-estimated SFR, $L_{\rm dust}$, AGN fraction are 3.31\,$M_\odot$\,yr$^{-1}$, 10\,$^{10.97}$\,$L_\odot$, $\sim$4.7\%, respectively.  }

\item{{\bf IR\,18293-3413}: a complex system, classified both as early-merger \citep[][]{Haan11} and mid-merger \citep[][]{Ricci17} galaxy system with no evidence of AGN activity \citep[X-ray spectrum;][]{Risaliti2000}, its radio SED covering $\sim$2.1\,decades in frequency (70\,MHz-8.4\,GHz) is best-fitted with two-component model representing two distinct star-forming regions with different ultrarelativistic CRe and one region without any turnover. The synchrotron spectral indices for the two populations are, $-$1.33 and $-$1.74, respectively. The turnover frequency is 1.06\,GHz. The $TF$ is $\sim$3.7\%. The CIGALE-estimated SFR, $L_{\rm dust}$, AGN fraction are 15.48\,$M_\odot$\,yr$^{-1}$, 10\,$^{11.7}$\,$L_\odot$, $\sim$2.5\%, respectively. 
}

\item{{\bf ESO\,593-IG008}: known as the `Bird' \citep[][]{Vaisanen17}, this system is a a well-known galaxy merger \citep[][]{Stierwalt13, Hung15}. Its radio SED covering 1.8 decades (70\,MHz-4.8\,GHz) is fitted with single-component star-forming region synchrotron (including absorption) with a turnover at $\sim$100\,MHz. Synchrotron spectral index, $\alpha$ =  $-$0.87. The computed $TF$ is 0.6\%. The CIGALE-estimated SFR, $L_{\rm dust}$, AGN fraction are 51.3\,$M_\odot$\,yr$^{-1}$, 10\,$^{11.18}$\,$L_\odot$, $\sim$3.9\%, respectively. 

}

\end{enumerate}

\begin{figure*}
 \includegraphics[width =  1\textwidth]{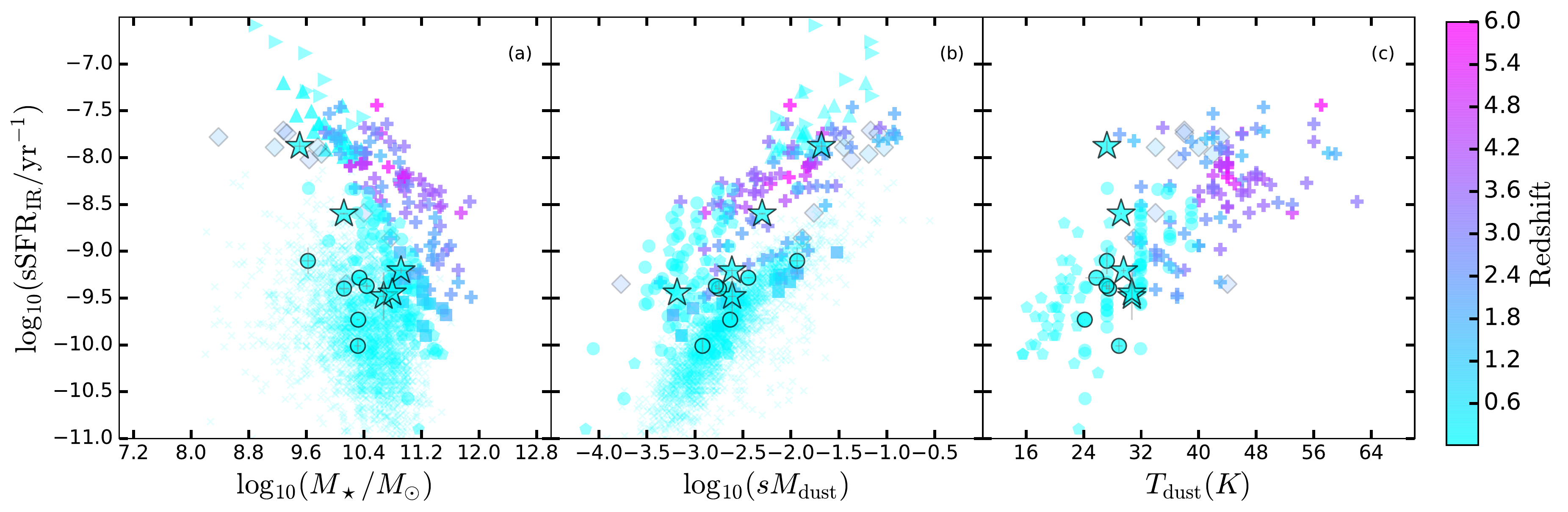}
 \caption{Distribution of sSFR$_{IR}$ with respect to stellar mass (panel a), specific dust mass (panel b) and dust temperature (panel c) for the 11 LIRGs studied here (star and \textbf{circle} symbols represent the merging and non-merging types). For comparison, we include data from literature for normal star forming galaxies \citep[cross;][]{Dacunha2010b}, LIRGs \citep[upward facing triangle;][]{Vega08}, \citep[circle;][]{Pasparialis2021}, \citep[diamond;][]{Dacunha2015}, \citep[square;][]{LoFaro2013}, and ULIRGs, \citep[right side facing triangle;][]{Vega08}, \citep[plus;][]{Dacunha2015}.}\label{fig:ssfr}

\end{figure*}

\begin{figure*}
 \includegraphics[width =  1\textwidth]{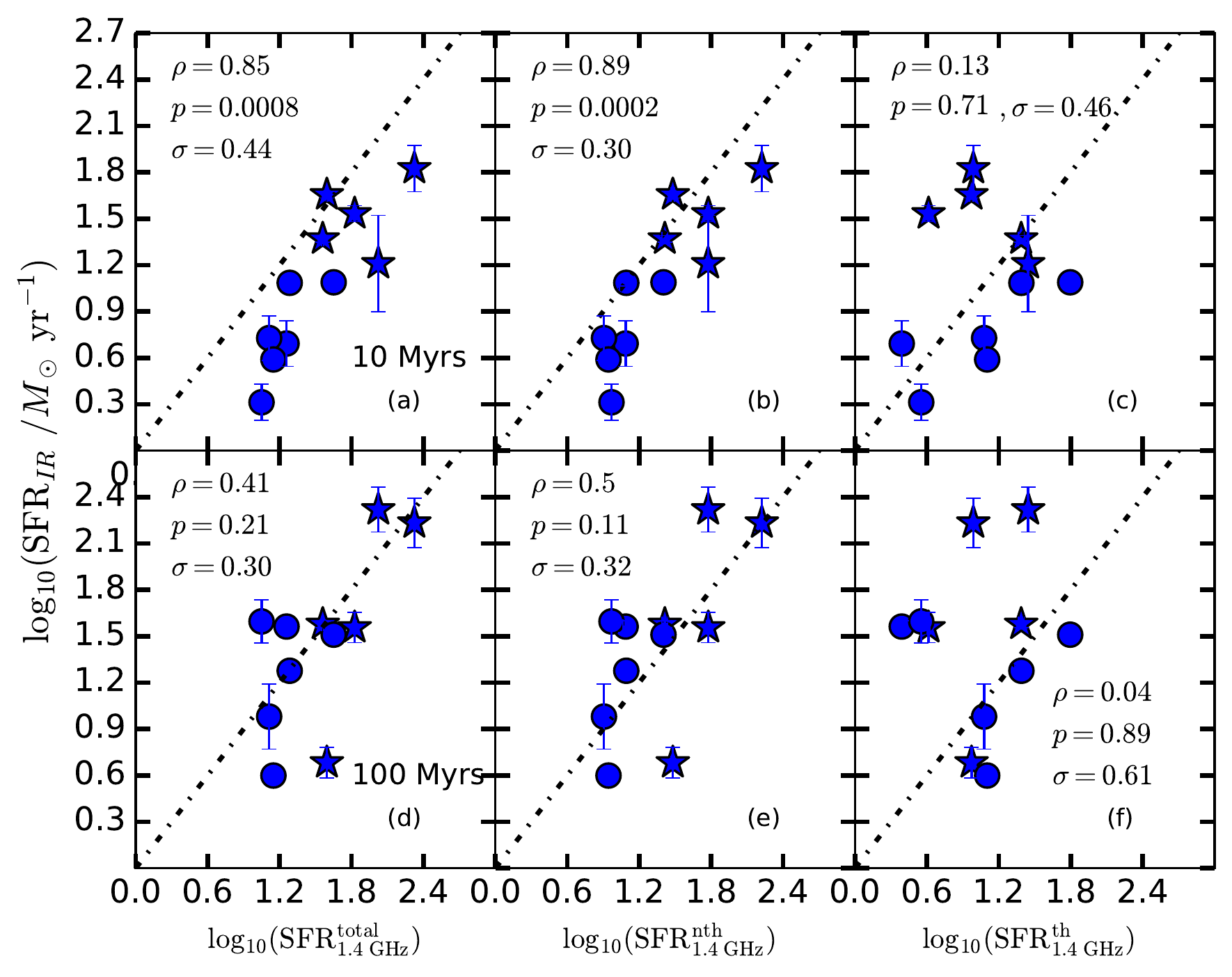}
 \caption{Comparison of the SFR$_{IR}$ obtained for two time intervals 10 and 100\,Myr ago from the CIGALE SED fitting and the 1.4\,GHz radio SFR (total, nonthermal, and thermal, respectively). The mergers and non-mergers are shown with star and circle symbols, respectively. $p-$ and $\rho-$ values are obtained from the application of the Spearman's rank correlation test while the scatter, $\sigma$, is the average deviation from the unity. }\label{fig:sfrir_rad}

\end{figure*}

\section{Discussion}\label{sec:discussion}

\subsection{$q_{IR}$ parameter}

%The $q_{IR}$ parameter is a discriminator for the the star-formation rates in galaxies \citep[][]{Bressan02, Murphy09, Ivison10}. 
We estimate $q_{IR}$ using far-infrared fluxes at 60\,$\mu$m and 100\,$\mu$m and 1.4\,GHz radio flux to compare with other samples in literature \citep[][]{Yun01, Galvin2018}. That is, we use the following expression given by \citet[][]{Yun01}:

\begin{equation}
    q_{IR} \equiv \log_{10}\left(\frac{\rm FIR}{3.75\times10^{12}\,{\rm W\,m^{-2}}}\right) -  \log_{10}\left(\frac{S_{\rm 1.4\,GHz}}{\rm W\,m^{-2}\,Hz^{-1}}\right)
    \label{eq:qir}
\end{equation}
The far infrared luminosity, FIR, is computed using ${\rm FIR} \equiv $1.26\,$\times$10$^{-14}$\,(2.58\,$S_{60\mu\rm m}$ + $S_{100\mu\rm m}$) W\,m$^{-2}$ where $S_{60\mu\rm m}$ and $S_{100\mu\rm m}$ are the 60\, and 100\,$\mu\rm m$ band flux densities, in Jy, from IRAS. The error on q$_{IR}$ parameter, computed using Eq.~\ref{eq:qir}, is derived following standard error propagation. Figure~\ref{fig:lirqir} shows the distribution of log$_{10}$\,$q_{IR}$ obtained based on Eq.~\ref{eq:qir} with respect to monochromatic infrared luminosity at 60\,$\mu\rm m$ ($=$4\,$\pi$\,$d_L^2$\,$\times\, S_{60\mu\rm m}$ where $d_L$ is the luminosity distance). To obtain the luminosity distance, we use the cosmological calculator\footnote{\url{http://www.astro.ucla.edu/~wright/CosmoCalc.html}} \citep[][]{Wright06} with the Hubble constant $ H_0 =  $ 69.6\,km\,s$^{-1}$\,Mpc$^{-1}$, $\Omega_{\rm M}  =  $ 0.286, and $\Omega_\lambda  =  $ 0.714 \citep{Bennett14}.  In this plot, the upper and lower dashed lines represent the 3$\sigma$ bounds of the mean value. The objects falling above and below the dashed lines are called the ``IR excess'' (upper dashed line) and the ``radio excess'' (lower dashed line), respectively. Our sample exhibits typical values of $q_{IR}$, indicating that radio emission is dominated by star-forming processes without any hint of abnormal behavior, such as AGN dominance or unusually high star-forming activity. This supports the results of our $UV$ to $IR$ fitting from CIGALE, where the median AGN fraction is $\sim$2.5\%, and only for one source ESO\,453-G005 it is close to $\sim$10\%. Furthermore, the distribution of $q_{IR}$, with respect to log$_{10}$\,$L_{60\mu\rm m}/(L_\odot)$ gives comparable results to those obtained for other LIRGs by \citet[][]{Yun01} and \citet[][]{Galvin2018}. According to the plot, the merging and non-merging galaxies take values from the main, most frequent range of $q_{IR}$ values and are indistinguishable in this aspect.

\subsection{Thermal fraction}\label{sec:tf}

From the radio--SED modeling, we computed the thermal fraction at 1.4\,GHz, $TF_{\rm 1.4 GHz}$, for our sample which range between 0.003$_{-0.002}^{+0.002}$ and 0.12$_{-0.04}^{+0.04}$ (Table~\ref{tab:rad_flux}). We plot the $TF_{\rm 1.4 GHz}$ against the rest frame infrared luminosity from CIGALE (Figure~\ref{fig:tf}). Our values are similar to those obtained for a different sample of starburst galaxies, including LIRGs and ULIRGs \citep[19 galaxies at median redshift equal to 0.09; ][]{Galvin2018}. These values are also compatible with those obtained from normal star-forming galaxy samples \citep[][]{Niklas97, Tabatabaei17}.

\subsection{Radio spectral indices and spectral curvature}

The integrated radio spectra for our sample of LIRGs show complex forms and are rarely described by single power-laws (Figure~\ref{fig:radiosed}). Eight out of 11 galaxies show bends in their spectra which we model as arising due to free-free absorption. Here we mention that curvatures in radio SEDs can also occur if the refractive index of the medium is less than unity \citep[Tsytovitch-Razin effect;][]{Israel90}, and due to synchrotron self-absorption \citep[SSA; ][]{Israel90}. For our galaxies, the radio surface brightness (=Peak flux/$\theta^2$, where flux is in Jy and $\theta$ is the size in arcsec of the source; for unresolved sources, we take the synthesized beam size as the upper limit for the size) is of the order of 10$^{-2}$ (Table~\ref{tab:gmrtobs}). For the SSA mechanism to be important at 100\,MHz frequencies for our sources, unreasonably high magnetic field strengths $>$1000\,G are needed \citep[][]{Kellermann69}. Therefore, SSA as a cause of turnover can be safely ruled out for our sample. The Razin turnover frequency is given as 20$\times\,n_e$/$B$(MHz), where $n_e$ is the electron density (cm$^{-3}$) and $B$ is the magnetic field strength in $\mu$G \citep[][]{Ginzburg65}. In ISM, for the typical CRe density $\sim$1 (cm$^{-3}$) \citep[][]{Ferriere01} and the typical magnetic field strength of 50$\mu$G \citep[][]{Crocker10}, the Razin turnover frequency turns out to be 0.4\,MHz. Therefore, we can safely rule out the Razin effect as a cause of low-frequency spectral turnover. 

The synchrotron spectral index ranges between $\sim$$-$0.5 and $\sim$$-$1.7. The galaxies IR\,16164$-$0746 and ESO\,453-G005 have relatively flatter spectra, $\sim-$0.5, compared to the canonical value of $-$0.75 \citep[][]{Condon92}. For these galaxies, the spectral flattening can be ascribed to a change in the CRe spectrum (electron energy index =2.1) because of ionization losses \citep[][]{Lacki10, Ramirez-Olivencia22}. Galaxies NGC\,3508, ESO\,440-IG058, NGC\,5135, IC\,4280, NGC\,6000, and ESO\,593-IG080 exhibit typical values of synchrotron spectral slope, ranging between $-$0.7 and $-$0.9. The remaining three galaxies, ESO\,500-G034,  ESO\,507-G070, and IR\,18293-3413, show steeper spectral slopes ranging between $-$1.17 and $-$1.74. Interestingly, galaxies showing stepper synchrotron spectral indices are fitted with two-component emission models. On the assumption that the injection indices of the electron energies are all similar but steepen due to physical processes, we expect the radiation spectral index to change by 0.5 due to synchrotron losses \citep[][]{Marvil15}. Therefore, within uncertainties, our spectral indices are consistent with a typical $\alpha$ $\sim$-0.75 and a value of $\alpha$ $\sim$ -1.25. The slightly flatter values are almost within the errors, which could be due to the complexity of the radio SED. We cannot evaluate it due to the poor spectral resolution achieved in our images. We speculate along the lines of \citet[][]{Galvin2018} who also modeled the radio SED LIRGs that the CRe energy spectrum is intrinsically steep for these galaxies. We remark that, unlike frequent assumptions, the non-thermal spectral index is not fixed, and it changes between $-$0.45 and $-$1.75, which could be due to the influence of star formation on the energetics of CRe as discussed in \citet[][]{Tabatabaei17}.

\subsection{Emission measure}

 The emission measure (EM) is an integral of the electron density along the line of sight. EM is calculated by assuming that the emission originates from a cylindrical geometry with constant temperature and electron density \citep[][]{Condon92}. From the spectral turnover frequency, one can estimate the EM from the formula of the optical depth, $\tau_\nu$: 
 
 \begin{equation}
     \tau_\nu  =   3.28\times\,10^{-7}\left(\frac{T_e}{10^4\,\rm K} \right)\,\left( \frac{\nu}{\rm GHz}\right)^{-2.1}\,\left( \frac{EM}{\rm pc\,cm^{-6}}\right)
 \end{equation}
 
 where $T_e$ is the electron temperature of the H$_\textsc{II}$ emitting region and $EM$ is the emission measure of depth $s$. $\tau_\nu$ is the optical depth, set at unity at turnover frequency measured from our SED modeling. The $EM$ is given by the integral of the electron density, $N_e$, along the line of sight of the $H_\textsc{II}$ region of depth $s$:
 
 \begin{equation}
     \frac{EM}{\rm pc\,cm^{-6}}  =   \int_{los}^{}\,\left(\frac{N_e}{cm^{-3}}\right)^2\,d\left(\frac{s}{pc}\right)
 \end{equation}
 
  We compute the EMs for our sources using the turnover frequencies obtained from our modeling (Table~\ref{tab:radiosed_pars}) and assuming a typical electron temperature 10$^4$\,K \citep[][]{Clemens10, Galvin2018}. Table~\ref{tab:em} provides the emission measures for our sample. Our $EM$ values cover the same range as that obtained by \citet[19 LIRGs; ][]{Galvin2018}. Furthermore, our EM values are also consistent with those obtained by \citet[20 LIRGs and ULIRGs; ][]{Clemens10}.               

\subsection{Thermal fraction vs. $q_{IR}$}

In star forming galaxies the radio spectral index is expected to be correlated with $TF$ and $q_{IR}$ with the star formation history playing a key role \citep[][]{Marvil15}. This is because in young star forming galaxies ($<10^8$\,yr; \citet{Condon92}) the radio emission is mainly due to free-free emission with a relatively flatter spectrum ($\alpha\sim-$0.1) leading to relatively higher values of $TF$ and $q_{IR}$ \citep[][]{Marvil15}. As the starburst ages, $TF$ and $q_{IR}$ are expected to decrease as the time-delayed nonthermal emission increases with time. This is mainly because the nonthermal emission originates from relatively old ($>10^7$\,yr; the average lifespan of $>8\,M_{\odot}$OB type stars) relativistic electrons considering the time taken for the diffusion ($\sim\,10^4$\,yr) of these electrons across the galaxy is negligible \citep{Clemens10, Galvin2018}. Figure~\ref{fig:tfsfh} shows the relation between $TF$ and $q_{IR}$ (computed using Eq.~\ref{eq:qir}) as a function of the SFR$_{IR}$ ratio obtained at time intervals 10\,Myr and 100\,Myr ago from the CIGALE analysis. This representation is informative in order to understand the evolution of the thermal fraction and $q_{IR}$ with the age of star formation. The large scatter observed in the plot can be explained by the following reasons: (i) ceasing of star formation before nonthermal emission commences, (ii) early contribution of nonthermal emission to the radio continuum while star formation is still ongoing, and (iii) due to limited sample size.

\subsection{FR$_{IR}$ vs. $M_{\rm dust}$}

In Figure~\ref{fig:sfrmdust} we show the relation of SFR$_{IR}$ with dust mass for our sample, including data from the literature for normal star-forming galaxies in the local Universe \citep{Dacunha2010b}, LIRGs at low ($z<0.5$) and high ($z>0.5$) redshifts \citep{Vega08, LoFaro2013, Pierra2015, Dacunha2015, Malek2018, Pasparialis2021}, and ULIRGs at low and high redshifts \citep{Vega08, Dacunha2015, Malek2017, Malek2018}. Our sample of LIRGs lies essentially in between the two sequences of normal and starburst galaxies defined by \citet{Rowlands2014}, in excellent agreement with the behavior shown by starburst galaxies. This is because our $IR$ selection $S_{60 \mu m}$\,$\geq$5\,Jy selects sources to have high SFRs. Moreover, our merger-type galaxies (filled star symbol) lie systematically higher in SFR$_{IR}$ than the nonmergers (filled circle) and at the same time lie systematically below the high redshift LIRGs (diamonds) and ULIRGs (plus), while the normal star-forming galaxies (cross symbol) occupy the region characterized by a lower star formation rate and dust mass. This can be understood as follows. With an increase in the SFR$_{IR}$, there is also an increase in the supernova rate, leading to a more efficient enrichment of the ISM with dust. Since SFR$_{IR}$ is correlated with dust mass, and the dust relates the gas content in galaxies (Kennicutt relation links SFR$_{IR}$ and gas mass), therefore, a positive trend in evolution of SFR$_{IR}$ with dust mass is expected with our LIRGs when compared with local star-forming galaxies and ULIRGs \citep[][]{Rowlands2014, Dovenski2020}. High SFR$_{IR}$ values are seen for interacting/merging LIRGs galaxies as compared to nonmergers \citep[see also, ][]{Pasparialis2021}. 

\subsection{SFR$_{IR}$ vs. stellar mass and nonthermal fraction}

Figure~\ref{fig:stellarmasssfr} explores the relation of stellar mass with the CIGALE estimated star formation rate (SFR$_{IR}$) for our galaxies, including the data for normal star-forming galaxies and the ULIRGs from the literature. As shown, our sample of LIRGs lies systematically above the main galaxy sequence represented by the parametric relation \citep[solid line with 0.4\,dex error denoted by dashed lines;][]{Saintonge2016} along with local star-forming galaxies from the sample of \citet{Dacunha2010b}. There is no evidence of a correlation between stellar mass and SFR$_{IR}$ for LIRGs and ULIRGs, including our sample (star symbol), which is explained by reaching saturation in SFR$_{IR}$ and stellar mass for these types of galaxies \citep[][]{Dacunha2010b, Pasparialis2021}. Figure~\ref{fig:ntf} explores the relationship between the 1.4\,GHz nonthermal fraction, $nTF$ (=nonthermal luminosity/total luminosity), and the SFR$_{IR}$ for our galaxy sample. All galaxies, including merger and non-merger types in our sample, exhibit high $nTF$ values at 1.4\,GHz, as expected. As we see a clear distinction between the SFR$_{ IR}$ values for merger- and non-merger-type galaxies, with the higher SFR$_{IR}$ values being exhibited by merger-type galaxies, one would expect stronger magnetic fields $B$ in these objects due to strong turbulence created by the merger process. For a given synchrotron lifetime of CRe, $\tau_{synch}$ $\propto B^{-1.5}$, the CRe have a good chance to efficiently lose their energy via synchrotron radiation before escaping the star-forming regions. 

In this regard, we note that resolved studies show that the radio spectral index is flatter in star-forming regions \citep[][]{Fletcher11, Tabatabaei07, Tabatabaei13, Hassani22} which contradicts the claimed argument of complete synchrotron cooling of CRes before escape from star-forming regions. Moreover, in star-forming regions, there is also a strong chance of CRe escape due to star formation feedback \citep[][]{Murphy08}. A fine balance between gas and magnetic fields/CRes can more realistically explain the nTF-SFR$_{IR}$ correlation observed in normal star-forming galaxies (\citet[][]{Tabatabaei13} and \citet[][]{Lacki10}).

\subsection{Specific star formation rate}
Figure~\ref{fig:ssfr} shows the evolution of specific star formation rate, (sSFR$_{IR}$ defined to be ratio of SFR$_{IR}$ and $M_\star$), with respect $M_\star$ (panel a), specific dust mass ($sM_{\rm dust}$ defined to be ratio of $M_{\rm dust}$ and $M_\star$; panel b), and $T_{\rm dust}$ (panel c), respectively, for our sample of LIRGs including the data from the literature for normal star-forming galaxies along with LIRGs and ULIRGs. The sSFR$_{IR}$ measures the recent star formation activity and is defined as the current star formation over the stellar mass of the galaxy \citep[][]{Dovenski2020, Pasparialis2021}. Panel (a) shows an anticorrelation between the sSFR$_{IR}$ and M$_\star$ for the three samples, indicating that massive galaxies are less efficient in star formation than less-massive ones. This is expected in the `downsizing' scenario of galaxy evolution, where massive galaxies formed most of their stars earlier and on shorter timescales. In comparison, less-massive galaxies evolve on longer timescales \citep[][]{Scodeggio09, Popesso11, Sobral11, Tatiana18}. 

Panel (b) shows a positive correlation between sSFR$_{IR}$ and $sM_{\rm dust}$ for the three samples, with LIRGs and ULIRGs occupying higher sSFR$_{IR}$ values per unit of specific M$_{\rm dust}$ compared to normal star-forming galaxies. In this plot, the contribution of M$_{\star}$ to SFR$_{IR}$ and M$_{\rm dust}$ is normalized, making it ideal for studying gas-to-dust behavior in galaxies \citep[][]{Smith2012, Hunt2014, Dovenski2020}. First, the figure shows that there is an increase in dust mass compared to stellar mass on a very short timescale for these galaxies with LIRGs and ULIRGs expected to occupy the top-right corner of the plot \citep[see, ][and the references therein]{Dovenski2020}. The relationship between sMdust and sSFR$_{IR}$ indicates an evolutionary scenario where sSFR$_{IR}$ decreases due to exhaustion of the gas reservoir and therefore causes an inefficient dust production, causing galaxies to occupy the bottom left corner of the plot \citep[][]{Burgarella2020}. As expected, our LIRGs follow the trend, and local and high redshift LIRGs and ULIRGs with normal star-forming galaxies are in a different branch. 

Panel (c) shows the distribution of sSFR$_{IR}$ with $T_{\rm dust}$ for our LIRGs and local and high redshift LIRGs and ULIRGs obtained from the literature. We note that the dust temperature obtained for our LIRGs falls in a narrow range. This is most likely due to the discrete nature of the parameter space used in the CIGALE modeling, including the values provided for the interstellar radiation field needed to heat the dust \citep[Eq.\,1 of; ][]{Pasparialis2021}. In general, a linear trend is observed between sSFR$_{IR}$ with $T_{\rm dust}$, extending to high$z$ galaxies, which is due to the heating of dust surrounding the young stellar population in highly star-forming galaxies \citep[][]{Magnelli2014, Liang2019}.
However, our sample has more homogeneous properties and indicates dust temperature within a limited range of 24 K to 32 K. 

This comparison indicates that our galaxies fall in the middle of the parameter space occupied by normal galaxies of the local Universe and distant starburst objects, exhibiting intermediate properties, as expected, between the two extremes of the galaxy evolution.

\subsection{Calibrating SFR in radio wavelength}\label{sec:calib}

    Our detailed radio SED modeling enabled us to decompose the nonthermal and thermal spectral luminosities and derive the respective SFRs. For this, we use the radio SFR calibration relations given by \citet[][]{Murphy11}. Thermal SFR, ${\rm SFR_{\nu}^{T}}$ is derived using Kroupa initial mass function \citep[IMF; ][]{Kroupa01} assuming solar metalicity and continuous star formation:

\begin{eqnarray}
\left(\frac{\rm SFR_{\nu}^{T}}{M_{\sun}~{\rm yr^{-1}}}\right) = 4.6\times10^{-28} \left(\frac{T_{\rm e}}{10^{4}~{\rm K}}\right)^{-0.45} \left(\frac{\nu}{\rm GHz}\right)^{0.1} \nonumber \\ \times \left(\frac{L_{\nu}^{\rm T}}{\rm erg~s^{-1}~Hz^{-1}}\right)\, \label{equ:sfrth}
\end{eqnarray}
where $L_{\nu}^{\rm T}$ is the thermal spectral luminosity and $T_e$=10$^4$\,K.

The nonthermal SFR, ${\rm SFR_{\nu}^{NT}}$, is derived using calibration between the supernova rate and the SFR using the output of Starburst99 model \citep[][]{Leitherer99}, and empirical relation between supernova rate and nonthermal spectral luminosity, $L_{\nu}^{\rm NT}$, of the Milky Way \citep[][]{Tammann82, Condon_yin90}:

\begin{eqnarray}
\left(\frac{\rm SFR_{\nu}^{NT}}{M_{\sun}~{\rm yr^{-1}}}\right) =6.64\times10^{-29}  \left(\frac{\nu}{\rm GHz}\right)^{\alpha^{\rm NT}} \left(\frac{L_{\nu}^{\rm NT}}{\rm erg~s^{-1}~Hz^{-1}}\right)
\label{equ:sfrnt}
\end{eqnarray}

where, $\alpha^{\rm NT}$=$\alpha$=$\alpha_{synch}$ is the synchrotron spectral index. For our calculations, we use the values obtained from the radio SED modeling (Table~\ref{tab:radiosed_pars}).

The total SFR at 1.4\,GHz, $\rm SFR_{\rm 1.4GHz}$ is derived using the relation between the IR emission and current SFR resulting from the integration of the output of Starburst99 spectrum over the wavelength range of 912\,$\AA$ -- 3646\,$\AA$, and the q$_{IR}$ relation (see eq.~\ref{eq:qir}). We use q$_{IR}$=2.64 \citep[][]{Bell03} to arrive at the total $\rm SFR_{\rm 1.4GHz}$ relation:
\begin{eqnarray}
\left(\frac{\rm SFR_{\rm 1.4GHz}}{M_{\sun}~{\rm yr^{-1}}}\right) = 6.35\times10^{-29}\left(\frac{L_{\rm 1.4GHz}}{\rm erg~s^{-1}~Hz^{-1}}\right) \label{equ:sfrtotal}
\end{eqnarray}

   Figure~\ref{fig:sfrir_rad} shows the comparison of the SFR$_{IR}$ obtained from the CIGALE SED fitting for two time intervals, 10 and 100 Myr ago, and the 1.4\,GHz radio SFR (total, nonthermal, and thermal, respectively). Upon visual inspection, we observe that SFR$_{IR}$ shows a one-to-one correlation with 1.4\,GHz radio SFR, albeit with a large scatter, $\sigma$, which measures average deviation from the unity. A striking result is that we obtain a much better correspondence of radio emission (total and synchrotron) with the young stellar population of about 10\,Myr than with the older population. To check this, we applied Spearman's rank correlation test, which measures the statistical dependence between the two variables \citep[][]{Spearman1904}. A null hypothesis of no correlation is tested against the alternate hypothesis of non-zero correlation at a certain significance level (=0.01, adopted by us). Typically, the $p-$value\footnote{$p-$value is the probability of obtaining the result as extreme as observed by chance} $<$0.01 means that the null hypothesis of no-correlation is rejected at a confidence level $>$99\%. For the star formation rate at 10\,Myr ago, the obtained $p-$values are 0.0008 and 0.0002, respectively, rejecting the null hypothesis that the radio (total, nonthermal, and thermal) and $IR$ SFRs are not correlated. At the same time, the $\rho$-values\footnote{$\rho-$value measures the strength and direction of association between two ranked variables} are positive, indicating a positive correlation between the two variables (panels a and b of Figure~\ref{fig:sfrir_rad}). The null hypothesis of no correlation is not rejected for a significance level of 0.01 for thermal SFR and SFR$_{IR}$ at 10\,Myr ago (panel c of Figure~\ref{fig:sfrir_rad}) and radio and SFR$_{IR}$ at 100\,Myr ago (panel d, e, f of Figure~\ref{fig:sfrir_rad}). This is probably due to the relatively short lifetime of the CRe synchrotron at the 1.4 GHz frequency. At this frequency, in a magnetic field of about 50$\mu$G \citep[][]{Crocker10}, the synchrotron lifetime is $\approx$ 3.3$\times$10$^5$\,yr. Therefore, synchrotron emission may be an effective indicator of recent SFR in galaxies. The thermal component appears to be less useful because of visible significant scatter in the predicted SFR. Since thermal emission is much weaker than synchrotron emission at this frequency, this may explain its weaker usefulness than the nonthermal component. Our analysis strengthens that 1.4\,GHz radio SFR measurements can be used as a diagnostic tool for high-$z$ galaxies \citep[][]{Murphy11, Tabatabaei17}.

  \subsection{Comparison of radio spectral indices from radio-only and CIGALE SED modeling}
  We emphasize that both radio-only and FUV to radio modeling includes synchrotron emission in the models, however, the direct comparison between the synchrotron spectral index obtained from the two SED fitting techniques is inadequate. This is because the CIGALE modeling (FUV to radio) uses a very simple formulation to compute the radio flux at 1.4\,GHz. This is done using the standard $q_{IR}$ relation at 1.4\,GHz and the integrated $IR$ luminosity between 8-1000$\mu$m. The radio flux at other frequencies is then computed using a single PL model with an assumed value of the spectral index. However, it is observed that the radio-only SEDs show complex morphologies (flattening or turnovers at lower frequencies and steepening at higher frequencies for some cases). Furthermore, such a direct comparison would only reveal the known scatter in the $q_{IR}$ relation \citep[][]{Wang19, Sinha22}. Furthermore, the limitations of this simple extrapolation in the CIGALE modeling are evident by the discontinuities in the radio wavebands (Figure~\ref{fig:cigalefitting}).

\section{Conclusion and Final remarks}\label{sec:conclusion}

In this study, we performed joint modeling of SEDs of 11 LIRGs, focusing in radio-only and FUV to radio bands where model parameters are estimated using state-of-the-art Bayesian (radio-only) and Bayesian-like (CIGALE modeling) inference techniques. The radio-only SED modeling allowed us to decompose nonthermal and thermal radio components while the CIGALE modeling allowed us to fit complex star formation history models, i.e., delayed star formation with an exponential burst (in our case), enabling us to estimate SFR$_{IR}$ at different time intervals. Our main finding are the following:

\begin{enumerate}
    \item The radio-only SED modeling shows that radio spectra have complex features, showing bends and turnovers. Unlike the frequent assumptions, this shows that the nonthermal spectral index is not fixed, and it changes between $-$0.45 and $-$1.75, which could be due to the influence of star formation on the energetics of CRe. The computed 1.4\,GHz $TF$ falls between $\sim$0.8\% and $\sim$0.12\% for our galaxies, similar to that obtained for other LIRGs in the literature.

    \item Although many studies have performed FUV-$IR$ SED modeling using CIGALE for star-forming galaxies, LIRGs, and ULIRGs (see Section~\ref{sec:intro}), only a handful include radio measurements in the SED fitting \citep[e.g.,][]{Vega08, Shen2020, Hamed2021}. The inclusion of radio measurements in the CIGALE modeling mostly serves to better constrain the derived dust luminosity. Our results indicate that, while the values of $L_{IR}$, fall in the range exhibited by galaxies of the same class, the uncertainties on them are improved by one order of magnitude compared to those given in the literature.

    \item The $M_\star$, $T_{\rm dust}$, SFR$_{IR}$, and AGN fraction derived using CIGALE modeling for our galaxies falls in the range exhibited by galaxies of the same class. The AGN is not energetically dominant in our sources, as the typical AGN fraction is $\leq$ 6\%, similar to those found in other samples. The bolometric AGN fraction obtained by CIGALE is similar to that obtained from spectroscopic methods for most of our galaxies, indicating the robustness of the CIGALE analysis.
    
    \item Comparison of the 1.4\,GHz radio SFRs obtained using the total, and nonthermal radio emissions shows a close correspondence with the CIGALE-obtained SFR$_{IR}$ at 10\,Myr ago as compared to 100\,Myr ago, strengthening the view that 1.4\,GHz SFR estimates is a good indicator of recent star formation.
    
    %\item The close correspondence between the $q_{IR}$ estimated from the CIGALE modeling and from Eq.~\ref{eq:qir} validates our analysis procedure.
    
\end{enumerate}

%The radio spectra of the LIRGs studied here show complex features. Moreover, the obtained $q_{IR}$, thermal fraction ($\leq$10\%) emission measure for our galaxies is comparable to other LIRGs in the literature. 

The comparison of astrophysical properties obtained by radio-only SED modeling and CIGALE modeling with other samples studied in the literature indicates that our sample belongs to a homogeneous population of LIRGs with respect to their astrophysical properties. Finally, we end with an obvious caveat that our findings are based on a relatively small sample comprising 11 LIRGs in total and five merger-type galaxies. Therefore, a larger sample with broadband SED coverage is needed to strengthen the tentative findings presented in this study.

\section*{Acknowledgements}
We thank the anonymous referee for carefully reading the manuscript and providing several constructive comments which substantially improved the content and presentation. S.D. acknowledges financial support from Jagiellonian University DSC grant N17/MNS/000014. S.D. and A.G. acknowledge support from the Polish National Science Centre (NCN) through the grant 2018/29/B/ST9/02298. K.M. acknowledges support from NCN UMO-2018/30/E/ST9/00082. S.D. thanks E. da Cunha for kindly providing the data presented in \citet[][]{Dacunha2010b} in the electronic form. Useful discussions with Ma{\l}gorzata Bankowicz and Alexander Herzig on CIGALE analysis and Marek Jamrozy on radio data are acknowledged. We thank Krzysztof~Chy$\dot{\rm z}$y, Agnieszka~Pollo, Micha\l~Ostrowski, and Nimisha~G.~ Kantharia for proofreading the manuscript and providing several valuable comments.

We thank the staff of the GMRT that made these observations possible. GMRT is run by the National Centre for Radio Astrophysics of the Tata Institute of Fundamental Research. 

This research has used the NASA/IPAC Extragalactic Database (NED),
which is operated by the Jet Propulsion Laboratory, California Institute of Technology, under contract with the National Aeronautics and Space Administration. This research has made use of the SIMBAD database,
operated at CDS, Strasbourg, France 2000,A\&AS,143,9, ``The SIMBAD astronomical database'', Wenger et al.This research has made use of the NASA/IPAC Infrared Science Archive, which is funded by the National Aeronautics and Space Administration and operated by the California Institute of Technology. This research has made use of the VizieR catalogue access tool, CDS,
 Strasbourg, France (DOI : 10.26093/cds/vizier). The original description 
 of the VizieR service was published in 2000, A\&AS 143, 23.

\facilities{SUMSS, GLEAM, GMRT, VLA, Effelsberg, ATCA, GALEX, XMM-OM, Swift, IRSA,  Spitzer, WISE, Herschel, IRAS, Gaia, AKARI}
\software{SKIRTOR \citep[][]{Stalevski2012, Stalevski2016}, UltraNest \citep[][]{Buchner21}, CIGALE \citep[][]{Noll2009, Serra11}}

%\bibliography{ref}
%\bibliographystyle{aasjournal}

\appendix
%\numberwithin{table}{section}
%\counterwithin{figure}{section}
\restartappendixnumbering

\section{Summary of radio observations gathered for radio SED modeling and optical--radio overlays}\label{app:A}

Tables ~\ref{tab:gmrtobs} and ~\ref{tab:radiosummary} provide the basic information on the GMRT and VLA continuum datasets analysed by us while the Table~\ref{tab:radioflux} provides the integrated flux densities used in radio-only SED modeling. Figure~\ref{fig:overlays} provides the overlay of the radio contours of DSS images for our galaxies. At the same time, the corner plots for the check on degeneracy on the model parameters are given in Figure~\ref{fig:corner_radio}.

\setcounter{table}{0}
\renewcommand{\thetable}{A\arabic{table}}

\begin{table}
%\centering
%\scriptsize
\small
\caption{Primary and phase calibrators used in our GMRT observations. \label{tab:gmrtobs}}
\begin{tabular}{ccccc}\\
\hline
Name  & Primary (325 MHz)  &  Phase (325 MHz) & Primary (610 MHz)  &  Phase (610 MHz) \\
\hline
 ESO\,500$-$G034 &  3C\,286, 3C\,468.1 & 1130$-$148 & 3C\,147, 3C\,286 & 1130$-$148 \\
 NGC\,3508 & 3C\,286,3C\,468.1 & 1130$-$148 & 3C\,147, 3C\,286 & 1130$-$148 \\
 ESO\,440$-$IG058 & 3C\,286, 3C\,468.1 & 1248$-$199 & 3C\,147, 3C\,286 & 1248$-$199 \\
 ESO\,507$-$G070  & 3C\,286, 3C\,468.1 & 1248$-$199& 3C\,147, 3C\,286 & 1248$-$199  \\
 NGC\,5135 & 3C\,286, 3C\,468.1 & 1248$-$199 & 3C\,147, 3C\,286 & 1248$-$199 \\
 IC\,4280 &  3C\,286, 3C\,468.1 & 1248$-$199 & 3C\,147, 3C\,286 & 1248$-$199 \\
 NGC\,6000 & 3C\,286, 3C\,468.1  & 1626$-$298 & 3C\,147, 3C\,286 & 1626$-$298\\
 IR\,16164$-$0746 & 3C\,286, 3C\,468.1 & 1626$-$298 & 3C\,147, 3C\,286 & 1626$-$298 \\
 ESO\,453$-$G005 &  3C\,286, 3C\,468.1 & 1626$-$298& 3C\,147, 3C\,286 & 1626$-$298 \\
 IR\,18293$-$3413 &  3C\,286, 3C\,468.1 & 1830$-$360& 3C\,147, 3C\,286 & 1830$-$360 \\
 ESO\,593$-$IG008 &  3C\,286, 3C\,468.1 & 1830$-$360 & 3C\,147, 3C\,286 & 1830$-$360\\
\hline
\end{tabular}
\end{table}

\startlongtable
\begin{deluxetable*}{ccccccccccccc}
%\tablenum{1}
\tablecaption{Summary of new GMRT and the archival VLA datasets analysed by us.\label{tab:radiosummary}}
\tablewidth{0pt}
\tabletypesize{\scriptsize}
\tablehead{
\colhead{ID} & \colhead{Date of obs.} & \colhead{Tel.} & \colhead{Arr.} & \colhead{Frequency} &  \colhead{$BW$} &  \colhead{\it IF} &  \colhead{Int.} & \colhead{rms} & \colhead{Synth. Beam} & \colhead{$PA$} & \colhead{S/N} & Area \\
\colhead{} & \colhead{}  & \colhead{} & \colhead{}  & \colhead{}  & \colhead{(MHz)}  & \colhead{} & \colhead{(min)} & \colhead{$(\frac{\rm mJy}{\rm beam}$)}  & \colhead{($^{\prime\prime} \times ^{\prime\prime}$)} & \colhead{($^\circ$)} & \colhead{} & \colhead{(\textbf{arcsec$^2$})} 
}
\decimalcolnumbers
\startdata
\multicolumn{13}{c}{{\bf ESO\,500-G034}}\\
23\_051 & 2013 Jun 11 & GMRT & & 325\,MHz & 32 & 1 & 32 & 0.78 &  12.49$\times$7.72  & 33.92  &   142 & 518.9\\
23\_051 & 2012 Oct 20  & GMRT & & 610\,MHz & 32 & 1&  32 & 0.18 &  6.04$\times$4.27  &  $-$21.00  & 394 & 330.4 \\
 AC345 & 1993 May 24 & VLA & CnB & 1.4\,GHz & 50 & 2 & 2.5 & 0.50 & 25.13$\times$14.66 & $-$21.90 & 22 & 1686.8 \\
AD229 & 1989 Feb 20 & VLA & BnA & 4.8\,GHz & 50 & 2 & 7.5 & 0.14 & 1.67$\times$0.91 & $-$61.11 & 79 & 11.1 \\
AV186 & 1991 Mar 26 & VLA & D & 14.9\,GHz & 50 & 2 & 10.5 & 0.47 & 9.49$\times$4.85 & $-$10.96 & 19 & 113.5 \\
\multicolumn{13}{c}{{\bf NGC\,3508}}\\
23\_051 &  2013 Jun 11 & GMRT & & 325\,MHz & 32 &1 & 32  & 0.84 &  10.94$\times$7.89  & 38.16  &   41 & 1703.1  \\
23\_051 & 2012 Oct 20  & GMRT & & 610\,MHz & 32 & 1 & 32  & 0.31 &  5.63$\times$4.32   & $-$34.77 &   42 & 997.5 \\
AC205 & 1988 Mar 23 & VLA & C & 1.4\,GHz & 50 & 3.9 & 2 &0.90 & 24.98$\times$16.46 & $-$9.73 & 26 & 1729.7\\
\multicolumn{13}{c}{{\bf ESO\,440-IG058}}\\
23\_051 &  2013 Jun 11 & GMRT & & 325\,MHz & 32 & 1 & 32  & 0.60 &  10.94$\times$7.89  & 38.16  &  41 & 705.9  \\
23\_051 & 2012 Oct 20  & GMRT &  & 610\,MHz & 32 &  1 & 32  & 0.17 &  7.79$\times$4.20   & $-$32.17 &   275 & 421.4 \\
AS412 & 1990 Oct 6 & VLA & CnB & 1.4\,GHz & 50 & 2 & 7.9 & 0.28 & 21.50$\times$11.44 & $-$48.59 & 30 & 1147.1 \\
AS412 & 1990 Oct 6 & VLA & CnB & 4.8\,GHz & 50 & 2 & 7.5 & 0.11 & 6.59$\times$4.09 & $-$44.93 & 92& 187.1 \\
\multicolumn{13}{c}{{\bf ESO\,507-G070}}\\
23\_051 &  2013 Jun 11 & GMRT &  & 325\,MHz & 32 & 1 & 32  & 1.4 &  10.73$\times$8.11 & 4.80  &   81 & 466.2 \\
23\_051 & 2012 Oct 20  & GMRT &  & 610\,MHz & 32 & 1 & 32  & 0.67 &  7.35$\times$4.11   & $-$43.75 &   190 & 174.5 \\
AR531 & 2003 Nov 1 & VLA & B & 1.4\,GHz & 50 & 2 & 10.5 & 0.45 & 8.18$\times$5.47 & 12.12 & 165 & 177.2 \\
AD215 & 1988 Jun 9 & VLA & DnC & 4.8\,GHz  &   50 & 2 & 9.3  &  0.15 &   13.69$\times$12.64 & $-$60.82 & 190 & 794.9 \\
AV186 & 1991 Mar 26 & VLA & D &  14.9\,GHz & 50 & 2 & 10 & 0.52 & 10.10$\times$4.67 &$-$16.23 & 39 & 142.5 \\
\multicolumn{13}{c}{{\bf NGC\,5135}}\\
23\_051 & 2013 Jun 11  & GMRT &  & 325\,MHz & 32 & 1 & 32  & 1.8 &  11.96$\times$7.97  & 4.75  &   558 & 775.0  \\
23\_051 &  2012 Oct 20 & GMRT &  & 610\,MHz & 32 & 1 & 32  & 0.35 &  7.84$\times$4.13   & $-$38.36 &   747 & 730.2 \\
AW126 & 1985 Apr 1 & VLA & BnA & 1.4\,GHz & 50 & 2 & 14.6 & 0.23 & 5.22$\times$3.28   & 49.38 & 10 & 156.1 \\
AC351 & 1993 Feb 4 & VLA & BnA & 4.8\,GHz & 50 & 2 & 18.3 & 0.1 & 1.53$\times$1.16 & $-$63.68 & 316 & 34.4\\
\multicolumn{13}{c}{{\bf IC\,4280}}\\
23\_051 & 2013 Jun 11  & GMRT &  & 325\,MHz & 32 & 1 & 32  & 0.59 &  11.95$\times$8.08  & 9.45  &   99 & 1226.3 \\
23\_051 &  2012 Oct 20 & GMRT &  & 610\,MHz & 32 & 1 & 32  & 0.23 &  7.21$\times$4.22   & $-$39.93 &   99 & 995.6\\
AC345 & 1993 May 24 & VLA & CnB &  1.4\,GHz & 50 & 2 & 3.9 & 0.55 &  18.83$\times$14.30  & $-$48.84 & 56 & 1114.6 \\
AD215 & 1988 Jun 14 & VLA & DnC & 4.8\,GHz & 50 & 2 & 8.5 & 0.17 & 13.15$\times$10.61 & $-$54.87 & 44  & 1251.0 \\
\multicolumn{13}{c}{{\bf NGC\,6000}}\\
23\_051 &  2012 Oct 20 & GMRT &  & 610\,MHz & 32 & 1 & 32  & 0.28 &  6.75$\times$4.36   & $-$17.18 &   600 & 1333.1 \\
AC345 & 1993 May 24 & VLA & CnB &  1.4\,GHz & 50 & 2 & 2 & 0.52 &  27.01$\times$12.43  & $-$44.71 & 207 & 2970.5 \\
AC326 & 1992 Jul 13 & VLA & D & 4.8\,GHz & 50 & 2 & 2 & 0.37 & 34.85$\times$16.08 & 18.22 & 156 & 2828.2 \\
\multicolumn{13}{c}{{\bf IR\,16164$-$0746}}\\
23\_051 &  2012 Oct 20 & GMRT &  & 610\,MHz & 32 & 1 & 32  & 1.3 &  5.54$\times$4.66   & $-$17.18 &   152& 127.8 \\
AT149 & 1993 Apr 26 & VLA & B &  1.4\,GHz & 50 & 2 & 2.2 & 0.43 & 6.60$\times$5.54   & $-$0.78 & 137 & 139.7 \\
\multicolumn{13}{c}{{\bf ESO\,453-G005}}\\
23\_051 & 2012 Oct 20  & GMRT &  &  610\,MHz & 32 & 1 & 32  & 0.35 &  8.10$\times$4.14   & $-$32.73 &   100 & 227.4 \\
AC345 & 1993 May 24 & VLA & CnB & 1.4\,GHz & 50 & 2 & 1.9 & 1.6 & 37.58$\times$11.48   &  $-$46.21 & 15 & 798.3 \\
\multicolumn{13}{c}{{\bf IR\,18293$-$3413}}\\
23\_051 &  2013 Jun 11 & GMRT &  & 325\,MHz & 32 & 1 & 32  & 1.7 &  14.20$\times$9.26 & $-$5.84  &   423 & 836.7  \\
23\_051 & 2012 Oct 20  & GMRT &  &  610\,MHz & 32 & 1 & 32  & 1.2 &  13.34$\times$3.98 & $-$47.24 &   200 & 687.9 \\
 AT0 & 1983 Nov 30 & VLA & BnA &  4.8\,GHz  & 50 & 2 & 8.5  &  0.21 &   1.03$\times$0.43 & 8.44 & 23 & 11.1 \\
AP534 & 2007 Jun 11 & VLA & A  & 8.4\,GHz & 50 & 2 & 80.3 & 0.038 & 0.60$\times$0.24 & 6.59 & 53 & 10.1 \\
\multicolumn{13}{c}{{\bf ESO\,593-IG008}}\\
23\_051 &  2013 Jun 11 & GMRT & & 325\,MHz & 32 & 1 & 32  & 0.62 &  11.15$\times$9.56 & $-$19.53  &   226 & 667.8 \\
23\_051 & 2012 Oct 20  & GMRT & & 610\,MHz & 32 & 1 & 32  & 0.34 &  10.19$\times$4.16   & $-$53.63 &   134 & 404.3 \\
 AT149 & 1993 May 23 & VLA & CnB & 1.4\,GHz  & 50 & 2 & 5.2  &  0.68 &   26.66$\times$12.93 & 33.36 & 75 & 1329.3\\
 AF446 & 2007 Mar 29 & VLA & D & 4.8\,GHz  & 50 & 2 & 3.6  &  1.1 &   31.22$\times$22.34 & $-$25.85 & 226 &1452.5 \\
\enddata
\tablecomments{
(1) proposal identifier of the observation, (2) date of observations, (3) telescope used, (4) configuration array used for VLA observations, (5) central frequency of observation, (6) bandwidth used, (7) number of sub bands used during the observation, (8) integration time used to image the target, (9) typical rms noise on image near to the target measured using \textsc{TVSTAT}, (10) synthesized beam achieved, (11) position angle of the restoring beam, measured counter-clockwise from the standard north direction, (12) signal to noise (S/N) ratio of the source detection (ratio of peak flux density of the source and the rms noise on the map), (12) integration area used in \textsc{TVSTAT}.}
\end{deluxetable*}

%\begin{multicols}{2}
\startlongtable
\begin{deluxetable}{ccccc}
\tablecaption{Integrated radio flux densities used for the SED fitting.\label{tab:radioflux}}
\tablewidth{0pt}
\tabletypesize{\scriptsize}
\tablehead{
\colhead{Name} &        \colhead{Obs. Freq.}    & \colhead{S }            & \colhead{error}  & \colhead{Ref.}  \\
\colhead{}   &         \colhead{(GHz)}        & \colhead{(mJy)}         & \colhead{(mJy)} & \colhead{}
}
 \decimalcolnumbers
\startdata
ESO 500-G034 & 0.091 &	174.25 &	33.06 & (a) \\
& 0.122 &	153.25 &	21.27 & (a) \\
& 0.150 & 155.9 & 24.19 & (b) \\
& 0.158 &	144.25 &	17.01 & (a) \\	
& 0.189 &	137.5 &	15.39 & (a) \\	
& 0.219 &	122.5 &	13.48 & (a) \\
& 0.325 &	139.74 & 7.28 & (c) \\
& 0.610 &	108.21 & 5.46 & (c) \\
& 1.4 &	62.61 &	2.23 & (d) \\ 
& 3.0 &	26.184 & 2.67 & (e) \\
& 4.8 & 19.32 &	0.72 & (d) \\
%& 10.0 & 15.7 & 0.15 & (8) \\ %not considered in fitting
& 14.9 & 9.67 &	0.88 & (d) \\
NGC 3508 & .074  &  560 & 70 & (f) \\
& 0.091  & 459.75 & 34.52 & (a) \\
& 0.122 & 339.75 	& 17.53 & (a) \\
& 0.150 & 230.0 & 32.8 & (b) \\
& 0.158 & 288.25 & 10.30 & (a) \\
& 0.189 & 274.25 & 7.41 & (a) \\
& 0.219 & 245.25 & 7.13 & (a) \\
& 0.325 & 195.80 & 10.66 & (c) \\
& 0.610 &	 135.92 & 7.15 & (c) \\
& 1.4 & 55.57 & 2.93 & (d) \\
& 3.0 & 3.13 & 0.41 & (e) \\
ESO440-IG058 & 0.091 &	260.75 & 40.56 & (a) \\
& 0.122 & 205.5 & 28.18 & (a) \\
& 0.150 & 156 & 23.07 & (b) \\
& 0.158 & 195.5 & 21.70 & (a) \\	
& 0.189 & 190.25 & 20.78 & (a) \\
& 0.219 & 172.75 & 18.21 & (a) \\
& 0.325 & 160.01 & 8.23 & (c) \\
& 0.610 & 97.36 & 4.92 & (c) \\
& 0.843 & 72.1 & 5.7 & (g) \\
& 1.4 & 49.795 & 1.64 & (d) \\
& 3.0 & 32.32 & 3.22 & (e) \\
& 4.8 & 19.71 & 0.68 & (d) \\
& 10.0 & 9.7 & 0.9 & (h) \\
ESO 507-G070 & 0.091 & 389.25 & 54.62 & (a) \\
& 0.122 &	344 &	40.52 & (a) \\
& 0.150 & 160.0 & 23.56 & (b) \\
& 0.158 &	276.5 &	30.61 & (a) \\
& 0.189 &	238.75 & 	29.15 & (a) \\
& 0.219 &	224.25 &	24.04 & (a) \\
& 0.325 &	117.14 &	6.90 & (c) \\
& 0.610 &	87.79 &	4.75 & (c) \\
& 1.4 &	55.12 &	1.93 & (d) \\
& 3.0 & 32.44 & 3.26 & (e) \\
& 4.8	 & 36.3 & 1.15 & (d) \\ %no error mentioned
%& 6.0 & 37 & 1 & \\ 
& 14.9 & 19.37 & 1.17 & (d) \\
NGC 5135 &	0.091 &	1011.25 &	173.16 & (a) \\
&	0.122 &	890.25 &	91.63 & (a) \\
&  0.150 &	826.8 &	83.7 & (b) \\
&	0.158 &	777.50 &	78.90 & (a) \\
&	0.189 &	649.75 &	67.38 & (a) \\
&	0.219 &	663.25 &	66.80 & (a) \\
& 0.325 &	533.40 &	27.29 & (c) \\
& 0.610 &	370.32 &	18.61 & (c) \\
& 1.40 &	165.69 &	5.03 & (d) \\
& 2.31 &	100.00 &	17.20 & (i) \\
& 3.0  &  95.766 &  9.794  & (e) \\
& 4.80 &	58.51 &	1.82 & (d) \\
& 6.70 &	48.00 &	7.20 & (j) \\
IC 4280 & 0.091 &	271.50 &	49.90 & (a) \\
& 0.122	& 283.75	& 35.56 & (a) \\
& 0.150	& 186.1 & 	27.54 & (b) \\
& 0.158	& 222.75	& 25.13 & (a) \\
& 0.189	& 189.15	& 22.93 & (a) \\
& 0.219	& 162.00	& 17.45 & (a) \\
& 0.325	& 162.4	& 8.46 & (c) \\
& 0.610 &  109.63	& 5.68 & (c) \\
& 1.4	& 48.04	& 1.92 & (d) \\
& 3.0     & 28.183  & 4.87 & (e) \\
NGC 6000 & 0.092	& 913	& 105.53 & (a) \\
& 0.122	& 727.75	& 77.61 & (a) \\
& 0.150 &  418.2           & 59.55 & (b) \\ 
& 0.158	& 586.25	& 61.00 & (a) \\
& 0.189	& 569.75	& 58.18 & (a) \\
& 0.220	& 494	& 50.05 & (a) \\
& 0.61 &	287.03 &	14.51 & (c) \\
& 1.4	& 153.32 & 	7.86 & (d) \\
& 4.8	& 62.70	& 2.10 & (d) \\
IR 16164-0746 & 0.091	& 191.25	& 58.92 & (a) \\
& 0.122	& 176.50	& 37.05 & (a) \\
& 0.150            & 151.5           &   21.88 &  (b) \\
& 0.158	& 183.75	& 26.98 & (a) \\
& 0.189	& 154.75	& 22.18 & (a) \\
& 0.219	& 200.25	& 30.21 & (a) \\
& 0.61 & 	88.17 & 	5.49 & (c) \\
& 1.4 & 	62.36 & 	3.26 & (d) \\
& 3.0 & 44.12 & 4.45 & (e) \\
ESO 453-G005 & 0.15	& 81.6	& 12.67 & (b)\\
& 0.61	& 38.21	& 2.17  & (c) \\
& 0.84	& 29.40	& 2.9 & (g) \\
& 1.40	& 23.39	& 2.55 & (d) \\
& 3.00	& 14.28	& 1.45 & (e) \\
IR 18293-3413 & 0.091	& 1390	& 152.79 & (a) \\ 
& 0.122	& 1006.75	& 107.71 & (a) \\
& 0.150            & 780.4             & 111.26 & (b) \\
& 0.158	& 907	& 92.79 & (a) \\
& 0.189	& 832.25	& 84.61 & (a) \\
& 0.219	& 710.25	& 72.27 & (a) \\
& 0.325	& 458.96	& 23.45 & (c) \\
& 0.610	& 405.18	& 20.84 & (c) \\
& 1.4	& 226.8	& 10.56 & (k) \\
& 3	& 106.53	& 10.81 & (e) \\
& 4.8	& 36.58	& 2.18 & (d) \\
& 8.4	& 22.09	& 1.16 & (d) \\
ESO 593-IG008 & 0.092	& 261.45	& 71.30  & (a) \\
& 0.122	& 379.5	& 53.73  & (a) \\
& 0.150    &  250.0         &  36.07 & (b) \\
& 0.158	& 289.5	& 34.07  & (a) \\
& 0.189	& 277.5	& 30.62 & (a) \\
& 0.220	& 240	& 25.86 & (a) \\
& 0.325	& 174.46	& 8.90 & (c) \\
& 0.61	& 116.29	& 5.93 & (c) \\ 
& 1.4	& 60.35 & 	2.36 & (d) \\
& 3.0     & 20.867  & 2.23 & (e) \\
 & 4.8	& 21.86	& 1.91 & (d) \\
\enddata
\tablecomments{ Columns:  (1) source name; (2) observing frequency; (3) integrated flux density of the source (4) the error in the flux density including the absolute calibration uncertainty (Sect~\ref{sec:datarad}),  (5) references - (a) GLEAM, (b)TGSS ADR1 \citep[][]{Intema17}, (c) current observations, (d)re-analysed VLA, (e) VLASS \citep[][]{Gordon21}, (f) VLSSr \citep[][]{lane14}, (g) SUMSS \cite, (h) ATCA, (i) S-PASS \citep[][]{Meyers17}, (j) Effelsberg \citep[][]{Impellizzeri08}, (k) NVSS.
}
\end{deluxetable}
%\end{multicols}{3}

%\setcounter{figure}{0} \renewcommand{\thefigure}{A.\arabic{figure}}
%\renewcommand\thefigure{\thesection.\arabic{figure}} 

\begin{figure*}
    \hbox{
    \hspace{0cm}{\includegraphics[width =  0.25\textwidth]{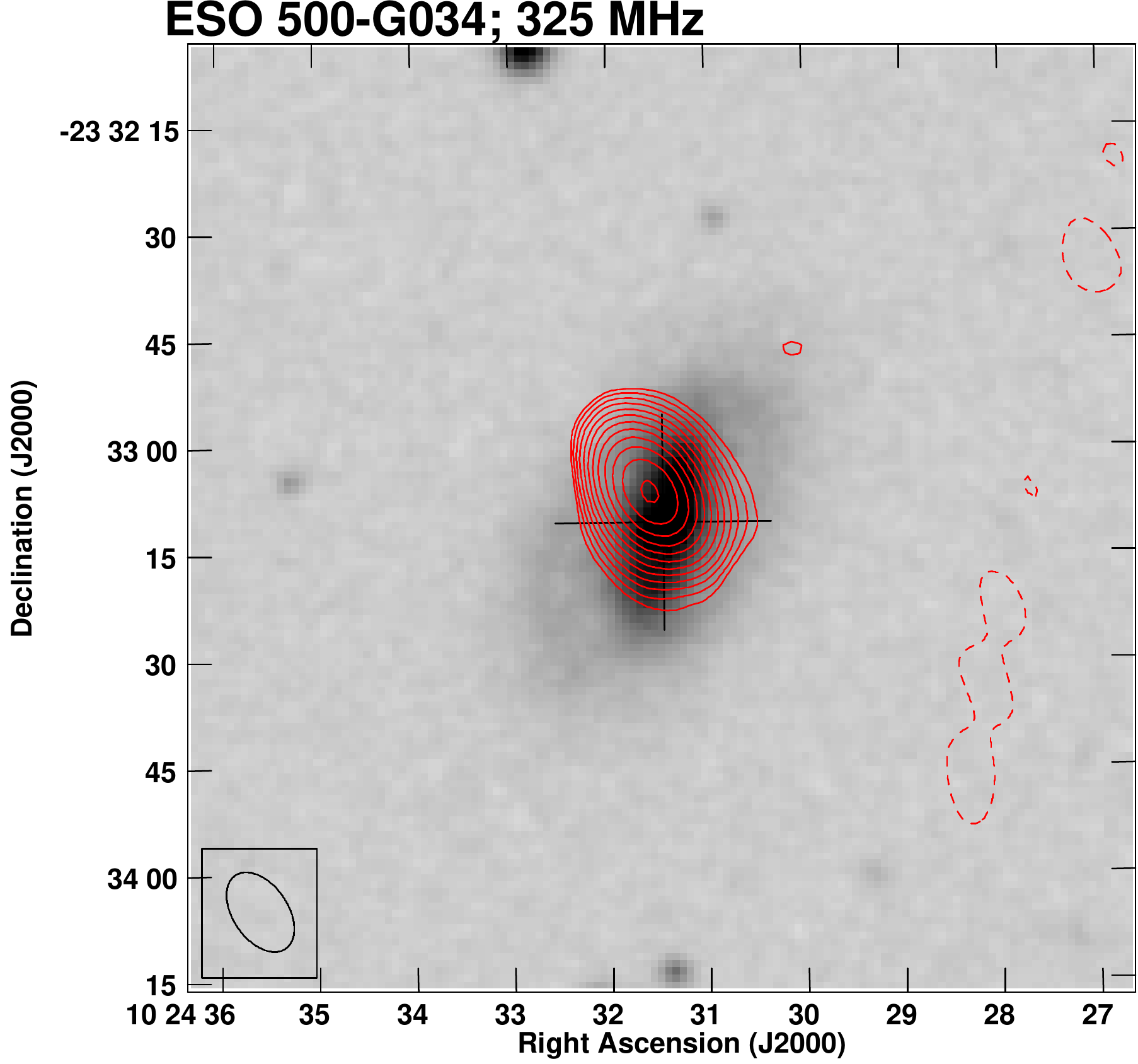}}
    \hspace{0cm}{\includegraphics[width =  0.25\textwidth]{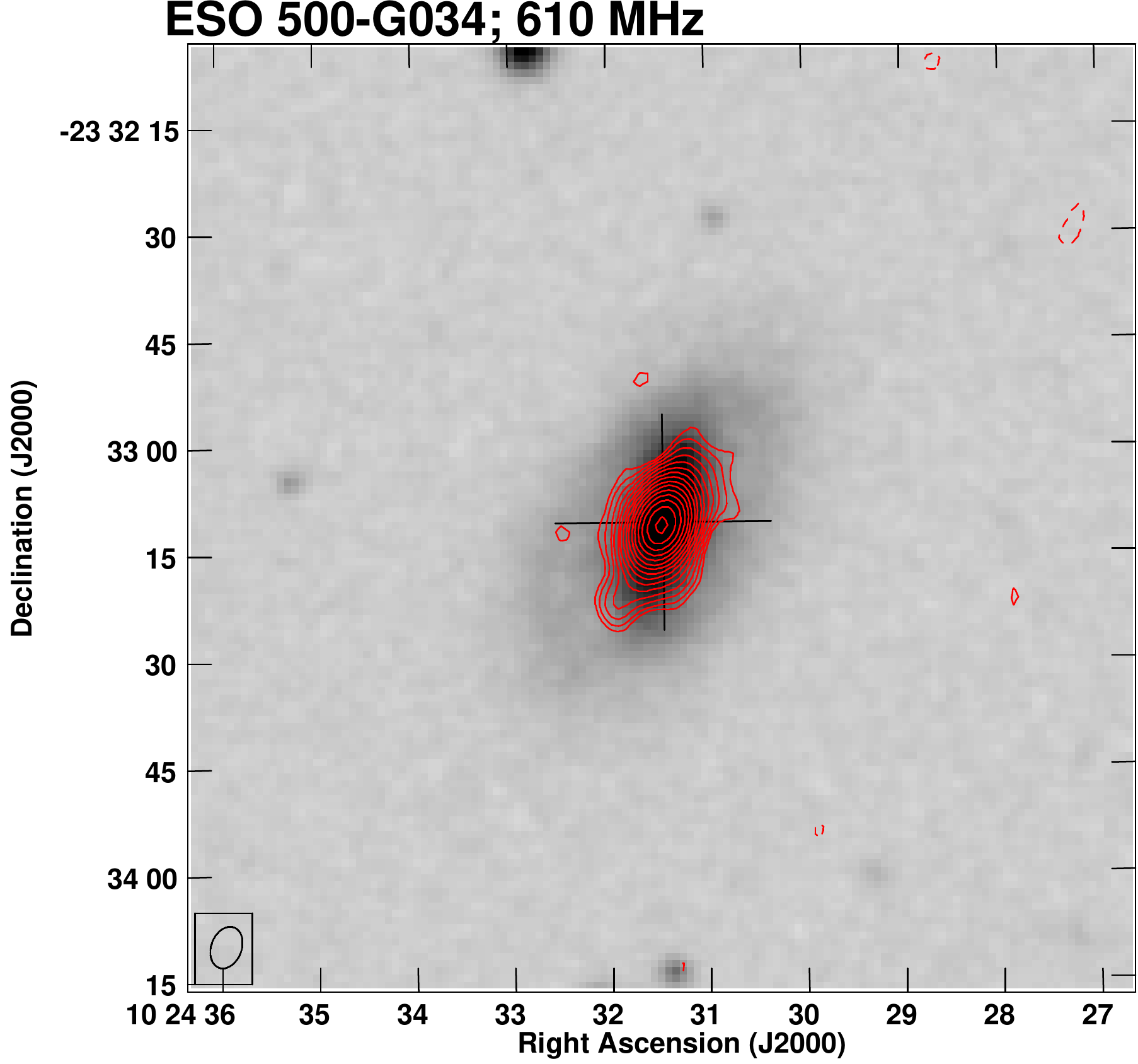}}
    \hspace{0cm}{\includegraphics[width =  0.25\textwidth]{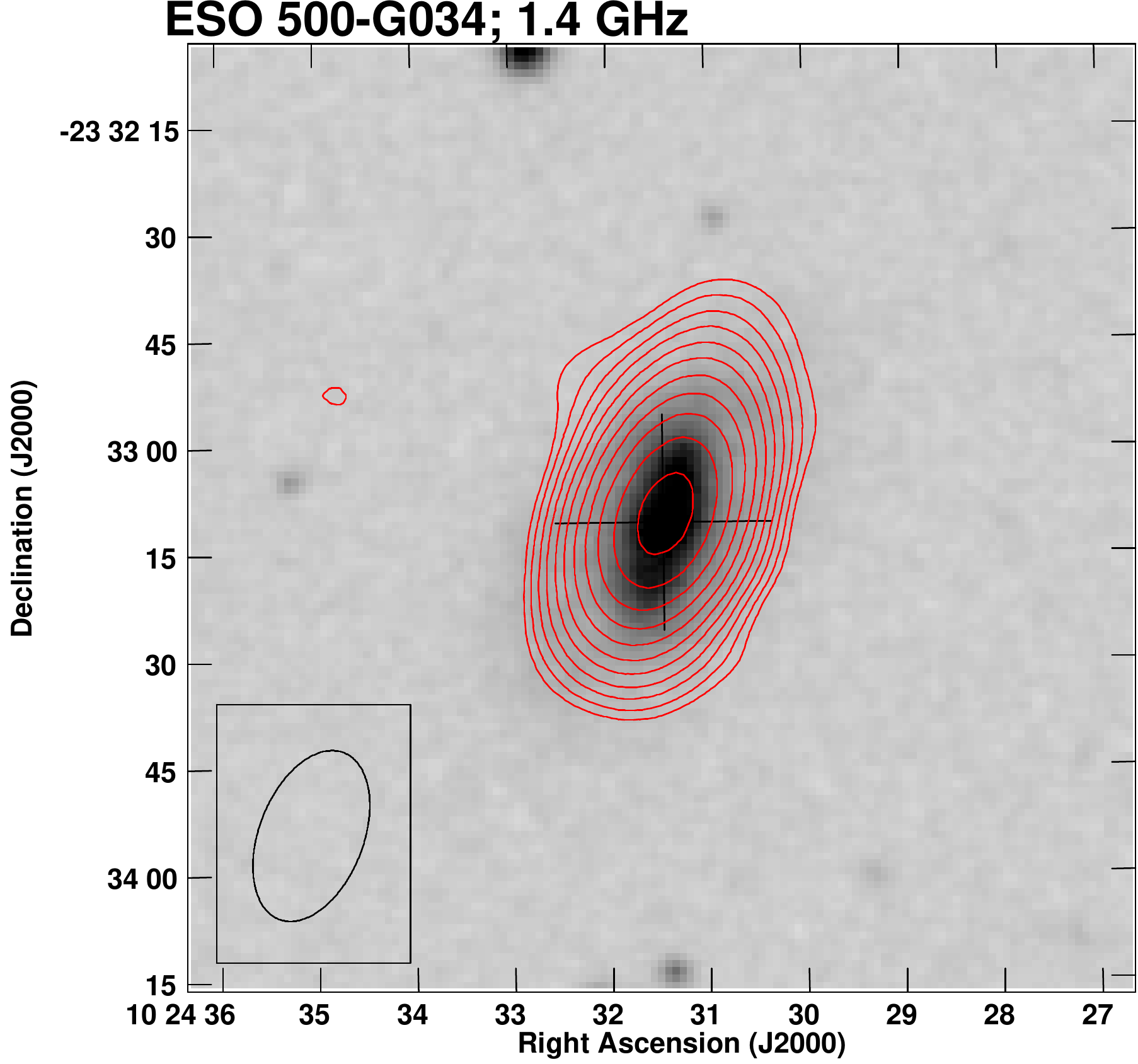}}
    \hspace{0cm}{\includegraphics[width =  0.25\textwidth]{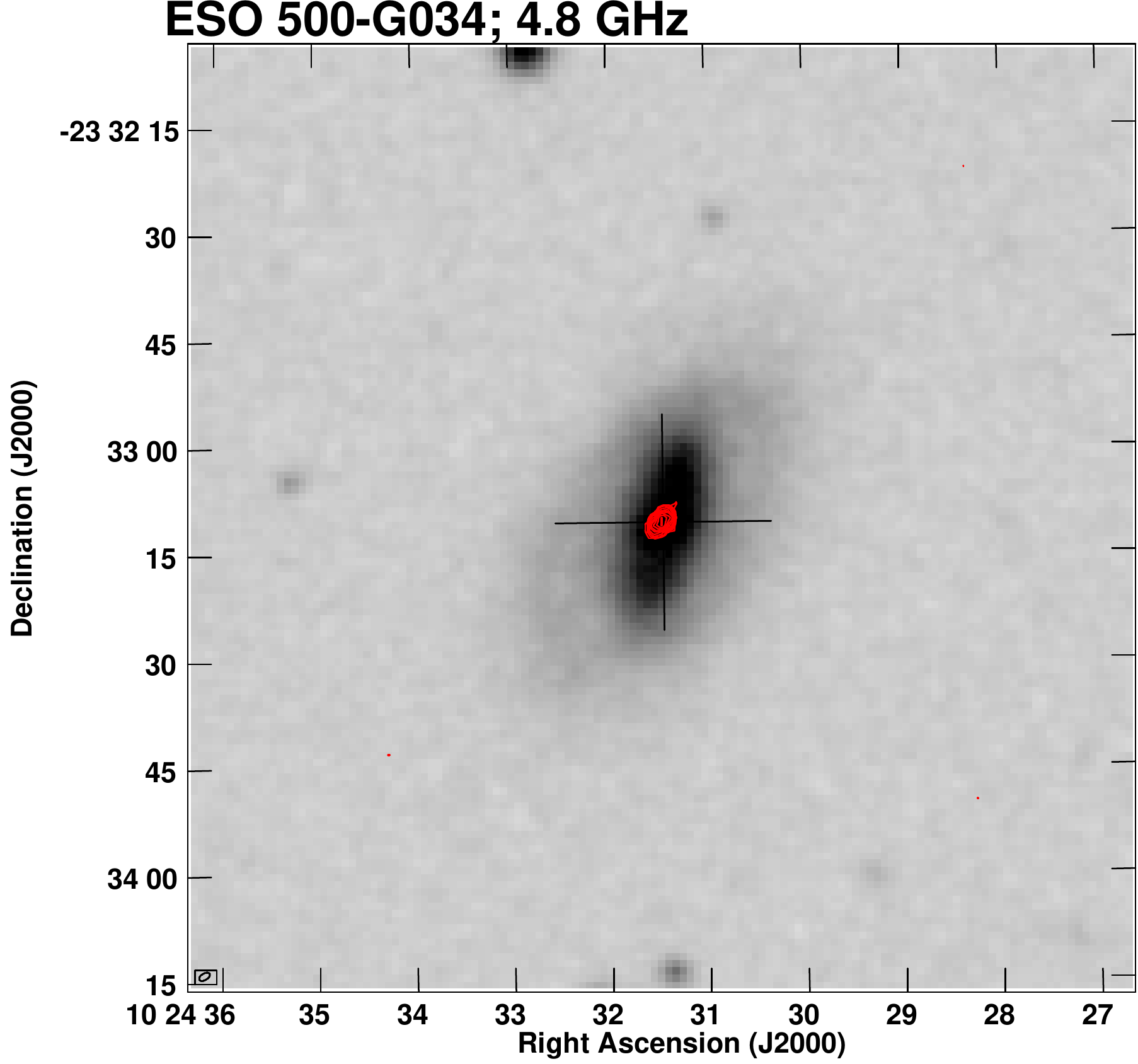}}
   }
   \hbox{
    \hspace{0cm}{\includegraphics[width =  0.25\textwidth]{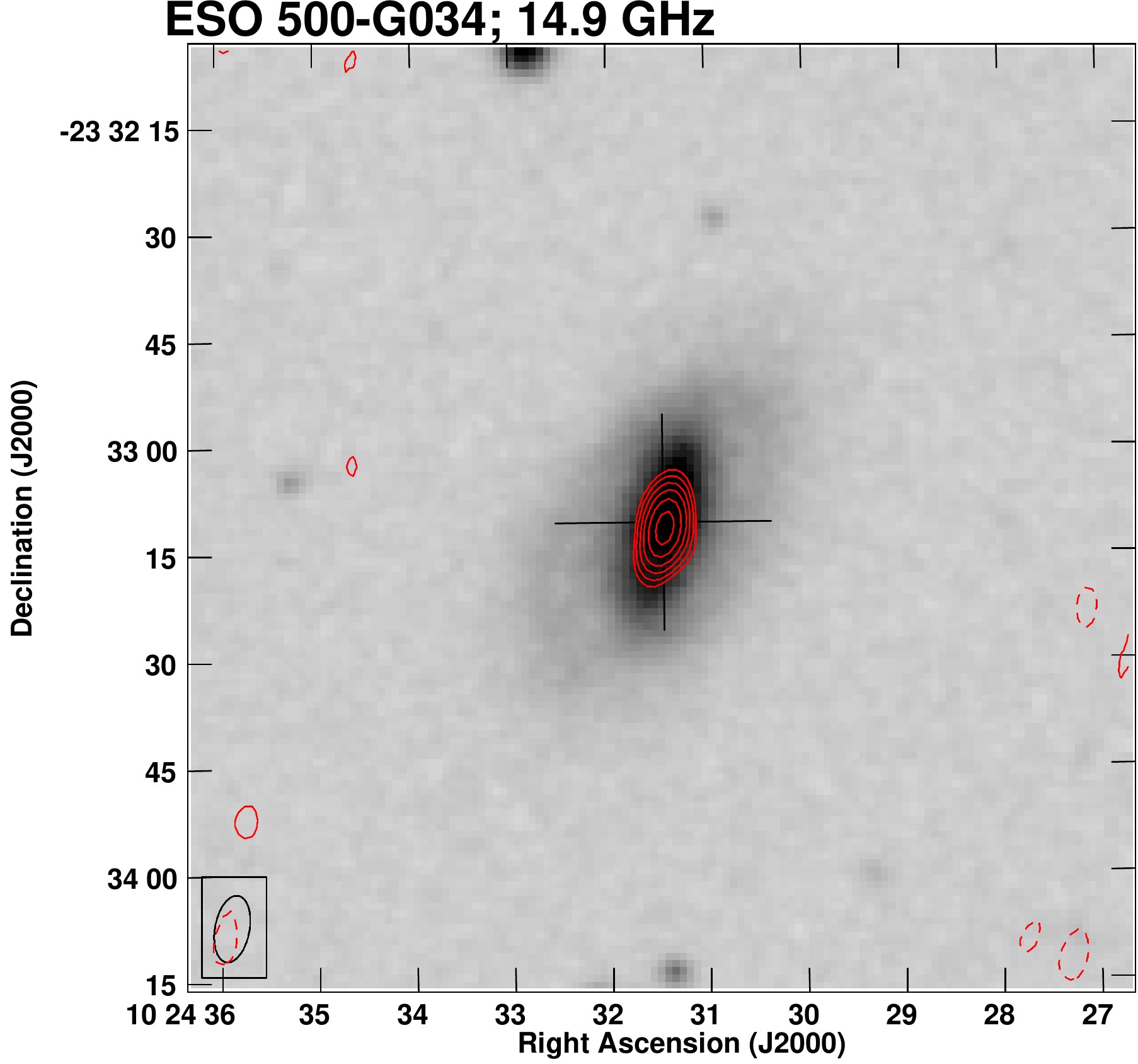}}
   \hspace{0cm}{\includegraphics[width =  0.25\textwidth]{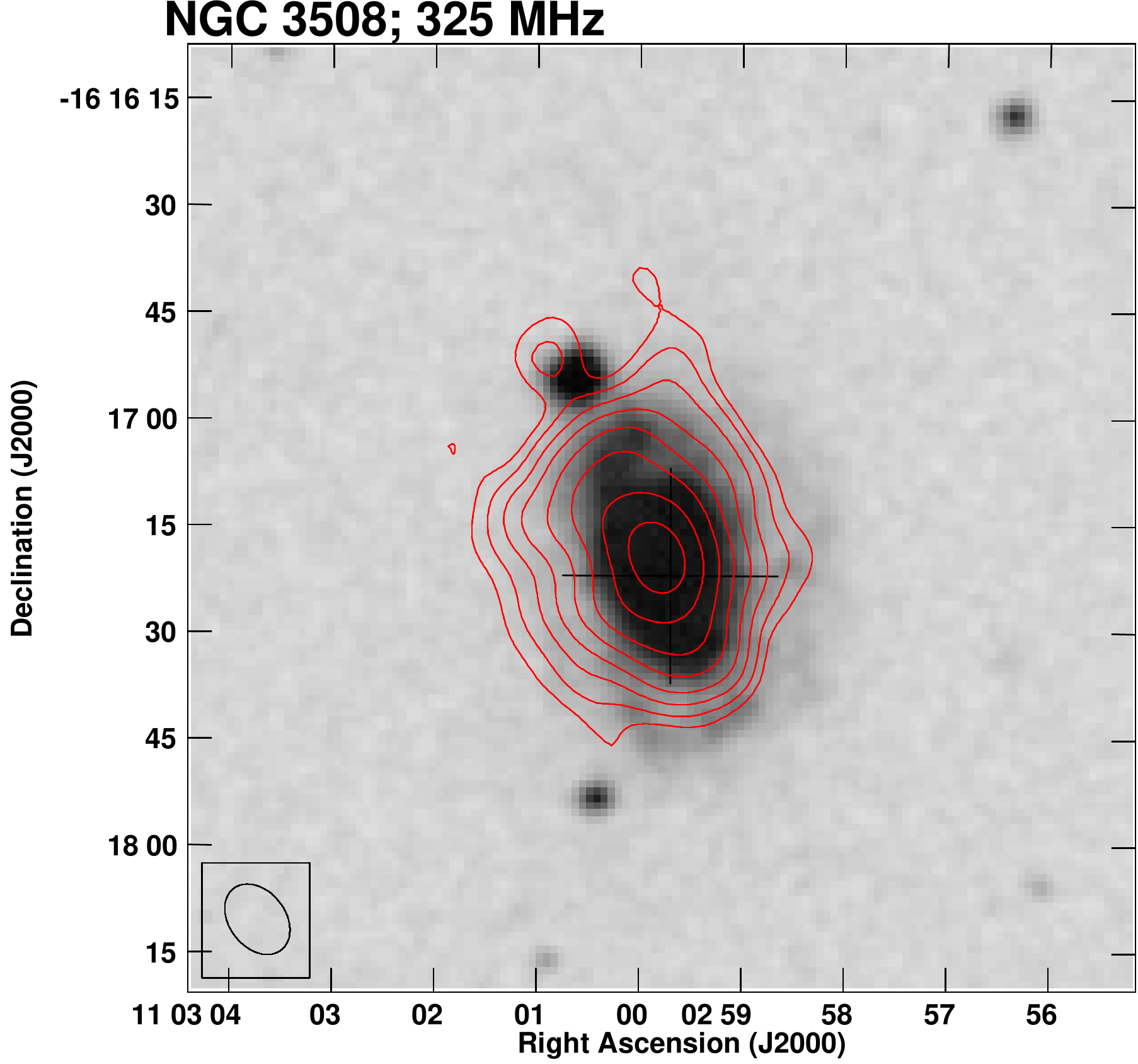}}
   \hspace{0cm}{\includegraphics[width =  0.25\textwidth]{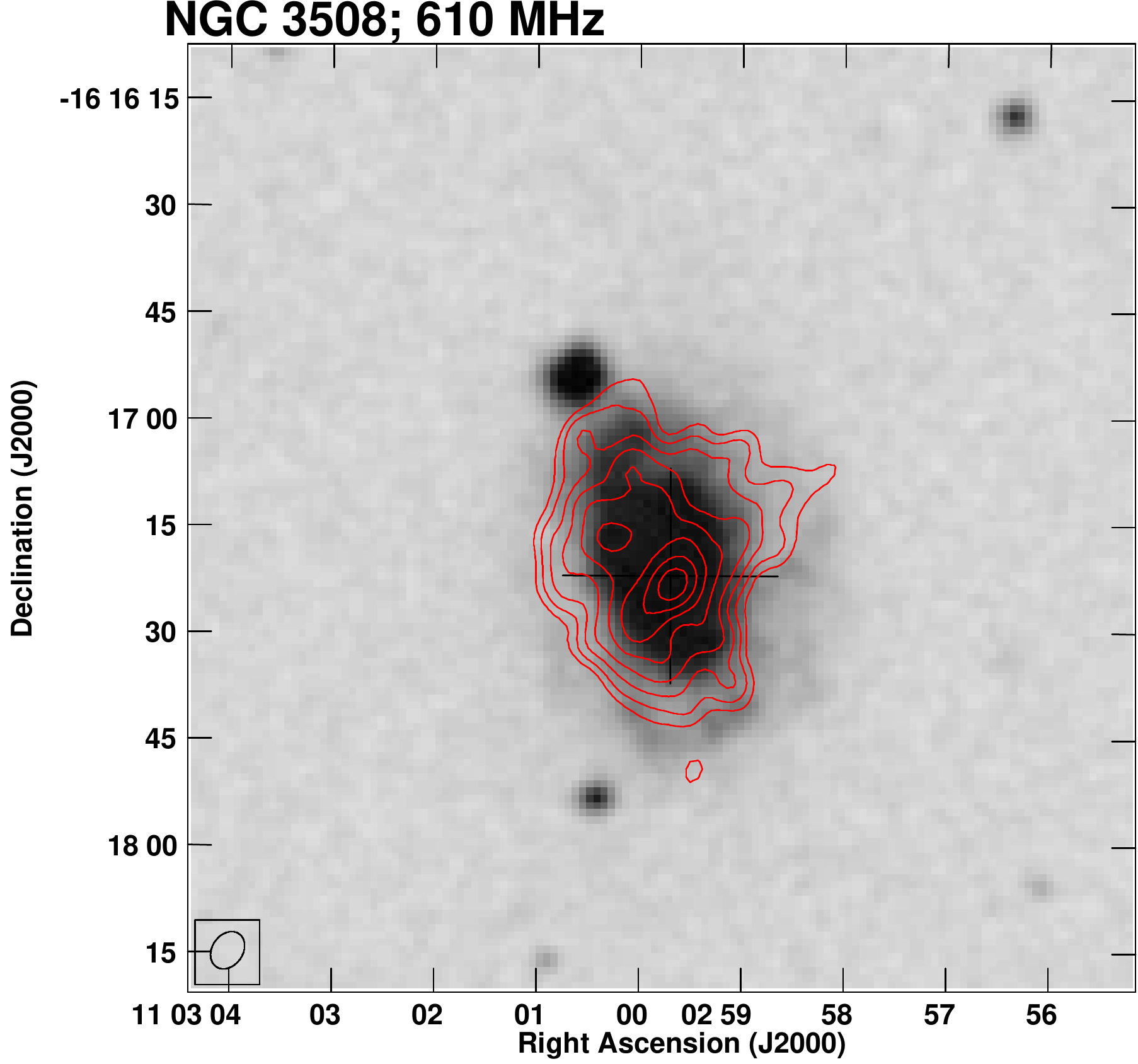}}
   \hspace{0cm}{\includegraphics[width =  0.25\textwidth]{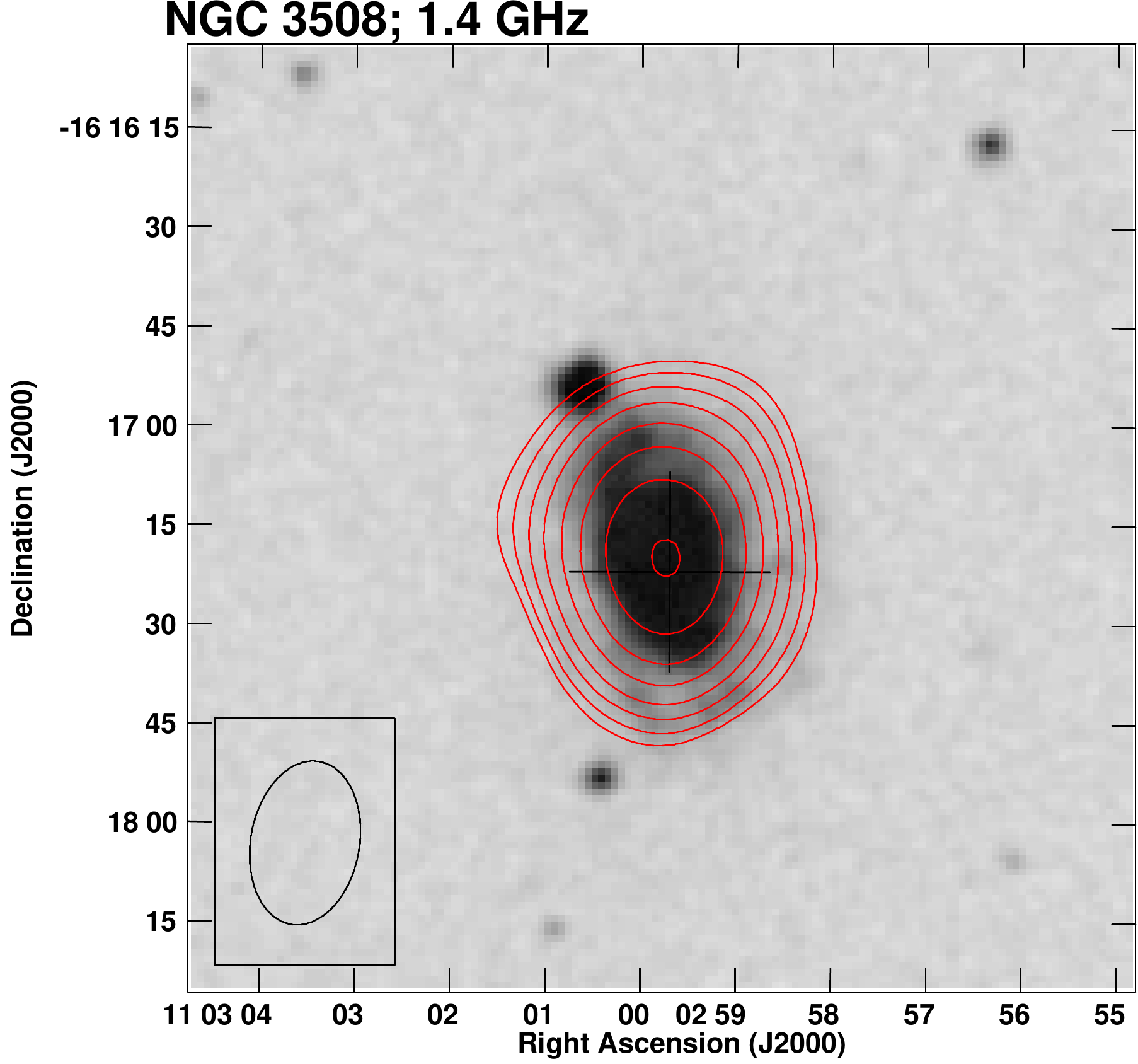}}
   }
   \hbox{
    \hspace{0cm}{\includegraphics[width =  0.25\textwidth]{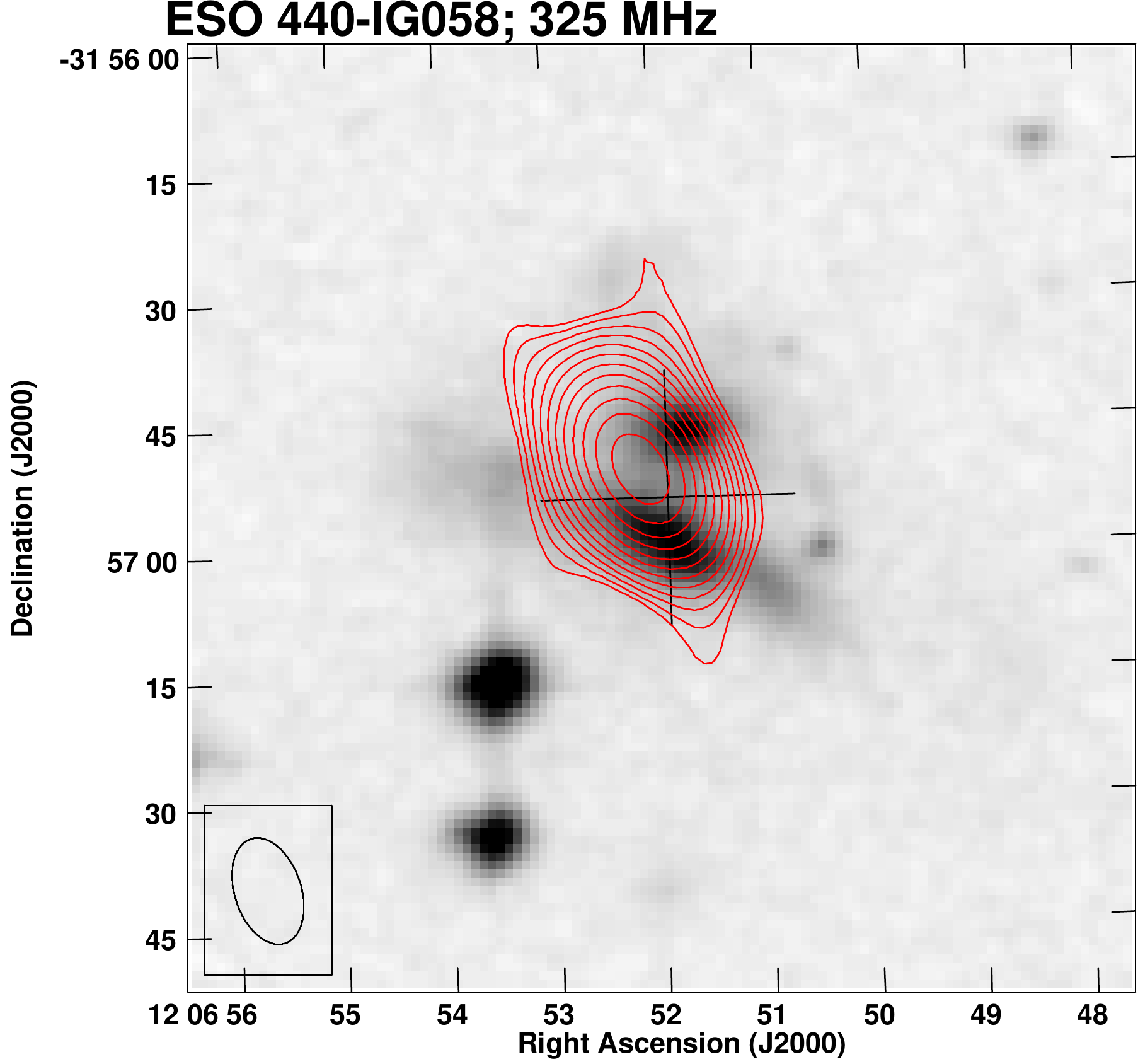}}
    \hspace{0cm}{\includegraphics[width =  0.25\textwidth]{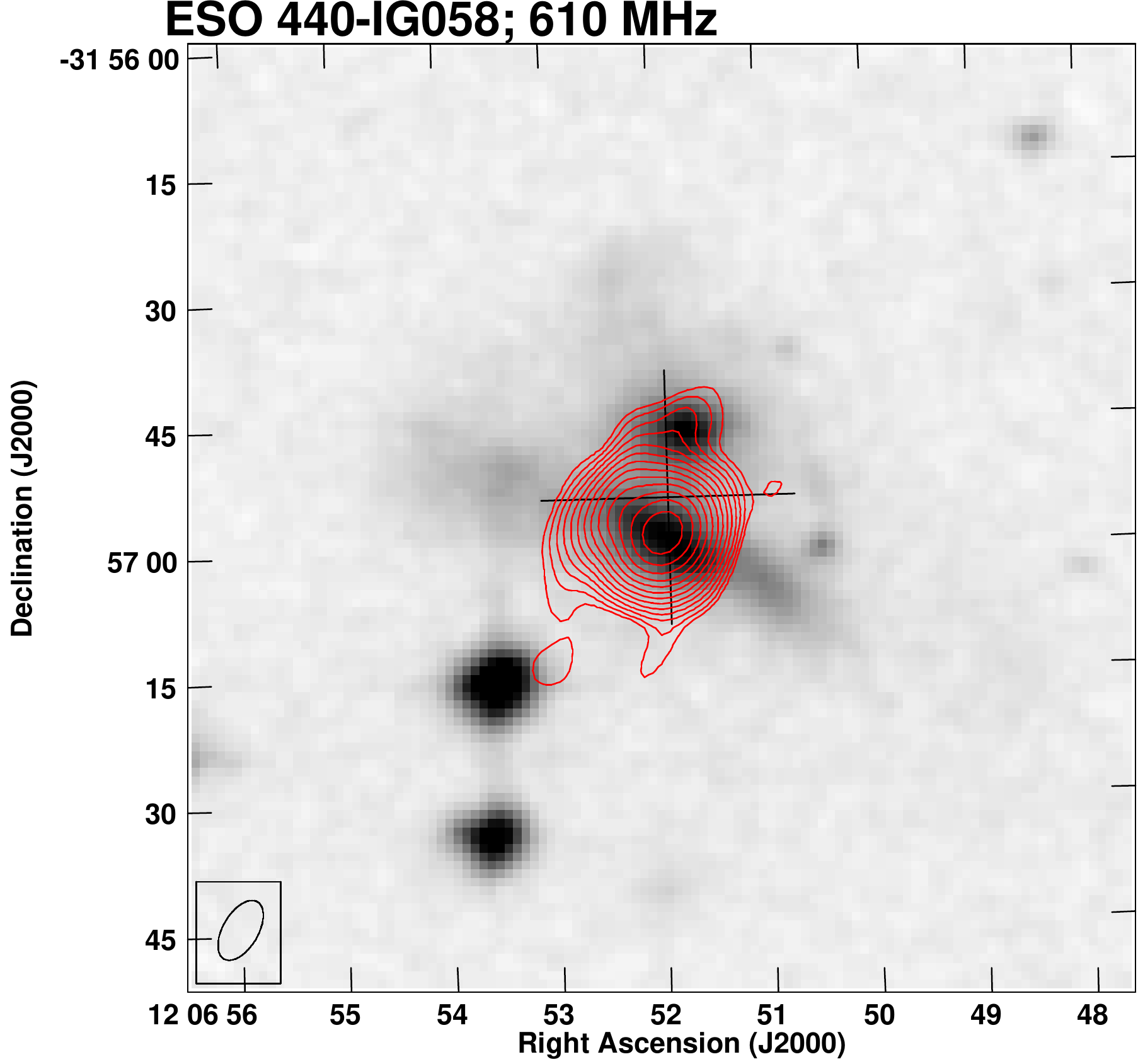}}
    \hspace{0cm}{\includegraphics[width =  0.25\textwidth]{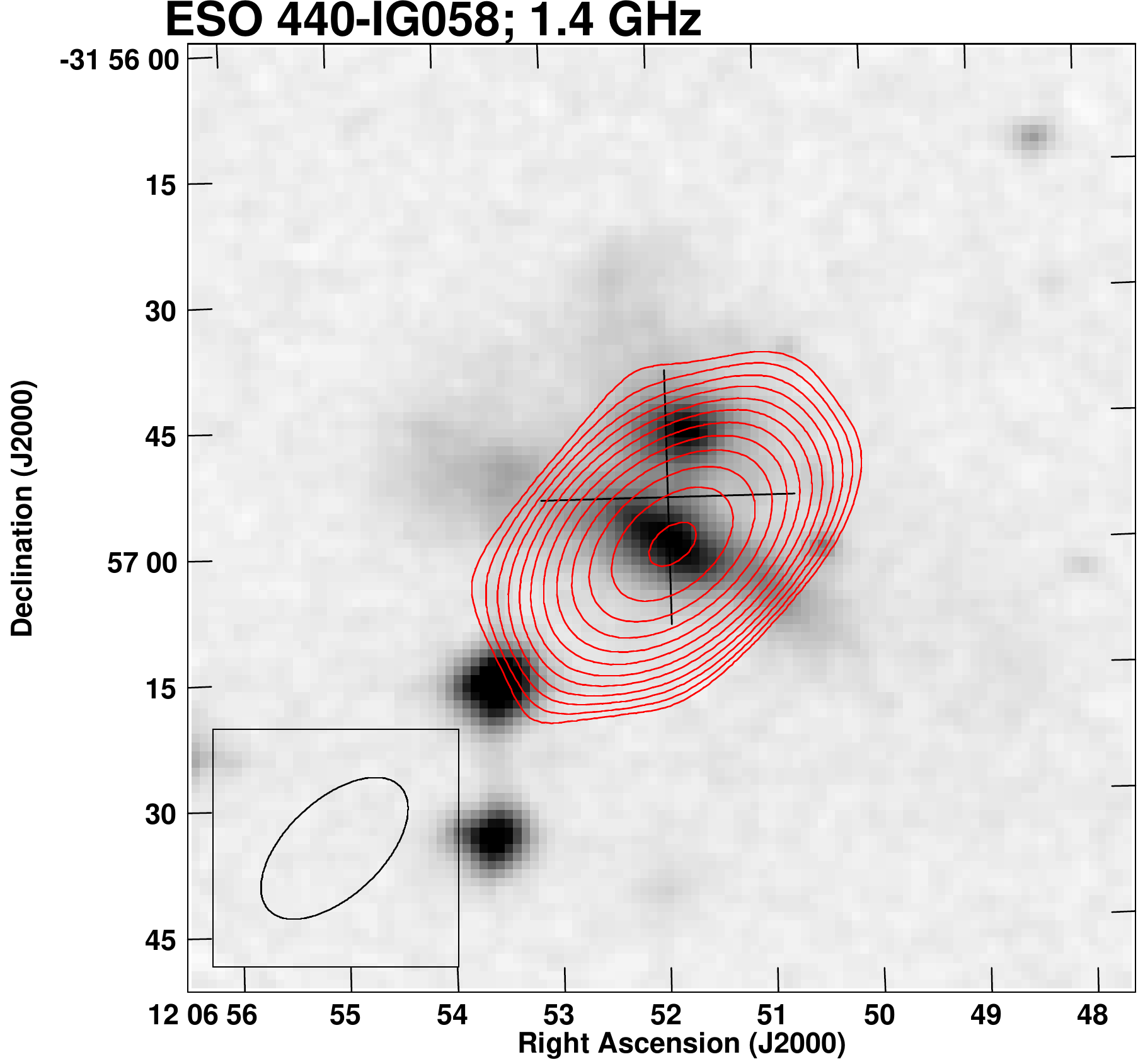}}
    \hspace{0cm}{\includegraphics[width =  0.25\textwidth]{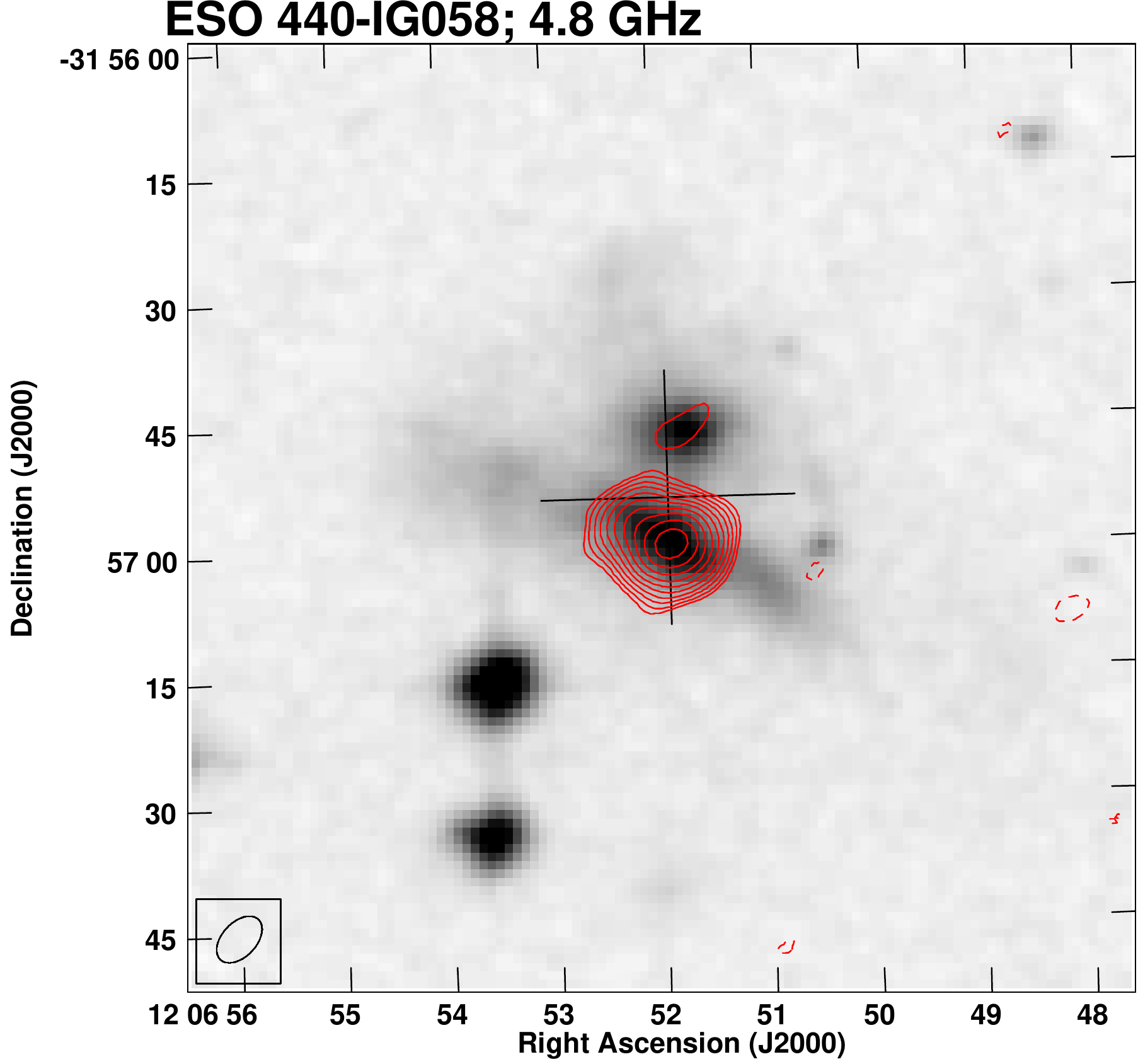}}
    }
    \hbox{
    \hspace{0cm}{\includegraphics[width =  0.25\textwidth]{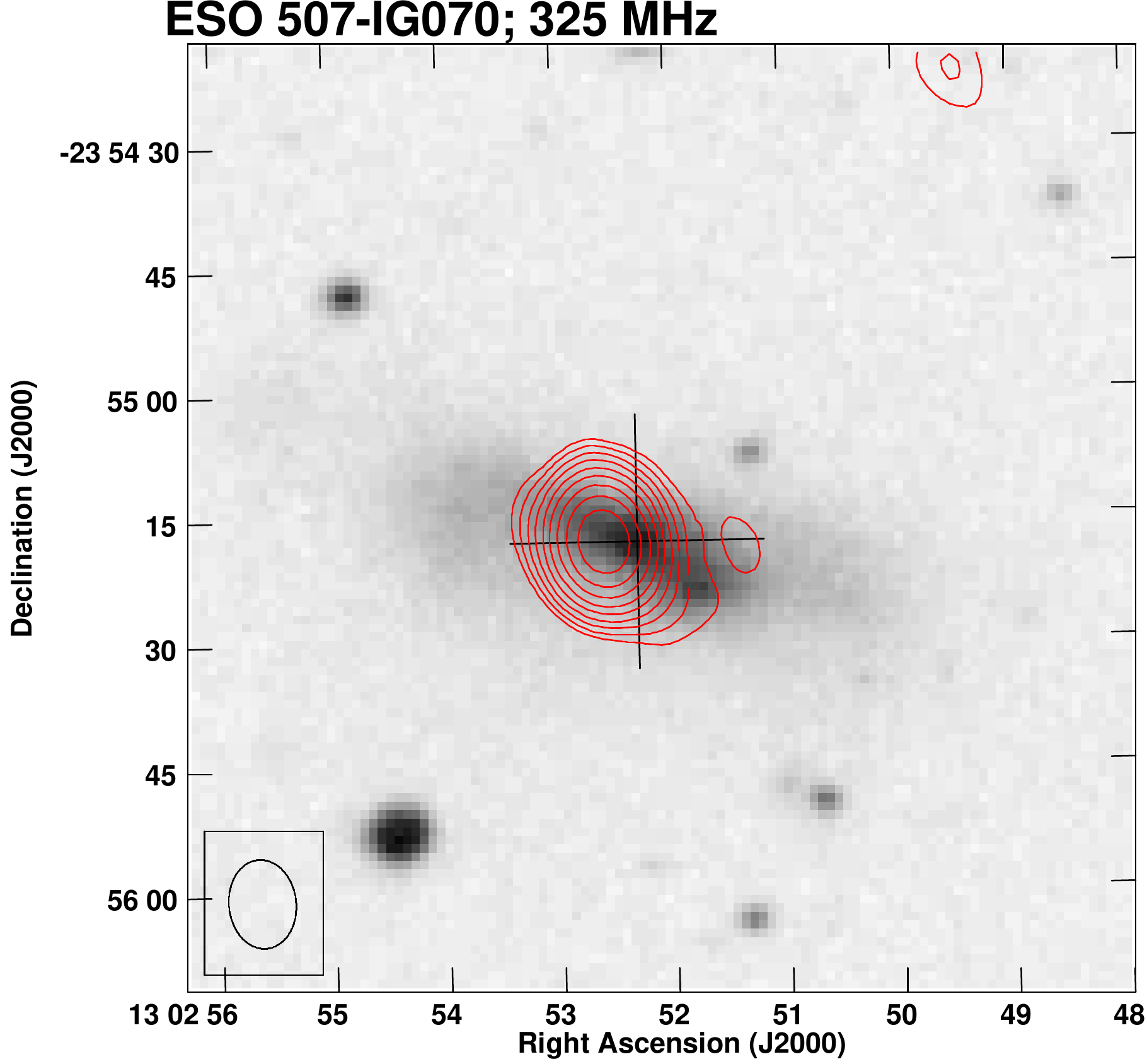}}
    \hspace{0cm}{\includegraphics[width =  0.25\textwidth]{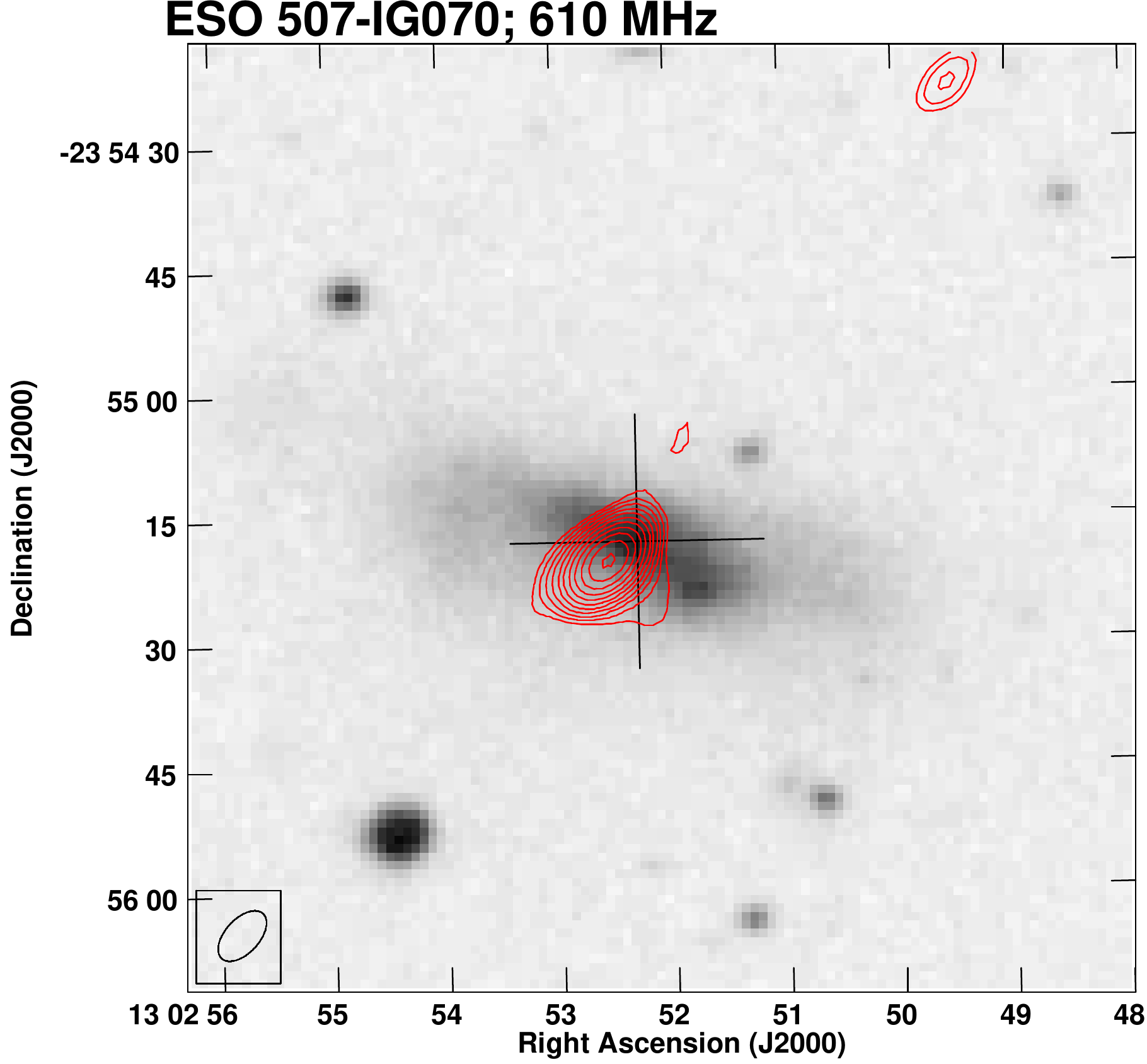}}
    \hspace{0cm}{\includegraphics[width =  0.25\textwidth]{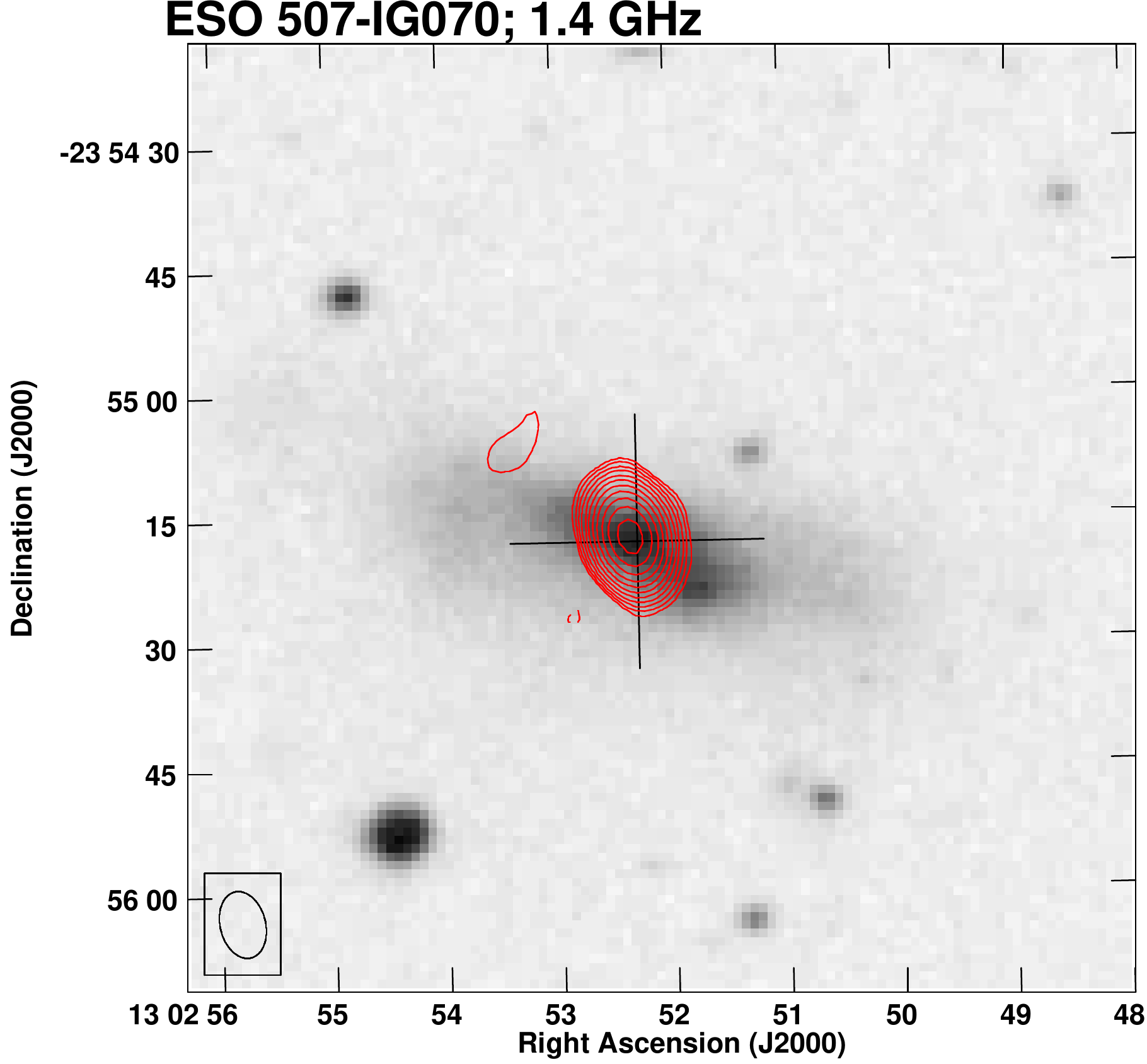}}
    \hspace{0cm}{\includegraphics[width =  0.25\textwidth]{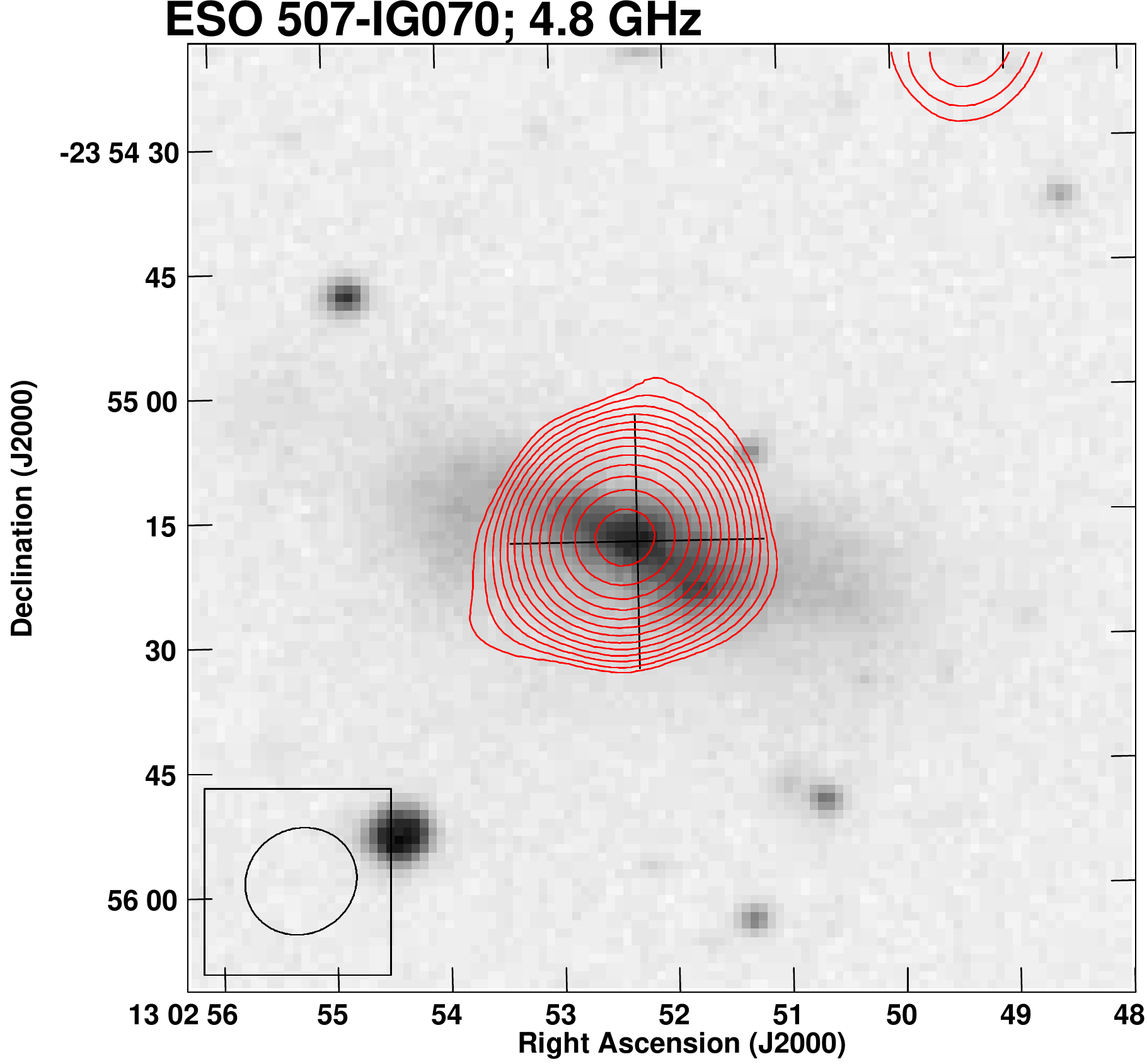}}
    }
    \hbox{
    \hspace{0cm}{\includegraphics[width =  0.25\textwidth]{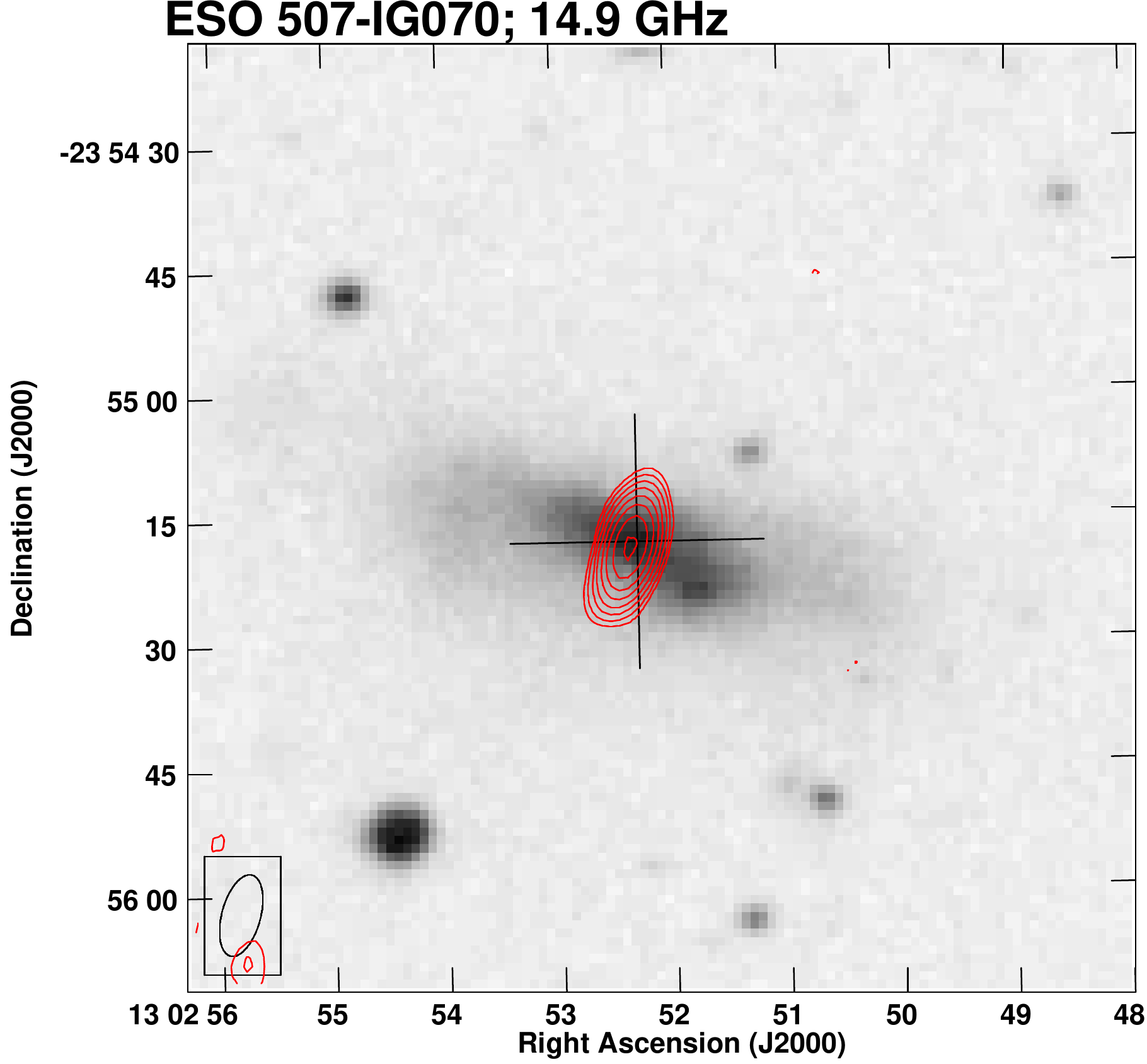}}
   \hspace{0cm}{\includegraphics[width =  0.25\textwidth]{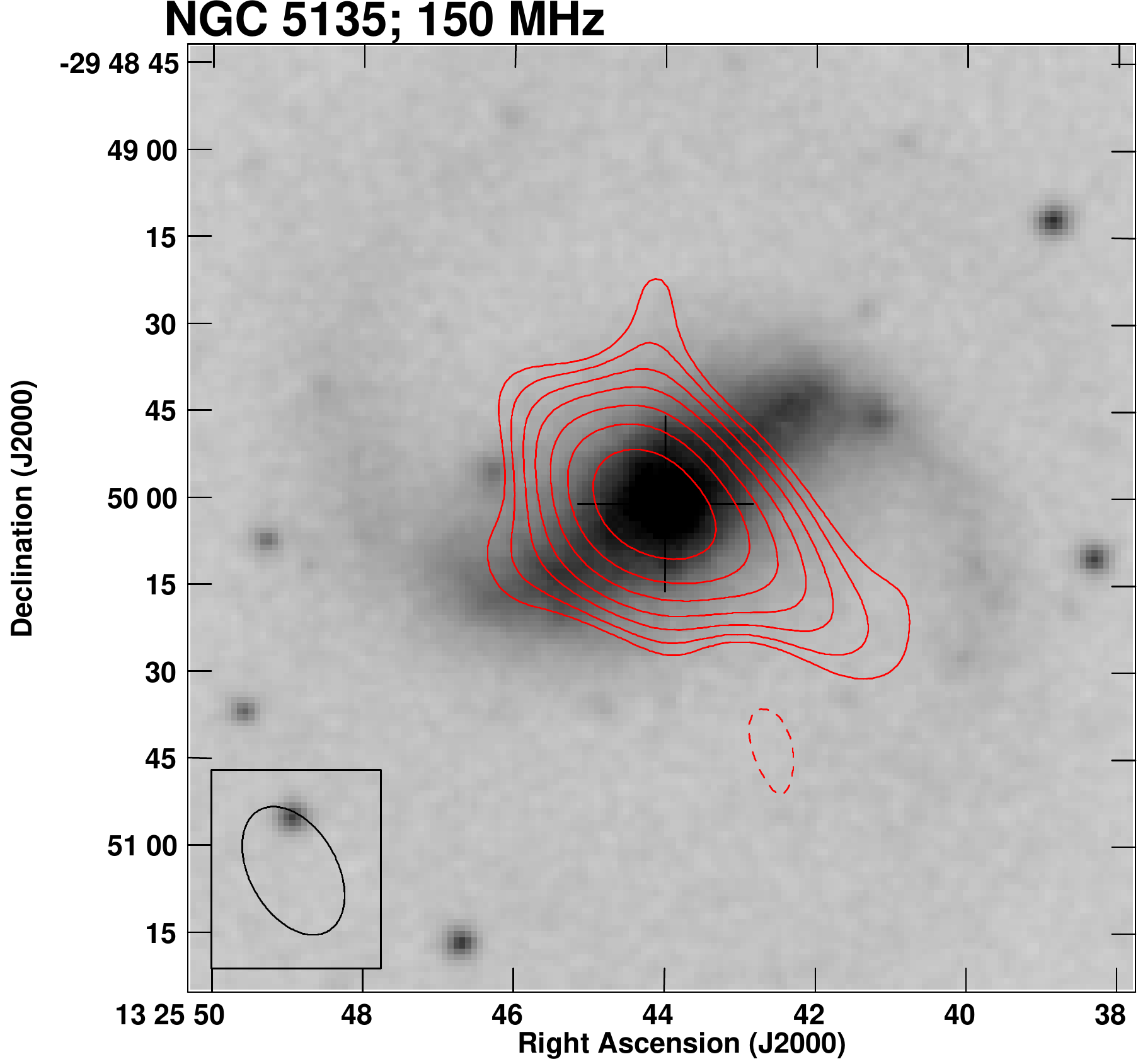}}
   \hspace{0cm}{\includegraphics[width =  0.25\textwidth]{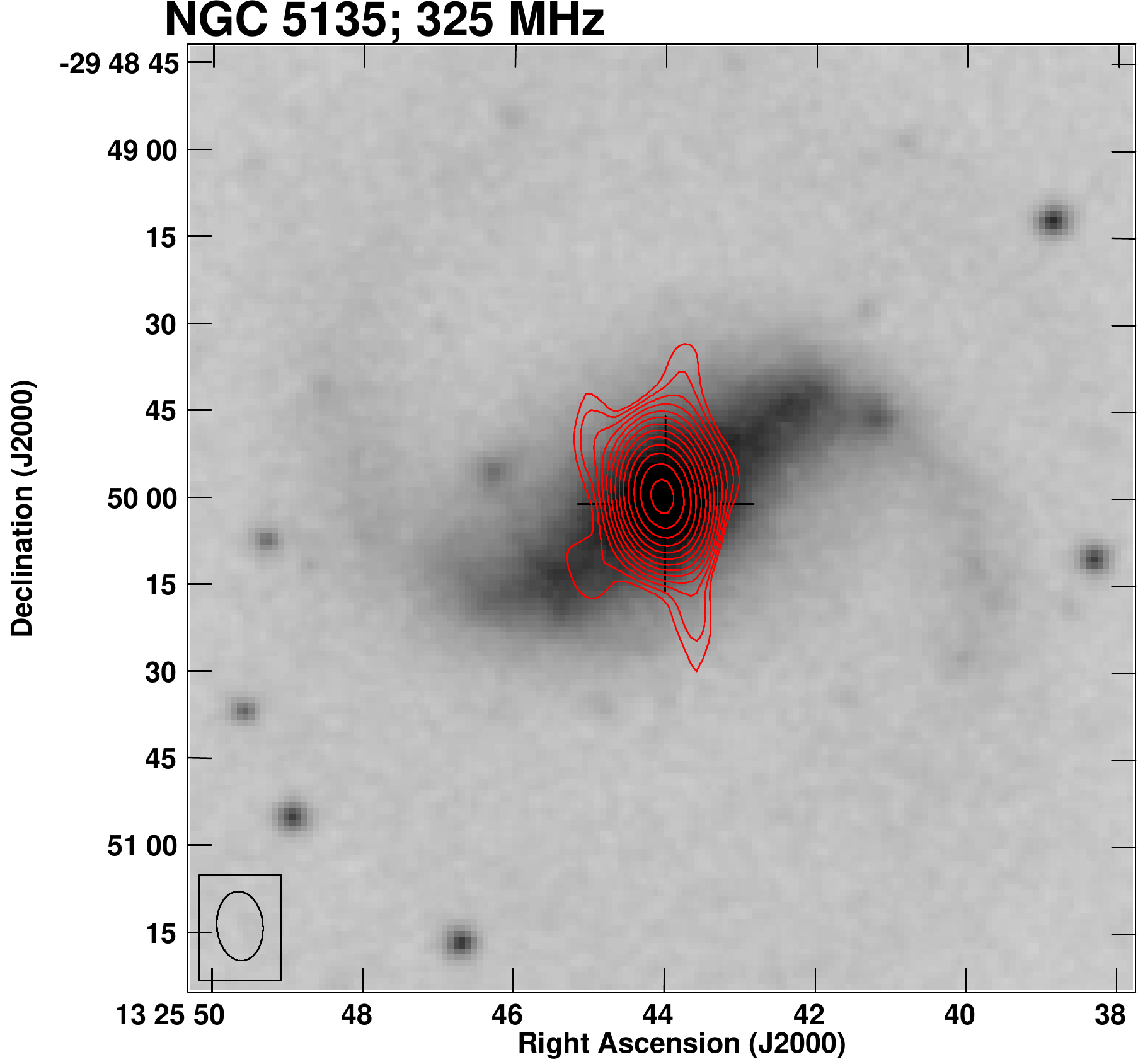}}
   \hspace{0cm}{\includegraphics[width =  0.25\textwidth]{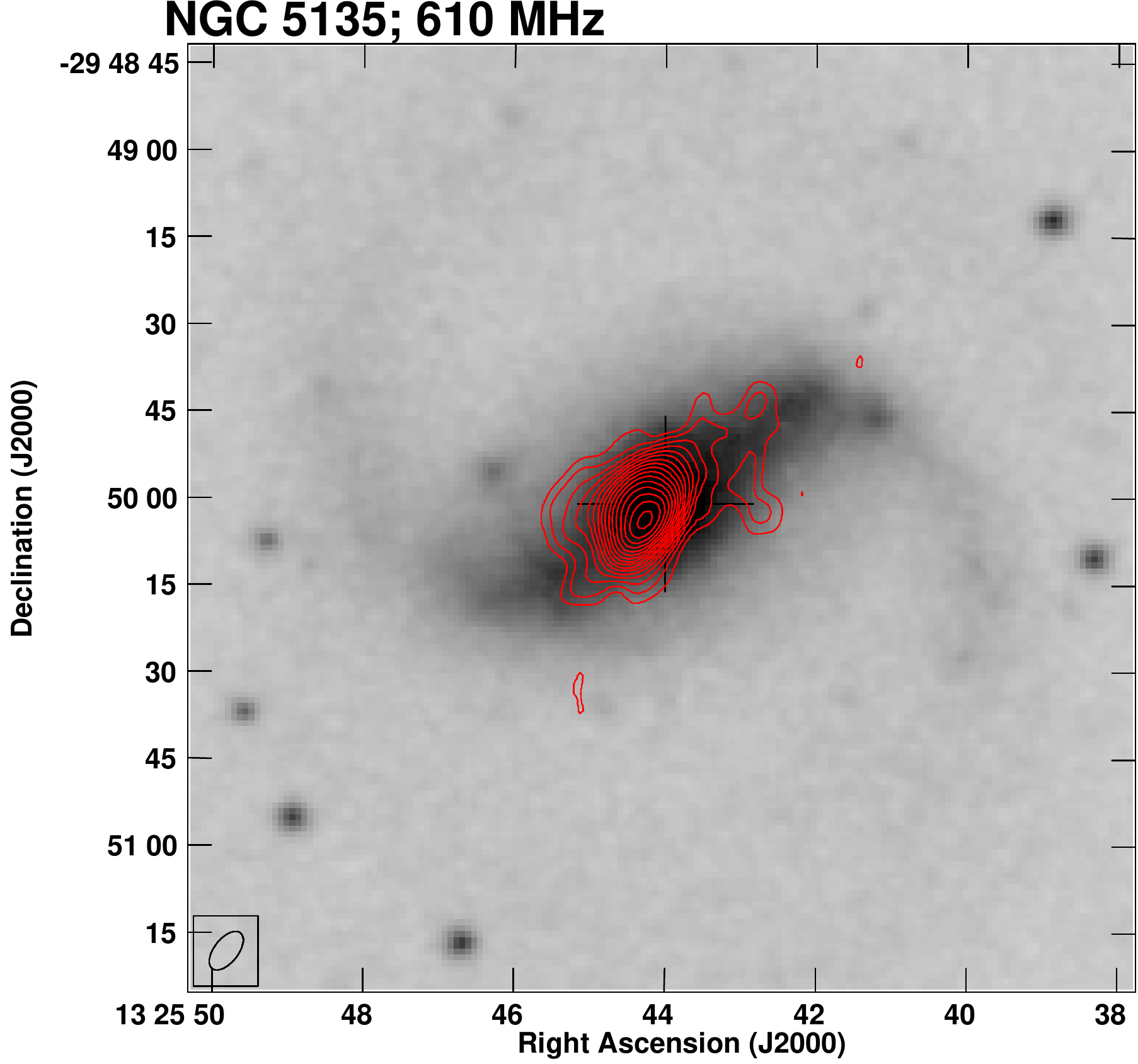}}
   }
    \caption{150\,MHz-15\,GHz radio contour images (red) overlaid on the optical DSS2 R-band images (grey scale) for the samples of 11 LIRGs analysed by us. The radio contours begin at 3$\sigma$ of the rms noise on the radio map (Column 8 of Tab.~\ref{tab:radiosummary}) and increase by $(\sqrt2)^n$ where $n$ ranges from 0,1,2,..20. The -3$\sigma$ radio intensity is shown by dashed contours and the synthesized beam achieved is given at the bottom left corner of the image. } 
    \label{fig:overlays}
\end{figure*}

\addtocounter{figure}{-1}
\begin{figure*}
   \hbox{
    \hspace{0cm}{\includegraphics[width =  0.25\textwidth]{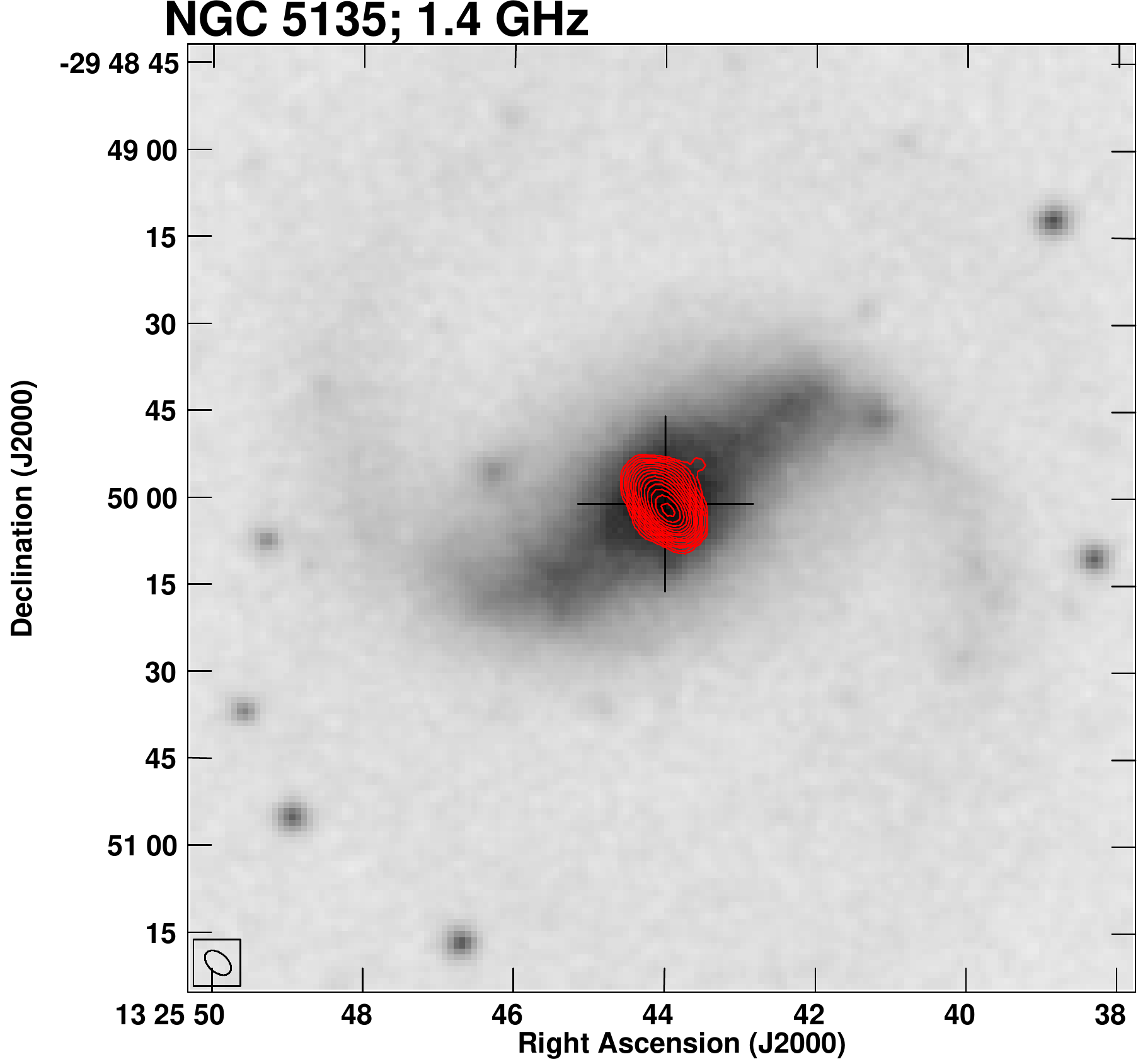}}
   \hspace{0cm}{\includegraphics[width =  0.25\textwidth]{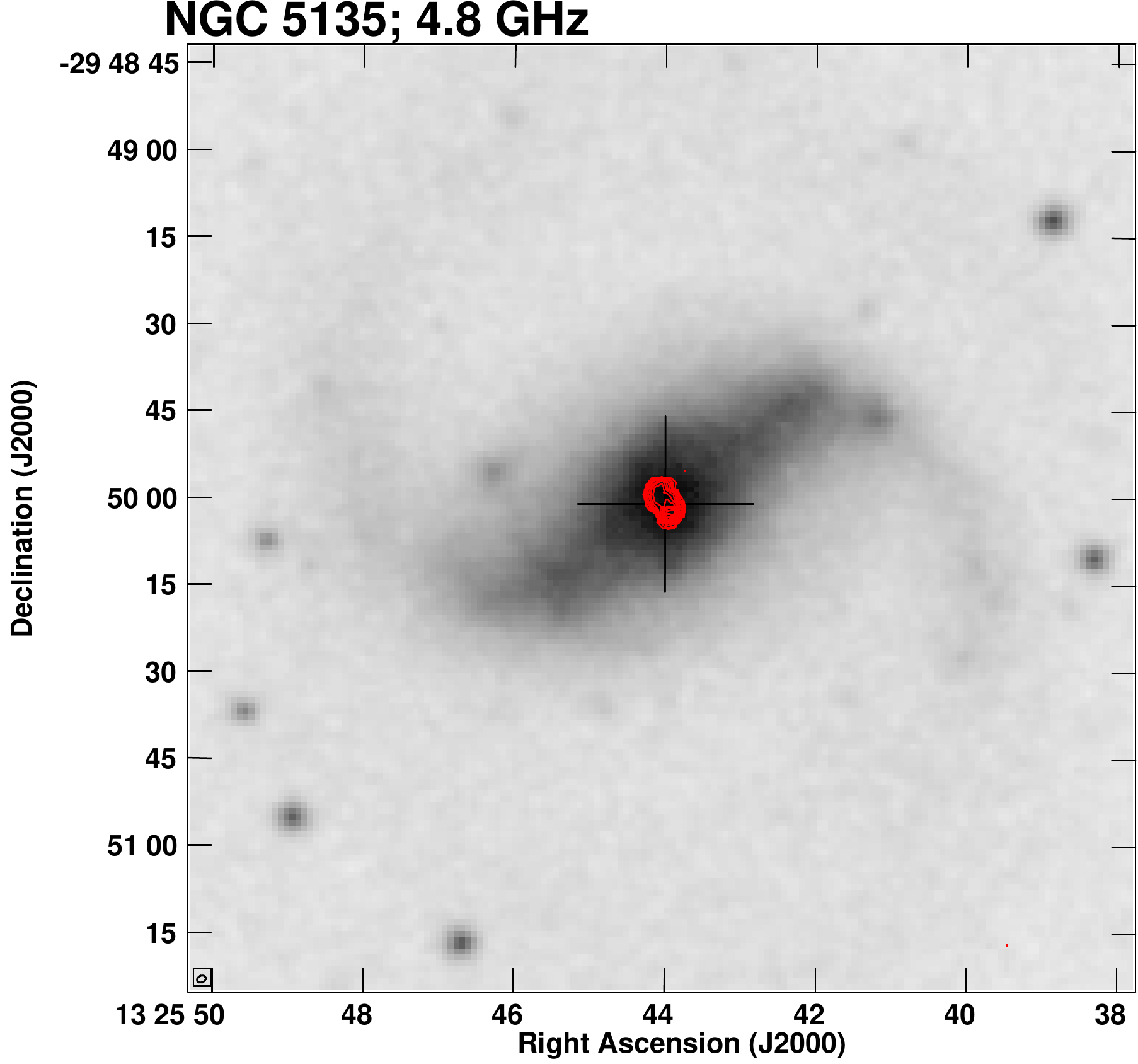}}
    \hspace{0cm}{\includegraphics[width =  0.25\textwidth]{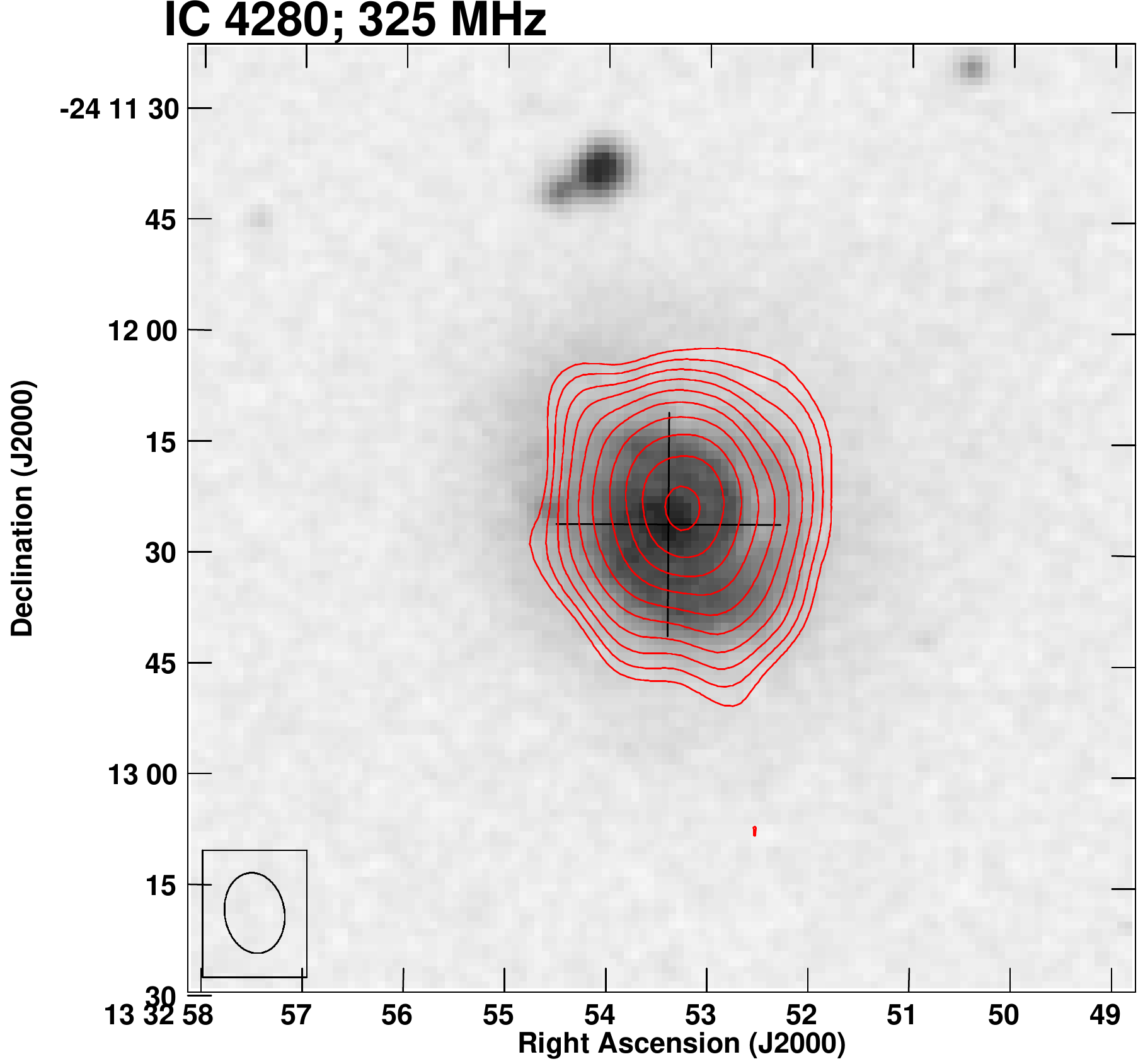}}
    \hspace{0cm}{\includegraphics[width =  0.25\textwidth]{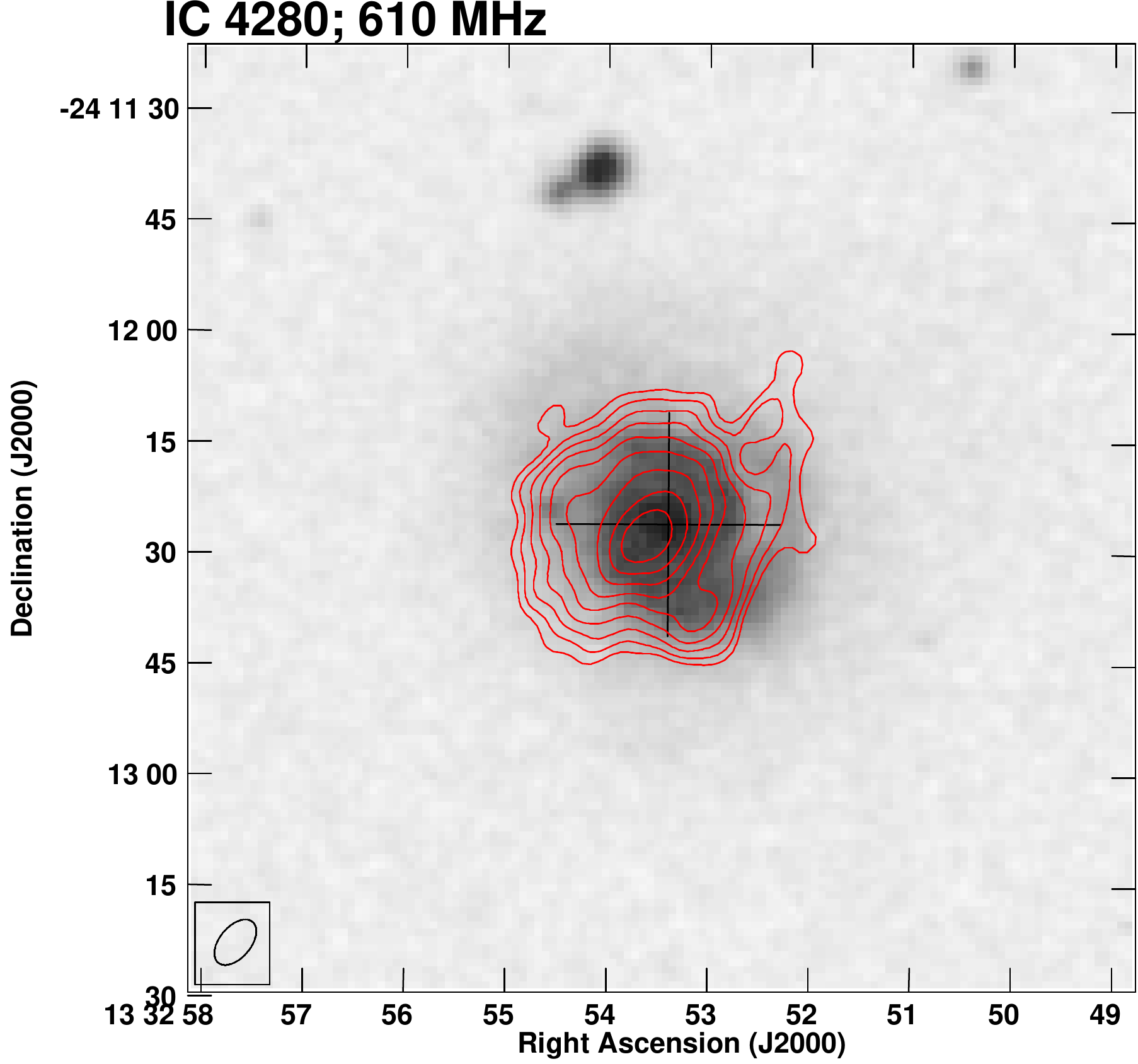}}
    }
    \hbox{
    \hspace{0cm}{\includegraphics[width =  0.25\textwidth]{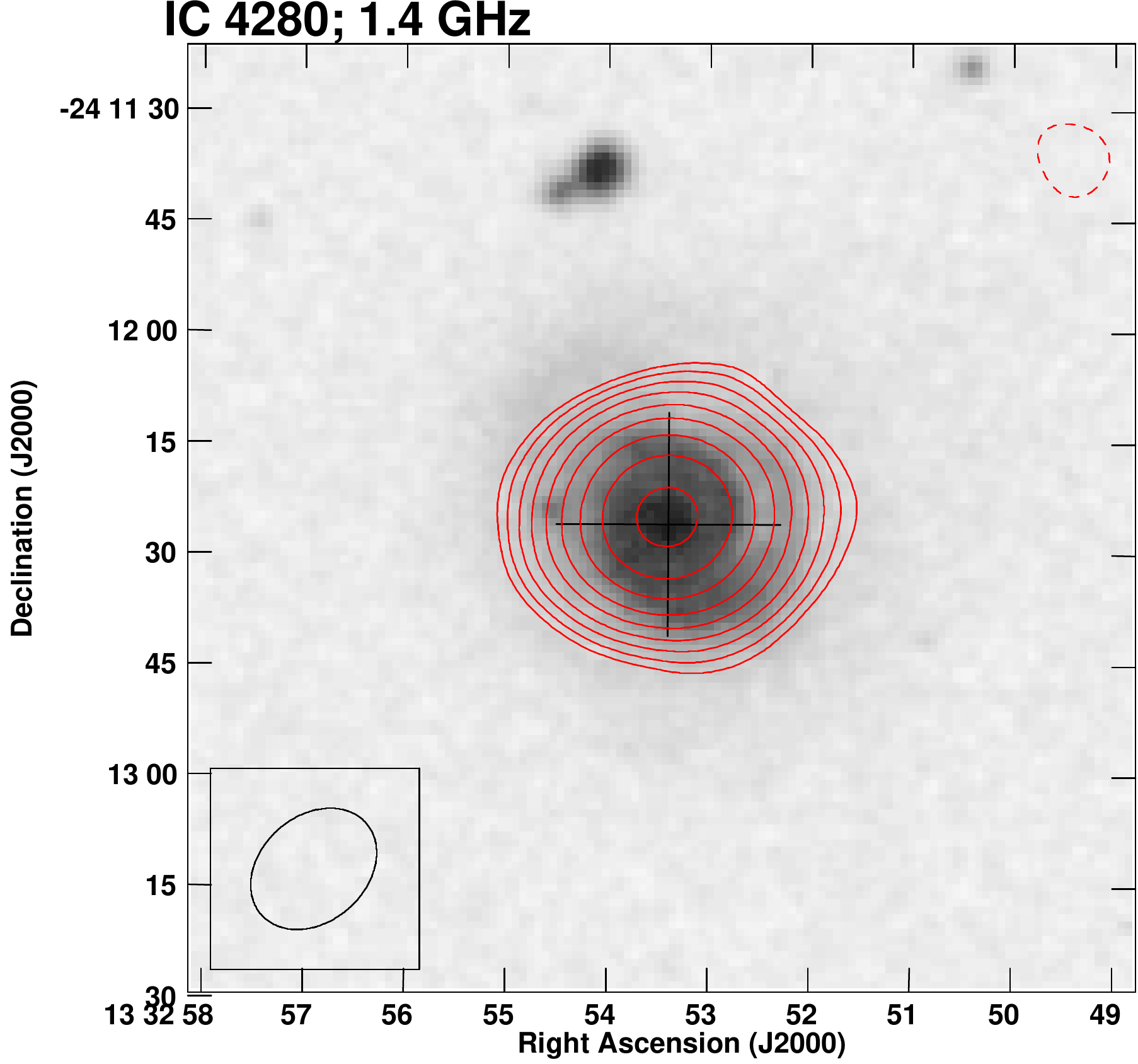}}
    \hspace{0cm}{\includegraphics[width =  0.25\textwidth]{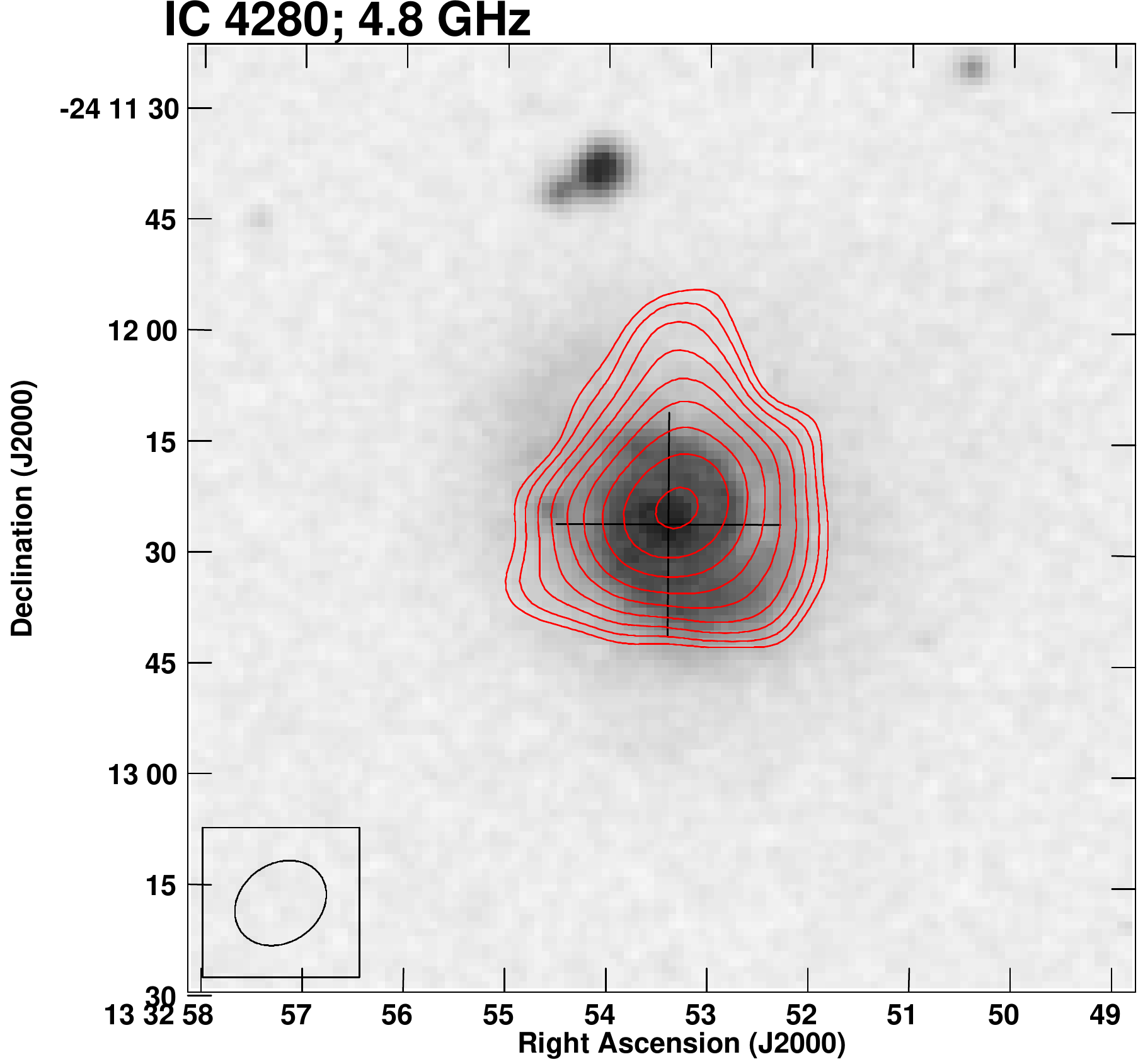}}
    \hspace{0cm}{\includegraphics[width =  0.25\textwidth]{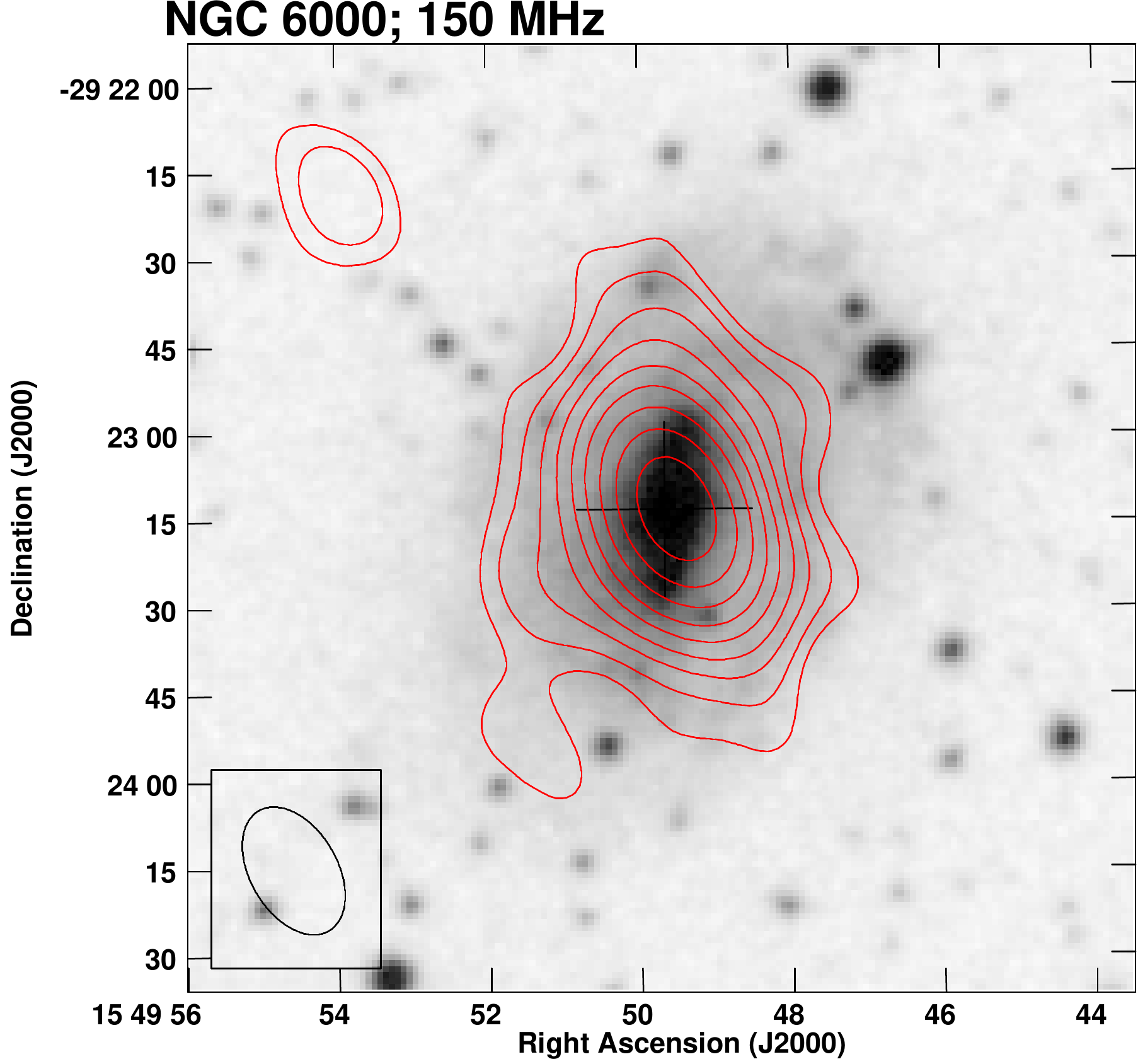}}
    \hspace{0cm}{\includegraphics[width =  0.25\textwidth]{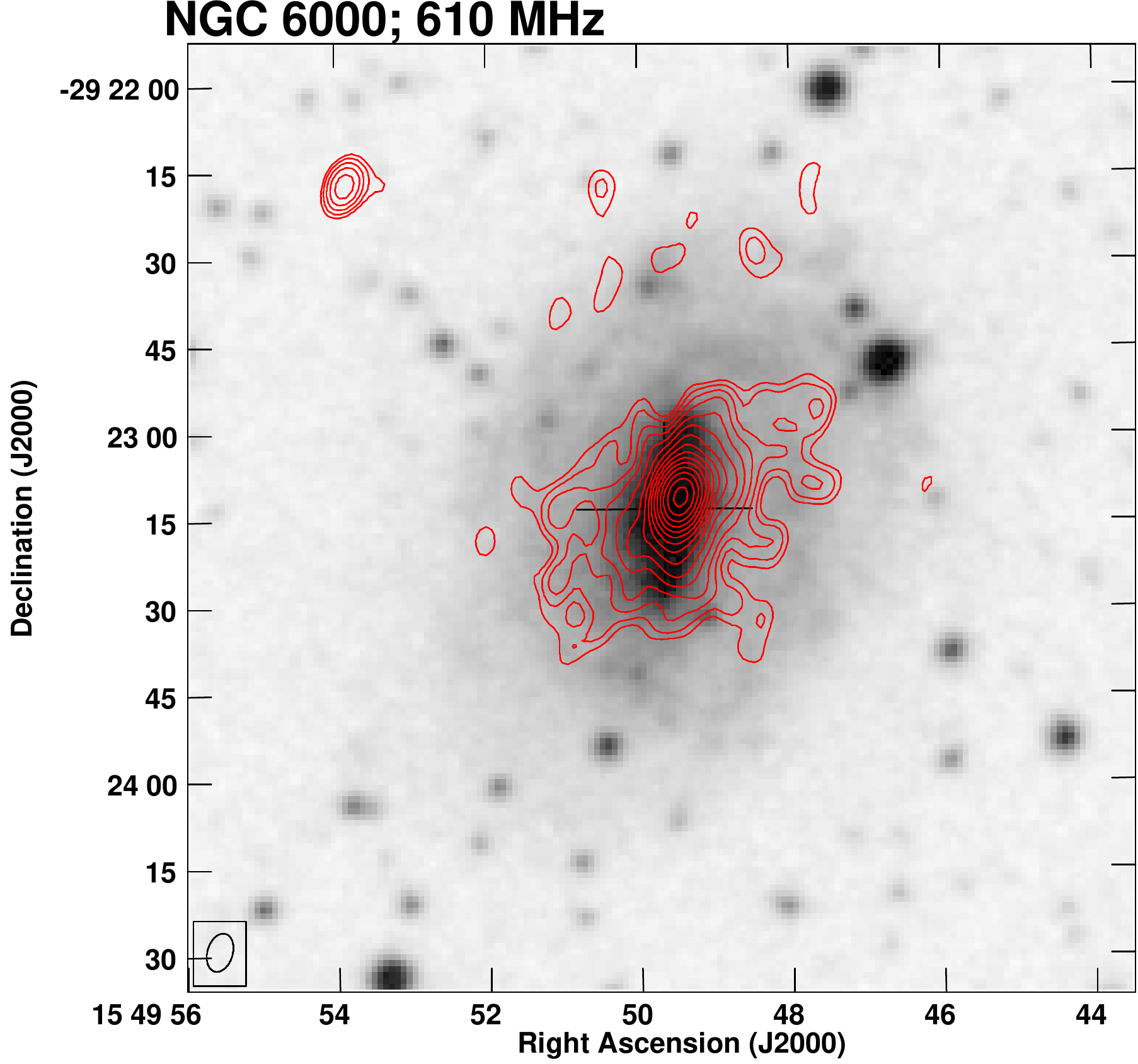}}
    }
    \hbox{
    \hspace{0cm}{\includegraphics[width =  0.25\textwidth]{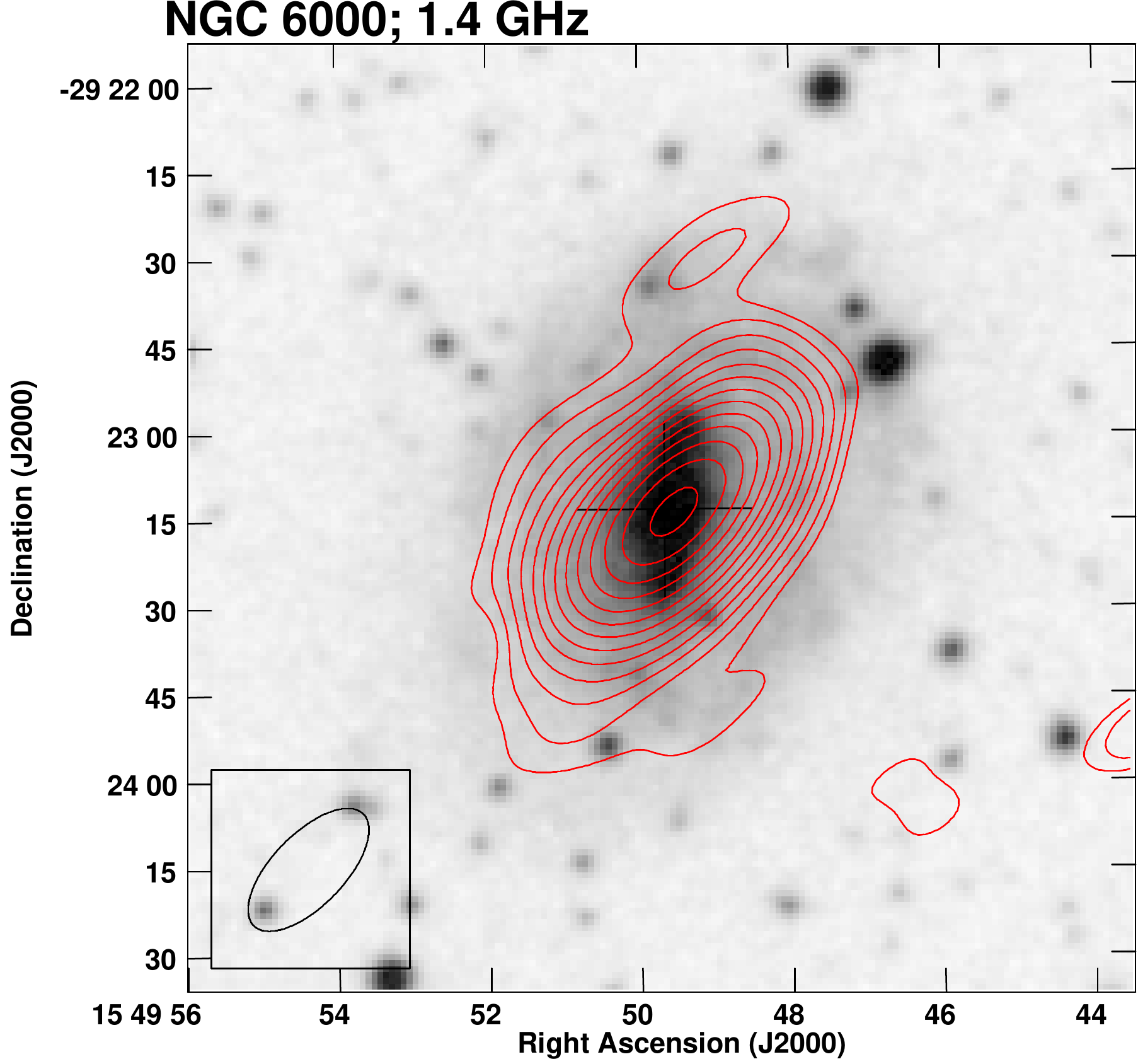}}
    \hspace{0cm}{\includegraphics[width =  0.25\textwidth]{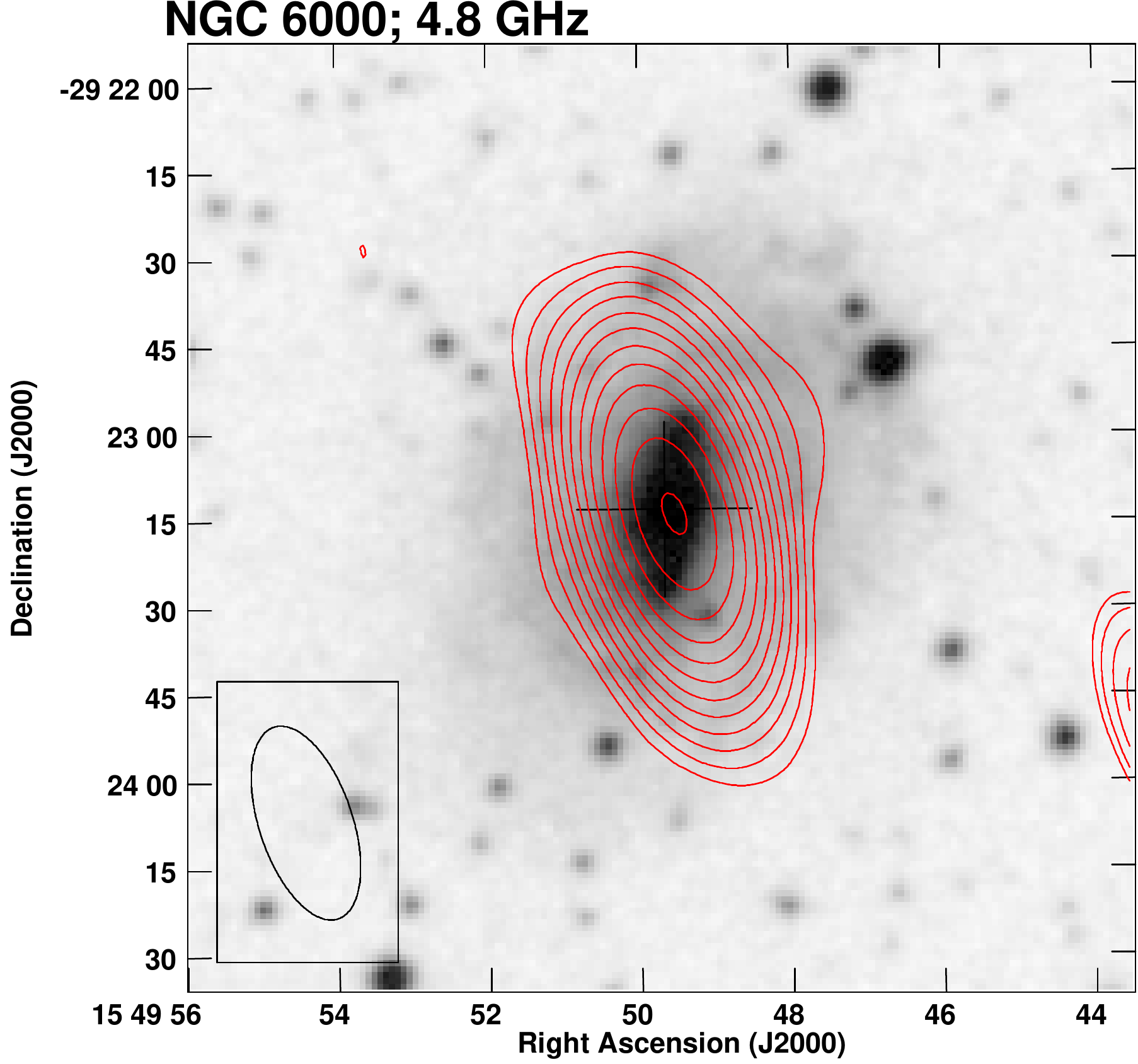}}
    \hspace{0cm}{\includegraphics[width =  0.25\textwidth]{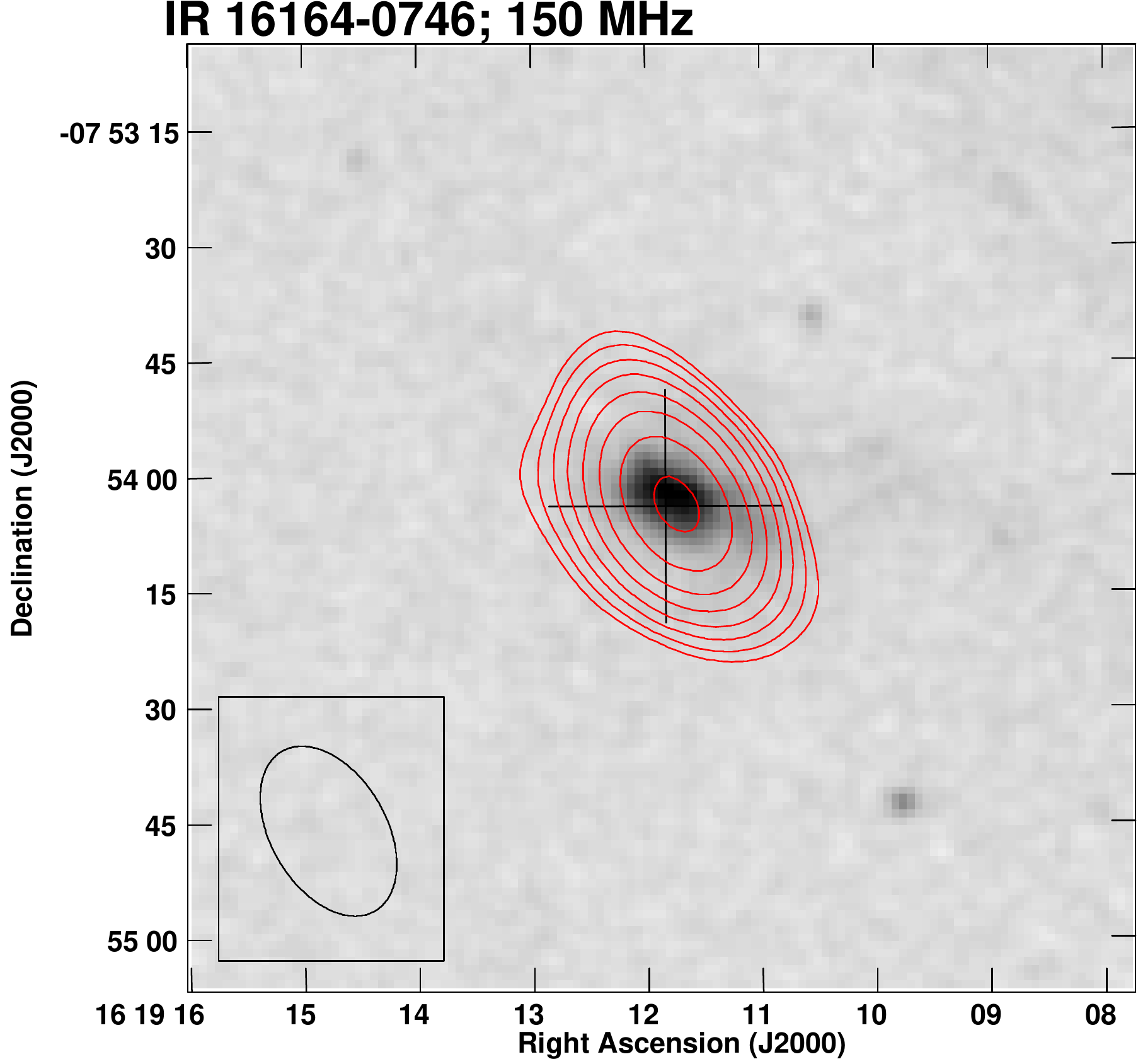}}
    \hspace{0cm}{\includegraphics[width =  0.25\textwidth]{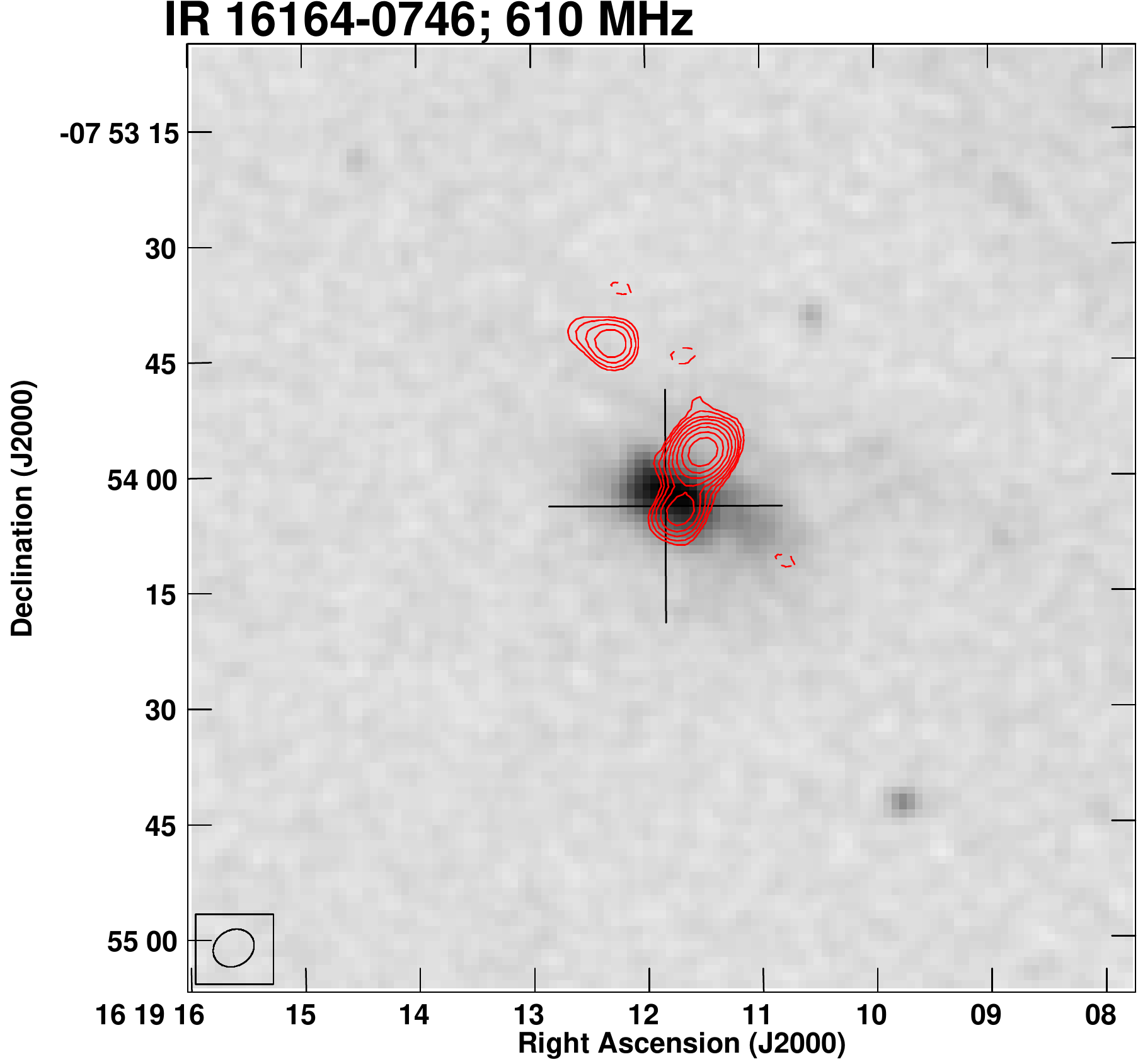}}
    }
    \hbox{
    \hspace{0cm}{\includegraphics[width =  0.25\textwidth]{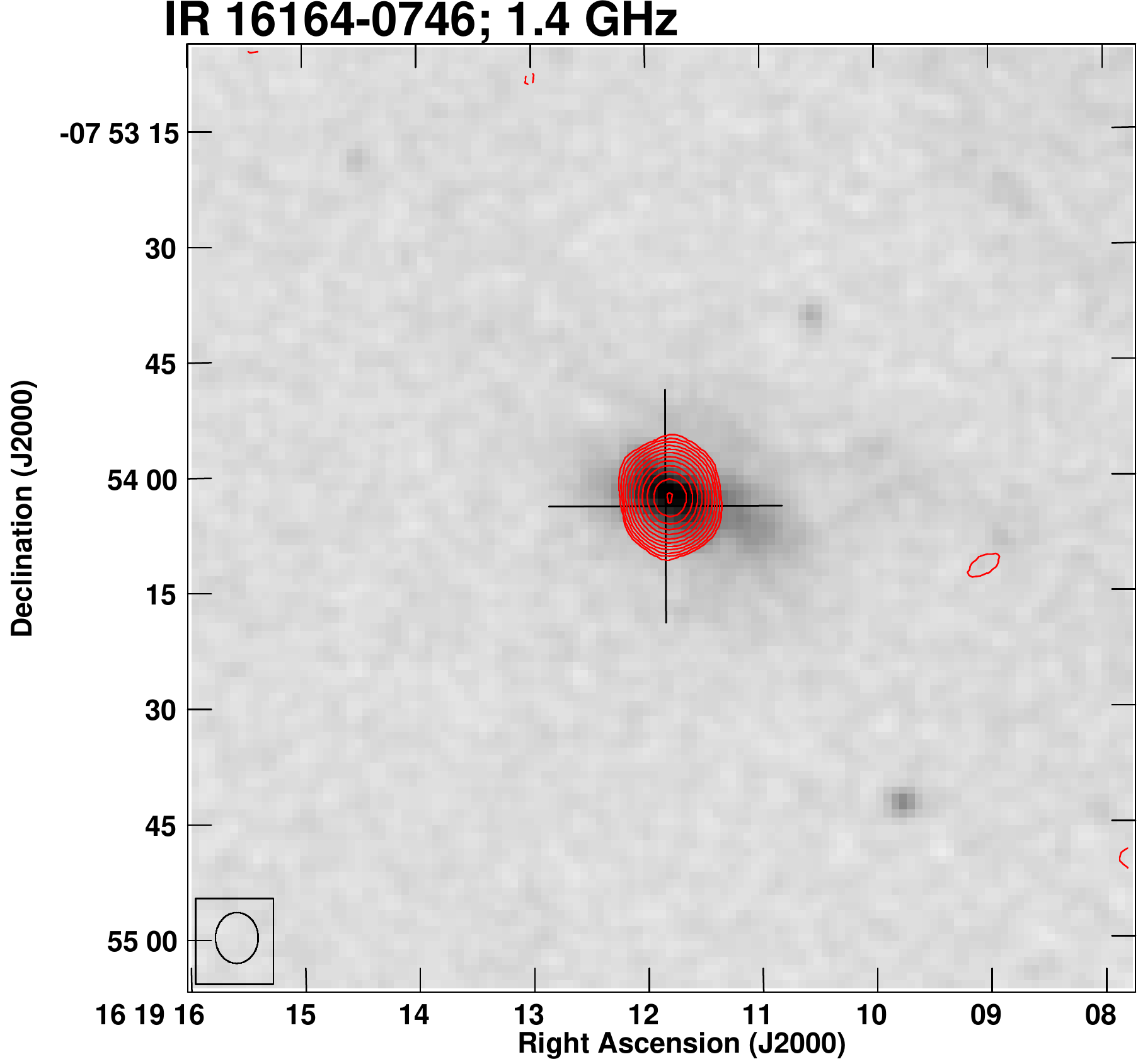}}
    \hspace{0cm}{\includegraphics[width =  0.25\textwidth]{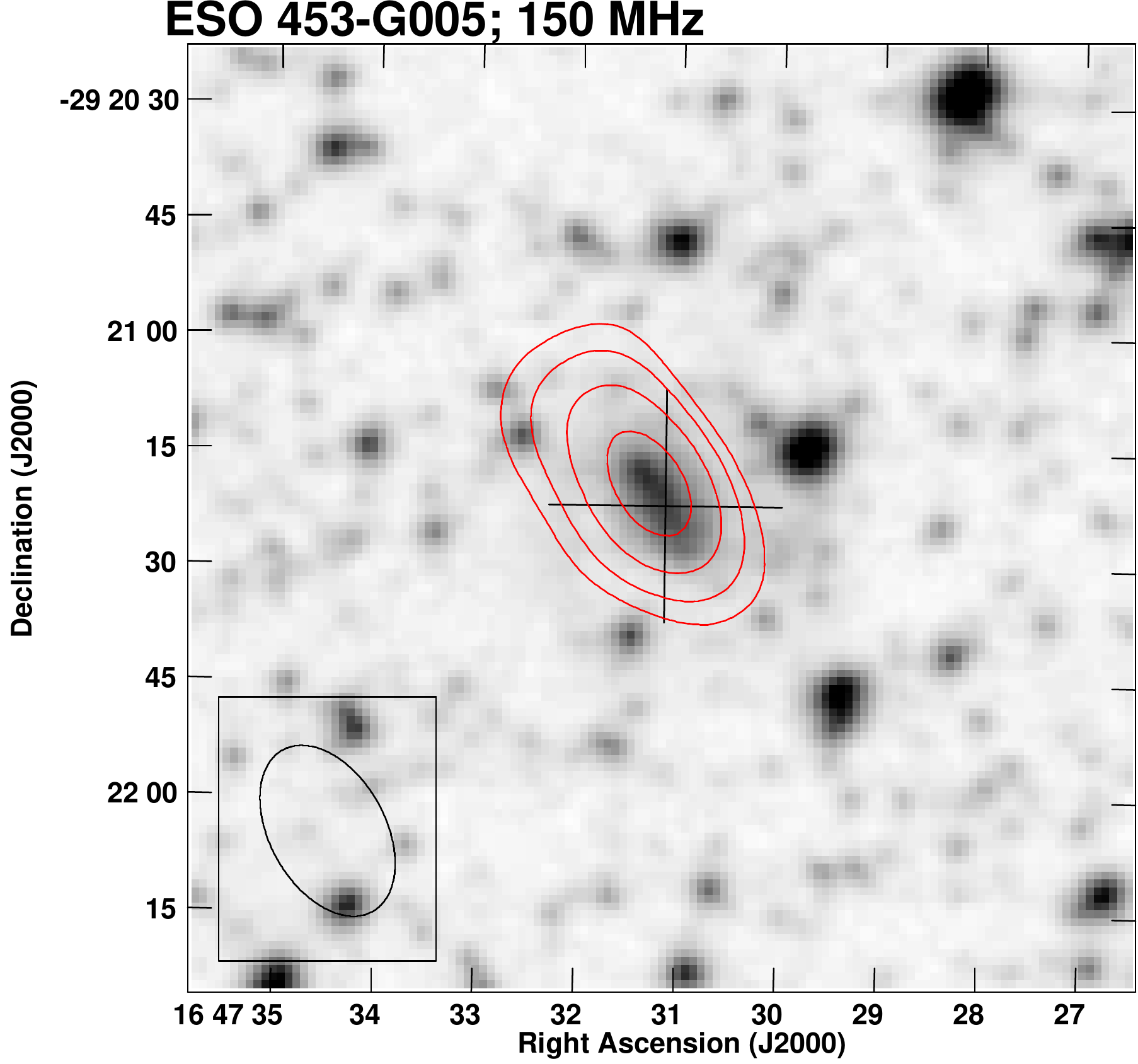}}
    \hspace{0cm}{\includegraphics[width =  0.25\textwidth]{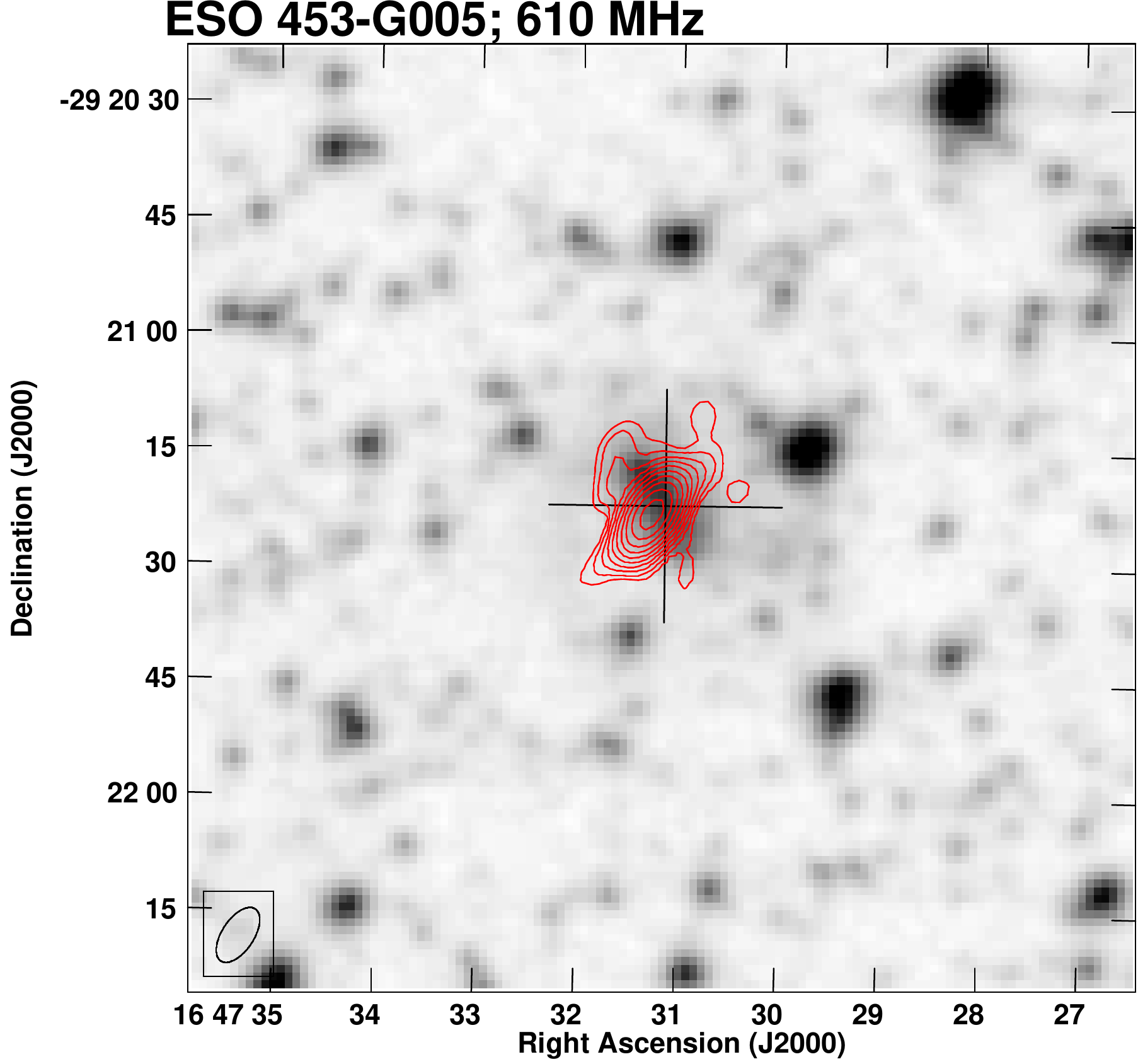}}
    \hspace{0cm}{\includegraphics[width =  0.25\textwidth]{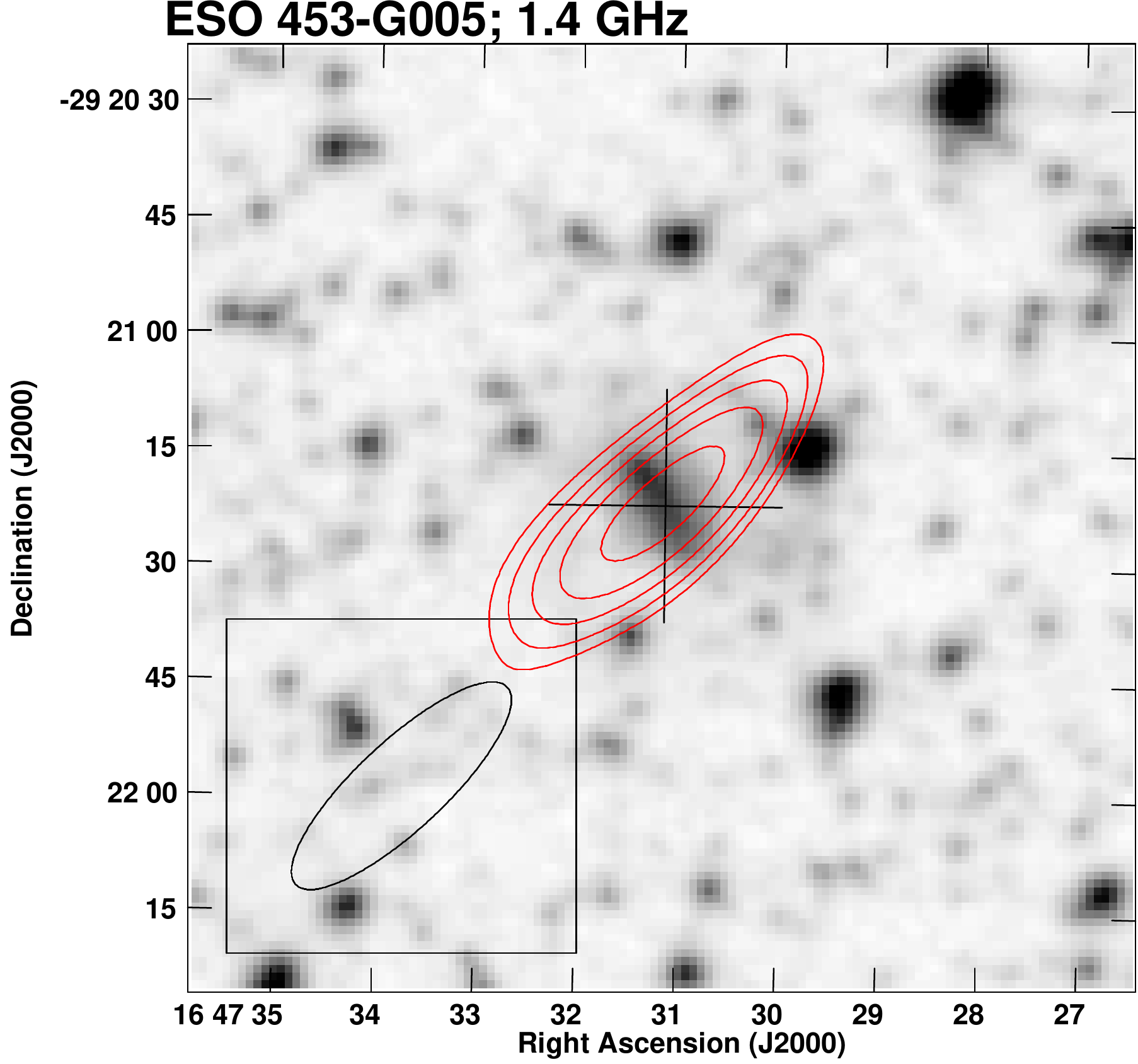}}
    }
    \hbox{
   \hspace{0cm}{\includegraphics[width =  0.25\textwidth]{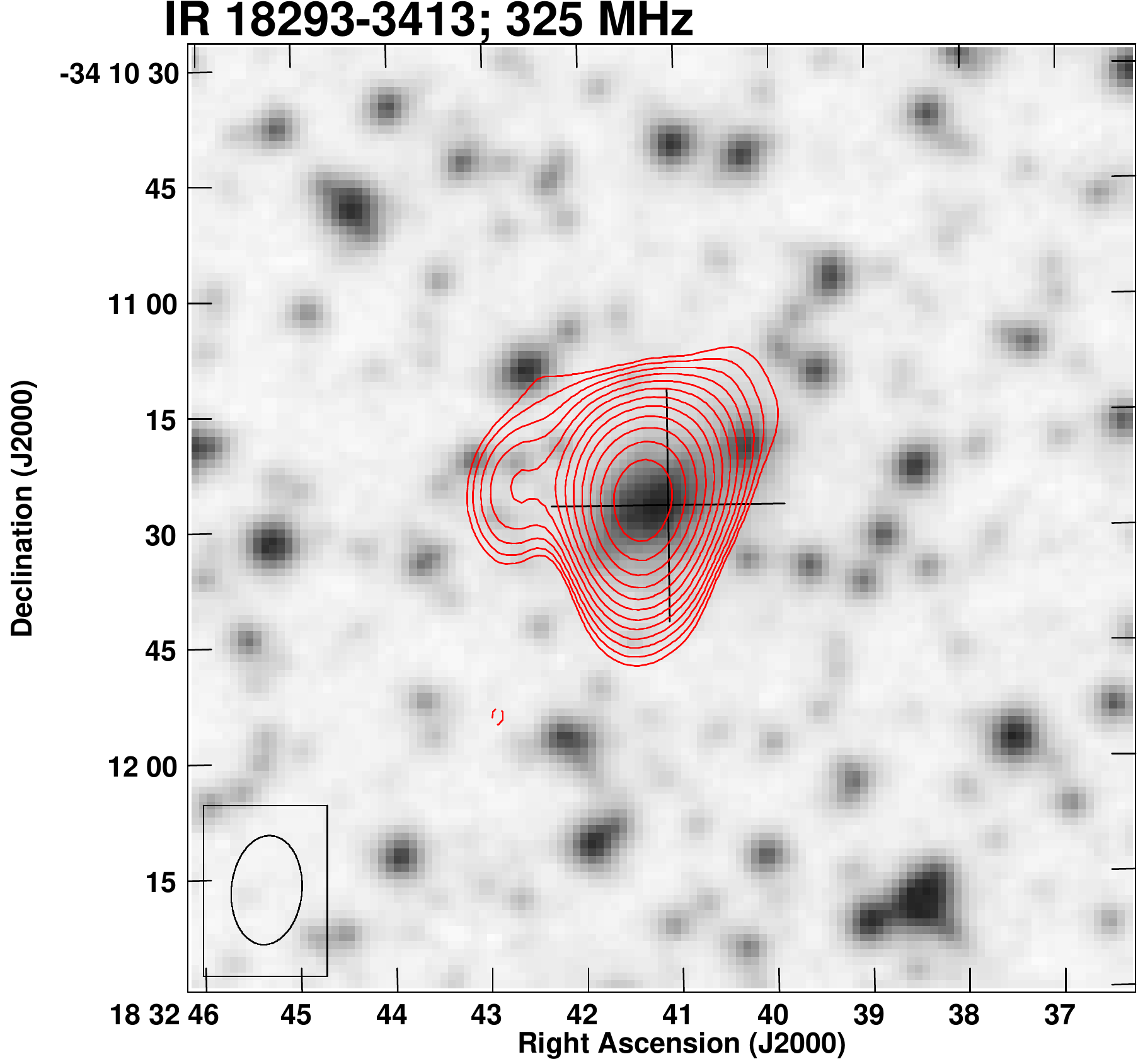}}
   \hspace{0cm}{\includegraphics[width =  0.25\textwidth]{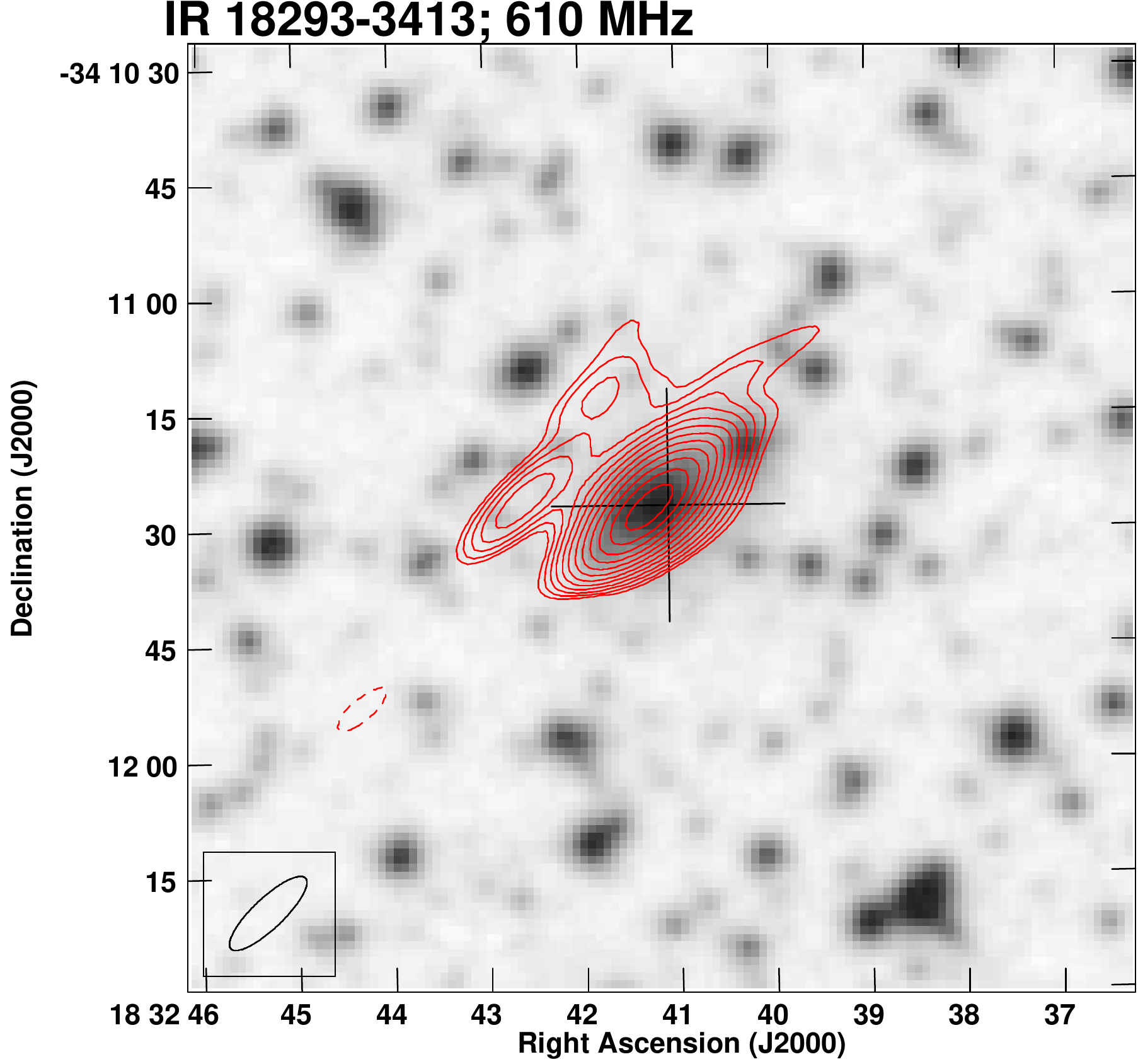}}
   \hspace{0cm}{\includegraphics[width =  0.25\textwidth]{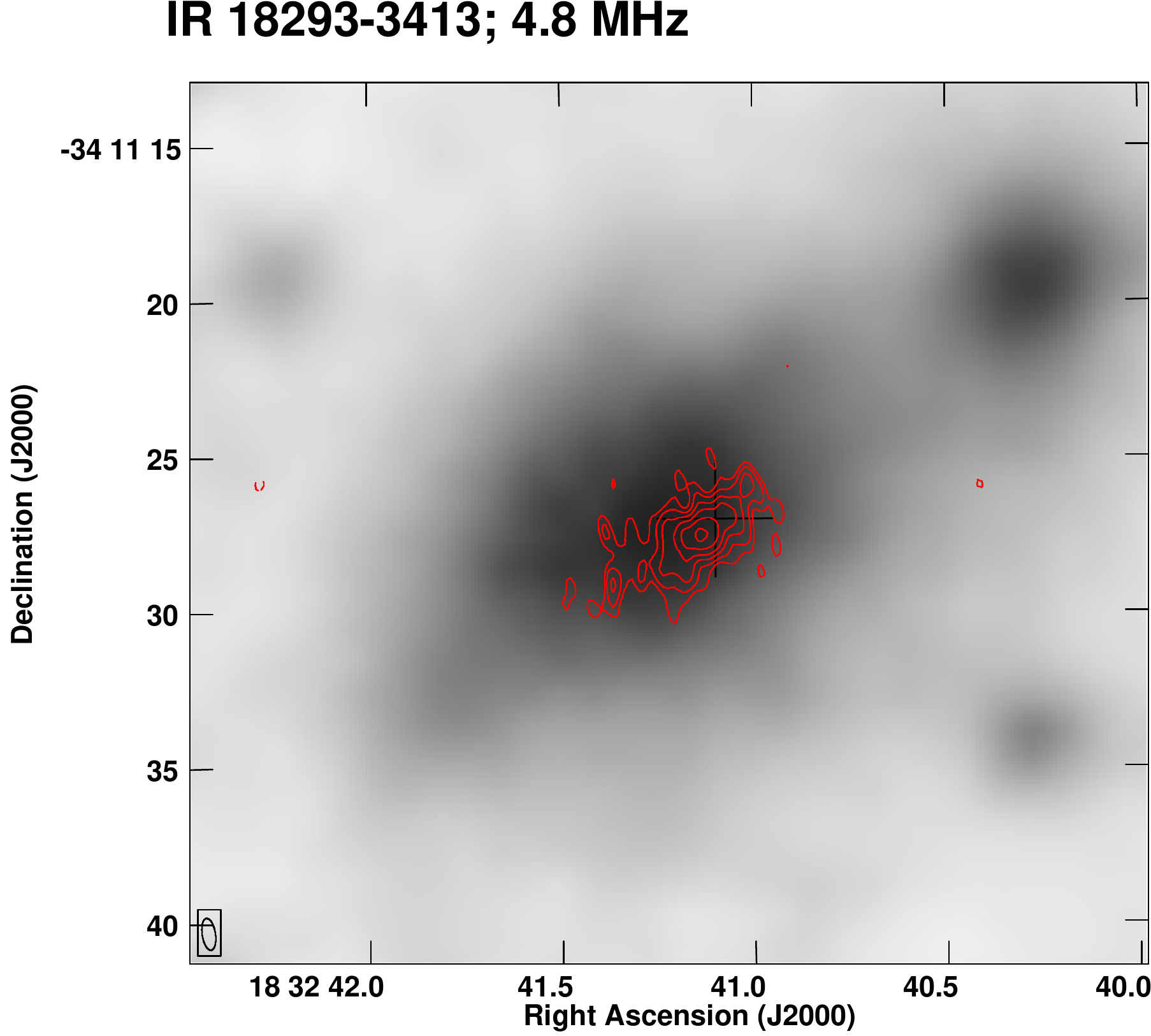}}
   \hspace{0cm}{\includegraphics[width =  0.25\textwidth]{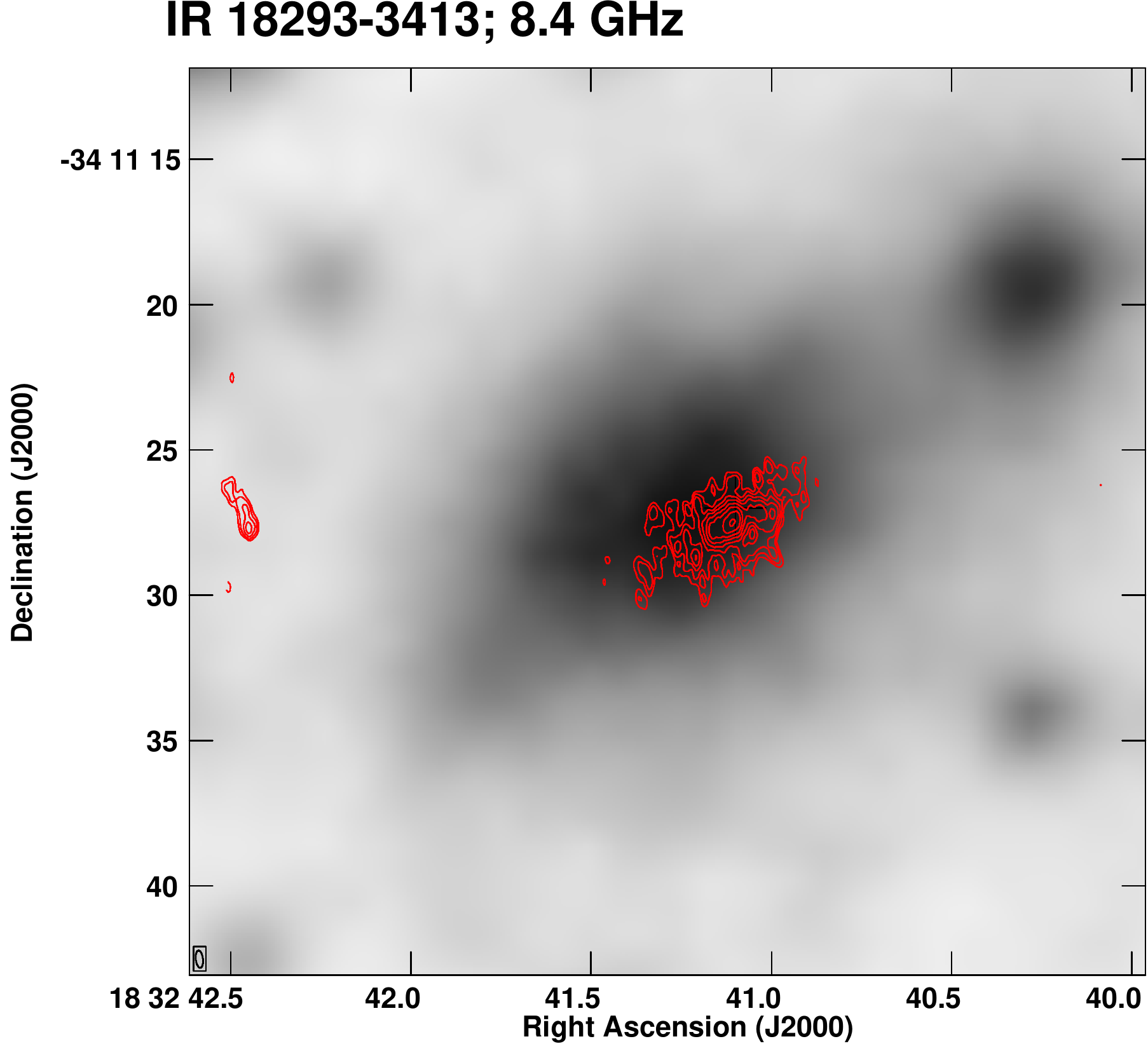}}
   
    }
    
\caption{continued}
\end{figure*}
\addtocounter{figure}{-1}

\begin{figure*}
    \hbox{
   
   \hspace{0cm}{\includegraphics[width =  0.25\textwidth]{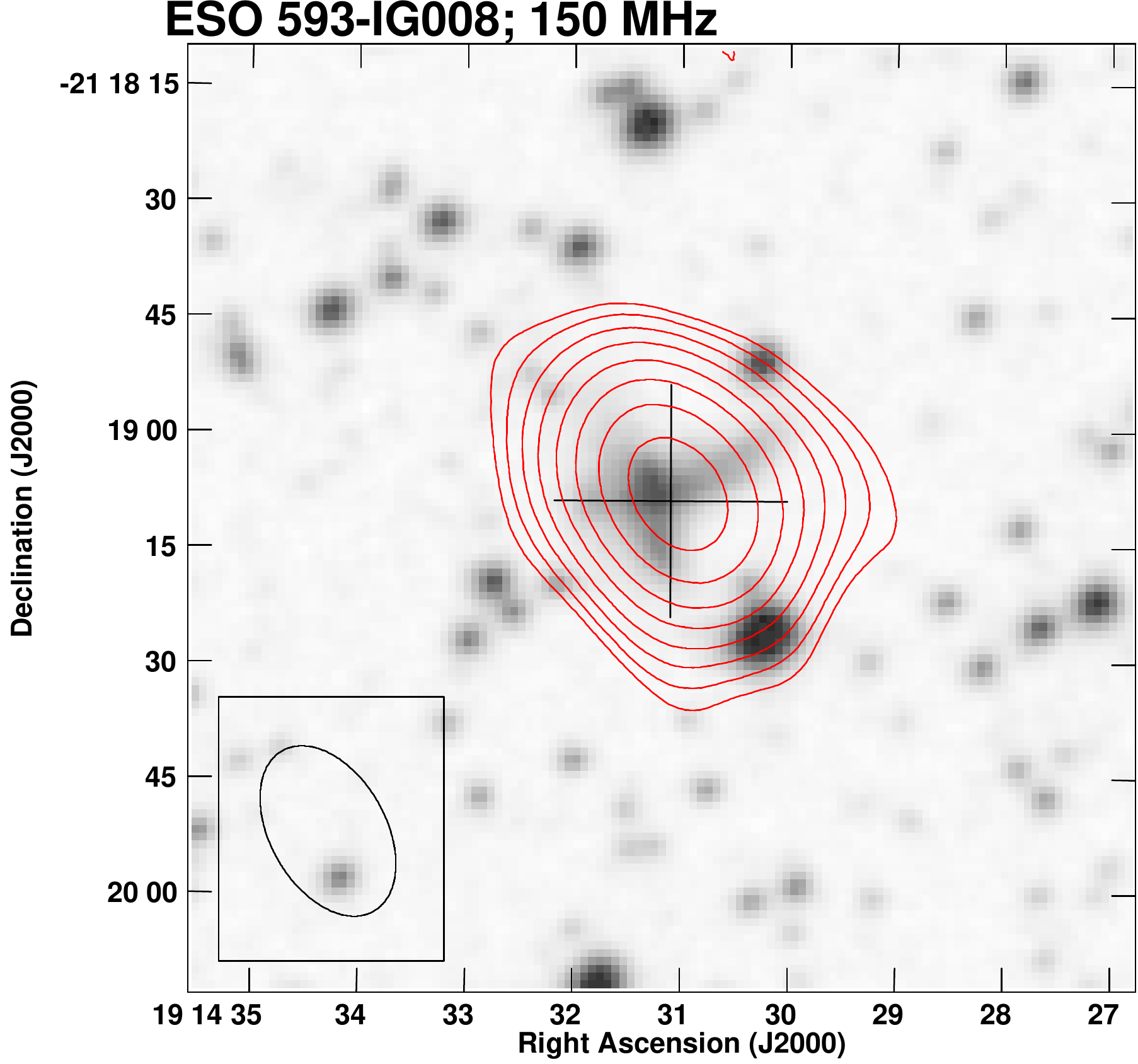}}
   \hspace{0cm}{\includegraphics[width =  0.25\textwidth]{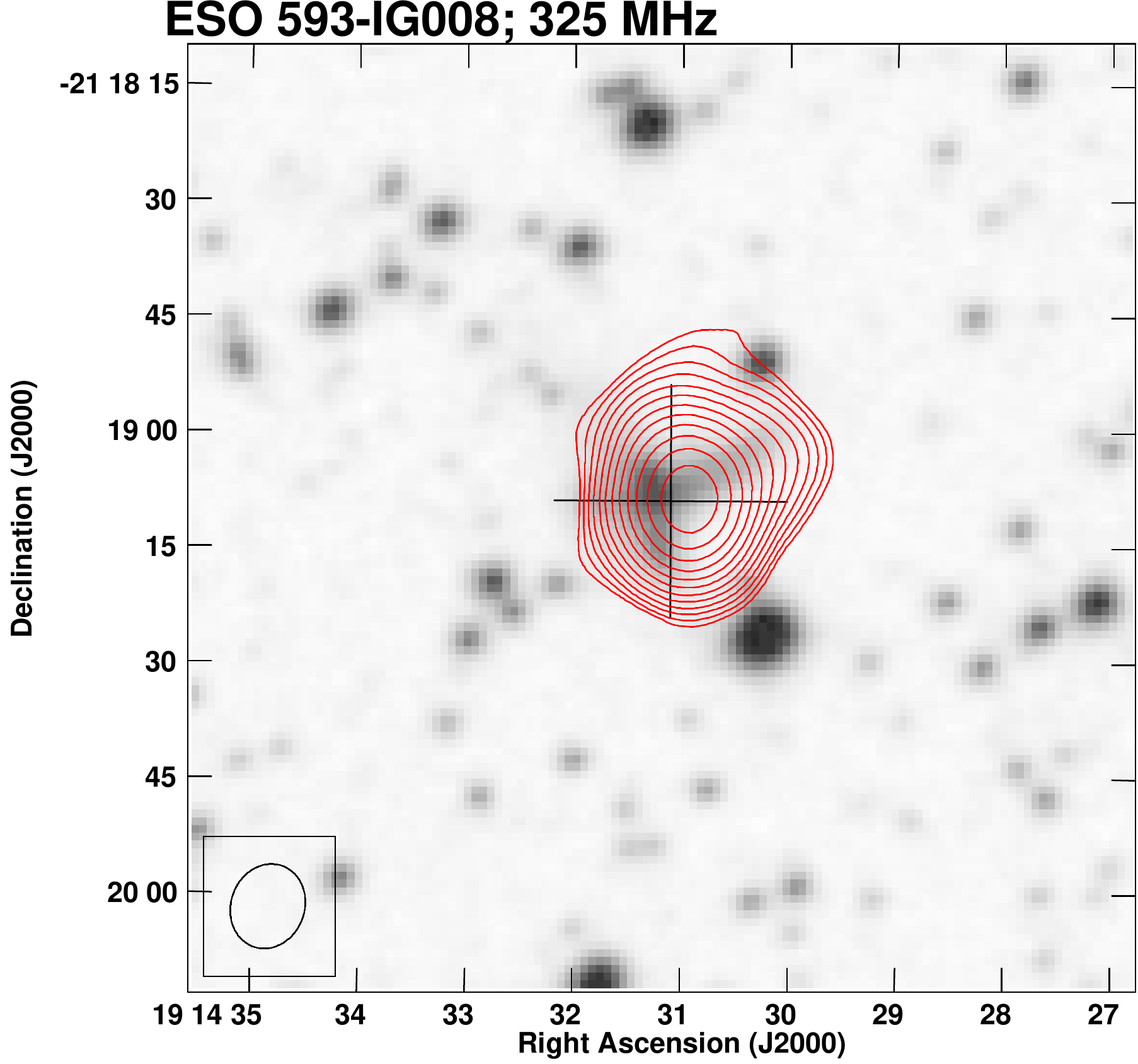}}
   \hspace{0cm}{\includegraphics[width =  0.25\textwidth]{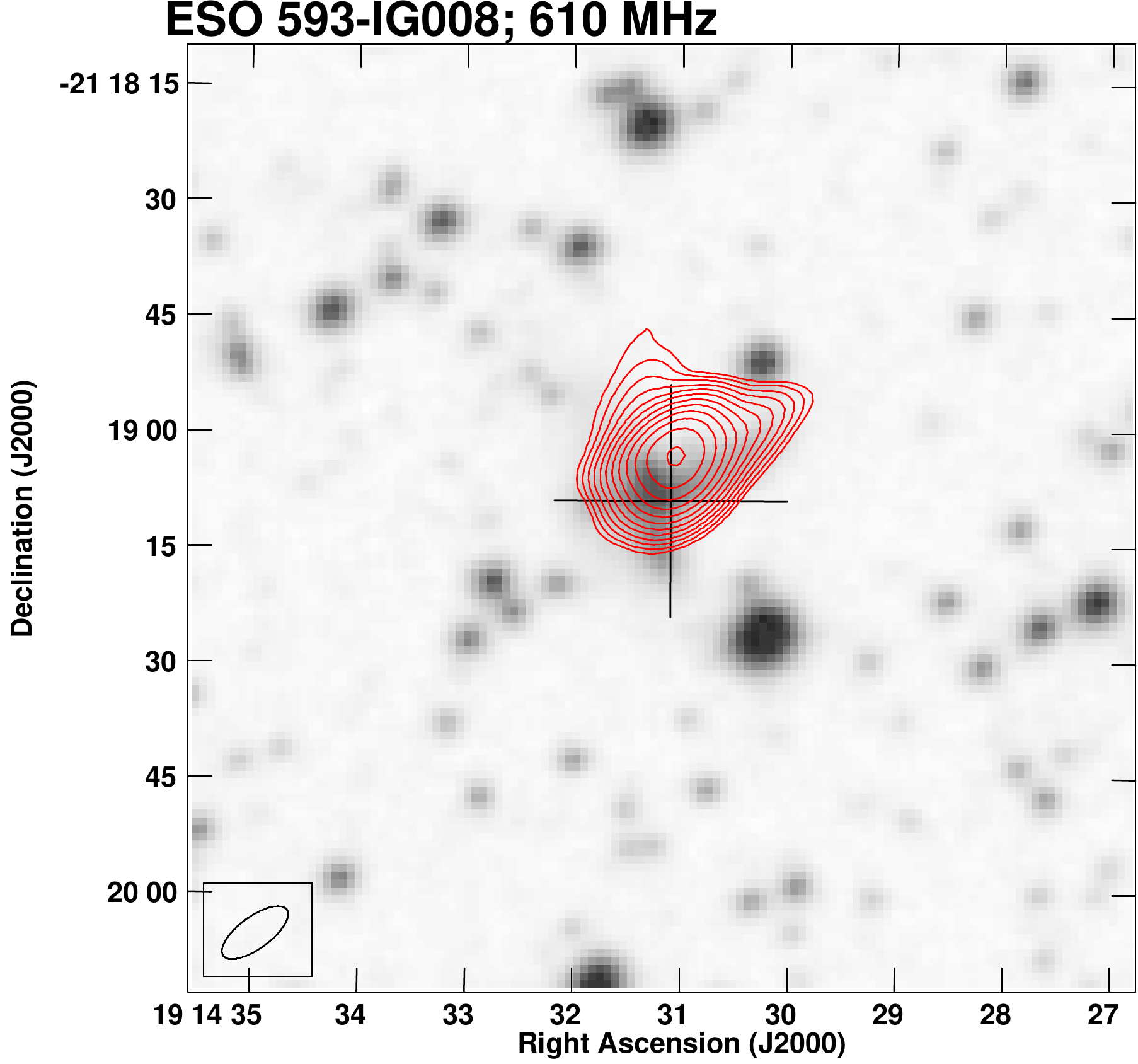}}
   \hspace{0cm}{\includegraphics[width =  0.25\textwidth]{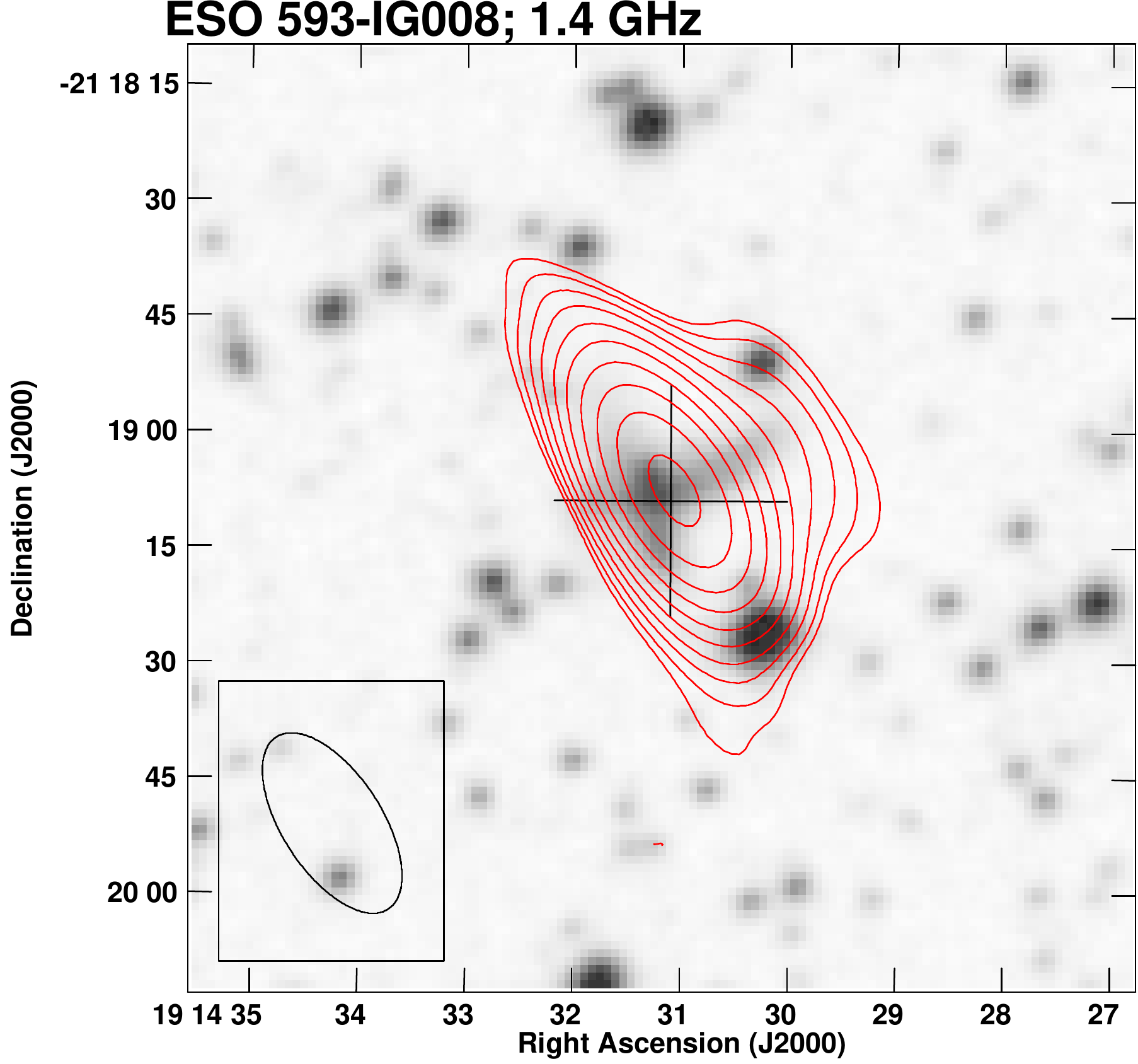}}
   }
   \hbox{
   
   \hspace{0cm}{\includegraphics[width =  0.25\textwidth]{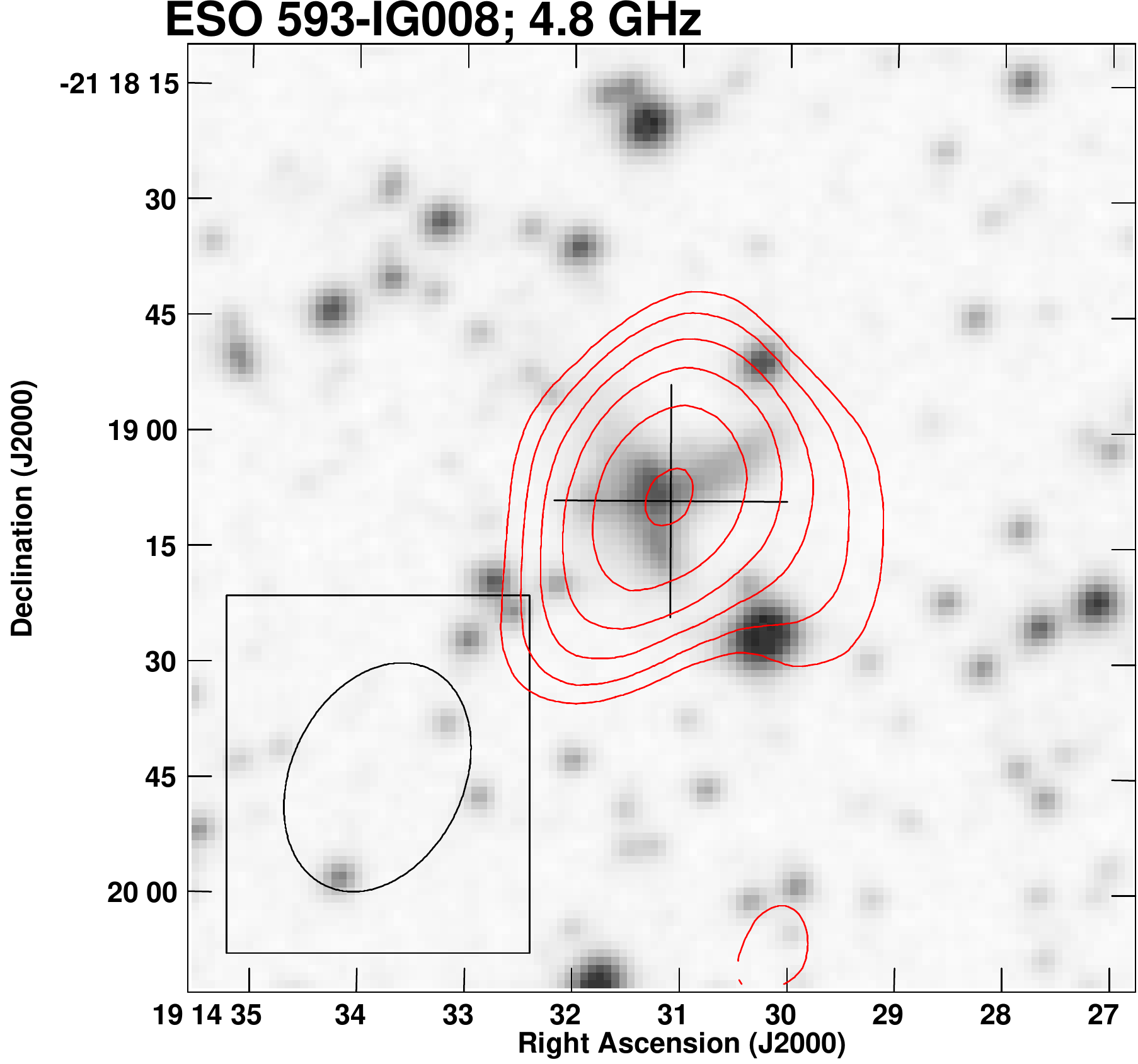}}
   }
\caption{continued}
\end{figure*}

\begin{figure*}
\centering
\hspace{0cm}{\includegraphics[width =  0.8\textwidth]{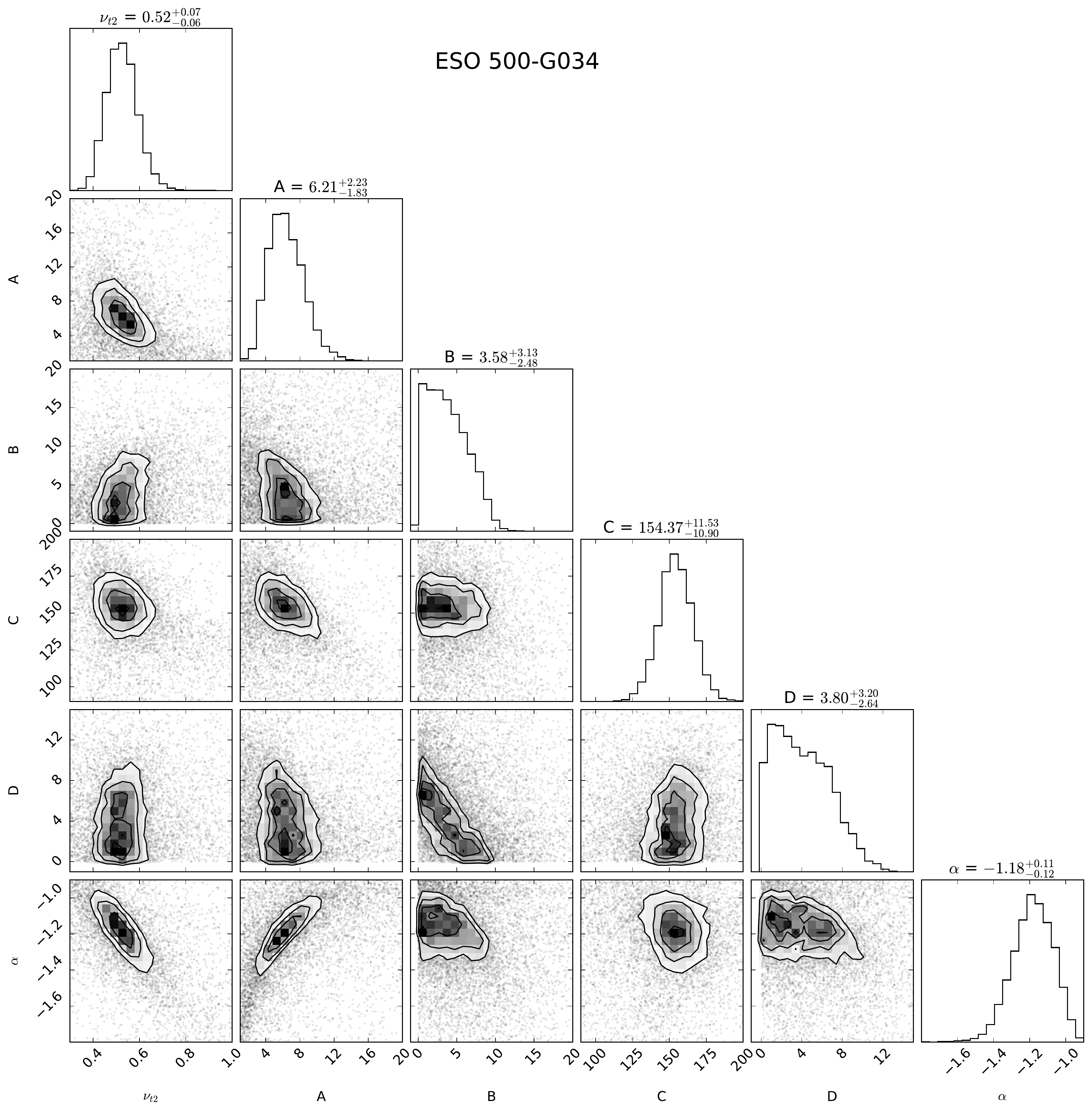}}
\caption{Corner plots showing the one and two-dimensional posterior probability distribution of the estimated parameters in the radio SED modeling for LIRG ESO 500-G034. The complete figure set (11 images) is available in the online journal.}\label{fig:corner_radio}
\end{figure*}

\clearpage

\figsetstart
\figsetnum{A2.1}
\figsettitle{ESO 500-G034}
\figsetgrpstart
\figsetgrpnum{2.2}
\figsetgrptitle{NGC 3508}
\figsetplot{./radio_seds_pdf/eso500_cornor.pdf}
\figsetgrpnote{for the description, see Figure~\ref{fig:corner_radio}}

\figsetnum{A2.2}
\figsettitle{NGC 3508}
\figsetgrpstart
\figsetgrpnum{2.2}
\figsetgrptitle{NGC 3508}
\figsetplot{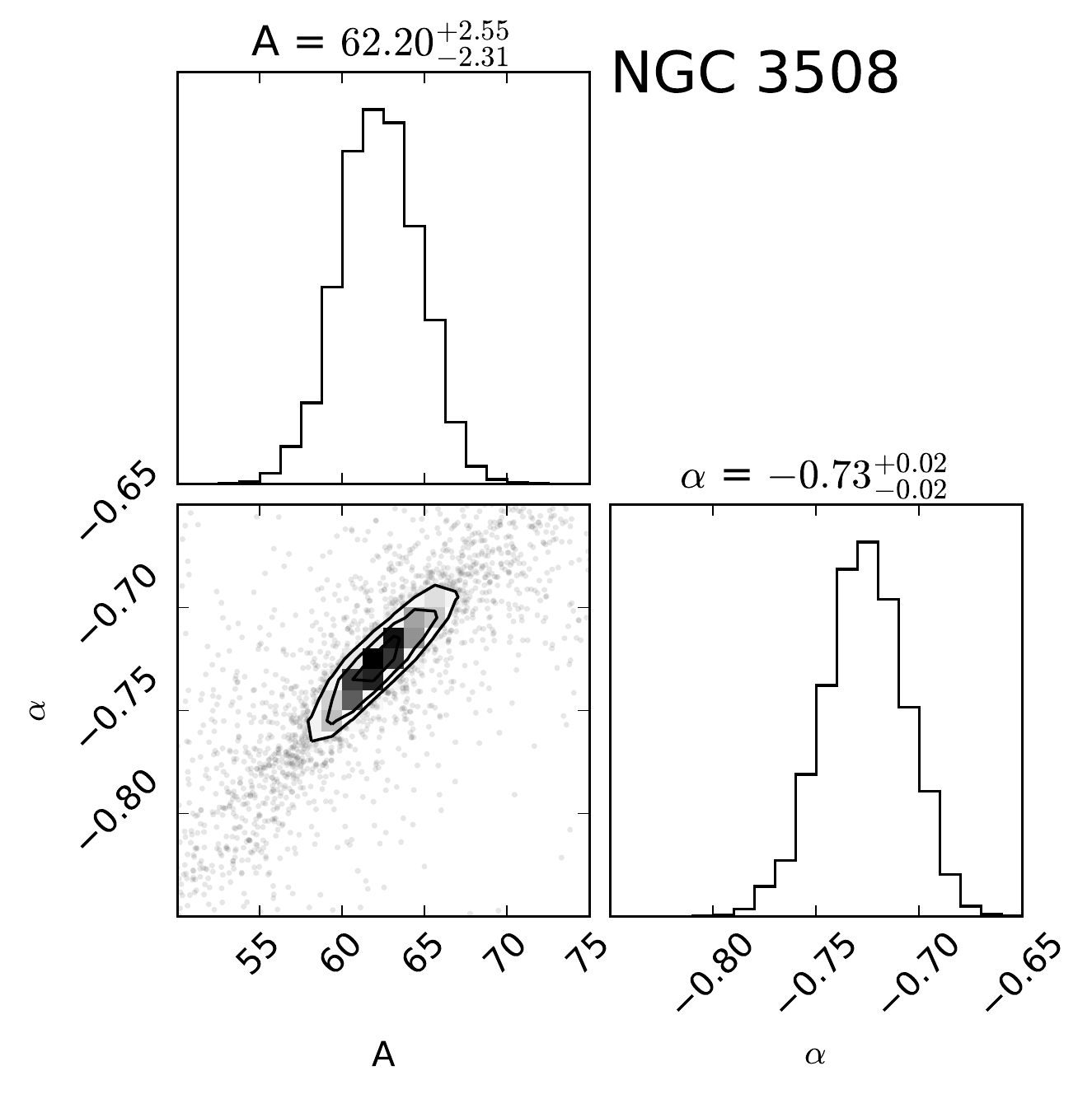}
\figsetgrpnote{for the description, see Figure~\ref{fig:corner_radio}}

\figsetnum{A2.3}
\figsettitle{ESO 440-IG058}
\figsetgrpstart
\figsetgrpnum{2.3}
\figsetgrptitle{ESO 440-IG058}
\figsetplot{./radio_seds_pdf/eso400_cornor.pdf}
\figsetgrpnote{for the description, see Figure~\ref{fig:corner_radio}}

\figsetnum{A2.4}
\figsettitle{ESO 507-G070}
\figsetgrpstart
\figsetgrpnum{2.4}
\figsetgrptitle{ESO 507-G070}
\figsetplot{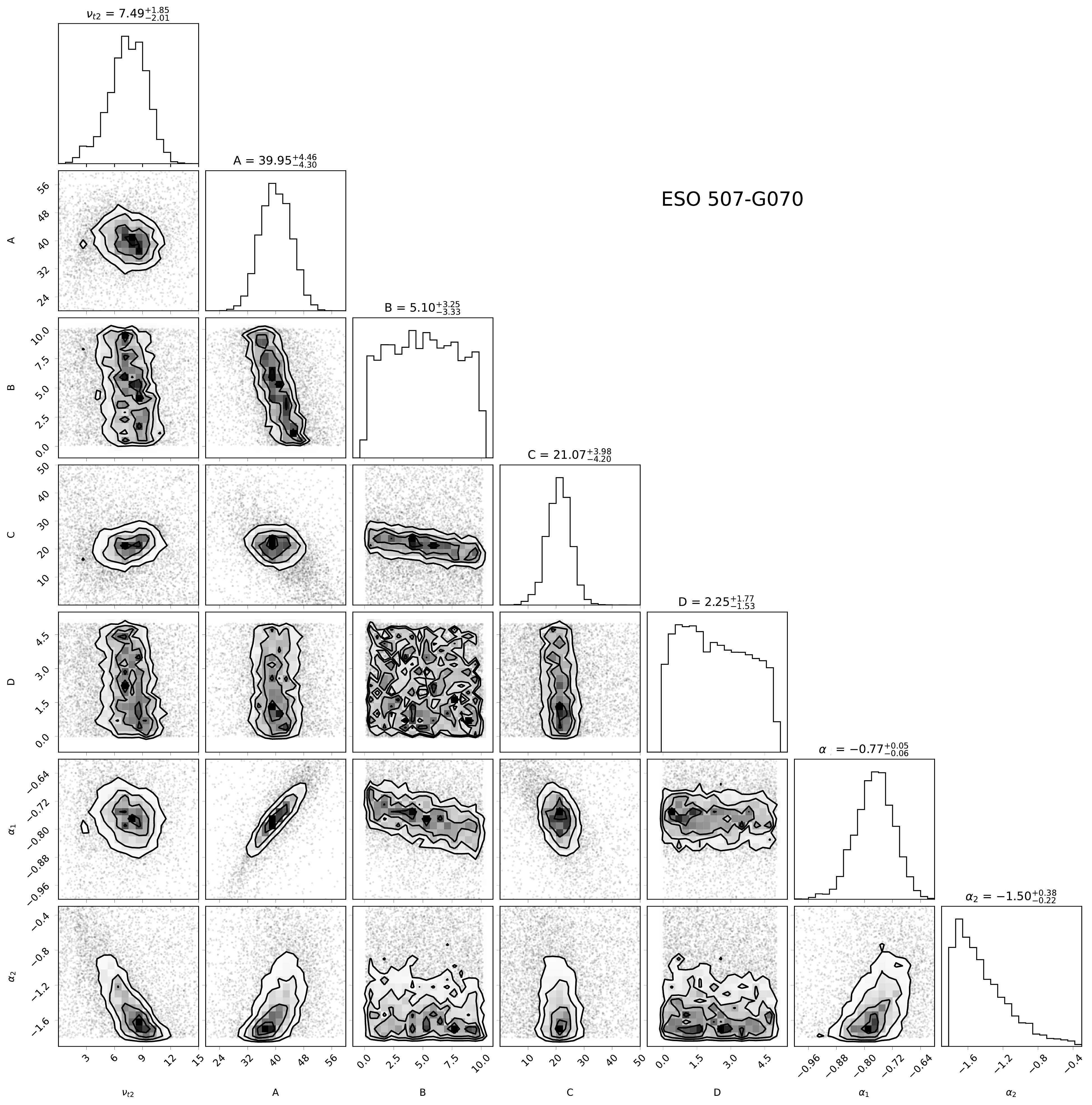}
\figsetgrpnote{for the description, see Figure~\ref{fig:corner_radio}}

\figsetnum{A2.5}
\figsettitle{NGC 5135}
\figsetgrpstart
\figsetgrpnum{2.5}
\figsetgrptitle{NGC 5135}
\figsetplot{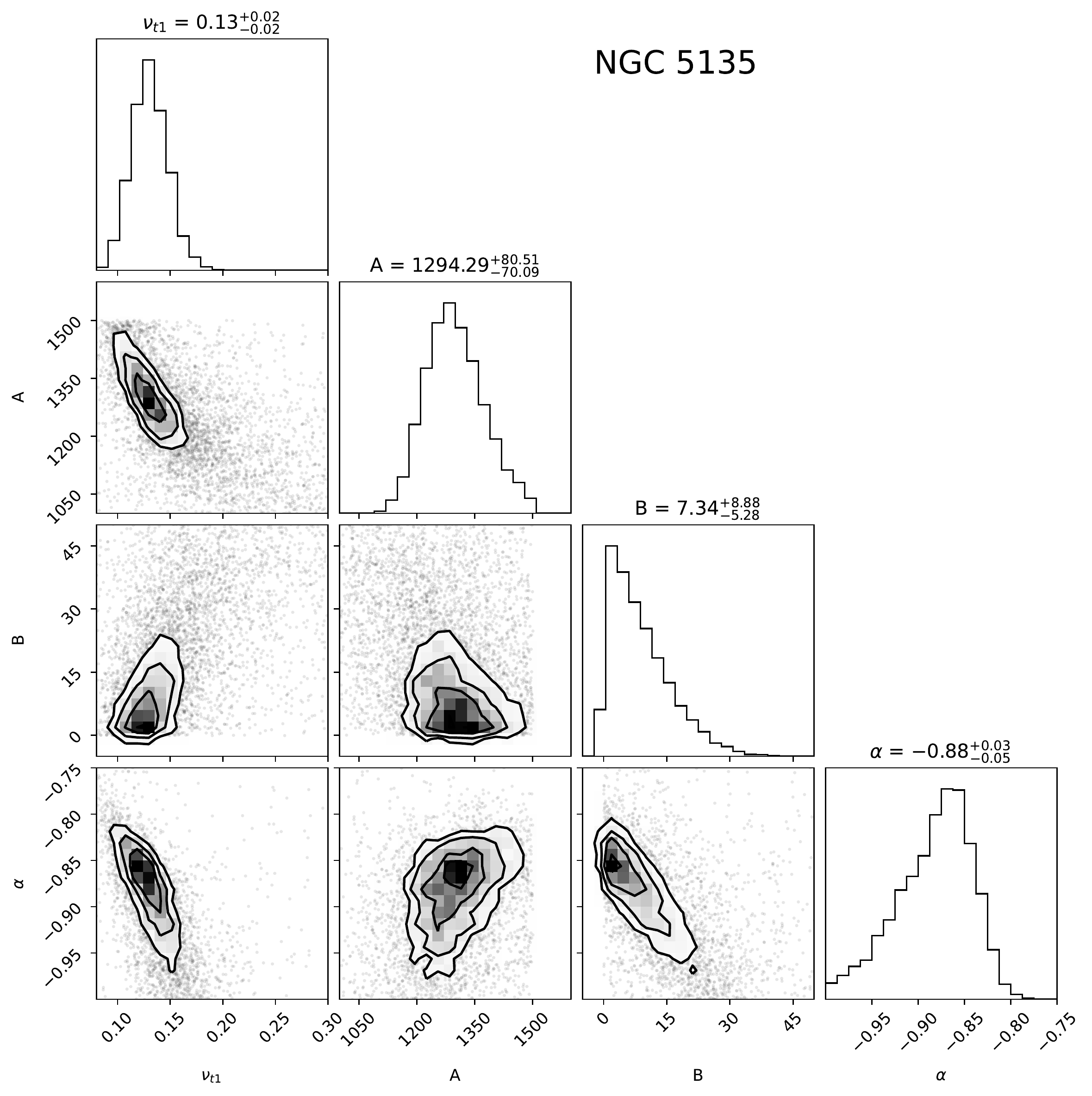}
\figsetgrpnote{for the description, see Figure~\ref{fig:corner_radio}}

\figsetnum{A2.6}
\figsettitle{IC 4280}
\figsetgrpstart
\figsetgrpnum{2.6}
\figsetgrptitle{IC 4280}
\figsetplot{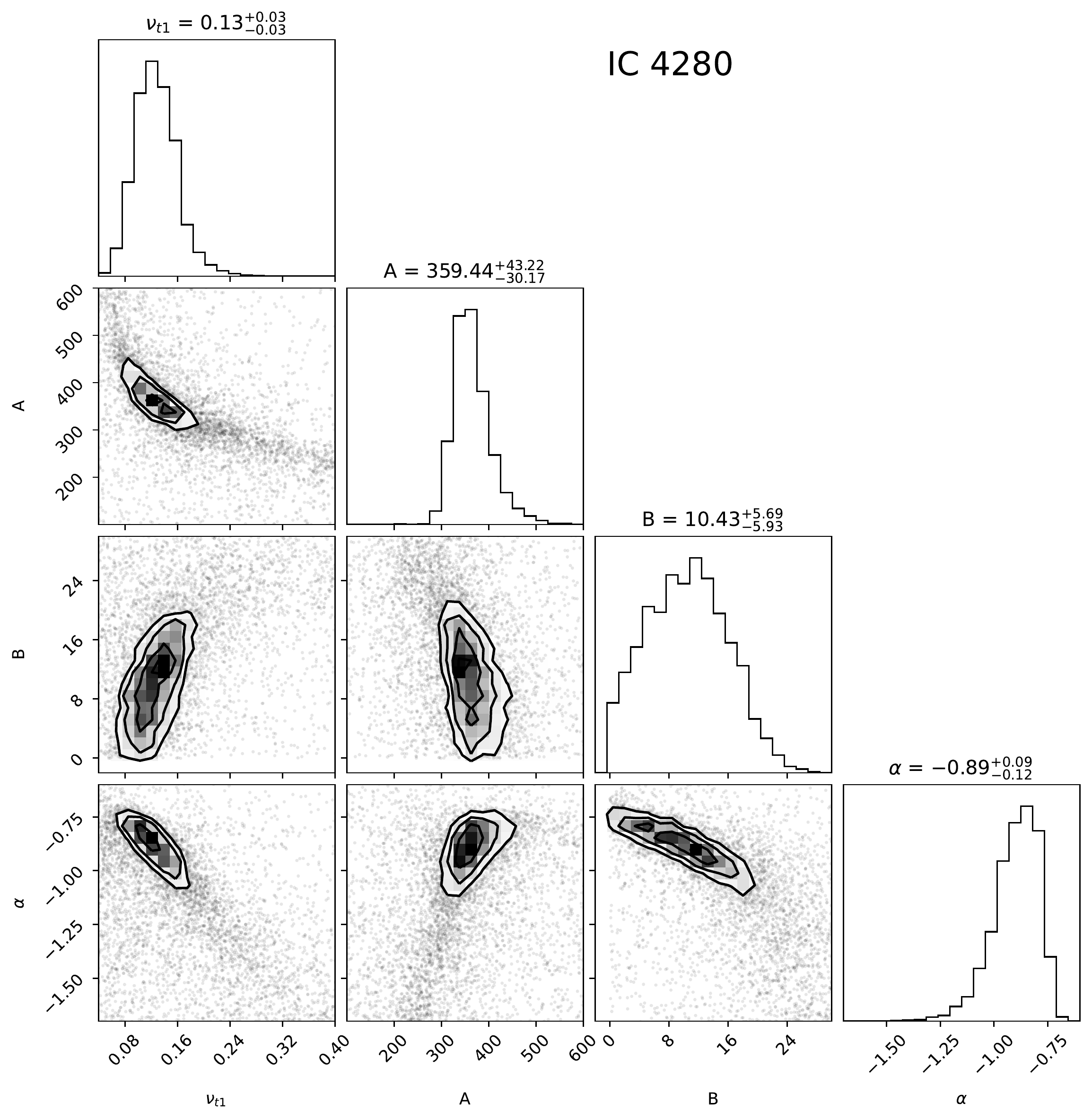}
\figsetgrpnote{for the description, see Figure~\ref{fig:corner_radio}}

\figsetnum{A2.7}
\figsettitle{NGC 6000}
\figsetgrpstart
\figsetgrpnum{2.7}
\figsetgrptitle{NGC 6000}
\figsetplot{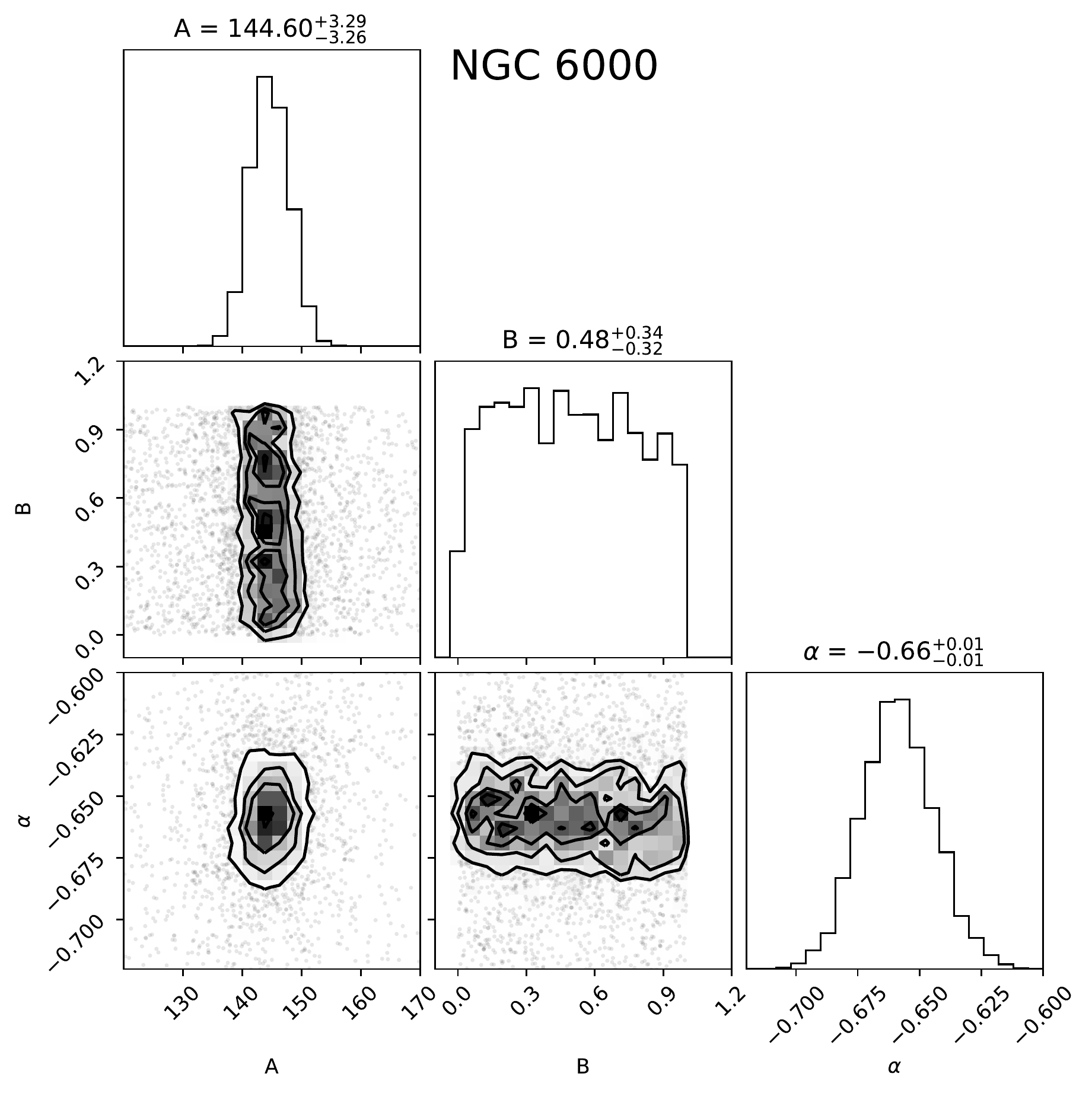}
\figsetgrpnote{for the description, see Figure~\ref{fig:corner_radio}}

\figsetnum{A2.8}
\figsettitle{IR 16164-0746}
\figsetgrpstart
\figsetgrpnum{2.8}
\figsetgrptitle{IR 16164-0746}
\figsetplot{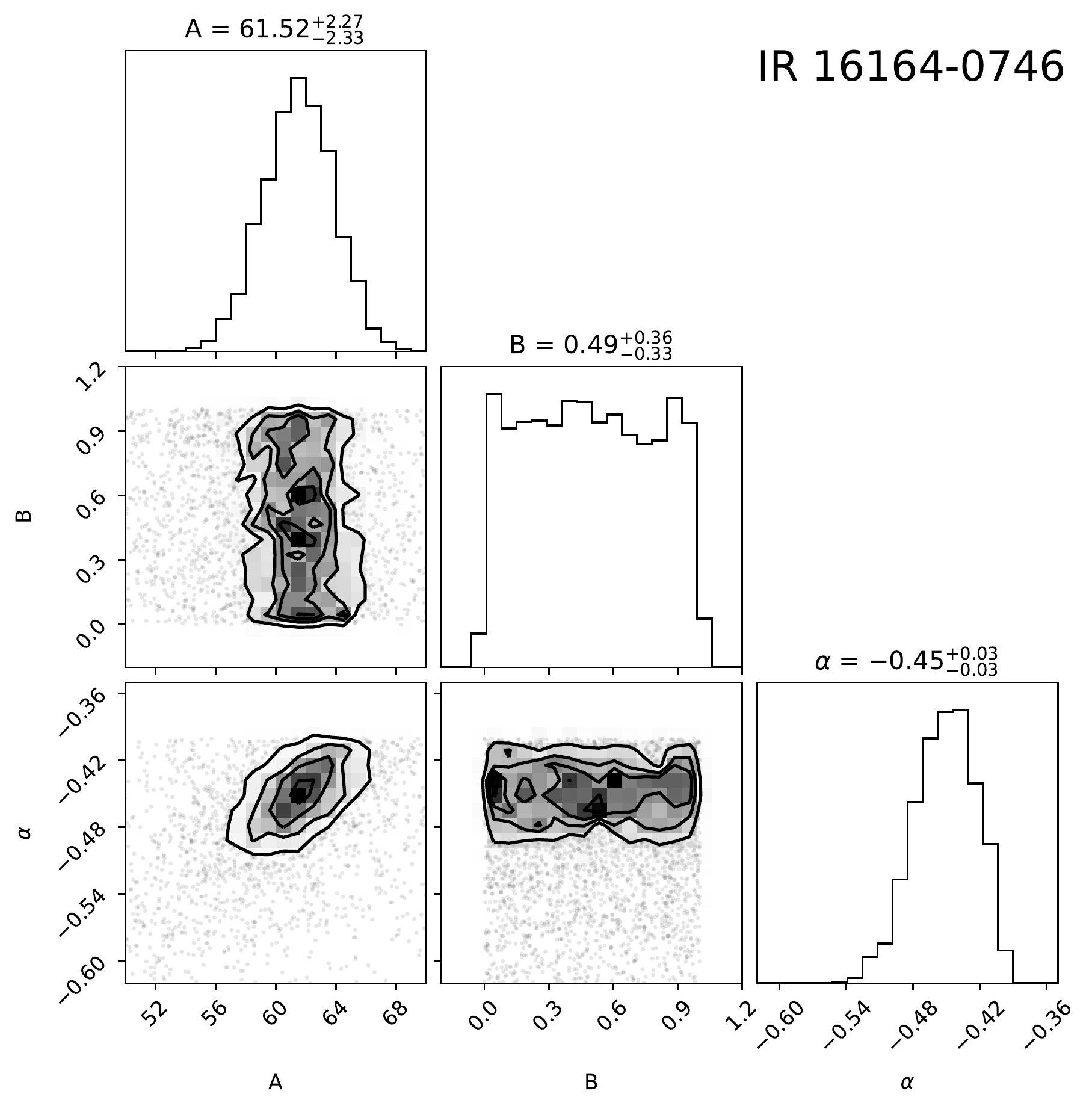}
\figsetgrpnote{for the description, see Figure~\ref{fig:corner_radio}}

\figsetnum{A2.9}
\figsettitle{ESO 453-G005}
\figsetgrpstart
\figsetgrpnum{2.9}
\figsetgrptitle{ESO 453-G005}
\figsetplot{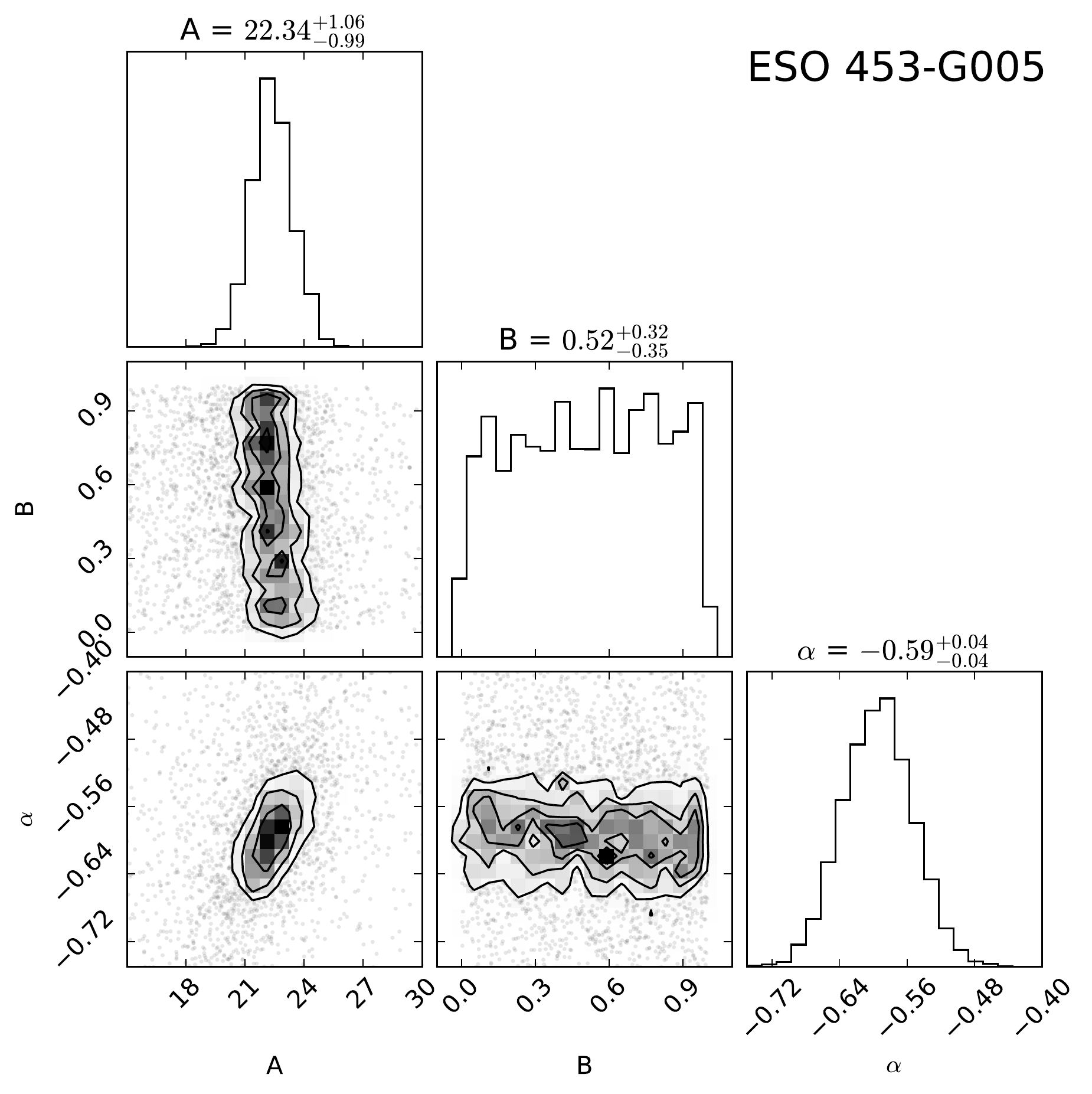}
\figsetgrpnote{for the description, see Figure~\ref{fig:corner_radio}}

\figsetnum{A2.10}
\figsettitle{IR 18293-3413}
\figsetgrpstart
\figsetgrpnum{2.10}
\figsetgrptitle{IR 18293-3413}
\figsetplot{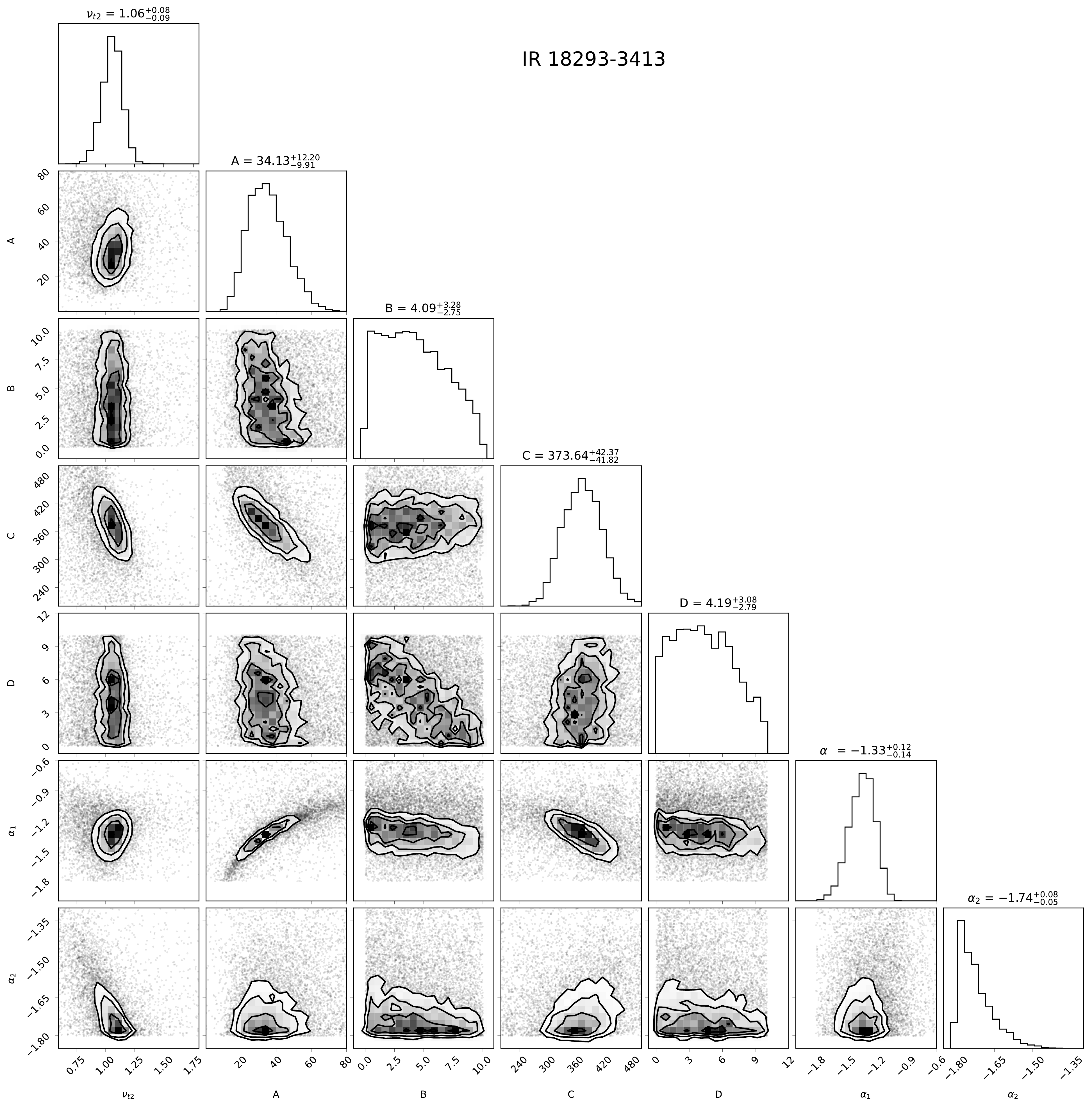}
\figsetgrpnote{for the description, see Figure~\ref{fig:corner_radio}}

\figsetnum{A2.11}
\figsettitle{ESO 593-IG008}
\figsetgrpstart
\figsetgrpnum{2.11}
\figsetgrptitle{ESO 593-IG008}
\figsetplot{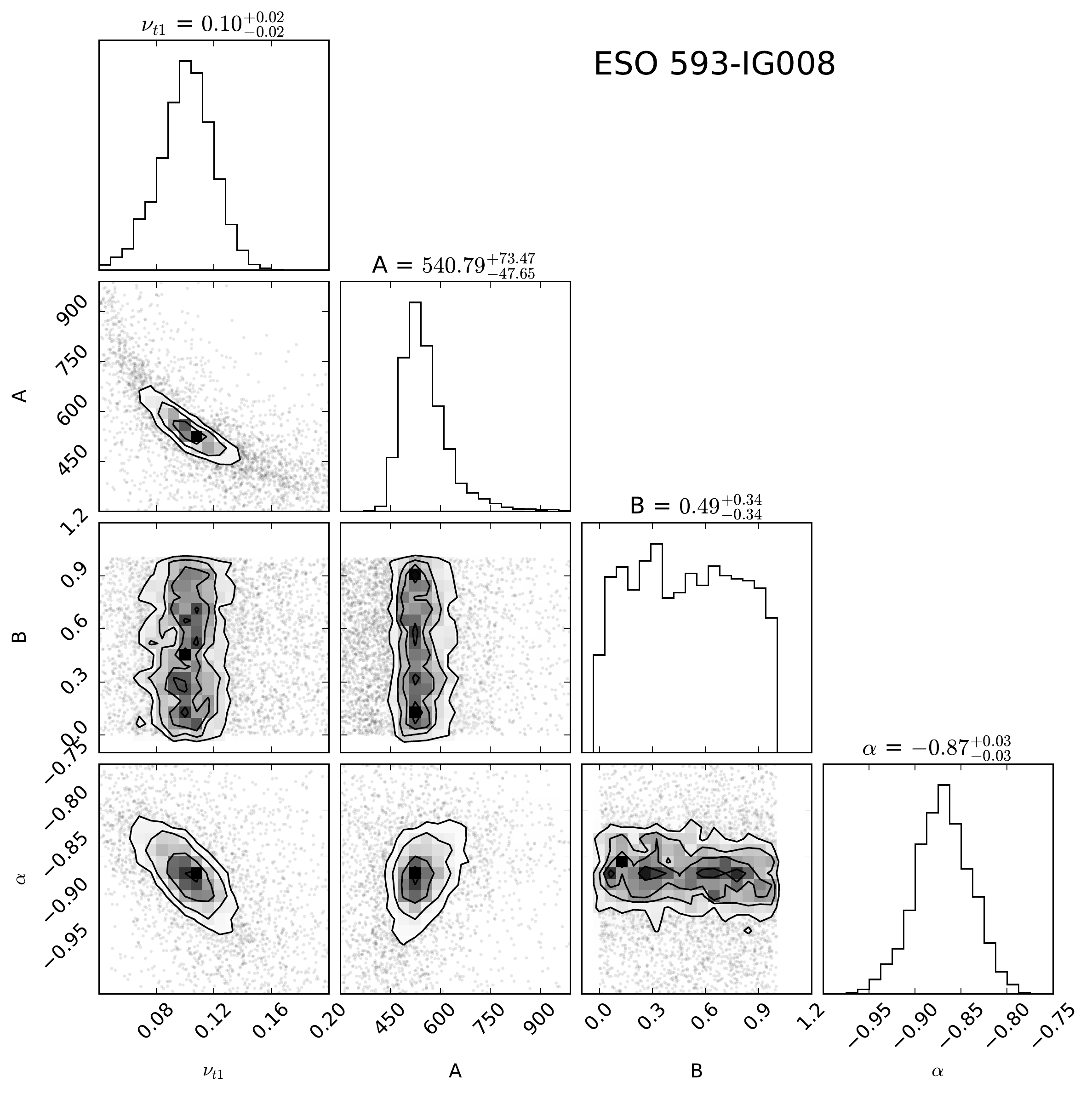}
\figsetgrpnote{for the description, see Figure~\ref{fig:corner_radio}}

\figsetgrpend

%
%\figsetgrpstart
%\figsetgrpnum{A2.3}
%\figsetgrptitle{ESO 440-IG058}
%\figsetplot{./radio_seds_pdf/eso440_cornor.pdf}
%\figsetgrpnote{for the description, see Figure~\ref{fig:corner_radio}}
%\figsetgrpend

\figsetend

\clearpage
\newpage

\section{Summary of UV-IR observations gathered for SED modeling via CIGALE 
}\label{app:B}
\restartappendixnumbering

Table~\ref{tab:uvdata} lists the basic information on the instruments used for gathering multiwavelength FUV--$IR$ datasets and Table~\ref{tab:UV-IRflux} lists the flux densities used in CIGALE modeling.

\begin{table*}[ht!]
\small
\caption{Basic information on the instruments/satellites used for UV-IR measurements in the CIGALE modeling\label{tab:uvdata}}
\begin{tabular}{cccc}
\hline
Instrument &  Passbands &  Central Wavelength   & Resolution  \\
         &       &         ($\mu$m)       &    ($\arcsec$)  \\ 
(1)  & (2)  & (3)          & (4)              \\ \hline

GALEX$^a$ & FUV, NUV & 0.15, 0.227 & 4.2, 5.3 \\
XMM-OM$^b$ & UVW2, UVW1  & 0.212, 0.291  & 1.98, 2   \\
Swift$^c$ & UVW2, UVM2, UVW1, & 0.193, 0.225, 0.260, & \\
Swift$^c$ &  u, b, v &  0.346, 0.439, 0.547  &  2.5 (at 350nm)$^\dag$ \\
SMSS$^d$ & u, v, g, r, i, z & 0.35, 0.38, 0.51, 0.62, 0.78, 0.92 & 3.1, 2.9, 2.6, 2.4, 2.3, 2.3 \\
SDSS DR16$^e$ &$u^\prime$, $g^\prime$, $r^\prime$, $i^\prime$, $z^\prime$ & 0.354, 0.477, 0.623, 0.762, 0.913 & 1.53, 1.44, 1.32, 1.26, 1.29  \\
2MASS$^f$ & J, H, Ks & 1.235, 1.662, 2.15 & 2 \\
WISE$^g$ & W1-W4  & 3.4, 4.6, 12, 22 & 6.1, 6.4, 6.5, 12 \\
{\it Spitzer}-IRAC$^h$ & IRAC1 - IRAC4  & 3.56,4.52,5.73,7.91 & 1.66, 1.72, 1.88, 1.98  \\
IRAS$^i$ & IRAS1 - IRAS4  &12, 25, 60, 100 & 45, 45, 90, 180  \\
AKARI$^{j,k}$ & N60, WIDE-S, WIDE-L, N160 & 65, 90, 140, 160 & 37,39,58,61 \\ 

{\it Herschel}-PACS$^l$ & Blue, Green, Red & 70, 100, 160 & 6, 7, 12  \\
{\it Herschel}-SPIRE$^{m,n}$ & PSW, PMW, PLW & 250, 350, 500 & 17.9, 24.2, 35.4\\ \hline
\end{tabular}
Columns: (1) name of the instrument/satellite (references for the instrument are given as superscript). $^a$\citet{Bianchi99}, $^b$\citet{mason2001}, $^c$\citep{Poole08}, $^d$\citep{Wolf18}, $^e$\url{https://www.sdss.org/dr16/imaging/other_info/}, $^f$ \citet{Jarrett2000},  $^g$\citet{Wright10}, $^h$\citet{Fazio2004},  $^i$\citet{Neugebauer84}, $^j$\citet{Murakami2007}, $^k$\citet{Doi2015}, $^l$\citet{Poglitsch2010}, $^m$\citet{Griffin2010}, $^n$ \citet[][]{Smith2017}, (2) name of the passband/filter used for observation, (3) central wavelength of the passband, (h) resolution (full width at half maximum). $^\dag$\url{https://swift.gsfc.nasa.gov/about_swift/uvot_desc.html}

\end{table*}

%\newpage
%\begin{multicols}{2}

\startlongtable
%\scriptsize
\begin{deluxetable}{ccccccc}
%\tablenum{1}
\tablecaption{Integrated UV-IR flux densities used for the SED fitting.\label{tab:UV-IRflux}}
\tablewidth{0pt}
\tabletypesize{\scriptsize}
\tablehead{
\colhead{Name} &        \colhead{Instrument} & \colhead{Passband}   & \colhead{S }            & \colhead{error}  & \colhead{Integration Area} & \colhead{Ref.}  \\
\colhead{}   &         \colhead{}    & \colhead{}  &   \colhead{(mJy)}         & \colhead{(mJy)} & \colhead{(arcsec$^{2}$)} & \colhead{}
}
 \decimalcolnumbers
\startdata
ESO 500-G034 & SMSS & $g^\prime$ & 10.9  & 3.2 & 3.3 $\times$ 3.3 \arcsec& (a)\\
 & &  $r^\prime$ & 16 & 0.9 & 3.3\arcsec $\times$ 3.3 \arcsec& (a) \\
 & & $i^\prime$ & 18.2 & 2.1 & 3.3\arcsec $\times$ 3.3 \arcsec& (a)\\
 & & $z^\prime$ & 21.3 & 1 &3.3\arcsec $\times$ 3.3 \arcsec & (a)\\
 & 2MASS & J & 52.6 & 0.88 & 55.6\arcsec $\times$ 26.7 \arcsec & (b)\\
 & & H & 72.4 & 1.69 & 55.6\arcsec $\times$ 26.7 \arcsec  & (b)\\
 & & Ks & 66.3 & 1.86 & 55.6\arcsec $\times$ 26.7 \arcsec  & (b) \\ 
 & WISE & W1 & 40.0 & 0.22 & 22\arcsec $\times$  22\arcsec  & (c)  \\
 & & W2 & 33.7 & 0.18 &  22\arcsec $\times$  22\arcsec  & (c)\\
 & & W3 & 223 & 1.23 & 22\arcsec $\times$ 22\arcsec & (c)\\
 & & W4 & 756 & 4.18 & 22\arcsec $\times$  22\arcsec  & (c)\\
% & IRAS & IRAS1 & 380 & 30 & $>$0.8\arcmin $\times$ $>$0.8\arcmin$^\dag$ & (4) \\
% & & IRAS2 & 1430 & 32 & $>$0.8\arcmin $\times$ $>$0.8\arcmin$^\dag$ & (4) \\
 & AKARI & N60 & 9430 & 350 & 40\arcsec $\times$ 50\arcsec$^\ddag$  & (d) \\
 & & WIDE-S & 13100 & 400 & 40\arcsec $\times$ 50\arcsec$^\ddag$ & (d) \\
 & & WIDE-L & 13700 & 500 & 70\arcsec $\times$ 90\arcsec$^\ddag$ & (d) \\
 & & N160 & 11400 & 600 & 70\arcsec $\times$ 90\arcsec$^\ddag$ & (d) \\
NGC3508 & GALEX & FUV & 1.44 & 0.04 & 3.5\arcsec $\times$ 3.5\arcsec   & (e)\\
 & & NUV & 2.8 & 0.02 & 3.5\arcsec $\times$ 3.5\arcsec & (e) \\
 & SDSS & $g^\prime$ & 13.7 & 1.08 & 22.9\arcsec $\times$ 22.9\arcsec & (f)\\
 %$u^\prime$ & 3.89 & 0.02 & 22.9\arcsec $\times$ 22.9\arcsec  & (f)\\
 & &  $r^\prime$ & 28 & 0.1 & 22.9\arcsec $\times$ 22.9\arcsec & (f)\\
 & & $i^\prime$ & 36.4 & 0.1 & 22.9\arcsec $\times$ 22.9\arcsec & (f)\\
 & & $z^\prime$ & 51.7 & 0.1 & 22.9\arcsec $\times$ 22.9\arcsec & (f)\\
 & 2MASS & J & 69.2 & 1.29 & 60.4\arcsec $\times$ 33.8 \arcsec  & (b)\\
 & & H & 81.2 & 2.06 &  60.4\arcsec $\times$ 33.8 \arcsec  & (b)\\
 & & Ks & 72.8 & 2.52 &  60.4\arcsec $\times$ 33.8 \arcsec  & (b) \\ 
  & WISE & W1 & 45.5 & 0.21 & 22\arcsec $\times$  22\arcsec & (c)  \\
 & & W2 & 31.8 & 0.175  &  22\arcsec $\times$  22\arcsec & (c)\\
 & & W3 & 254 & 1.41 & 22\arcsec $\times$  22\arcsec & (c)\\
 & & W4 & 495 & 3.19 & 22\arcsec $\times$  22\arcsec & (c)\\
 & AKARI & N60 & 6347 & 493 & 42\arcsec $\times$ 42\arcsec$^\ddag$ & (d)\\
 & & WIDE-S & 10690 & 486 & 42\arcsec $\times$ 42\arcsec$^\ddag$ & (d) \\
 & & WIDE-L & 14229 & 1450 & 60\arcsec $\times$ 60\arcsec$^\ddag$ & (d) \\
 & & N160 & 14740 & 1620 & 60\arcsec $\times$ 60\arcsec$^\ddag$ & (d) \\
 ESO 440-IG058  & GALEX & FUV & 0.18 & 0.02 & 4.63\arcsec $\times$ 4.63\arcsec  & (e) \\
 & & NUV & 0.45 & 0.01 & 3.5\arcsec $\times$ 3.5\arcsec & (e) \\
& SMSS & g & 1.88 & 0.43 & 7.5\arcsec $\times$ 7.5\arcsec & (a)\\
 & &  r & 2.00 & 0.11 & 7.5\arcsec $\times$ 7.5\arcsec & (a)\\
 & & i & 2.74 & 0.09 & 7.5\arcsec $\times$ 7.5\arcsec & (a)\\
 & & z & 2.83 & 0.08 & 7.5\arcsec $\times$ 7.5\arcsec & (a)\\
 & IRAC & IRAC1 & 18.75 & 1.8 & & (g) \\
 & & IRAC2 & 14.62 & 1.4 & & (g) \\
& & IRAC3 & 56.63 & 5.6 & & (g) \\
 & & IRAC4 & 182.56 & 18 & & (g) \\
 & IRAS & IRAS1 & 200 & 38 & 0.8\arcmin $\times$ $>$0.8\arcmin$^\dag$ & (h)\\
 & & IRAS2 & 760 & 35 & 0.8\arcmin $\times$ $>$0.8\arcmin$^\dag$ & (h) \\
& {\it Herschel}-PACS & Blue & 7972 & 363 & 35\arcsec $\times$ 35\arcsec  & (i)\\
& & Green & 10620 & 480 &  35\arcsec $\times$ 35\arcsec  & (i) \\
& & Red & 8597 & 380 &  35\arcsec $\times$ 35\arcsec  & (i)\\
 & {\it Herschel}-SPIRE & PSW & 3371 & 203 & 50\arcsec $\times$ 50\arcsec  & (i) \\
 & & PMW & 1278 & 79 & 50\arcsec $\times$ 50\arcsec & (i) \\
 & & PLW & 489 & 30 & 50\arcsec $\times$ 50\arcsec & (i) \\
 ESO 507-G070 & GALEX & FUV & 0.09 & 0.01 & 3.5\arcsec $\times$ 3.5\arcsec& (e) \\
 & & NUV & 0.21 & 0.01 & 3.5\arcsec $\times$ 3.5\arcsec & (e) \\
 & SMSS & $g^\prime$ & 2.72  & 0.14 & 3.55\arcsec $\times$  3.55\arcsec & (a)\\
 & &  $r^\prime$ & 3.98 & 0.05 & 3.55\arcsec $\times$  3.55\arcsec & (a)\\
 & & $i^\prime$ & 6.42 & 0.09 & 3.55\arcsec $\times$  3.55\arcsec& (a) \\
  & 2MASS & J & 16.7 & 0.18 & 14\arcsec $\times$  14\arcsec & (b)\\
 & & H & 23.6 & 0.31 & 14\arcsec $\times$  14\arcsec  & (b)\\
 & & Ks & 23.4 & 0.30 & 14\arcsec $\times$  14\arcsec  & (b) \\ 
  & WISE & W1 & 24.7 & 0.14 & 22\arcsec $\times$  22\arcsec & (c)  \\
 & & W3 & 104 & 0.58 & 22\arcsec $\times$  22\arcsec & (c)\\
 & {\it Herschel}-PACS & Blue & 16320 & 750 & 43\arcsec $\times$ 43\arcsec & (i)\\
& & Green & 16900 & 770 & 43\arcsec $\times$ 43\arcsec & (i) \\
& & Red & 10660 & 470 & 43\arcsec $\times$ 43\arcsec & (i)\\
 & {\it Herschel}-SPIRE & PSW & 3747 & 228 & 55\arcsec $\times$ 55\arcsec & (i) \\
 & & PMW & 1405 & 87 & 55\arcsec $\times$ 55\arcsec & (i) \\
 & & PLW & 451 & 29 & 55\arcsec $\times$ 55\arcsec & (i) \\
 NGC 5135 & GALEX & FUV & 0.97 & 0.01 & 3.5\arcsec $\times$ 3.5\arcsec & (e) \\
 & & NUV & 3.16 & 0.01 & 3.5\arcsec $\times$ 3.5\arcsec & (e) \\
 & SMSS & $g^\prime$ & 27.9  & 1.7 & 3.83\arcsec $\times$ 3.83\arcsec &(a) \\
   & & $r^\prime$ & 40.6 & 3.9 & 3.83\arcsec $\times$ 3.83\arcsec &(a) \\
 & &  $i^\prime$ & 48.7 & 3.6 & 3.83\arcsec $\times$ 3.83\arcsec & (a)\\
 & & $z^\prime$ & 54.5 & 1.5 & 3.83\arcsec $\times$ 3.83\arcsec&(a) \\
& 2MASS & J & 59.8 & 0.832 & 14\arcsec $\times$ 14\arcsec & (b)\\
 & & H & 79.9 & 1.11 & 14\arcsec $\times$ 14\arcsec  & (b)\\
 & & Ks & 76.7 & 1.14 & 14\arcsec $\times$ 14\arcsec  & (b) \\ 
 & IRAC & IRAC1 & 131.9 & 1.3 & & (g)\\
 & & IRAC2 & 103.9 & 1.3 & & (g)\\
& & IRAC3 & 219.9 & 2.1 & & (g)\\
 & & IRAC4 & 521.9 & 5.2 & & (g)\\
 & IRAS & IRAS1 & 630 & 35 & 0.8\arcmin $\times$ $>$0.8\arcmin$^\dag$ &(h) \\
& {\it Herschel}-PACS & Blue & 21440 & 1070 & 70\arcsec $\times$ 70\arcsec & (i) \\
& & Green & 31120 & 1560 & 70\arcsec $\times$ 70\arcsec & (i) \\
& & Red & 26860 & 1340 & 70\arcsec $\times$ 70\arcsec & (i) \\
& {\it Herschel}-SPIRE & PSW & 12370 & 810 & 100\arcsec $\times$ 100\arcsec & (i) \\
 & & PMW & 5058 & 333 & 100\arcsec $\times$ 100\arcsec & (i) \\
  & & PLW & 1577 & 106 & 100\arcsec $\times$ 100\arcsec & (i)\\
 IC 4280 & GALEX & FUV & 0.71 & 0.02 & 3.5\arcsec $\times$ 3.5\arcsec & (e) \\
 & & NUV & 1.62 & 0.02 & 3.5\arcsec $\times$ 3.5\arcsec & (e) \\
  & SMSS & $g^\prime$ & 18.9  & 3.1 & 3.36\arcsec $\times$ 3.36\arcsec &(a) \\
   & & $r^\prime$ & 25.3 & 0.4 & 3.36\arcsec $\times$ 3.36\arcsec &(a) \\
 & &  $i^\prime$ & 32.4 & 0.7 & 3.36\arcsec $\times$ 3.36\arcsec & (a)\\
 & & $z^\prime$ & 37.5 & 0.8 & 3.36\arcsec $\times$ 3.36\arcsec&(a) \\
  & 2MASS & J & 33.8 & 0.19 & 14\arcsec $\times$ 14\arcsec & (b)\\
 & & H & 44.1 & 0.244 & 14\arcsec $\times$ 14\arcsec  & (b)\\
 & & Ks & 39.7 & 0.33 & 14\arcsec $\times$ 14\arcsec  & (b) \\ 
 & WISE & W1 & 58.9 & 0.27 & 22\arcsec $\times$ 22\arcsec & (c)  \\
 & & W2 & 39.3 & 0.22  &  22\arcsec $\times$ 22\arcsec & (c)\\
 & & W3 & 281 & 1.55 & 22\arcsec $\times$ 22\arcsec & (c)\\
 & & W4 & 365 & 2.35 & 22\arcsec $\times$ 22\arcsec & (c)\\
 & {\it Herschel}-PACS & Blue & 7647 & 383 & 45\arcsec $\times$ 45\arcsec & (i) \\
& & Green & 12970 & 650 & 45\arcsec $\times$ 45\arcsec & (i) \\
& & Red & 12520 & 630 & 45\arcsec $\times$ 45\arcsec & (i) \\
 & {\it Herschel}-SPIRE & PSW & 5631 & 369 & 65\arcsec $\times$ 65\arcsec& (i)\\
 & & PMW & 2431 & 150 & 65\arcsec $\times$ 65\arcsec& (i) \\
 & & PLW & 771 & 48 & 65\arcsec $\times$ 65\arcsec& (i) \\
 NGC 6000 & SWIFT-UVOT & W2 & 0.432 & 0.02 &4.85\arcsec $\times$ 4.85\arcsec & (j)\\
 & & M2 & 0.578 & 0.02 &5.00\arcsec $\times$ 5.00\arcsec& (j)\\
 & & W1 & 0.959 & 0.04 & 6.25\arcsec $\times$ 6.25\arcsec & (j) \\
 & & U & 3.13 & 0.12 &4.35\arcsec $\times$ 4.35\arcsec  & (j) \\
 & & B & 6.41 & 0.26 & 5.40\arcsec $\times$ 5.40\arcsec& (j) \\
 & & V & 11.3 & 0.45 & 5.85\arcsec $\times$ 5.85\arcsec &  (j)\\
  & SMSS & $g^\prime$ & 24.3  & 0.9 & 3.58\arcsec $\times$ 3.58\arcsec& (a)\\
 & &  $r^\prime$ & 44.6 & 1.6 & 3.58\arcsec $\times$ 3.58\arcsec& (a)\\
 & & $i^\prime$ & 65.8 & 1.1 & 3.58\arcsec $\times$ 3.58\arcsec& (a)\\
  & & $z^\prime$ & 82.5 & 10.2 & 3.58\arcsec $\times$ 3.58\arcsec& (a)\\
   & WISE & W1 & 145 & 0.80 &  22\arcsec $\times$ 22\arcsec & (c)  \\
 & & W2 & 106 & 0.58  &  22\arcsec $\times$ 22\arcsec & (c)\\
 & & W3 & 827 & 4.57 & 22\arcsec $\times$ 22\arcsec & (c)\\
 & & W4 & 3010 & 16.6& 22\arcsec $\times$ 22\arcsec & (c)\\
  & AKARI & N60 & 41889 & 2010 & 42\arcsec $\times$ 42\arcsec$^\ddag$ & (d) \\
 & & WIDE-S & 43500 & 963 & 42\arcsec $\times$ 42\arcsec$^\ddag$ & (d) \\
 & & WIDE-L & 36169 & 4030 & 60\arcsec $\times$ 60\arcsec$^\ddag$ & (d) \\
 & & N160 & 47470 & 4950 & 60\arcsec $\times$ 60\arcsec$^\ddag$ & (d) \\
 IR 16164-0746  & GALEX & FUV & 0.024 & 0.002 & 3.5\arcsec $\times$ 3.5\arcsec & (e) \\
 & & NUV & 0.07 & 0.01 & 4.8\arcsec $\times$ 3.5\arcsec  & (e) \\
   & SMSS & $g^\prime$ & 1.91  & 0.33 & 3.2\arcsec $\times$ 3.2\arcsec& (a)\\
 & &  $r^\prime$ & 3.27 & 0.39 & 3.2\arcsec $\times$ 3.2\arcsec& (a)\\
 & & $i^\prime$ & 4.44 & 0.28 & 3.2\arcsec $\times$ 3.2\arcsec& (a)\\
  & & $z^\prime$ & 85.35 & 0.09 & 3.2\arcsec $\times$ 3.2\arcsec& (a)\\
  & 2MASS & J & 10.30 & 0.19 & 14\arcsec $\times$ 14\arcsec & (b)\\
 & & H & 14.90 & 0.276 & 14\arcsec $\times$ 14\arcsec  & (b)\\
 & & Ks & 15.1 & 0.34 & 14\arcsec $\times$ 14\arcsec  & (b) \\ 
  & WISE & W1 & 58.9 & 0.27 & 22\arcsec $\times$ 22\arcsec & (c)  \\
 & & W2 & 39.3 & 0.22  &  22\arcsec $\times$ 22\arcsec & (c)\\
 & & W3 & 281 & 1.55 & 22\arcsec $\times$ 22\arcsec & (c)\\
 & & W4 & 365 & 2.35 & 22\arcsec $\times$ 22\arcsec & (c)\\
 & {\it Herschel}-PACS & Blue & 13290 & 610 & 40\arcsec $\times$ 40\arcsec & (i) \\
& & Green &15030 & 680 & 40\arcsec $\times$ 40\arcsec & (i) \\
& & Red & 10210 & 450 & 40\arcsec $\times$ 40\arcsec& (i) \\
& {\it Herschel}-SPIRE & PSW & 3581 & 220 & 60\arcsec $\times$ 60\arcsec & (i) \\
 & & PMW & 1266 & 80 & 60\arcsec $\times$ 60\arcsec& (i) \\
  & & PLW & 407 & 28 & 60\arcsec $\times$ 60\arcsec & (i) \\
 ESO 453-G005 & GALEX & NUV & 0.131 & 0.01 & 3.7\arcsec $\times$ 3.7\arcsec  & (e) \\
 & SMSS & r & 4.06 & 0.19 & 3.8\arcsec $\times$ 3.8\arcsec & (a)\\
 & &  z & 4.68 & 0.12 & 3.8\arcsec $\times$ 3.8\arcsec & (a)\\
 & 2MASS & J & 11.1 & 0.29 & 14\arcsec $\times$ 14\arcsec & (b)\\
 & & H & 14.7 & 0.45 & 14\arcsec $\times$ 14\arcsec  & (b)\\
 & & Ks & 12.9 & 0.40 & 14\arcsec $\times$ 14\arcsec  & (b) \\ 
 & WISE & W1 & 27.6 & 0.25 & 22 $\times$ 22\arcsec & (c)  \\
 & & W2 & 16.3 & 0.19  &  22  $\times$ 22\arcsec & (c)\\
 & & W3 & 63.3 & 0.7 & 22  $\times$ 22\arcsec & (c)\\
 & & W4 & 209 & 1.7 & 22  $\times$ 22\arcsec & (c)\\
 & {\it Herschel}-PACS & Blue & 12190 & 550 & 35\arcsec $\times$ 35\arcsec & (i) \\
& & Green & 13950 & 630 & 35\arcsec $\times$ 35\arcsec &  (i) \\
& & Red & 10040 & 440 & 35\arcsec $\times$ 35\arcsec &  (i) \\
 & {\it Herschel}-SPIRE & PSW & 4150 & 252 & 55\arcsec $\times$ 55\arcsec &  (i) \\
 & & PMW & 1635 & 101 & 55\arcsec $\times$ 55\arcsec &  (i)\\
 & & PLW & 575 & 35 & 55\arcsec $\times$ 55\arcsec &  (i)\\
 IR 18293-3413 & XMM-OM & UVW1 & 0.02 & 0.003 & 5.7\arcsec x 5.7\arcsec & (k) \\ 
 & SMSS & g & 4.06 & 1.21 & 3.9\arcsec $\times$ 3.9\arcsec & (a)\\
 & &  r & 9.19 & 0.43 & 3.9\arcsec $\times$ 3.9\arcsec & (a)\\
 & & z & 19.9 & 0.2 & 3.9\arcsec $\times$ 3.9\arcsec & (a)\\
 & & Ks & 12.9 & 0.40 & 14\arcsec $\times$ 14\arcsec  & (b) \\ 
 & 2MASS & J & 49.8 & 0.28 & 14\arcsec $\times$ 14\arcsec & (b)\\
 & & H & 80.70 & 0.37 & 14\arcsec $\times$ 14\arcsec  & (b)\\
 & & Ks & 88.4 & 0.33 & 14\arcsec $\times$ 14\arcsec  & (b) \\ 
  & WISE & W1 & 88.3 & 0.50 & 22\arcsec $\times$ 22\arcsec  & (c)  \\
 & & W2 & 78.2 & 0.43  &  22\arcsec $\times$ 22\arcsec & (c)\\
 & & W3 & 761 & 3.51 & 22\arcsec $\times$ 22\arcsec & (c)\\
 & & W4 & 2270 & 126 & 22\arcsec $\times$ 22\arcsec & (c)\\
  & {\it Herschel}-PACS & Blue & 45710 & 2110 & 50\arcsec $\times$ 50\arcsec & (i) \\
& & Green & 59130 & 2710 & 35\arcsec $\times$ 35\arcsec &  (i) \\
& & Red & 45840 & 2070 & 35\arcsec $\times$ 35\arcsec &  (i) \\
 & {\it Herschel}-SPIRE & PSW & 17220 & 1070 & 70\arcsec $\times$ 70\arcsec &  (i) \\
 & & PMW & 6454 & 400 & 70\arcsec $\times$ 70\arcsec &  (i)\\
 & & PLW & 1987 & 122 & 70\arcsec $\times$ 70\arcsec &  (i)\\
 ESO 593-IG008 & GALEX & FUV & 0.166 & 0.01 & 3.5\arcsec $\times$ 3.5\arcsec & (e) \\
 & & NUV & 0.30 & 0.01 & 3.5\arcsec $\times$ 3.5\arcsec & (e) \\
 & SMSS & g & 1.69 & 0.12 & 2.8\arcsec $\times$ 2.8\arcsec  & (a)\\
 & &  r & 4.07 & 0.1 & 2.8\arcsec $\times$ 2.8\arcsec & (a)\\
 & & i & 6.23 & 0.15 & 2.8\arcsec $\times$ 2.8\arcsec & (a)\\
 & & z & 6.15 & 0.07 & 2.8\arcsec $\times$ 2.8\arcsec & (a)\\
 & 2MASS & J & 14 & 0.5 & 24.5\arcsec $\times$ 24.5\arcsec & (b)\\ %https://irsa.ipac.caltech.edu/cgi-bin/2MASS/PubGalPS/nph-galps?locstr=ESO+593-IG+008&radunits=arcsec&radius=30
 & & H & 25.2 & 1.0 & 24.5\arcsec $\times$ 24.5\arcsec & (b)\\
 & & Ks & 27.1 & 1.2 & 24.5\arcsec $\times$ 24.5\arcsec  & (b) \\ 
 & IRAS & IRAS1 & 180 & 25 & 0.8\arcmin $\times$ $>$0.8\arcmin$^\dag$& (h) \\
 & & IRAS2 & 510 & 38 & 0.8\arcmin $\times$ $>$0.8\arcmin$^\dag$ & (h) \\
   & {\it Herschel}-PACS & Blue & 7972 & 363 & 40\arcsec $\times$ 40\arcsec & (i) \\
& & Green & 10620 & 480 & 40\arcsec $\times$ 40\arcsec &  (i) \\
& & Red & 8597 & 380 & 40\arcsec $\times$ 40\arcsec &  (i) \\
 & {\it Herschel}-SPIRE & PSW & 3371 & 203 & 50\arcsec $\times$ 50\arcsec &  (i) \\
 & & PMW & 1278 & 79 & 50\arcsec $\times$ 50\arcsec &  (i)\\
 & & PLW & 489 & 30 & 50\arcsec $\times$ 50\arcsec &  (i)\\
 \enddata
\tablecomments{Columns:  (1) source name; (2) Instrument; (3) Passband; (4) total flux densities of the source (5) the error in the flux densities; (6) integration area used in aperture photometry in terms of length of semi-major axis a $\times$ length of semi-minor axis b (for circular apertures, semi-major axis = semi-minor axis), $^\dag$ source is flagged as resolved or marginally resolved in \citep[][]{Sanders03}, $^\ddag$ full-width at half maximum of the 2-D Gaussian used in profile fitting photometry; (7) references - (a) \citet[][]{Wolf18},  (b) 2MASS Extended Mission Data Release \citep[][]{Skrutskie06} (c) \citet[][]{Wright10,Mainzer11}
, (d) \citet[][]{Kawada07}  (e) \url{https://galex.stsci.edu/GR6/?page=mastform},(f) The 16th Data Release of the Sloan Digital Sky Surveys \citep[][]{Ahumada20}, (g) For discussion on IRAC flux estimation, see Section~\ref{sec:datauv}, (h) \citep[][]{Sanders03}, (i) \citet[][]{Chu17}, (j) \citet[][]{Yershov14}, (k) \citet[][]{Page12} }
%(7) The Galaxy Evolution Explorer (GALEX) Source Catalogs \citep[][]{Seibert12}, (8) \citep[][]{Gallimore10}, (9) \citet[][]{Henden2015}, (10) \citep[][]{Poole08},(11)  (12) \citet[][]{Dole2004}, (13), (14) \citet[][]{Hirashita08}
\end{deluxetable}

%---------
%\setcounter{table}{0}

\end{document}